\documentclass[12pt]{article}
\usepackage[margin=1in]{geometry} 

\usepackage{authblk}
\usepackage{hhline}

\usepackage{graphicx,epstopdf}              
\usepackage{mathtools}                        
\usepackage{amsfonts,amsthm,amssymb,latexsym,amsmath}
\usepackage{mathrsfs}
\usepackage{bbm}
\usepackage{units}  
\usepackage[font=small,labelfont=bf,skip=0pt]{caption}
\usepackage{subcaption}
\usepackage{textcomp} 
\usepackage{hhline}
\usepackage{multirow}      
\usepackage{tabularx}
\usepackage{booktabs}
\usepackage[backref=page]{hyperref}
\usepackage{float}
\usepackage{cleveref}
\usepackage{bbm}
\makeatletter
\AddToHook{cmd/appendix/before}{\def\cref@section@alias{appendix}}
\makeatother
\usepackage{pdflscape}
\usepackage{braket}
\usepackage[framemethod=TikZ]{mdframed} % fancy boxes
\usepackage{dsfont}

\usepackage[overleftharpoonup, overrightharpoonup]{overarrows}

\usepackage{tikz}
\usepackage{tikz-cd}
\usepackage{zx-calculus}

\usetikzlibrary{external}
\usetikzlibrary{zx-calculus}

\usepackage[export]{adjustbox}

\usepackage{array, makecell}

\DeclareFontFamily{U}{mathx}{}
\DeclareFontShape{U}{mathx}{m}{n}{<-> mathx10}{}
\DeclareSymbolFont{mathx}{U}{mathx}{m}{n}
\DeclareMathAccent{\widehat}{0}{mathx}{"70}
\DeclareMathAccent{\widecheck}{0}{mathx}{"71}
\newcommand{\Rep}{{\mathrm{Rep}}}
\newcommand{\GHZ}{{\mathrm{GHZ}}}
\newcommand{\GHX}{{\mathrm{GHX}}}
\newcommand{\Sym}{{\mathrm{Sym}}}
\newcommand{\SSB}{{\mathrm{SSB}}}
\newcommand{\op}{{\mathrm{op}}}
\newcommand{\ZZ}{{\mathbb Z}}
\newcommand{\RR}{{\mathbb R}}
\newcommand{\CC}{{\mathbb C}}

\newcommand{\wh}{\widehat}
\newcommand{\wc}{\widecheck}
\newcommand{\counith}{{\wh\counit}}
\newcommand{\haarc}{{\wc\haar}}
\newcommand{\bbphi}{\bar{\phi}}
\newcommand{\Fiso}{\Phi_\Theta}
\DeclareMathOperator{\Irr}{Irr}
\newcommand{\FF}{\mathsf{F}}
\newcommand{\drep}{r}
\newcommand{\dchi}{x}
\newcommand{\dsi}{\mathds{1}}
\newcommand{\id}{\mathrm{id}}
\newcommand{\DD}{\mathsf{D}}
\newcommand{\ofcoeff}{\kappa }

\def\cC{\mathcal C}

\def\cH{\mathcal H}
\def\cV{\mathcal V}
\def\cO{\mathcal O}

\def\Xr{{\overleftarrow{X}}}
\def\Xl{{\overrightarrow{X}}}
\def\Zr{{\overleftarrow{Z}}}
\def\Zl{{\overrightarrow{Z}}}

\def\Al{{\overrightarrow{A}}}
\def\Br{{\overleftarrow{B}}}

\def\ZXgate{{\overleftrightarrow{ZX}}}
\def\XZgate{{\overleftrightarrow{XZ}}}
\def\includegraphicsr#1#2{\raisebox{-.5\height}{\includegraphics[]{#2}}}
\newcommand{\includegraphicstable}[1]{\includegraphics[valign=c]{#1}}

\newcommand{\secref}[1]{Sec.\,\ref{#1}}
\newcommand{\appref}[1]{Appendix\,\ref{#1}}

\newcommand{\Hom}{\mathrm{Hom}}
\newcommand{\Cocom}{\mathrm{Cocom}}
\newcommand{\VEC}{\mathrm{Vec}}
\newcommand{\TY}{\mathrm{TY}}
\newcommand{\Aut}{\mathrm{Aut}}

\def\mult{m}
\def\comult{\Delta}
\def\haar{\ell}
\def\cohaar{\phi}
\def\unit{e}
\def\counit{\epsilon}

\newcommand{\dl}[1]{{#1}^*}

\newcommand{\lact}[1]{\overrightarrow{#1}}
\newcommand{\ract}[1]{\overleftarrow{#1}}

\newcommand{\gaugemap}{\mathrm{KW}}
\newcommand{\kw}{\mathsf{D}_\mathrm{KW}}
\newcommand{\aso}[1]{\widetilde{#1}}

\theoremstyle{definition}
\newtheorem{definition}{Definition}[section]

\usepackage{environ}
\NewEnviron{eqs}{%
    \begin{equation}\begin{split}
            \BODY
    \end{split}\end{equation}
}

\newcommand{\gel}[1]{\langle #1 \rangle}

\usepackage{enumerate}

\numberwithin{equation}{section}

\usepackage{dcolumn}
\usepackage{bm}

\usepackage[normalem]{ulem}

\usepackage{amstext}
\usepackage{amsbsy} 

\usepackage[all,matrix,cmtip]{xy}
\usepackage{color}
\definecolor{red}{rgb}{1,0,0}
\definecolor{blue}{rgb}{0,0,1}
\definecolor{dblue}{rgb}{0,0,0.4}
\definecolor{green}{rgb}{0,1,0}
\definecolor{black}{rgb}{0,0,0}
\definecolor{white}{rgb}{1,1,1}
\definecolor{pastelblue}{RGB}{20,93,160}

\definecolor{brn}{rgb}{.8,.4,.0}
\definecolor{redo}{rgb}{1,.5,.0}
\definecolor{ddgrn}{rgb}{0,0.4,0}
\definecolor{dgrn}{rgb}{0,0.55,0}
\definecolor{dbl}{rgb}{0,0,0.5}

\hypersetup{colorlinks,citecolor=pastelblue,linkcolor=pastelblue,urlcolor=pastelblue}

\usepackage[nosort]{cite}

\newcommand{\two}{\mathbf{2}}

\renewcommand{\t}[1]{\widetilde{#1}} 

\newcommand{\ee}{\hspace{1pt}\mathrm{e}}

\newcommand{\Rf}[1]{Ref.~\cite{#1}}
\newcommand{\Rfs}[1]{Refs.~\cite{#1}}

\newcommand{\Tr}{{\rm Tr}}

\newcommand{\bpm}{\begin{pmatrix}}
    \newcommand{\epm}{\end{pmatrix}}
\newcommand{\bmm}{\begin{matrix}}
    \newcommand{\emm}{\end{matrix}}
\newcommand{\bvm}{\begin{vmatrix}}
    \newcommand{\evm}{\end{vmatrix}}

\newcommand{\cA}{ {\cal A} } 

\newcommand{\cD}{ {\cal D} } 
 
\newcommand{\cF}{ {\cal F} }

\newcommand{\cL}{ {\cal L} } 
 
\newcommand{\cN}{ {\cal N} }

\usepackage{euscript}

\newcommand\sH         {\mathsf{H}}

\newcommand{\al}{\alpha} 
\newcommand{\bt}{\beta}

\newcommand{\Ga}{\Gamma}

\newcommand{\Th}{\Theta} 
\newcommand{\si}{\sigma}

\makeatletter
\newsavebox{\@brx}
\newcommand{\llangle}[1][]{\savebox{\@brx}{\(\m@th{#1\langle}\)}%
    \mathopen{\copy\@brx\kern-0.5\wd\@brx\usebox{\@brx}}}
\newcommand{\rrangle}[1][]{\savebox{\@brx}{\(\m@th{#1\rangle}\)}%
    \mathclose{\copy\@brx\kern-0.5\wd\@brx\usebox{\@brx}}}
\makeatother

\newcommand{\ie}{\begin{equation}\begin{aligned}[]}
        \newcommand{\fe}{\end{aligned}\end{equation}}

\newcommand\Tstrut{\rule{0pt}{3.0ex}}  % "top" strut
\newcommand\Bstrut{\rule[-1.75ex]{0pt}{0pt}}   % "bottom" strut
\newcommand\TBstrut{\Tstrut\Bstrut}           % "top and bottom" strut

\usetikzlibrary{arrows,snakes,shapes.arrows,decorations.markings}
     \tikzset{>=triangle 90}
     \tikzstyle{bbc}=[draw,circle,fill=black,scale=.75]
     \tikzstyle{rc}=[circle,fill=red,scale=.6]
     \tikzstyle{wc}=[draw,circle,scale=.75]

\tikzset{snake it/.style={decorate, decoration=snake}}

\tikzset{zxarrow/.style={->, >=stealth, rounded corners}}
\tikzset{zxarrowr/.style={<-, >=stealth, pos=0.4, rounded corners}}
\tikzset{zxdrrow/.style={>-, >=stealth, rounded corners}}

\tikzset{
    on each segment/.style={
        decorate,
        decoration={
            show path construction,
            moveto code={},
            lineto code={
                \path [#1]
                (\tikzinputsegmentfirst) -- (\tikzinputsegmentlast);
            },
            curveto code={
                \path [#1] (\tikzinputsegmentfirst)
                .. controls
                (\tikzinputsegmentsupporta) and (\tikzinputsegmentsupportb)
                ..
                (\tikzinputsegmentlast);
            },
            closepath code={
                \path [#1]
                (\tikzinputsegmentfirst) -- (\tikzinputsegmentlast);
            },
        },
    },
    mid arrow/.style={postaction={decorate,decoration={
                markings,
                mark=at position .5 with {\arrow[#1]{stealth}}
    }}},
}

\usetikzlibrary{decorations.markings}

\tikzset{line/.style={line width=0.25mm},
curve/.style={line,smooth,tension=1},
->-/.style={decoration={
  markings,
  mark=at position #1 with {\arrow[>=stealth]{>}}},postaction={decorate}},
-<-/.style={decoration={
  markings,
  mark=at position #1 with {\arrow[>=stealth]{<}}},postaction={decorate}},
}

\tikzset{bg/.style={opacity=.5}}

        \zxNewNodeFromPic{TriNodeTxt}[
        rotate=-\zxCurrentRotationMode,
        ]{
        \node[
        regular polygon, regular polygon sides=3, draw=black,fill=colorZxZ, font=\footnotesize,
        inner sep=1pt, zx main node,
        shape border rotate=90+\zxCurrentRotationMode, % Rotates the shape but not the text
        ] {\tikzpictext};
        }
        \zxNewNodeFromPic{ZTri}[
        every node/.append style={transform shape},
        scale=0.7
        ]{
        \node[regular polygon, regular polygon sides=3, shape border
        rotate=90, draw=black,fill=colorZxZ, inner sep=0.6mm, zx main node] {};
        }
        \zxNewNodeFromPic{XTri}[
        every node/.append style={transform shape},
        scale=0.7
        ]{
        \node[regular polygon, regular polygon sides=3, shape border
        rotate=90, draw=black,fill=colorZxX, inner sep=0.6mm, zx main node] {};
        }
\zxNewNodeFromPic{rep}{
\node[draw,fill=yellow!30,draw=black,inner sep=1mm, zx main node, alias=mynode] at (0,0) {\tikzpictext};
}

\zxNewNodeFromPic{amult}{
    \node[draw,circle,fill=pink!70,draw=black,inner sep=0.6mm, zx main node, alias=mynode2] at (0,0) {\tikzpictext};
}

\zxNewNodeFromPic{Had}{
\node[draw,draw=black,fill=colorZxH, inner sep=1mm, zx main node, alias=mynode4] at (0,0) {\tikzpictext};
}

\zxNewNodeFromPic{end}{
\node[draw,circle,fill=blue!30,draw=black,inner sep=0.5mm, zx main node, alias=mynode66] at (0,0) {\tiny\tikzpictext};
}

\newcommand{\zxbase}[1]{\raisebox{0pt}[0pt][0pt]{$ #1$}}

\newcommand{\myzx}[1]{\begin{ZX}#1\end{ZX}}

\zxConfigureExternalSystemCallAuto
\zxExternalSuffix{}

\newcommand{\tikzname}[1]{\tikzsetnextfilename{#1}}
\DeclareRobustCommand{\authoricon}[1]{%
  \raisebox{.15ex}{\includegraphics[scale=0.7]{author-icons/#1.pdf}}%
}

\newcommand{\NT}[1]{}
\newcommand{\dcl}[1]{}
\newcommand{\AC}[1]{}
 \newcommand{\fixme}[1]{}

\title{Generalized Kramers-Wannier Self-Duality\\
in Hopf-Ising Models
}
\author{\centering Da-Chuan Lu$^{\authoricon{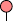},\authoricon{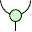},\authoricon{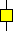}}$, 
Arkya Chatterjee$^{\authoricon{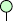},\authoricon{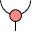},\authoricon{hadamard0}}$, 
Nathanan Tantivasadakarn$^{\authoricon{comultr}}$\\\vspace{-10pt}\small
$^{\authoricon{brar}}$Department of Physics, Harvard University, Cambridge, Massachusetts 02138, USA\\\vspace{-10pt}
$^{\authoricon{comultg}}$Department of Physics and Center for Theory of Quantum Matter, University of Colorado,
Boulder, Colorado 80309, USA\\\vspace{-8pt}
$^{\authoricon{brag}}$Department of Physics, Massachusetts Institute of Technology,  Cambridge, Massachusetts 02139, USA\\\vspace{-10pt}
$^{\authoricon{comultr}}$C.N. Yang Institute for Theoretical Physics, Stony Brook University, Stony Brook, New York 11794, USA}

\begin{document}

\maketitle

\renewcommand*{\thefootnote}{${\authoricon{hadamard0}}$}
\footnotetext[1]{These authors contributed equally to this work.}
\renewcommand{\thefootnote}{\arabic{footnote}}

\begin{abstract}

The Kramers–Wannier transformation of the 1+1d transverse-field Ising model exchanges the paramagnetic and ferromagnetic phases and, at criticality, manifests as a non-invertible symmetry. Extending such self-duality symmetries beyond gauging of abelian groups in tensor-product Hilbert spaces has, however, remained challenging. In this work, we construct a generalized 1+1d Ising model based on a finite-dimensional semisimple Hopf algebra $H$ that enjoys an anomaly-free non-invertible symmetry $\mathrm{Rep}(H)$. We provide an intuitive diagrammatic formulation of both the Hamiltonian and the symmetry operators using a non-(co)commutative generalization of ZX-calculus built from Hopf-algebraic data. When $H$ is self-dual, we further construct a generalized Kramers–Wannier duality operator that exchanges the paramagnetic and ferromagnetic phases and becomes a non-invertible symmetry at the self-dual point. This enlarged symmetry mixes with lattice translation and, in the infrared, flows to a weakly integral fusion category given by a $\mathbb{Z}_2$ extension of $\mathrm{Rep}(H)$.
Specializing to the Kac–Paljutkin algebra $H_8$, the smallest self-dual Hopf algebra beyond abelian group algebras, we numerically study the phase diagram and identify four of the six $\mathrm{Rep}(H_8)$-symmetric gapped phases, separated by Ising critical lines and meeting at a multicritical point. We also realize all six $\mathrm{Rep}(H_8)$-symmetric gapped phases on the lattice via the $H$-comodule algebra formalism, in agreement with the module-category classification of $\mathrm{Rep}(H_8)$. Our results provide a unified Hopf-algebraic framework for non-invertible symmetries, dualities, and the tensor product lattice models that realize them.
\end{abstract}

\tableofcontents

\section{Introduction}
Symmetry is a central organizing principle in quantum many-body physics and quantum field theory, shaping kinematics, classifying phases of matter, and constraining low-energy dynamics. Beyond symmetry, \textit{duality} provides an additional organizing principle by relating distinct gapped phases and rendering otherwise complicated phase diagrams tractable under duality transformations \cite{KramersWannier,Wegner71,aasen2016duality,Aasen:2020jwb,lootens_dualities_2022}. Such dualities are typically realized through \textit{gauging}, and a necessary condition for a self-duality is that the symmetry obtained after gauging coincides with the original one~\cite{Ginsparg:1988ui,Fendley:1989vt,Petkova:2000ip,Frohlich:2004ef,Ruelle:2005nk,Lootens24,Chen:2023qst,linhao2024dual,Lu:2024ytl,vanhove2025duality,bram2025gauging,Tantivasadakarn:2025txn}. At the self-dual point, this mechanism can lead to additional exact \textit{self-duality symmetries}\footnote{It is also dubbed as ``auto-orbifold'' in the literature.} that go beyond ordinary group symmetries and are naturally described by fusion categories, providing powerful constraints on critical points and multicritical behavior. More generally, non-invertible symmetries, realized by topological defects with non-group fusion rules, are ubiquitous in lattice models, conformal field theories, and gauge theories \cite{Fuchs:2002cm,gaiotto_generalized_2015,tachikawa_gauging_2020,Bhardwaj:2017xup,Chang19,Thorngren2019iar,thorngren_fusion_2021} (see \cite{mcgreevy2023generalized,Cordova22,Schafer-Nameki:2023review,Brennan:2023review,Bhardwaj:2023review,Shao23,Carqueville:2023review,davi2024review} for reviews of recent developments).

The prototypical example of a duality is the Kramers--Wannier transformation, which exchanges the paramagnetic and ferromagnetic phases of the 1+1d transverse-field Ising model \cite{KramersWannier}. The transformation can be understood as gauging the $\mathbb{Z}_2$ spin-flip symmetry to yield another Ising chain with dual $\mathbb{Z}_2$ symmetry, but with an inverted coupling constant \cite{Fendley:1989vt}. At a critical value of the coupling, the Kramers--Wannier transformation commutes with the Hamiltonian, and in the continuum, gives rise to a topological defect that exchanges the order and disorder operators. The same lattice construction extends to finite abelian groups, for example the $N$-state clock and Potts models by gauging the $\mathbb{Z}_N$ symmetry, as well as the Ashkin-Teller family \cite{fradkin1980disorder,berkcan1983order,fateev1985parafermionic,Delfino:2003ia}. This is because in 1+1d, gauging a finite abelian group $A$ produces a theory with dual global symmetry $\widehat{A}$, which is isomorphic to the original symmetry \cite{Vafa:1989ih,Bhardwaj:2017xup}.

In contrast, for a finite non-abelian group $G$, gauging no longer returns a group symmetry but instead produces a dual non-invertible symmetry $\mathrm{Rep}(G)$, whose simple objects correspond to the irreducible representations of $G$ \cite{Bhardwaj:2017xup}. Since $\mathrm{Rep}(G) \ncong G$ as a symmetry structure, a Kramers--Wannier-type self-duality cannot arise from gauging $G$. As a result, genuinely non-abelian self-dualities are far more subtle and require additional structure beyond ordinary group symmetries. 

The generalization of abelian self-duality symmetries requires generalizing from groups to finite-dimensional \textit{self-dual} semisimple Hopf algebras \cite{choi2023self,Diatlyk:2023fwf,perez2023notes}. One might ask why (if at all) must we use this more complicated machinery of Hopf algebras. It turns out that the goal of defining a lattice model on a tensor product Hilbert space makes this structure very natural. First, in the group case, the local Hilbert space is already a group algebra, with a notion of multiplication. Second, in order to act with a symmetry operator, we need to have a notion of ``copying" the symmetry action so that it can act identically across all sites. This is the notion of comultiplication. These two structures, their compatibility conditions, along with the notion of an inverse for a group, is exactly what is encoded in the structure of a Hopf algebra.

An additional reason why Hopf algebras are important for symmetries can also be seen from the continuum perspective. In 1+1d quantum field theories, non-invertible symmetries are described by unitary fusion categories \cite{Bhardwaj:2017xup,Choi2023xjw,Thorngren2019iar}. 
When such a non-invertible symmetry admits a symmetric gapped phase with a unique ground state (also called anomaly-free\footnote{we will use this as the definition of anomaly-free throughout the paper, rather than an algebra object that is gaugeable. See \Rf{Choi2023xjw} for subtleties.}) the corresponding fusion category $\cC$ is said to be equipped with a fiber functor ${\cF:\cC\to \VEC}$~\cite{Thorngren2019iar,Choi2023xjw}. By the Tannaka--Krein reconstruction theorem, any (unitary) fusion category admitting a fiber functor is equivalent to a representation category of the form $\Rep(H)$ for some finite-dimensional Hopf ($C^\star$-)algebra $H$. Ordinary group symmetries arise as the special case where $H$ is the function algebra $\mathrm{Fun}(G)$ on a finite group $G$.

In the continuum, one can gauge the full $\Rep(H)$ symmetry and obtain a dual non-invertible symmetry $\Rep(\dl{H})$, where $\dl{H}$ denotes the linear dual Hopf algebra. A necessary condition for self-duality is therefore that $H$ be isomorphic to its dual, ${H \cong \dl{H}}$. We refer readers to \cite{choi2023self,Diatlyk:2023fwf,perez2023notes} for detailed discussions of gauging non-invertible symmetries and the resulting self-duality structures. More precisely, the resulting self-duality symmetry in the continuum can be described by a $\mathbb{Z}_2$ extension of $\Rep(H)$. Namely, it is a $\mathbb{Z}_2$-graded fusion category $\mathcal{C}=\mathcal{C}_0\oplus\mathcal{C}_1$, with ${\mathcal{C}_0=\Rep(H)}$ and ${\mathcal{C}_1=\{\mathcal{D}\}}$, whose fusion rules are
\begin{equation}
    \mathcal{D}\times \mathcal{D} = \sum_{a\in \Irr(H)} d_a \, a, \qquad
    a\times \mathcal{D} = \mathcal{D}\times a = d_a\,\mathcal{D}.
\end{equation}
When $H$ is an abelian group algebra, this fusion category reduces to the well-studied Tambara--Yamagami fusion categories \cite{tambara1998tensor}. For a general self-dual Hopf algebra $H$, the above fusion category has been studied in \Rf{davydov2013z}. The smallest self-dual Hopf algebra that is both non-commutative and non-cocommutative is the Kac--Paljutkin algebra $H_8$. The corresponding $\mathbb{Z}_2$ extension of $\Rep(H_8)$ describing its self-duality has been analyzed explicitly in \Rf{choi2023self}. \Cref{tab:catdata} summarizes the correspondence between physical, categorical, and Hopf-algebraic data.
\begin{table}[h]
    \centering
    \renewcommand{\arraystretch}{1.2}
    \begin{tabular}{|c|c|c|}
        \hline
        \textbf{Physics data} & \textbf{Category description} & \textbf{Algebraic description} \\ \hline
        Non-invertible symmetry & Fusion category & weak Hopf algebra\\ \hline
        \makecell{Anomaly-free\\ non-invertible symmetry} & \makecell{Fusion category \\ with fiber functor $(\cC,\cF)$}  & Hopf algebra $H$ \\ \hline
        Topological defect line & (Simple) object in $\cC$ & (Irreducible) representation of $H$ \\ \hline
        Symmetric gapped phase & Module categories over $\cC$ & Comodule algebra over $H$ \\ \hline
        Dual symmetry &\makecell{Category of $\cA$-$\cA$-bimodule \\ objects in $\cC$, ${}_\cA \cC_\cA$} & dual Hopf algebra $\dl{H}$\\\hline
        Self-duality symmetry & $\ZZ_2$-extension of $\cC$ & Hadamard form of self-dual $H$\\ \hline
        Bulk TFT 
        & Drinfeld center $Z(\mathcal{C})$ 
        & Drinfeld double $D(H)$ \\ \hline
    \end{tabular}
    \caption{Correspondence between physical, categorical, and Hopf-algebraic data. The dual symmetry refers to the symmetry under gauging the whole anomaly-free non-invertible symmetry, i.e. gauging the algebra object $\cA=\bigoplus_{a\in \mathrm{Irr}(\cC)} d_a a$.}
    \label{tab:catdata}
\end{table}

To complement our current understanding of such symmetries in the continuum, we aim to provide a realization of this general self-duality symmetry in a lattice model with a tensor product Hilbert space. On the lattice, non-invertible symmetries of the form $\Rep(H)$ can be realized by matrix product operators (MPOs) acting on a tensor-product Hilbert space, which flow to the corresponding categorical symmetry in the infrared (IR) limit~\cite{buerschaper2013qdalgebra,inamura_lattice_2022,molnar2022MPO,zhian2024chain,jia2024generalized}. The structure of these symmetry MPOs, their symmetric matrix product states (MPS), and the associated gauging map closely parallels the familiar $\Rep(G)$ construction for finite groups, while exhibiting important new features specific to Hopf algebras. Lattice realizations of $\Rep(G)$ symmetries in group-valued quantum spin chains, as well as their gauging, have been studied previously in \cite{tantivasadakarn_hierarchy_2022,nat2025repG,sakura2024reps3,apoorv2024reps3,arkya2024reps3dual,alison2024repd8s,Pace:2024acq,blanik2025}. Our contributions here are to generalize the gauging to the general case of $\Rep(H)$, and further construct the self-duality operator when $H$ is self-dual.

We remark that there have been previous works, which also construct non-invertible symmetries in spin chains that can also realize self-duality symmetries. Notably the anyon chain in \Rf{Feiguin07} can be constructed for any fusion category symmetry. Nevertheless, the Hilbert space is not a tensor product Hilbert space. Another work is \Rf{molnar2022MPO}, which uses a weak Hopf algebra to construct such symmetries in a tensor product Hilbert space. However, in this construction the symmetry action corresponding to the unit object acts as a global projector, rather than identity. This means that the space in which this symmetry acts is also secretly a constrained (non-tensor-product) subspace. In contrast, our construction is on a fully tensor product Hilbert space without any extra projections (but at the price of mixing with lattice translation). 
Generally, MPOs provide a framework for realizing non-invertible symmetries on the lattice, but at this level of generality they form only a pre-bialgebra structure, making properties such as anomalies, gapped phases, and gauging difficult to analyze \cite{molnar2022MPO,yuhan2025mpo}. By focusing on a particular class of anomaly-free non-invertible symmetries with a Hopf algebra structure, these features become more rigid and tractable. 

A separate contribution we make is to use a generalized ZX calculus for Hopf algebras\cite{collins:2019qdw,majid2022zxhopf,collins2024hopf} in our work. Within this framework, one can define generalized Hopf Pauli operators and construct matrix product states and local operators in direct analogy with the ZX calculus for Abelian group algebras \cite{coecke2011zxreview,coecke2018picturing,backens2014zxcal,duncan2020zxreview}. The resulting graphical language allows derivations to proceed by diagrammatic rewriting rather than explicit tensor contractions, greatly simplifying calculations. It also yields a basis-independent description of operators and their algebraic relations, in contrast to explicit tensor representations, which depend on a choice of basis. Recent applications of ZX calculus to non-invertible symmetries and their gapped phases can be found in \Rfs{nat2024zxKW,lu2025strange}.

The rest of the paper is organized as follows. In Section~\ref{sec:hopf-pauli}, we review the fundamental properties of finite-dimensional semisimple Hopf algebras and introduce the generalized Hopf ZX calculus, which establishes the diagrammatic toolkit for Hopf-Frobenius algebras, which we  use throughout this work. Section~\ref{sec:Hopfqudit} defines the lattice kinematics of Hopf algebra-valued spin chains, including generalized Hopf Pauli operators, and shows how they can be used to realize MPO representations of both $\Rep(H)$ and $\Rep(\dl H)$ symmetries, together with their corresponding paramagnetic and symmetry-breaking fixed-point Hamiltonians.
In Section~\ref{sec:KW}, we detail the gauging procedure of $\Rep(H)$ and use it to construct an effective gauging map that takes a system with $\Rep(\dl H)$ symmetry and outputs a system with $\Rep( H)$ symmetry on a tensor-product Hilbert space. Then, for self-dual Hopf algebras, we further introduce the Hopf Hadamard gate and combine it with gauging to realize the non-invertible Hopf Kramers--Wannier duality defect $\kw$, and analyze its fusion rules and categorical data. Section~\ref{sec:HopfIsing} introduces a family of self-dual Hopf-Ising models and presents a numerical study of the phase diagram for the model corresponding to the Kac--Paljutkin algebra $H_8$, identifying its critical lines and a multicritical point. Section~\ref{sec:H-gapped-phases} provides a general construction of $\Rep(H)$-symmetric gapped phases using the formalism of $H$-comodule algebras, and applies this framework to explicitly construct lattice Hamiltonians for all six symmetric gapped phases of the $H_8$ model. Finally, Section~\ref{sec:conclusion} concludes with a summary and outlook on future directions.

\section{Review of Hopf algebras and Hopf ZX calculus}
\label{sec:hopf-pauli}

In this section, we will review Hopf algebras and how it can be further equipped with an extra Frobenius structure, giving rise to a Hopf-Frobenius algebra. Throughout, we will introduce the corresponding graphical notation of ZX calculus\footnote{The normalization in the ZX calculus literature differs from those of the Hopf algebras at certain places. In particular, the normalization of (co)units, the (co)unit copy, bialgebra, and Hopf rules. Here, we follow the Hopf algebra normalization. See e.g. Ref.~\cite{nat2024zxKW} for the former convention in the context of $\ZZ_2$.}

ZX calculus provides a graphical framework for reasoning about linear maps and tensor networks, offering a diagrammatic language that is both rigorous and intuitively transparent (see ~\cite{coecke2011zxreview,vandewetering2020zxcalculus,duncan2020zxreview} for a review of the $\ZZ_2$ case). It captures the algebraic structure of quantum observables through two fundamental types of tensors (often called spiders), corresponding to complementary commutative Frobenius algebras \cite{coecke2018picturing,quanlong2019zx}. This formalism is particularly powerful because it encodes algebraic equalities as local diagrammatic rewrite rules, thereby enabling computations to be carried out via graphical manipulations rather than symbolic algebra \cite{backens2014zxcal,quanlong2019zx}. When adapted to Hopf-Frobenius algebras, the ZX calculus naturally represents comultiplication, counit, antipode, and related structures, providing a unified language to visualize and manipulate algebraic relations \cite{collins:2019qdw,majid2022zxhopf,collins2024hopf}. This makes it an especially suitable tool for constructing and analyzing models with non-invertible symmetries, such as the Hopf-Ising model we discuss in this paper (see, e.g., \cref{sec:HopfIsing}).

\subsection{Hopf \texorpdfstring{$C^\star$}{C*}-algebra}
\begin{table}[ht!]
    \centering
    \begin{tabular}{|c|c|c|c|c|}
        \hline
        &     Unit & Counit & Integral & Cointegral\\
        \hline
        Circle   &  \tikzsetnextfilename{ketr} $\unit= \myzx{ \zxN{} \\ \zxX{}\uar}$  & \tikzsetnextfilename{brag} $\widehat{\counit}= \myzx{\zxZ{}  \dar \\ \zxN{} }$  & \tikzsetnextfilename{ketg} $\haar = \myzx{\zxN{}\\ \zxZ{}\uar}$   & \tikzsetnextfilename{brar} $\widehat{\cohaar}= \myzx{\zxX{}  \dar \\\zxN{} }$ 
        \TBstrut \\
        \hline
        Triangle & \tikzsetnextfilename{ketrtri} $\wc{\unit}= \myzx{ \zxN{}\\ \zxXTri.{}\uar}$  & \tikzsetnextfilename{bragtri} $\counit= \myzx{\zxZTri'{}  \dar \\\zxN{} }$   & \tikzsetnextfilename{ketgtri} $\wc{\haar} = \myzx{ \zxN{}\\ \zxZTri.{}\uar}$   & \tikzsetnextfilename{brartri} $\cohaar= \myzx{\zxXTri'{}  \dar \\\zxN{} }$ 
        \TBstrut\\
        \hline
    \end{tabular}
    \caption{Summary of states. Normalized states are the ones without hats or checks (circles for kets, and triangles for bras). The hatted version is multiplied by a factor of $\sqrt{|H|}$ and the checked version is divided by $\sqrt{|H|}$ (here, $|H|$ is the dimension of the vector space underlying $H$). 
    % The hat and check versions 
    The triangular (co)integral satisfies the idempotence condition under (co)multiplication. The normalization is such that $\counit(\haar) = \cohaar(\unit)=1$. Moreover, the circle tensor is always $\sqrt{|H|}$ times the corresponding triangle tensor. }
    \label{tab:statesconvention}
\end{table}

In this subsection, we review Hopf $C^\star$-algebras.\footnote{The star will be denoted using $^\star$ while, the dual Hopf algebra (defined in \cref{sec:dualHopf}) will be denoted $\dl{H}$. Note that throughout this paper, we will work exclusively with finite-dimensional Hopf $C^\star$-algebras.} 
Before describing the $\star$-structure, let us first define what a Hopf algebra is.
A Hopf algebra over $\CC$ is a sextuple $(H, \mult, \unit, \comult, \counit, S)$, defined as follows. $H$ is a finite dimensional vector space over $\CC$. Elements $a\in H$ can be denoted diagrammatically as $a= \myzx{ \zxN{} \\ \zxX{a}\uar}$, while linear maps $\alpha: H \rightarrow \CC$ are denoted $\alpha=\myzx{\zxZ{\alpha}  \dar \\ \zxN{} }$. These diagrams are to be read from bottom to top\footnote{All the diagrams in this paper are read from bottom to top and from right to left, i.e. elements at the bottom (right) are mapped to those at the top (left). We will use Latin letters for elements in $H$ and Greek letters for elements in the dual Hopf algebra, $\dl{H}$ (see \cref{sec:dualHopf}).}.
The multiplication and comultiplication $\mult,\comult$ are linear maps,
\begin{align}
    \mult: H \otimes H& \rightarrow H
    \hspace{62pt} 
    \tikzsetnextfilename{multr}
    \begin{ZX}
        & \zxN{}  &\\
        & \zxX{} \ar[u] &  \\
        \zxN{} \ar[ru, bend left]  & 
        &  \zxN{} \ar[lu, bend right]
    \end{ZX}
    \\
    \mult(a_i\otimes a_j) &= \mult_{ij}^k a_k\\
    \comult: H &\rightarrow H \otimes H
    \hspace{36.5pt}
    \tikzsetnextfilename{comultg}
    \begin{ZX}
        \zxN{} \ar[rd, bend right]  & 
        &  \zxN{} \ar[ld, bend left] \\
        & \zxZ{} &  \\
        & \zxN{} \ar[u] &  
    \end{ZX}
    \\
    \comult(a_i) &= \comult_{i}^{jk} a_j \otimes a_k
\end{align}
For the comultiplication, it is convenient to use the Sweedler notation which is standard in the Hopf algebra literature, $\comult(a) = \sum_{(a)} a_{(1)} \otimes a_{(2)} \equiv \sum_{i} a_{(1),i} \otimes a_{(2),i}$. 
The multiplication and comultiplication are associative and coassociative respectively, i.e.
\begin{align}\label{eq:assoc}
    \tag{ASC}
\text{(co)associativity:} &&
    \tikzname{assoc1}
    \begin{ZX}
        &   &   &\zxN{}\dar  & &    & \\
        &   &   &[\zxWRow]  & &    & \\
        &   &   &\zxX{} \uar & &  &  \\
        &   &   & &   &\zxX{} 
        \ar[llu,(]  & \\
        &   &   & &  &    & \\
        &\zxN{} \ar[uuurr,)] &  & &\zxN{} \ar[uur,)]  &  &\zxN{} \ar[uul,(] 
    \end{ZX}
    =
    \tikzname{assoc2}
    \begin{ZX}
        &   &   &   &\zxN{} \dar &    & \\
        &   &   &   &[\zxWRow] &    & \\
        &   &   &   &\zxX{} \uar &  &  \\
        &   &\zxX{} 
        \ar[urr,)]  & &   &   & \\
        &   &   & &  &    & \\
        &\zxN{} \ar[uur,)]  &    &\zxN{} \ar[uul,(] &  & &\zxN{} \ar[uuull,(]
    \end{ZX}
    \ 
    \,,
    \qquad 
    \tikzname{assoc3}
    \begin{ZX}
        &\zxN{} \ar[dddrr,(] &  & &\zxN{} \ar[ddr,(]  &  &\zxN{} \ar[ddl,)]\\
        &   &   & &  &    & \\
        &   &   & &   &\zxZ{} 
        \ar[lld,)]  & \\
        &   &   &\zxZ{} \dar & &  &  \\
        &   &   &[\zxWRow]  & &    & \\
        &   &   &\zxN{}\uar  & &    & 
    \end{ZX}
    =
    \tikzname{assoc4}
    \begin{ZX}
        &\zxN{} \ar[ddr,(]  &    &\zxN{} \ar[ddl,)] &  & &\zxN{} \ar[dddll,)] \\
        &   &   & &  &    & \\
        &   &\zxZ{} 
        \ar[drr,(]  & &   &   & \\
        &   &   &   &\zxZ{} \dar &  &  \\
        &   &   &   &[\zxWRow] &    & \\
        &   &   &   &\zxN{} \uar &    & 
    \end{ZX}
    \ .
\end{align}
The unit $\unit$, depicted \includegraphics{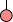}, and counit $\wh\counit$,\footnote{We use the hat to denote the fact that the counit is not a normalized state with respect to the inner product. See Table \ref{tab:statesconvention}.} 
depicted \includegraphics{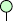}, are linear maps
\begin{align}
    \unit:\CC \to H \,,\qquad \wh\counit:H\to \CC\,.
\end{align}

They satisfy the following defining properties:
\tikzsetfigurename{unit}
\begin{align}
\label{eq:Id}
\tag{Id}
    \begin{ZX}
        &   & \zxN{} \dar \\
        &   & [\zxWRow] \\
        &   & \zxX{} \uar  \\
        & \zxX{} \ar[ru, bend left]  & 
        \zxN{} \uar\dar\\
        & [\zxWRow] &
    \end{ZX}
    =
    \begin{ZX}
        & \zxN{} \dar &   \\ 
        & [\zxWRow] &   \\ 
        & \zxX{} \uar  &   \\ 
        & \zxN{} \uar\dar    & \zxX{} \ar[lu, bend right]  \\
        & [\zxWRow] &
    \end{ZX}
    =
    \begin{ZX}
        \zxN{} \dar \\
        [\zxWRow] \\
        \zxN{} \uar \dar \\
        [\zxWRow] \\
        \zxN{} \uar 
    \end{ZX} 
    \ ,
    \qquad 
    \begin{ZX}
        &    [\zxWRow] &\\
        &\zxZ{} \ar[rd, bend right]  & 
        \zxN{} \uar\dar\\
        &    & \zxZ{} \dar  \\
        &    & [\zxWRow] \\
        &    & \zxN{} \uar 
    \end{ZX}
    =
    \begin{ZX}
        &    [\zxWRow] &\\
        & \zxN{} \uar\dar & \zxZ{} \ar[ld, bend left]   \\
        & \zxZ{} \dar      & \\
        & [\zxWRow]     & \\
        & \zxN{} \uar   &    
    \end{ZX}
    =
    \begin{ZX}
        \zxN{} \dar \\
        [\zxWRow] \\
        \zxN{} \uar \dar \\
        [\zxWRow] \\
        \zxN{} \uar 
    \end{ZX}
    \ .
\end{align}

The unital associative algebra formed by $ (H,\mult,\unit) $  and the coalgebra formed by $ (H,\comult,\wh\counit) $ together form a bialgebra. This means that $ \comult $ and $ \wh\counit $ are algebra homomorphisms, so that they satisfy ${\comult(ab) = \comult(a)\comult(b)}$, ${\comult(\unit) = \unit\otimes\unit}$ and ${\wh{\counit}(ab) = \wh{\counit}(a)  \wh{\counit}(b)}$, ${\wh{\counit}(\unit) = 1}$. Diagrammatically,
\tikzsetfigurename{bialgebra}
\begin{align}\label{eq:hopfbialgebra}
\tag{B}
\text{Bialgebra:} &&
    \begin{ZX}
        &\zxN{}\ar[rdd,(]&&\zxN{}\ar[ldd,)] \\
        &&& \\
        &&\zxZ{}\dar&\\
        &&&\\
        &&\zxX{}\uar&\\
        &&&\\
        &\zxN{}\ar[ruu,)]&&\zxN{}\ar[luu,(]
    \end{ZX}
    &=
    \begin{ZX}
        &\zxN{}\ar[rdd,(]   &&&& \zxN{}\ar[ldd,)] \\
        &&&&&   \\
        &&\zxX{}\dar && \zxX{}\dar& \\
        &&&&&\\
        &&\zxZ{}\uar\ar[rruu]&& \zxZ{}\uar\ar[lluu, 3d above, very thick]&\\
        &&&&&\\
        &\zxN{}\ar[ruu,)]&&&&\zxN{}\ar[luu,(]
    \end{ZX}
    \  , \\
    \label{eq:copy}
    \tag{UC}
    \text{(Co)unit copy:} &&
    \tikzsetfigurename{copy}
    \begin{ZX}
        &&\zxZ{}\dar&\\
        &&\zxX{}\uar&\\
        &&&\\
        &\zxN{}\ar[ruu,)]&&\zxN{}\ar[luu,(]
    \end{ZX}
    &= 
    \begin{ZX}
        &\zxZ{}\dar&&\zxZ{}\dar\\
        &&&\\
        &\zxN{}\uar&&\zxN{}\uar\\
    \end{ZX}
    \   , \qquad \tikzsetfigurename{copy2}
    \begin{ZX}
        & &&\\
        & &&\\
        &\zxZ{} \ar[luu,)] \ar[ruu,(]&&\\
        & \zxX{} \ar[u] &&
    \end{ZX}
    = 
    \begin{ZX}
        & &&\\
        & &&\\
        &\zxX{}\ar[uu]&&\zxX{}\ar[uu]
    \end{ZX} \ ,\\
    \label{eq:scalar}
    \tag{Sc}
\text{Scalar:} &&
    \tikzsetfigurename{counit-unit}
    \begin{ZX}
        \zxZ{}\dar &\\
        \zxX{}\uar &
    \end{ZX} &= 1 \, .
\end{align}

Finally, we have the antipode, which is a linear map ${S:H\to H}$, depicted \tikzsetnextfilename{antipode} \begin{ZX}
    \\
    \zxrep{$S$}\ar[u]\ar[d]\\
\end{ZX},
which satisfies the Hopf rule
\tikzsetfigurename{Hopfrule}
\begin{align}\label{eq:Hopf-rule}
    \tag{Hf}
\text{Hopf:} &&
    \begin{ZX}
        &\zxN{}  & \\
        & & \\
        &\zxX{} \ar[uu] \ar[dr,)] \ar[dd,C] & \\
        &   & \zxrep{$S$} \\
        &\zxZ{} \ar[dd] \ar[ur,(]  &   \\
        &  &  \\
        &\zxN{}  &  
    \end{ZX} =   
    \begin{ZX}
        &\zxN{}  & &    & \\
        & &  &    & \\
        &\zxX{} \ar[uu] \ar[dl,(]  &   &\\
        \zxrep{$S$} &   &  \\
        &\zxZ{} \ar[ul,)] \ar[dd] \ar[uu,C-]  & &  &  \\
        &  & &    & \\
        &\zxN{}  & &    & 
    \end{ZX} = 
    \begin{ZX}
        \zxN{}   \\
        \\
        \zxX{}  \ar[uu] \\
        \\
        \zxZ{} \ar[dd]    \\
        \\
        \zxN{} 
    \end{ZX}
\end{align}
and the antipode rule
\begin{align}\label{eq:antipoderule}
    \tag{S}
\text{Antipode:} &&
    \tikzsetfigurename{antipoderulem}
    \begin{ZX}
        &&\\
        &\zxrep{$S$} \ar[u] \ar[d]&\\
        &\zxX{} \ar[ld,(] \ar[rd,)]&\\
        \zxrep{$S$}\ar[d]&&\zxrep{$S$} \ar[d]\\
        &&
    \end{ZX}=\begin{ZX}
        &&& \\
        &\zxX{} \ar[u] \ar[d,C] &&\\
        &\zxN{} \ar[rd,)]\ar[u,C-,3d above,very thick]\ar[ld,(,3d above,very thick] &&\\
        &&&
    \end{ZX},\quad 
    \begin{ZX}
        &&\\
        \zxrep{$S$} \ar[u]&&\zxrep{$S$}\ar[u]\\
        &\zxZ{} \ar[lu,)], \ar[ru,(] \ar[d]&\\
        &\zxrep{$S$} \ar[d]&\\
        &&
    \end{ZX}=\begin{ZX}
        \zxN{}\ar[rd,(]&&\\
        &\zxN{} \ar[ru,(,3d above,very thick]\ar[d,C-] &\\
        &\zxZ{} \ar[d] \ar[u,C,3d above,very thick] &\\
        &\zxN{}&
    \end{ZX} \ .
\end{align}
All Hopf algebras we consider throughout this paper are assumed to be semi-simple and cosemi-simple,\footnote{Note that any finite dimensional $C^\star$-algebra is isomorphic to a direct sum of matrix algebras (see e.g.~Prop.~7.1.5 of \Rf{rordam2000introduction}) and, thus, is semisimple. Since we only work with Hopf $C^\star$-algebras in this paper, the semi-simplicity is automatic rather than an added assumption.} for which the antipode is involutive, ${S^2=\mathbbm{1}}$~\cite{larson1988semisimple,etingof1998finite}.

A left integral ${l \in H}$ is an element that satisfies ${a l = \wh\counit(a) l}$ for any ${a\in H}$. Similarly, a right integral ${r\in H}$ satisfies ${ra = \wh\counit(a) r }$. A Haar integral is defined as a left and right integral.  For a semi-simple Hopf algebra, the Haar integral is unique (up to normalization). We will often drop the word Haar when referring to these integrals.

In the Hopf algebra literature, the normalization of the Haar integral is usually chosen such that it is idempotent under multiplication.
We will denote the Haar integral with this normalization as $\wc\haar$ and depict it as \includegraphics{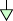}. $\wc\haar$ satisfies ${\wc\haar^2 =\wc\haar}$, i.e.~${\wh{\counit}(\wc\haar) =1}$. 
For a left integral $l$, one can show that $S(l)$ is a right integral, and vice versa. Therefore, $S(\wc\haar)$ is also a Haar integral. But since the Haar integral is unique for semisimple Hopf algebras (up to normalization), we conclude that ${S(\wc\haar) = \wc\haar}$.

Similarly, we will refer to the Haar integral of the dual Hopf algebra $\dl{H}$ (see \cref{sec:dualHopf}) as the Haar cointegral $\cohaar$ \footnote{The Haar cointegral is also called Haar measure or Haar functional.}. 
Similar to the normalization of $\wc\haar$, it is conventional to normalize $\cohaar$ such that ${\cohaar(\unit) = 1}$. This implies ${(\phi \otimes \phi) \circ \Delta = \phi}$. In diagrams, we depict $\cohaar$ as \includegraphics{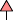}.
Together, these properties are diagrammatically represented as
\tikzsetfigurename{haar-cohaar}
\begin{equation} \label{ref:integral}
\text{(Co)integral Copy}:  \quad
\tag{IC}
\begin{aligned}
%\tikzsetfigurename{haar}
    \begin{ZX}
        & &  & \zxN{} \dar \\
        & &  &  \\
        & &  & \zxX{} \uar  \\
        &  &  & 
        \zxN{} \uar\dar \\ 
        &\zxZTri.{}  \ar[rruu, bend left] &  & 
    \end{ZX}
    &= 
    \begin{ZX}
        &   & \zxN{} \dar \\
        &  &  \\
        & \zxZ{}\dar  & \zxN{} \uar\dar \\
        & \zxN{} \uar\dar   & 
        \zxN{} \uar\dar \\ 
        &   &\zxZTri.{} 
    \end{ZX}
    =
    \begin{ZX}
        & \zxN{} \dar &  & \\
        &  &  & \\
        & \zxX{} \uar  &  & \\
        & \zxN{} \uar\dar & &   \\ 
        & &  & \zxZTri.{}  \ar[lluu, bend right]  
    \end{ZX}
    \ , 
    &
    \begin{ZX}
        & \zxZ{} \dar  \\
        & \zxN{} \uar\dar \\ 
        & \zxZTri.{}
    \end{ZX}
    &= 1 \, ,\\
%\tikzsetfigurename{cohaar-identity}
\begin{ZX}
    & \zxXTri'{}  \ar[rrdd, (] &  &\\
    & &  & \zxN{} \uar\dar \\ 
    & &  & \zxZ{} \dar  \\
    & &  &  \\
    & &  & \zxN{} \uar \\
\end{ZX}
&= 
\begin{ZX}
    &   &\zxXTri'{} \\
    & \zxN{} \uar\dar   & \zxN{} \uar\dar \\
    & \zxX{}\uar  & \zxN{} \uar\dar \\
    &  &  \\
    &   & \zxN{} \uar 
\end{ZX}
=
\begin{ZX}
    & &  &\zxXTri'{} \ar[lldd, )]  \\
    & \zxN{} \uar\dar & &   \\
    & \zxZ{} \dar  &  & \\
    &  &  & \\
    & \zxN{} \uar &  & 
\end{ZX}
\ ,&
\qquad 
\begin{ZX}
    & \zxXTri'{} \\
    & \zxN{} \uar\dar  \\
    & \zxX{} \uar
\end{ZX}
&= 1 \, .
\end{aligned}
\end{equation}

To promote a Hopf algebra over $\CC$ to a Hopf $C^\star$-algebra~\cite{nill1997hopfchain,vaes2001hopf}, one must equip $H$ with a $\star$-structure\footnote{All semisimple Hopf algebras over $\CC$ with ${|H|< 24}$ can be promoted to Hopf $C^\star$-algebras \cite{abella2019some}. However, it remains an open question whether every semisimple Hopf algebra over $\CC$ admits such $\star$-structure.}.
The $\star$-structure will allow us to define Hilbert spaces, adjoints, and unitarity \cite{klimyk2012quantum,buerschaper2013qdalgebra}. The conjugate-linear map $\star:H\rightarrow H$ is an involutive anti-homomorphism, i.e.~it satisfies
\tikzsetfigurename{starmult}
\begin{equation}
    (a^\star)^\star = a,\qquad (ab)^\star = b^\star a^\star, \begin{ZX}
        &&\\
        &\zxrep{$\star$} \ar[u]&\\
        &\zxX{} \ar[ld,(] \ar[rd,)] \ar[u]&\\
        &&
    \end{ZX} = \begin{ZX}
    &&& \\
    &\zxX{} \ar[u] \ar[d,C] &&\\
    &\zxN{} \ar[rd,)]\ar[u,C-,3d above,very thick]\ar[ld,(,3d above,very thick] &&\\
    \zxrep{$\star$} \ar[d]&&\zxrep{$\star$} \ar[d]&\\
    &&&
    \end{ZX}.
\end{equation}
and it follows that $\unit^\star = \unit$. The $\star$ is compatible with comultiplication\footnote{Here we take the convention $(a\otimes b)^\star = a^\star \otimes b^\star$. There is an alternative convention in the literature using $(a\otimes b)^\star = b^\star \otimes a^\star$.},
\tikzsetfigurename{starcomult}
\begin{equation}
    \comult(a^\star) = \comult(a)^\star,\begin{ZX}
        &&\\
        &\zxZ{} \ar[lu,)], \ar[ru,(] \ar[d]&\\
        &\zxrep{$\star$} \ar[d]&\\
        &&
    \end{ZX} = \begin{ZX}
    &&\\
    \zxrep{$\star$} \ar[u]&&\zxrep{$\star$}\ar[u]\\
    &\zxZ{} \ar[lu,)], \ar[ru,(] \ar[d]&\\
    &&
    \end{ZX} , \qquad \counit(a^\star) = \overline{\counit(a)}
\end{equation}
the interplay between antipode and $\star$ is given by
\begin{equation}
    S(S(a^\star)^\star) = a.
\end{equation}
Due to the uniqueness of the Haar integral, we have ${\wc{\haar}^\star = \wc{\haar}}$.

The cointegral, along with the $\star$-structure, allows us to define an inner product on $H$ as
\begin{equation}\label{eq:innerproduct}
    \braket{a,b}_\cohaar\equiv\phi(a^\star b)
\end{equation}
The unit is normalized under this inner product, ${\langle \unit,\unit\rangle_\cohaar = 1}$. However, the idempotent integral $\wc{\haar}$ is not. 
Instead, we can define a rescaled integral, $\haar = \sqrt{|H|}\wc{\haar}$, depicted as \includegraphics{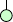}. %\tikzsetnextfilename{ketg} $\myzx{\zxN{}\dar \\ \zxZ{}\uar}$. 
This state is normalized with respect to the inner product $\braket{\cdot,\cdot}_\cohaar$. Indeed,
\tikzsetnextfilename{haarnormalized}
\ie 
\braket{{\haar},{\haar}}_\cohaar = |H|\cohaar(\wc\haar^\star \wc\haar) 
=|H|\wh{\counit}(\wc\haar) \cohaar(\wc\haar)  =|H| \cdot 1 \cdot \frac{1}{|H|} = 1 
\,, \quad \text{i.e.}\quad 
\begin{ZX}
    & \zxXTri'{}  &\\
    & \zxX{} \ar[u] &  \\
    \zxrep{$\star$} \ar[ru, bend left]  & & \zxZ{} \ar[lu, bend right]\\
    \zxZ{} \ar[u] & & 
\end{ZX} 
= 1 
\fe
where we used the fact that $\cohaar(\wc\haar)= \frac{1}{|H|}$ for any semisimple Hopf algebra~\cite{kodiyalam2003planar}. 

The underlying complex vector space of the Hopf $C^\star$-algebra $H$ along with the above inner product forms a Hilbert space. We will consider spin chains with a tensor product Hilbert space, where each factor will be precisely $H$ itself.
An arbitrary state in this local Hilbert space will be denoted as $\ket{a}$, with ${a\in H}$. As we will discuss in \cref{sec:HopfPauli}, the inner product also allows us to define the adjoint of the generalized Hopf Pauli operators. When we consider multi-site operators, we will also make use of the natural extension of the above inner product on $H$ to one on $H\otimes H$, defined via $\braket{a\otimes b, c\otimes d}_\cohaar = \braket{a, c}_\cohaar \braket{b, d}_\cohaar $.

\subsection{Dual Hopf algebra}\label{sec:dualHopf}
Given a finite dimensional Hopf algebra $H$, the linear dual space ${\dl{H} = \Hom(H,\CC)}$, which is isomorphic to $H$ as a complex vector space, can also be equipped with a Hopf algebra structure, called the dual Hopf algebra. Given $\alpha,\beta \in \dl{H}$ and $a\in H$, the multiplication on $\dl{H}$ is defined as,
\begin{equation}
    m^*( \alpha, \beta )(a) = (\alpha\otimes \beta)(\comult(a)) = \sum_{(a)} \alpha(a_{(1)})\beta(a_{(2)}) 
\end{equation}
and diagrammatically~\footnote{Note that our diagrammatic convention of the dual Hopf algebra is different from that of \Rfs{davydov2013z,collins:2019qdw,majid2022zxhopf}, whose counterparts of the r.h.s.~of \cref{eq:dl-mult,eq:dl-comult} look planar. On the other hand, our convention matches the algebraic definition in \appref{app:lineartrans}. A similar convention is used for the definition of the associated Hopf algebra constructed by the Frobenius form in \secref{sec:hopf-frob}.}, 
\tikzsetfigurename{dualmult}
\begin{equation}\label{eq:dl-mult}
        \mult^* : \dl{H} \times \dl{H} \rightarrow \dl{H}, \begin{ZX}
            &&\\
            &\zxX{} \ar[u] \ar[ld,(] \ar[rd,)]&\zxN|[xshift=2pt]{\text{*}}\\
            &&
        \end{ZX} \equiv \begin{ZX}
            &&&&&\\
            &&&&&\\
            &&&&&\\
            \zxN{} \ar[dd]&&\zxZ{} \ar[rr,C',3d above,very thick] \ar[ll,C'] \ar[r,C.] &\zxN{}\ar[luuu,N]&\zxN{} \ar[dd]&\\
            &&&&&\\
            &&&&&
        \end{ZX},
    \end{equation}
and the unit in $\dl{H}$ is defined by $\begin{ZX}
    &\\
    \zxX[phase in label={label position=0, text=black,fill=none}]{\text{*}} \ar[u] &
\end{ZX} = \begin{ZX}
    &&\\
    &&\\
    \zxZ{} \ar[r,C.]&\zxN{} \ar[luu,N]&\\
    &&
\end{ZX}$. 
The comultiplication $\Delta^*: \dl{H} \rightarrow \dl{H} \otimes \dl{H}$ is defined as
\tikzsetfigurename{dualcomult}
    \begin{equation}\label{eq:dl-comult}
        \Delta^*(\alpha)(a\otimes b) = \alpha(\mult(a, b)),\quad \begin{ZX}
            &&\\
            &\zxZ{} \ar[lu,)], \ar[ru,(] \ar[d]&\zxN|[xshift=2pt]{\text{*}}\\
            &&
        \end{ZX} = \begin{ZX}
            &&&&&\\
            &&&&&\\
            \zxN{} \ar[uu]&&\zxX{} \ar[r,C'] \ar[rr,C.,3d above,very thick] \ar[ll,C.]&\zxN{} \ar[lddd,N]&\zxN{} \ar[uu]&\\
            &&&&&\\
            &&&&&\\
            &&&&&
        \end{ZX}
    \end{equation}
where ${\alpha\in \dl{H}}$ and ${a,b\in H}$,
while the counit is defined by 
\tikzsetfigurename{dualcounit}
$\begin{ZX}
    \zxZ[phase in label={label position=0, text=black,fill=none}]{\text{*}} \ar[d]  \\
\end{ZX} = \begin{ZX}
 \zxX{} \ar[r,C'] & \zxN{} \ar[ldd,N]\\
 &\\ &
\end{ZX}$. 
As with $H$, we will also employ Sweedler notation for the dual Hopf algebra as
\begin{align}
   \Delta^*(\alpha) =  \alpha \circ m  \equiv \sum_{(\alpha)} \alpha_{(1)} \otimes \alpha_{(2)}
\end{align}
Finally, the dual antipode is defined as $\dl{S}(\alpha) = \alpha \circ S$, which can be depicted diagrammatically as
\tikzsetfigurename{dualanti}
\begin{equation}
    \begin{ZX}
        &&\\
        &&\\
        \zxN{} \ar[ruu,N]&\zxrep{$S$} \ar[l,C.] \ar[r, C']& \zxN{} \ar[ldd,N]\\
        &&\\
        &&
    \end{ZX}\ .
\end{equation}
The dual of a Hopf $C^\star$-algebra $H$ is again a Hopf $C^\star$-algebra, with the $\star$ operation given by,
\begin{equation}\label{eq:dualstar}
    \alpha^\star(a) = \overline{\alpha(S(a)^\star)}.
\end{equation}
It follows that $\counit^\star = \counit$, similar for Haar cointegral $\cohaar$. The inner product on $H$ \eqref{eq:innerproduct} induces a natural inner product $\braket{\cdot,\cdot}^*: \dl{H}\times \dl{H} \rightarrow \CC$ on $\dl{H}$, defined by 
\ie 
\braket{\alpha,\beta}^* \equiv \braket{\alpha^*,\beta^*}_{\cohaar}\,.
\fe 
However, compared to the inner product defined using the Haar cointegral of $\dl H$, namely $\haar$, we find that
\ie 
\braket{\alpha^*,\beta^*}_{\cohaar} = |H| \dl{m}(\alpha^\star, \beta)(\haar) = |H| \braket{\alpha,\beta}_\haar \,.
\fe 
That is, these two inner products differ by a factor of $|H|$.

\paragraph{Example: Group algebra \texorpdfstring{$\CC[G]$}{C[G]} and its dual \texorpdfstring{$\CC^G$}{C^G}}
For elements in the group algebra $g,h\in \CC[G]$, $m$ is the usual group multiplication, while comultiplication duplicates the group value:
\begin{align}
    \mult: g&\otimes h \mapsto gh\,,& \comult: g&\mapsto g\otimes g
\end{align} 
Both the antipode and $\star$ inverts group elements:
\begin{align}
\quad S(g) &= g^{-1}\,,& \star:&g\mapsto g^{-1}
\end{align} 
The unit is the group identity $e_G$. We define the dual basis elements to be $\delta_g$, which act as $\delta_g(h) = \delta_{g,h}$. The counit $\wh{\counit}$ evaluates any group element to $1$ and so is an equal sum of all dual basis elements. To summarize,
\begin{align}
    \unit&=e_G\,, &  \wh{\counit} &= \sum_{g\in G} \delta_g\,, & \haar &= \frac{1}{\sqrt{|G|}}\sum_{g\in G} g\,, &  \wh{\cohaar}&=\sqrt{|G|}\delta_{e_G}\,,\\      \wc{\unit} &= \frac{1}{\sqrt{|G|}} e_G \,, &  \counit&=\frac{1}{\sqrt{|G|}}\sum_{g\in G}
     \delta_g\,,  &  \wc{\haar}&=\frac{1}{|G|}\sum_{g\in G} g\,, & \cohaar &= \delta_{e_G}\,.
\end{align}
We see that the idempotent Haar integral $\wc{\haar}$ and the cointegral $\phi$ are projectors. The cointegral allows us to define the inner product
\begin{align}
\langle g,h \rangle_\cohaar = \delta_{g,h}
\end{align}
Under this inner product, we see that $\wc{\haar}$ is not properly normalized as a state. Instead, $\haar$ is. This normalized state is also sometimes called the $\ket{+}$ state in the literature.

Dual to the group algebra is the Hopf algebra $\CC^G$, also called the function algebra $\text{Fun}(G)$, with the basis $\delta_g \in \CC^G$. We have,
\begin{align}\label{eq:FunG}
    \mult^*(\delta_g \otimes \delta_h)=\delta_{g,h}\delta_g\,,\quad \comult^*(\delta_g) =\sum_{hk=g} \delta_h\otimes \delta_k\,,\quad S^*(\delta_g) = \delta_{g^{-1}} \,,\quad \star: \delta_g \mapsto \delta_g\,.
\end{align}

\subsection{Hopf-Frobenius algebra}\label{sec:hopf-frob}
It is known that a finite dimensional Hopf algebra is integral (i.e.~has a non-zero Haar integral)  \cite{swedler1969hopf}, and admits a Frobenius structure. Therefore it can be turned into a Hopf-Frobenius algebra \cite{coecke2011zxreview,collins:2019qdw,majid2022zxhopf,Freed:2018cec}. Indeed, given a Hopf algebra with Haar (co)integral, one can define the following red and green Frobenius forms, 
\tikzsetfigurename{Frobeniusform}
\begin{align}\label{eq:FrobeniusForm}
    \begin{ZX}
        &\zxX{} \ar[ld,(] \ar[rd,)]&\\
        &&
    \end{ZX}
    = \begin{ZX}
        &\zxX{} \ar[d]&\\
        &\zxX{} \ar[ld,(] \ar[rd,)]&\\
        \zxrep{$S$}\ar[d]&&\zxN{}\ar[d]\\
        &&
    \end{ZX}\ , 
    \qquad
    \begin{ZX}
        &&\\
        &\zxX{} \ar[lu,)] \ar[ru,(]&
    \end{ZX}
    =\begin{ZX}
        &&\\
        &\zxZ{} \ar[lu,)] \ar[ru,(] \ar[d]&\\
        &\zxZ{}&
    \end{ZX}\ ,
    \qquad 
    % \\
    \begin{ZX}
        &\zxZ{} \ar[ld,(] \ar[rd,)]&\\
        &&
    \end{ZX}
    = \begin{ZX}
        &\zxX{} \ar[d]&\\
        &\zxX{} \ar[ld,(] \ar[rd,)]&\\
        &&
    \end{ZX}\ , 
    \qquad 
    \begin{ZX}
        &&\\
        &\zxZ{} \ar[lu,)] \ar[ru,(]&
    \end{ZX}
    =\begin{ZX}
        &&\\
        \zxN{}\ar[u]&&\zxrep{$S$}\ar[u]\\
        &\zxZ{} \ar[lu,)] \ar[ru,(] \ar[d]&\\
        &\zxZ{}&
    \end{ZX}
    \ .
\end{align}
They generalize the notion of (unnormalized) Bell pairs in the qubit case.\footnote{In the qubit case, i.e.~${H=\CC[\ZZ_2]}$, the Bell pair is unique due to the trivial action of antipode, and the red or green node does not need to be present because of the identity removal axiom.}

Since we assumed the semisimple and cosemisimple Hopf algebra with $S^2=\mathbbm{1}$, the red Frobenius form is symmetric, as one can check,
\tikzsetfigurename{FrobeniusformCom}
\begin{equation}\label{eq:FrobeniusFormCocom}
    \begin{ZX}
        &&\\
        &\zxX{} \ar[ld,(] \ar[rd,)]&\\
        &&
    \end{ZX}=\begin{ZX}
        &&& \\
        &\zxX{}\ar[d,C] &&\\
        &\zxN{} \ar[rd,)]\ar[u,C-,3d above,very thick]\ar[ld,(,3d above,very thick] &&\\
        &&&
    \end{ZX},\quad \quad 
    \begin{ZX}
    &&\\
    &\zxX{} \ar[lu,)], \ar[ru,(] &\\
    &&
    \end{ZX}
    =
    \begin{ZX}
    \zxN{}\ar[rd,(]&&\\
    &\zxN{} \ar[ru,(,3d above,very thick]\ar[d,C-] &\\
    &\zxX{}  \ar[u,C,3d above,very thick] &\\
    &&
    \end{ZX},
\end{equation}
and similarly for the green Frobenius form. One can also check that these Frobenius forms have the following normalization,
\tikzsetfigurename{straightening}
\begin{equation}\label{eq:straightening}
    \begin{ZX}
        &\zxX{} \ar[ld,(] \ar[rd,s] &&\\
        &&\zxX{} \ar[ru,(]&
    \end{ZX}=\begin{ZX}
        &&\zxX{} \ar[ld,s] \ar[rd,)] &\\
        &\zxX{} \ar[lu,)]&&
    \end{ZX}=\begin{ZX}
        &\zxN{}\ar[d]\\[8pt]
        &
    \end{ZX}\ ,\quad \begin{ZX}
        &\zxX{} \ar[d,C] \ar[d,C-]&\\
        &\zxX{}&
    \end{ZX}=|H|
\end{equation}
Similarly for the green Frobenius form. Moreover, we have,
\tikzsetfigurename{straighteningG}
\begin{equation}\label{eq:rcgc=S}
    \begin{ZX}
        &\zxX{} \ar[ld,(] \ar[rd,s] &&\\
        &&\zxZ{} \ar[ru,(]&
    \end{ZX} = \begin{ZX}
        &\zxZ{} \ar[ld,(] \ar[rd,s] &&\\
        &&\zxX{} \ar[ru,(]&
    \end{ZX} = \begin{ZX}
        &\\
        &\zxrep{$S$}\ar[u]\ar[d]\\
        &
    \end{ZX} \ .
\end{equation}
The Frobenius forms are compatible with the $\star$-structure, i.e.
\ie 
    \begin{ZX}
        &&\\
        &\zxX{} \ar[ld,(] \ar[rd,)]&
        \\
        \zxrep{\tiny$\star$}\dar&&
        \\
        &&
    \end{ZX}
    =
    \begin{ZX}
        &&\\
        &\zxX{} \ar[ld,(] \ar[rd,)]&
        \\
        &&\zxrep{\tiny$\star$}\dar
        \\
        &&
    \end{ZX}
    \ , \quad 
    \begin{ZX}
    &&\\
    \zxrep{\tiny$\star$}\uar&&\\
    &\zxX{} \ar[lu,)], \ar[ru,(] &\\
    &&
    \end{ZX}
    =
    \begin{ZX}
    &&\\
    &&\zxrep{\tiny$\star$}\uar\\
    &\zxX{} \ar[lu,)], \ar[ru,(] &\\
    &&
    \end{ZX}
    \ ,
\fe 
and similarly for the green Frobenius forms.

With the Frobenius form, we can define the associated Hopf algebra $\aso{H}$,
\tikzsetfigurename{assoH}
\begin{align}\label{eq:assoH}
    \aso{\mult}: \aso{H}\times \aso{H} \rightarrow \aso{H},\quad \tikzsetnextfilename{multg} \begin{ZX}
        &&\\
        &\zxZ{} \ar[u] \ar[ld,(] \ar[rd,)]&\\
        &&
    \end{ZX} &\equiv \tikzsetnextfilename{multgdef} \begin{ZX}
        &&\ar[ddd,)]&&&\\
        &\zxZ{} \ar[ldd,(]&&&\zxZ{}\ar[rdd,)]&\\
        &&\zxZ{} \ar[lu] \ar[rru,3d above,very thick]\ar[d,(]&&&\\
        &&\zxZ{} &&&
        &&&&&
    \end{ZX}
    \\
% \end{equation}
% \tikzsetfigurename{assoHco}
% \begin{align}
    \aso{\comult}: \aso{H}\rightarrow \aso{H}\times \aso{H} ,\quad  \tikzsetnextfilename{comultr}\begin{ZX}
        &&\\
        &\zxX{} \ar[d] \ar[lu,)] \ar[ru,(]&\\
        &&
    \end{ZX} &\equiv \tikzsetnextfilename{comultrdef} \begin{ZX}
        &&&&&&\\
        &&&\zxX{}\ar[ddd,(]&&&\\
        &&&\zxX{} \ar[lld,3d above,very thick] \ar[u,(] \ar[rd]&&&\\
        &\zxX{}\ar[luu,)]&&&\zxX{}\ar[ruu,(]&&\\
        &&&\zxN{}&&&
    \end{ZX}
\end{align}
The associated Hopf algebra is isomorphic to $H^{*}$ \cite{majid2022zxhopf,collins2024hopf,Freed:2018cec}.

Moreover, one can show that the Frobenius forms relate the counit to the integral and the unit to the cointegral:
\tikzsetfigurename{flipped}
\begin{equation}\label{eq:frob-haar}
    \includegraphicsr{}{tikz-figs/ketg.pdf}=\begin{ZX}
        &&&\\
        &\zxZ{} &&\\
        &&\zxZ{}\ar[lu,)]   \ar[ruu,(]&
    \end{ZX},\quad 
    \includegraphicsr{}{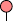} =\begin{ZX}
        &\zxX{} \ar[rdd,)]&&\\
        \zxX{}\ar[ur,)]&&&\\
        &&&
    \end{ZX}.
\end{equation}
Thus, the integral $\haar \in H$ serves as the unit for $\aso{H}$ and the cointegral $\wh{\cohaar} \in \dl{H}$ serves as the counit of $\aso{H}$. Similarly,
the integral and cointegral in $\aso{H}$ match the unit and counit of $H$, respectively. 

To summarize, the Hopf algebra and its associated Hopf algebra corrspond to the following data:
\begin{align}
    \text{Hopf algebra $H$: }&\left(H, \includegraphicsr{}{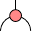}, \includegraphicsr{}{tikz-figs/ketr.pdf}, \includegraphicsr{}{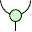}, \includegraphicsr{}{tikz-figs/brag.pdf}, \includegraphicsr{}{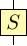}\right)
    , \\
    \text{associated Hopf algebra $\aso{H}:$ }&\left(H,\includegraphicsr{}{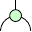}, \includegraphicsr{}{tikz-figs/ketg.pdf}, \includegraphicsr{}{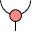}, \includegraphicsr{}{tikz-figs/brar.pdf},\includegraphicsr{}{tikz-figs/antipode.pdf}\right)
\end{align}
In addition, the red(green) algebra and red(green) coalgebra together define a red(green) Frobenius algebra, meaning that they satisfy the following Frobenius conditions:
\tikzsetfigurename{Frobenius}
\begin{equation}\label{eq:hopf-frob-cond}
\tag{Fr}
    \begin{ZX}
        &&\\
        \zxX{} \ar[u] \ar[dd] \ar[rd]&&\\
        &\zxX{} \ar[uu] \ar[d]&\\
        &&
    \end{ZX} = \begin{ZX}
        &&\\
        &\zxX{} \ar[u] \ar[dd]&\\
        \zxX{} \ar[uu] \ar[d] \ar[ru]&&\\
        &&\\
        &&
                    \end{ZX} = \begin{ZX}
        &&\\
        &\zxX{} \ar[lu] \ar[ru] \ar[d]&\\
        &\zxX{} \ar[ld] \ar[rd]&\\
        &&
    \end{ZX}\,,
    \qquad
    \begin{ZX}
        &&\\
        \zxZ{} \ar[u] \ar[dd] \ar[rd]&&\\
        &\zxZ{} \ar[uu] \ar[d]&\\
        &&
    \end{ZX} = \begin{ZX}
        &&\\
        &\zxZ{} \ar[u] \ar[dd]&\\
        \zxZ{} \ar[uu] \ar[d] \ar[ru]&&\\
        &&\\
        &&
    \end{ZX} = \begin{ZX}
        &&\\
        &\zxZ{} \ar[lu] \ar[ru] \ar[d]&\\
        &\zxZ{} \ar[ld] \ar[rd]&\\
        &&
    \end{ZX}
    \,.
\end{equation}
With our normalization, the red and green Frobenius algebras are special,
\tikzsetfigurename{Frob-special}
\begin{equation}
\begin{ZX}
    &\zxN{}  & \\
    & & \\
    &\zxZ{} \ar[uu] \ar[dd,C-] \ar[dd,C] & \\
    &   &\\
    &\zxZ{} \ar[dd]   &   \\
    &  &  \\
    &\zxN{}  &  
\end{ZX} = \sqrt{|H|}\ \begin{ZX}
\zxN{} \ar[ddddddd]\\ \zxN{} \\ \zxN{} \\ \zxN{} \\  \zxN{} \\ \zxN{} \\ \zxN{} \\
\zxN{}\end{ZX},\quad \begin{ZX}
&\zxN{}  & \\
& & \\
&\zxX{} \ar[uu] \ar[dd,C-] \ar[dd,C] & \\
&   &\\
&\zxX{} \ar[dd]   &   \\
&  &  \\
&\zxN{}  &  
\end{ZX} = \sqrt{|H|}\ \begin{ZX}
\zxN{} \ar[ddddddd]\\ \zxN{} \\ \zxN{} \\ \zxN{} \\  \zxN{} \\ \zxN{} \\ \zxN{} \\
\zxN{}\end{ZX}
\end{equation}
The data of the Hopf-Frobenius algebra is summarized in Table~\ref{tab:Hopf-Frobenius}.

\begin{table}[ht!]
    \centering
    \begin{tabular}{|c|c|c|c|c|c|}
        \hline
        &   Green Frobenius & Red Frobenius & Antipode & Integral & Cointegral  \\
        \hline
        $H$          & \includegraphicstable{tikz-figs/comultg.pdf} \quad  \includegraphicstable{tikz-figs/brag.pdf}   & \includegraphicstable{tikz-figs/multr.pdf} \quad \includegraphicstable{tikz-figs/ketr.pdf}  & \includegraphicstable{tikz-figs/antipode.pdf} & \includegraphicstable{tikz-figs/ketg.pdf} &  \includegraphicstable{tikz-figs/brar.pdf}\TBstrut\\
        \hline
        $\aso{H}$   & \includegraphicstable{tikz-figs/multg.pdf} \quad  \includegraphicstable{tikz-figs/ketg.pdf}   & \includegraphicstable{tikz-figs/comultr.pdf} \quad  \includegraphicstable{tikz-figs/brar.pdf}  & \includegraphicstable{tikz-figs/antipode.pdf} & \includegraphicstable{tikz-figs/ketr.pdf} &  \includegraphicstable{tikz-figs/brag.pdf}\TBstrut\\
        \hline
    \end{tabular}
    \caption{The Hopf-Frobenius algebra consists of a Hopf algebra $H$ and an associated Hopf algebra $\aso{H}$, sharing the same antipode, in which the (co)unit of $\aso{H}$ is the (co)integral of $H$ and vice versa. Moreover, the coalgebra of $H$ and algebra of $\aso{H}$ forms a green Frobenius algebra and the algebra of $H$ and coalgebra of $\aso{H}$ forms a red Frobenius algebra.}
    \label{tab:Hopf-Frobenius}
\end{table}

\paragraph{Example: Hopf-Frobenius algebra of \texorpdfstring{$\CC[G]$}{C[G]}}
The red and green Frobenius forms correspond to maximally entangled states. Namely, 
\begin{align}
    \includegraphics[]{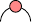} &=\sqrt{|G|}\sum_{g\in G} \delta_g \otimes \delta_{g},  &     \includegraphics[]{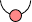} &= \frac{1}{\sqrt{|G|}}\sum_{g\in G} g \otimes g,\\
     \includegraphics[]{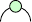} &=\sqrt{|G|} \sum_{g\in G} \delta_g \otimes \delta_{g^{-1}},  &     \includegraphics[]{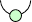} &= \frac{1}{\sqrt{|G|}}\sum_{g\in G} g \otimes g^{-1}
\end{align}
Using this, the multiplication and comultiplication of the associated Hopf algebra are
\begin{align}
    \t{m} (g,h) &= \sqrt{|G|} \delta_{g,h} g \,, & \t{\Delta}(g) &= \frac{1}{\sqrt{|G|}} \sum_{hk=g} h \otimes k \,.
\end{align}

\section{Lattice models from Hopf qudits and symmetries}
\label{sec:Hopfqudit}

In this section, we generalize the group-based Pauli operators to Hopf algebras, and proceed to use them to construct the MPO for $\Rep(H)$ symmetry, and their fixed point paramagnetic and ferromagnetic wavefunctions. We note that some of these constructions have appeared previously in the literature \cite{buerschaper2013qdalgebra,inamura_lattice_2022,molnar2022MPO,zhian2024chain,jia2024generalized,zhiyuan2025hopfcircuit}, here we fix notation and present a unified discussion.

\subsection{Hopf Pauli operators}

\label{sec:HopfPauli}
The generalized Hopf Pauli operators share similar structure with the non-abelian group case, which are used in constructing quantum double models  \cite{buerschaper2013qdalgebra,kitaev2003quantumdouble,bombin2008qd,cui2021qdhopf,zhian2023double} and Hopf algebra valued spin chains~\cite{Albert:2021vts,nat2025repG,jia2024generalized,zhian2024chain}. In both cases, the operators naturally separate into ``\(X\)''-type (or $L$) maps, associated with left and right multiplication by algebra elements, and ``\(Z\)''-type (or $T$) maps, associated with characters or dual actions.
However, there are certain subtleties that one must keep in mind when generalizing from groups. First, since we no longer assume cocommutativity, there will now be two distinct $Z$ operators, corresponding to left and right comultiplication. Second, there is no longer a canonical basis for the algebra where the multiplication is invertible.

We begin by constructing the generalized Hopf Pauli operator, taking the local Hilbert space to be $H$ itself.\footnote{In the most general case, the local Hilbert space corresponds to an $H$-Hopf module (we will postpone a more detailed discussion to \Cref{sec:H-gapped-phases}; cf.~\Rf{buerschaper2013qdalgebra}), which has both compatible $H$-action and $H$-coaction. For simplicity, we take the $H$-Hopf module to be $H$ itself, so that the action and coaction follow from the Hopf algebra multiplication and comultiplication.}
We label elements of the Hilbert space as $\ket{a}$ for $a\in H$. 
First, we can define the following single site Hopf Pauli operators,\footnote{\Rf{jia2024generalized} also defines Hopf Pauli operators. Our notation is slightly different from theirs.}
\tikzsetfigurename{HopfPauli}
\begin{align}
    \Xl^a\ket{b} &= \ket{ab} = \tikzsetnextfilename{Xl}\begin{ZX}
        &&&\\
        &\zxX{} \ar[u] \ar[d] \ar[ld,(]&&\\
        \zxX{a}&&&
    \end{ZX} &  \Xr^a\ket{b} &= \ket{bS(a)} = \tikzsetnextfilename{Xr}\begin{ZX}
        &&&\\
        &\zxX{} \ar[u] \ar[dd] \ar[rd,)]&&\\
        &&\zxrep{$S$} \ar[rd,(]&\\
        &&&\zxX{a}
    \end{ZX}\\
    \Zl^\alpha\ket{a} &= \sum_{(a)} \alpha(S(a_{(1)})) \ket{a_{(2)}} = \tikzsetnextfilename{Zl} \begin{ZX}
        \zxZ{\alpha}&&&\\
        &\zxrep{$S$}\ar[lu,(]&&\\
        &&\zxZ{} \ar[uu] \ar[d] \ar[lu,)]&\\
        &&&
    \end{ZX}    &  \Zr^\alpha\ket{a} &= \sum_{(a)} \ket{a_{(1)}} \alpha(a_{(2)})  = \tikzsetnextfilename{Zr} \begin{ZX}
        &&&\\
        &\zxZ{\alpha}&&\\
        \zxZ{} \ar[u] \ar[d] \ar[ru,(]&&&\\
        &&&
    \end{ZX}
\end{align}
Using the multiplication in the Hopf algebra, we have the expected algebra of the $X$ operators. Indeed, if we fix a basis, ${\{a_i \, |\,  i=1,\dots, |H|\}}$, we have the following algebra
\begin{equation}
    \Xl^{a_i} \Xl^{a_j} = \sum_k \mult_{ij}^k \Xl^{a_k} \,,
    \qquad \Xr^{a_i} \Xr^{a_j} = \sum_k \mult_{ij}^k \Xr^{a_k} \,,
\end{equation}
where $\mult_{ij}^k$ is the multiplication tensor in this basis. Similarly, in terms of the dual basis for $\dl H$, ${\{\alpha_i \, | \, i=1,\dots, |H|\}}$ and the comultiplication tensor $\comult_k^{ij}$, we find the algebra of the $Z$ operators,
\ie 
\Zl^{\alpha_i} \Zl^{\alpha_j} = \sum_k \comult_k^{ij} \Zl^{\alpha_k}\,, 
\qquad 
\Zr^{\alpha_i} \Zr^{\alpha_j} = \sum_k \comult_k^{ij} \Zr^{\alpha_k} \,.
\fe

The algebra involving both $X$ and $Z$ type operators is more subtle. In general, we find
\ie \label{eq:ZalphaXa}
\Zr^{\alpha} \Xl^{a} = \sum_{(a)}\sum_{(\alpha)} \alpha_{(1)}(a_{(2)}) \Xl^{a_{(1)}} \Zr^{\alpha_{(2)}} \,.
\fe 
For the case where ${\al =\Ga_{pq}}$ for some ${\Ga \in {\rm Irr}(H)\subseteq \dl{H}}$, we can simplify \cref{eq:ZalphaXa} using the fact that the irreducible representation satisfies ${\Ga(\mult(a,b))_{pq} = \Ga(a)_{ps} \Ga(b)_{sq}}$\footnote{\label{footnote:Hmodule-rep}The irreducible representations of $H$, denoted collectively by the notation ${\rm Irr}(H)$, are more precisely described as simple finite-dimensional (left) $H$-modules; see \Cref{sec:Hmod-Hcomod} for further discussion on $H$-modules.}, where the subscript indices ${p,q,s\in \{1,\dots,d_\Ga\}}$, where $d_\Ga$ is the dimension of $\Ga$. With this, we find that
the operators $X^a$ and $Z^{\Ga_{pq}}$ fail to commute up to a matrix multiplied in the virtual dimension, i.e.
\begin{equation} \label{eq:ZGaXa}
    \Zr^{\Gamma_{pq}} \Xl^{a} = \sum_{(a)}\sum_{s = 1}^{d_\Ga} \Xl^{a_{(1)}} \left(\Gamma(a_{(2)})_{ps} \Zr^{\Gamma_{sq}} \right) \,.
\end{equation}
For some choices of $\Ga$ and $a$, this algebra can be simplified even further.
If we require the operators $\Xl^a$ and $\Zr^\Ga$ commute up to a \textit{phase} when  
$\Gamma(a_{(2)})$ is a fixed phase factor times the $d_\Ga$-dimensional identity matrix for each $a_{(2)}$ appearing in $\comult(a)$, meaning that $a_{(2)}$ are the central elements of $H$ (elements that commute with any other element of $H$). 
A sufficient condition for this is for $a$ to be a central group-like element, i.e.~${\comult(a)=a\otimes a}$ and ${\mult(a,\cdot) = \mult(\cdot, a)}$. In this case, we have
\begin{equation} \label{eq:ZGaXacentral}
    \Zr^{\Gamma_{pq}} \Xl^{a} =\frac{\chi_\Gamma(a)}{d_\Gamma} \Xl^{a}  \Zr^{\Gamma_{pq}} 
    \,  .
\end{equation}
Note that $\frac{\chi_\Gamma(a)}{d_\Gamma}$ has norm 1 for central group-like element $a$ and arbitrary unitary representation $\Gamma$ (see \appref{app:def} for definitions).

Using the inner product defined in \eqref{eq:innerproduct}, we can define the adjoint of any operator $A$ via ${\langle x,A(y) \rangle = \langle A^\dagger(x),y \rangle}$, where ${x,y\in H}$. For the single-site Hopf Pauli operators defined above, we find their adjoints are given by
\begin{equation}\label{eq:pauli-dag}
        (\Xl^a)^\dagger = \Xl^{a^\star},\quad (\Xr^a)^\dagger = \Xr^{a^\star},\quad (\Zl^\alpha)^\dagger = \Zl^{\alpha^\star}, \quad (\Zr^\alpha)^\dagger = \Zr^{\alpha^\star} \,.
\end{equation}
For example, the last identity of \cref{eq:pauli-dag} is expressed diagrammatically as
\tikzsetfigurename{HopfPauliadjoint}
\begin{equation}
\left(
    \begin{ZX}
        &&&\\
        &&\zxZ{\alpha}&\\
        &\zxZ{} \ar[u] \ar[d] \ar[ru,(]&&\\
        &&&
    \end{ZX} 
    \right)^\dagger
    =
    \begin{ZX}
    &\zxZ{\overline{\alpha}}&&\\
    &\zxrep{$S$} \ar[u]&&\\
    &\zxrep{$\star$} \ar[u]&&\\
    \zxZ{} \ar[uuu] \ar[d] \ar[ru,(]&&&\\
    &&&
    \end{ZX} 
    \stackrel{\eqref{eq:dualstar}}{=} \begin{ZX}
    &&&\\
    &\zxZ{\alpha^\star}&&\\
    \zxZ{} \ar[u] \ar[d] \ar[ru,(]&&&\\
    &&&
    \end{ZX} \,.
\end{equation}
In the intermediate step, the notation $\overline{\al}$ is used to denote complex conjugation. In symbols, the above equation can be re-written as\footnote{Note that ZX calculus diagrams do not effectively differentiate linear maps from anti-linear ones, which means that any complex conjugation of scalars needs to be tracked separately.}:
\ie 
(\Zr^{\al})^\dagger \ket{a} = \sum_{(a)} \overline{\al(S(a_{(2)})^\star) } \ket{a_{(1)}} = \sum_{(a)} \al^\star(a_{(2)}) \ket{a_{(1)}} \equiv \Zr^{\al^\star} \ket{a} \,.
\fe 
Note that when the Hopf Pauli operator corresponding to the idempotent (co)integral takes a simple form as the following projectors
\begin{align}
\label{eq:singlesiteprojector}
Z^\cohaar &\equiv \Zl^\phi = \Zr^\phi \stackrel{\eqref{ref:integral}}{=} \begin{ZX}
        &\\
        \zxX{} \ar[u]&\\
        \zxXTri'{} \ar[d]& \\
        &
    \end{ZX}, & X^{\wc\haar} &\equiv \Xl^{\wc\haar} = \Xr^{\wc\haar} \stackrel{\eqref{ref:integral}}{=} \begin{ZX}
        &\\
        \zxZTri.{} \ar[u]&\\
        \zxZ{} \ar[d]& \\
        &
    \end{ZX},
\end{align}
where we removed the arrow to emphasize the fact that the left and right versions are equal.

Next, we define a shorthand to denote products of $X$ and $Z$ operators acting on different Hopf qudits. 
\begin{align}
    (\prod_{j=1}^n X_j)^a &\equiv \sum_{(a)} \prod_j X_j^{a_{(j)}}, &(\prod_{j=1}^n Z_j)^\alpha &\equiv \sum_{(\alpha)} \prod_j Z_j^{\alpha_{(i)}} \label{eq:prodXprodZ}, \\
    (\prod_{j=1}^n X_j)^a_\op &\equiv \sum_{(a)} \prod_j X_j^{a_{(n-j)}}, &  (\prod_{j=1}^n Z_j)^\alpha_\op &\equiv \sum_{(\alpha)} \prod_j Z_j^{\alpha_{(n-j)}},
    \label{eq:prodXprodZop}
\end{align}
where $X$ can be either $\Xl$ or $\Xr$ and $Z$ can be either $\Zl$ or $\Zr$. The subscript ``op" is to denote that the opposite comultiplication is used in splitting $a$ (likewise, the opposite multiplication is used in splitting $\alpha$).

For example, we can define the following two-body operators
\tikzsetfigurename{HopfdefineZZ}
\begin{align}
    (\Xr \Xl)^{a}_\op &=\sum_{(a)}\Xr^{a_{(2)}} \Xl^{a_{(1)}} &&=\begin{ZX}
        &&&\\
\zxX{}\ar[u]\ar[dddd]&&&\zxX{}\ar[u]\ar[dddd]\\
        & \zxrep{$S$}\ar[rd,s]\ar[lu,s]&&\\
        &&\zxN{}\ar[ruu,s]&\\
        &&\zxZ{}\ar[u,C,3d above,very thick] \ar[u,C-]\ar[d]&\\
        \zxN{}&&\zxX{a}&\zxN{}
    \end{ZX} \\
    &=\sum_{(Sa)} \Xr^{S((Sa)_{(1)})} \Xl^{(S((Sa)_{(2)})} &&=   \tikzsetnextfilename{XXterm}
    \begin{ZX}
        &&&\\
    \zxX{}\ar[u]\ar[ddd]&&&\zxX{}\ar[u]\ar[ddd]\\
        &&\zxrep{$S$}\ar[ru,s]&\\
        &\zxZ{}\ar[luu,s]\ar[ru,s] \ar[d]&&\\
        &\zxX{Sa}&&\\
        &&&
    \end{ZX}
\end{align}
\begin{align}
    (\Zr \Zl)^{\alpha}_\op &=\sum_{(\alpha)}\Zr^{(\alpha)_{(2)}} \Zl^{(\alpha)_{(1)}} &&=\tikzsetnextfilename{ZZterm0}
    \begin{ZX}
        &\zxZ{\alpha}&&\\
        &\zxX{}\ar[d,C,3d above,very thick] \ar[d,C-] \ar[u]&&\\
        &\zxN{} \ar[ldd,s] &&\\
        &&\zxrep{$S$}\ar[lu,s] \ar[rd,s]&\\
\zxZ{}\ar[uuuu]\ar[d]&&&\zxZ{}\ar[uuuu]\ar[d]\\
        &&&
    \end{ZX}\\
    &=\sum_{(\alpha S)} \Zr^{(\alpha S)_{(1)}S} \Zl^{(\alpha S)_{(2)}S} &&=   \tikzsetnextfilename{ZZterm}
    \begin{ZX}
        &&\zxZ{\alpha S}&\\
        &&\zxX{}\ar[rdd,s]\ar[ld,s] \ar[u]&\\
        &\zxrep{$S$}\ar[ld,s]&&\\
    \zxZ{}\ar[uuu]\ar[d]&&&\zxZ{}\ar[uuu]\ar[d]\\
        &&&
    \end{ZX}
\end{align}
which will generalize the $XX$ and $ZZ$ terms appearing in different presentations of the usual Ising model for qubits. We elect to draw the diagrams in the second presentation moving forward, since they are planar.

We note that the $ZZ$ operator corresponding to the cointegral takes a particularly convenient diagrammatic form. Namely, $(\ract{Z} \lact{Z})^{\phi} $ term only invokes the structure of the green Frobenius algebra.

\tikzsetfigurename{HopfPauliStr22}
\begin{equation}
    (\ract{Z} \lact{Z})^{\phi}_\op =  \frac{1}{|H|}\sum_{\Gamma \in \Irr(H)} d_\Gamma (\ract{Z} \lact{Z})_\op^{\chi_\Gamma} = \frac{1}{\sqrt{|H|}}\begin{ZX}
        &&\zxX{}&\\
        &&\zxX{}\ar[rdd,s]\ar[ld,s] \ar[u]&\\
        &\zxrep{$S$}\ar[ld,s]&&\\
        \zxZ{}\ar[uuu]\ar[d]&&&\zxZ{}\ar[uuu]\ar[d]\\
        &&&
    \end{ZX} = \frac{1}{\sqrt{|H|}}\begin{ZX}
        &&&&\\
        &\zxZ{} \ar[dd] \ar[lu,)] \ar[ru,(]&&&\\
        &&&&\\
        &\zxZ{} \ar[ld,(] \ar[rd,)]&&&\\
        &&&&
    \end{ZX}
\end{equation}
where $d_\Ga$ is the dimension of the irreducible representation $\Ga$ of $H$ and $\chi_\Ga$ is the corresponding character. The characters of ${\Gamma\in\mathrm{Irr}(H)}$ form a basis for the sub-algebra of cocommutative elements of $\dl{H}$, denoted $\Cocom(\dl{H})$\footnote{\label{footnote:characters-cocom}
Since the characters of ${\Ga\in{\rm Irr}(H)}$ form a basis of $\Cocom(\dl H)$, one can think of the latter as a lift from the character ring to an algebra over $\CC$. 
In this sense, $\Cocom(\dl H)$ is the ``algebra of characters of $H$''. Each character of $H$ corresponds to a cocommutative element in $\dl H$. This can be understood as follows. Since $\chi$ is the character of some ${\Ga\in\Irr(H)}$, from the cyclicity of trace, we have ${\chi(ab) = \chi(ba)}$ for ${a,b \in H}$. Upon taking the dual of this identity, following \Cref{sec:dualHopf}, we get the dual identity, ${\dl{\Delta}(\chi) =  (\dl{\Delta})^{cop}(\chi)} $, i.e.~$\chi$ is a cocomutative element of $\dl H$.
}~\cite{Yammine2021}. 
 Similarly, $(\ract{X} \lact{X})^{\wc\haar}$ only invokes the structure of the red Frobenius algebra.

\tikzsetfigurename{HopfPauliStrS}
\begin{equation}\label{eq:Pauli-XX}
    (\ract{X} \lact{X})_\op^{\wc\haar} = \frac{1}{|H|}\sum_{\drep\in \Irr(\dl{H})} d_\drep(\ract{X} \lact{X})_\op^{\dchi_\drep} =  \frac{1}{\sqrt{|H|}}\begin{ZX}
        &&&\\
        \zxX{}\ar[u]\ar[ddd]&&&\zxX{}\ar[u]\ar[ddd]\\
        &&\zxrep{$S$}\ar[ru,s]&\\
        &\zxZ{}\ar[luu,s]\ar[ru,s] \ar[d]&&\\
        &\zxZ{}&&\\
        &&&
    \end{ZX} = \frac{1}{\sqrt{|H|}}\begin{ZX}
        &&&&\\
        &\zxX{} \ar[dd] \ar[lu,)] \ar[ru,(]&&&\\
        &&&&\\
        &\zxX{} \ar[ld,(] \ar[rd,)]&&&\\
        &&&&
    \end{ZX}.
\end{equation}
where $\drep\in{\rm Irr}(\dl{H})$ is an irreducible representation of $\dl{H}$ of dimension $d_\drep$, and its character is denoted by $\dchi_\drep$.

We define a few more two-site operators for later convenience. These are the generalizations of the familiar CNOT gates of qubits:\footnote{We choose to call these operators $ZX$ operators, because they are precisely made from contracting the Hopf Pauli $Z$ and $X$ tensors together. We avoid using the term ``control'' in the general case, because it is only well-defined if the comultiplication is group like, i.e. $\comult(g) = g \otimes g$.} 
\begingroup
\allowdisplaybreaks
\begin{align}
\ZXgate &=  \begin{ZX}
        \zxN{}&&\zxN{}\\
        &&\zxX[]{}\ar[ddd]\ar[u]\\
        &\zxN{}\ar[ld]\ar[ru]&\\
        \zxZ[]{}\ar[uuu]\ar[d]&&\\
        \zxN{}&&\zxN{} \\
    \end{ZX}:
    &   \ZXgate_{ve}\ket{a}_v\ket{b}_e &=\sum_{(a)} \ket{a_{(1)}}_v\ket{a_{(2)}b}_e 
     \,,\\
\ZXgate^\dagger &= \begin{ZX}
        \zxN{}&&\zxN{}\\
        &&\zxX[]{}\ar[ddd]\ar[u]\\
        &\zxrep{$S$}\ar[ld]\ar[ru]&\\
        \zxZ[]{}\ar[uuu]\ar[d]&&\\
        \zxN{}&&\zxN{} \\
    \end{ZX}: &
    \ZXgate_{ve}^\dagger\ket{a}_v\ket{b}_e &=\sum_{(a)} \ket{a_{(1)}}_v\ket{S(a_{(2)})b}_e     \,,\\
    \XZgate &=  \begin{ZX}
        \zxN{}&&\zxN{}\\
        \zxX[]{}\ar[ddd]\ar[u]&&\\
        &\zxN{}\ar[lu]\ar[rd]&\\
        &&\zxZ[]{}\ar[uuu]\ar[d]\\
        \zxN{}&&\zxN{} \\
    \end{ZX}
    :&\XZgate_{ev}\ket{b}_e\ket{a}_v &=\sum_{(a)} \ket{ba_{(1)}}_e \ket{a_{(2)}}_v   \,, \\ 
    \XZgate^\dagger &=  \begin{ZX}
        \zxN{}&&\zxN{}\\
        \zxX[]{}\ar[ddd]\ar[u]&&\\
        &\zxrep{$S$}\ar[lu]\ar[rd]&\\
        &&\zxZ[]{}\ar[uuu]\ar[d]\\
        \zxN{}&&\zxN{} \\
    \end{ZX}
    :&\XZgate^\dagger_{ev}\ket{b}_e\ket{a}_v &=\sum_{(a)}\ket{bS(a_{(1)})}_e\ket{a_{(2)}}_v   \,.
\end{align}
\endgroup
One can show that these are unitary using \eqref{eq:Hopf-rule}. 
In the group case, these reduce to the controlled left-multiplication operator $C\Xl  = \sum_{g,h} \ket{g,gh}\bra{g,h}$, controlled right multiplication operator $C\Xr  = \sum_{g,h} \ket{g,hg^{-1}}\bra{g,h}$ and their inverses.
The single-site Hopf Pauli operators map under the unitaries defined above as
\begin{align}
    \Xr^a_v \otimes \mathbbm 1_e &\xrightarrow{\ZXgate_{ve}^\dagger}  \sum_{(a)} \Xr_v^{a_{(2)}} \otimes \Xl_e^{a_{(1)}} \,, 
    & \mathbbm 1_e \otimes  \Xr^a_v  &\xrightarrow{\XZgate_{ev}}  \sum_{(a)} \Xr_e^{a_{(2)}} \otimes \Xr_v^{a_{(1)}} \,, 
    \\
    \mathbbm 1_v \otimes \Zr^{\alpha}_e &\xrightarrow{\ZXgate_{ve}^\dagger} \sum_{(\alpha)} \Zr^{\alpha_{(1)}}_v \otimes\Zr^{\alpha_{(2)}}_e \,, 
    & \Zr^{\alpha}_e \otimes \mathbbm 1_v   &\xrightarrow{\XZgate_{ev}}  \sum_{(\alpha)} \Zr_e^{\alpha_{(1)}} \otimes \Zl_v^{\alpha_{(2)}} \,,
\end{align}
while some of the two-site Hopf Pauli operators map as
\begin{align}
    \sum_{(a)} \Xl^{a_{(1)}}_v \otimes \Xl^{a_{(2)}}_e &\xrightarrow{\ZXgate_{ve}^\dagger}  \Xl^a_v \otimes \mathbbm 1_e \,,  
    & \sum_{(a)} \Xr_e^{a_{(1)}} \otimes \Xl_v^{a_{(2)}}  &\xrightarrow{\XZgate_{ev}} \mathbbm 1_e \otimes \Xl_v^{a} \,,
    \\
    \sum_{(\alpha)} \Zr^{\alpha_{(2)}}_v \otimes \Zl^{\alpha_{(1)}}_e &\xrightarrow{\ZXgate_{ve}^\dagger}  \mathbbm 1_v \otimes \Zl^\alpha_e  \,, 
    & \sum_{(\alpha)} \Zl_e^{\alpha_{(2)}} \otimes \Zl_v^{\alpha_{(1)}}   &\xrightarrow{\XZgate_{ev}} \Zl^{\alpha}_e \otimes \mathbbm 1_v \,.
\end{align}
In these equations, the notation ${\mathcal O \xrightarrow{U} \mathcal O '}$ should be understood as shorthand for ${U \mathcal O  U^\dagger  = \mathcal O '}$.
As before, we used Sweedler's notation, ${\comult(a) = \sum_{(a)} a_{(1)} \otimes a_{(2)}}$ and ${\alpha\circ\mult = \sum_{(\alpha)} \alpha_{(1)}\otimes\alpha_{(2)} }$.

\subsection{\texorpdfstring{$\Rep(H)$}{Rep(H)} symmetry and SSB}
\label{sec:RepH-symm}

The $\Rep(H)$ symmetry MPO can be constructed similar to the $\Rep(G)$ symmetry MPO
~\cite{JiWen20,Albert:2021vts,tantivasadakarn_hierarchy_2022,nat2025repG} (see also the generalization to (weak) Hopf algebras \cite{inamura_lattice_2022,molnar2022MPO,zhian2024chain,jia2024generalized,zhiyuan2025hopfcircuit}). Given a representation $\Gamma \in \Rep(H)$, the corresponding symmetry operator is
\tikzsetfigurename{rephmpo11}
\begin{equation}\label{eq:rephmpo}
    \Br^{\Gamma}=
    \begin{ZX}
        && && && && && \\
        \zxN{}\ar[rrrrr,blue]&& && &&... && &&\zxN{} \ar[lll,blue] \\
        \zxN{} \ar[u,C,blue]&&\zxrep{$\Gamma$} \ar[ll,blue] \ar[rr,blue] \ar[zxarrowr,rrdd,N] && \zxrep{$\Gamma$}\ar[r,blue] \ar[zxarrowr,rrdd,N] && ... &&\zxrep{$\Gamma$} \ar[rr,blue] \ar[zxarrowr,rrdd,N] \ar[l,blue] && \zxN{} \ar[u,C-,blue]\\
        && && && && && \\
        && && && && && \\
        &&& \zxZ{}\ar[dd] \ar[uuuuu,3d above,very thick] \ar[ru,(]   && \zxZ{} \ar[dd] \ar[uuuuu,3d above,very thick] \ar[ru,(]&& \zxN{} && \zxZ{}\ar[dd] \ar[uuuuu,3d above,very thick]\ar[ru,(] &\\
        && && && && && \\
        && && && && && \\
    \end{ZX} = 
    \begin{ZX}
        && && && && && \\
        &\zxrep{$\Gamma$} \ar[u,C,blue] \ar[u,C-,blue]& && && && && \\
        \zxN{} &&\zxX{} \ar[lu,)] \ar[rr] \ar[zxarrowr,rrdd,N] && \zxX{}\ar[r] \ar[zxarrowr,rrdd,N] && ... &&\zxN{}  \ar[rrdd,N] \ar[l] && \\
        && && && && && \\
        && && && && && \\
        &&& \zxZ{}\ar[dd] \ar[uuuuu,3d above,very thick] \ar[ru,(]   && \zxZ{} \ar[dd] \ar[uuuuu,3d above,very thick] \ar[ru,(]&& \zxN{} && \zxZ{}\ar[dd] \ar[uuuuu,3d above,very thick]\ar[ru,(] &\\
        && && && && && \\
        && && && && && \\
    \end{ZX}
    \stackrel{\eqref{eq:assoc}}{=}\begin{ZX}
        && && && && && \\
        && && && && &&\zxZ{\chi_\Gamma} \ar[ld,)] \\
        &&\zxN{}  \ar[rr] \ar[ld,s] && \zxX{}\ar[r] \ar[ld,)] && ... &&\zxX{} \ar[r] \ar[ld,)] \ar[l] &&  \\
        & \zxZ{}\ar[dd] \ar[uuu,3d above,very thick]  && \zxZ{} \ar[dd] \ar[uuu,3d above,very thick]&& \zxN{} && \zxZ{}\ar[dd] \ar[uuu,3d above,very thick] && \zxN{} &\\
        && && && && && \\
        && && && && && \\
    \end{ZX}= \left( \prod \ract{Z} \right)^{\chi_\Gamma}
    \,.
\end{equation}
Here, $\chi_\Gamma=\mathrm{tr}(\Gamma)$ is the character of the representation $\Gamma$ \cite{larson1971characters,cohen2014character}.

We can understand this symmetry operator as a generalization of the group case. In the group algebra case, the group elements get ``copied", multiplied together and the character is computed. In the general Hopf algebra, the copy operation is replaced with comultiplication. Also note that in general, the comultiplication can either act from the right or the left. This is analogous to how a non-abelian group symmetry can be defined with either a left or a right action. Our convention here uses the right comultiplication, hence the arrow acting from the right in $\Br^{\Gamma}$.

\medskip

It is also instructive to define the symmetry MPO corresponding to the regular representation of $H$, whose character is nothing but $|H|\cohaar$, 
\tikzsetfigurename{rephmporeg11}
\begin{equation}
    \Br^{\text{Reg}}= |H|\begin{ZX}
        && && && && && \\
        && && && && &&\zxXTri'{} \ar[ld,)] \\
        &&\zxN{}  \ar[rr] \ar[ld,s] && \zxX{}\ar[r] \ar[ld,)] && ... &&\zxX{} \ar[r] \ar[ld,)] \ar[l] &&  \\
        & \zxZ{}\ar[dd] \ar[uuu,3d above,very thick]  && \zxZ{} \ar[dd] \ar[uuu,3d above,very thick]&& \zxN{} && \zxZ{}\ar[dd] \ar[uuu,3d above,very thick] && \zxN{} &\\
        && && && && && \\
        && && && && && \\
    \end{ZX} 
\end{equation}
since \tikzsetfigurename{char-sum}
\begin{equation}\label{eq:chardecompose}
\sum_{\Gamma \in \Irr(H)} d_\Gamma \begin{ZX}
    \zxZ{\chi_\Gamma} \ar[d] \\
\end{ZX} = |H| \begin{ZX}
    \zxXTri'{} \ar[d]&\\ &\end{ZX} =\begin{ZX}
    &\\
\zxX{} \ar[u,C] \ar[u,C-] \ar[rd,(]&\\ &
    \end{ZX}.
\end{equation}
Note that when defined on a plaquette in a 2D lattice, $\Br^{\text{Reg}} $ is exactly the plaquette term in the Hopf quantum double model~\cite{buerschaper2013qdalgebra}.

Since $\chi_\Gamma$ is a character, it satisfies,
\tikzsetfigurename{cocom}
\begin{equation}\label{eq:cocom-cond}
\tag{cocom}
    \begin{ZX}
        &\zxZ{\chi_\Gamma}&& \\
        &\zxX{} \ar[u] \ar[d,C] &&\\
        &\zxN{} \ar[rd,)]\ar[u,C-,3d above,very thick]\ar[ld,(,3d above,very thick] &&\\
        &&&
    \end{ZX} = \begin{ZX}
        &\zxZ{\chi_\Gamma}&& \\
        &\zxX{} \ar[u] \ar[ld,(]  \ar[rd,)]&&\\
        &&&
    \end{ZX}, \quad \chi_\Gamma(ba)=\chi_\Gamma(ab),\  \forall a,b\in H
 \end{equation}
 meaning that $\chi_\Gamma \in \text{Cocom}(\dl{H})$. 
In fact, for finite dimensional semisimple Hopf algebra, $\chi_\Gamma$, with ${\Gamma \in \Irr(H)}$, form a complete basis for $\text{Cocom}(\dl{H})$ (see related discussion in footnote \ref{footnote:characters-cocom}). Consequently, the symmetry MPO commutes with translation symmetry on a periodic chain: 
\tikzsetfigurename{rephmpotranslation}
\begin{equation}\label{eq:translation-inv}
    T\Br^{\Gamma}T^{-1}=\begin{ZX}
        && && && && && \\
        &\zxN{} \ar[u] \ar[rrrrrrd,N]& &\zxN{} \ar[u]& &\zxN{} \ar[u]& && && \\
        &\zxN{} \ar[rru,N,3d above,very thick]& &\zxN{} \ar[rru,N,3d above,very thick]& && &\zxN{} & &&\zxZ{\chi_\Gamma} \ar[ld,)] \\
        &&\zxN{}  \ar[rr] \ar[ld,s] && \zxX{}\ar[r] \ar[ld,)] && ... &&\zxX{} \ar[r] \ar[ld,)] \ar[l] &&  \\
        & \zxZ{}\ar[d] \ar[uu,3d above,very thick]  && \zxZ{} \ar[d] \ar[uu,3d above,very thick]&& \zxN{} && \zxZ{}\ar[d] \ar[uu,3d above,very thick] && \zxN{} &\\
        &\zxN{} \ar[rrd,N]& &\zxN{} \ar[rrd,N]& && &\zxN{} & &&\\
        &\zxN{} \ar[d] \ar[rrrrrru,N,3d above,very thick]& &\zxN{} \ar[d]& &\zxN{} \ar[d]& && && \\
        && && && && && \\
    \end{ZX} = 
    \begin{ZX}
        && && && && && &&\\
        &&\zxN{}  \ar[rrrrrr] \ar[ldd,s]  && && && \zxN{}\ar[rrdd,s] && &&\zxZ{\chi_\Gamma} \ar[ld,)]\\
        &&&& \zxN{}  \ar[rr] \ar[ld,s] && ... &&\zxX{} \ar[rrr,3d above,very thick] \ar[ld,)] \ar[l] &&  &\zxX{} \ar[ld,)]&\\
        & \zxZ{}\ar[dd] \ar[uuu,3d above,very thick]  && \zxZ{} \ar[dd] \ar[uuu,3d above,very thick]&& \zxN{} && \zxZ{}\ar[dd] \ar[uuu,3d above,very thick] && &\zxN{}  &&\\
        && && && && && &&\\
        && && && && && &&\\
    \end{ZX} \stackrel{\text{\eqref{eq:cocom-cond}}}{=}\begin{ZX}
        && && && && && \\
        && && && && &&\zxZ{\chi_\Gamma} \ar[ld,)] \\
        &&\zxN{}  \ar[rr] \ar[ld,s] && \zxX{}\ar[r] \ar[ld,)] && ... &&\zxX{} \ar[r] \ar[ld,)] \ar[l] &&  \\
        & \zxZ{}\ar[dd] \ar[uuu,3d above,very thick]  && \zxZ{} \ar[dd] \ar[uuu,3d above,very thick]&& \zxN{} && \zxZ{}\ar[dd] \ar[uuu,3d above,very thick] && \zxN{} &\\
        && && && && && \\
        && && && && &&
    \end{ZX} = \Br^{\Gamma}
\end{equation}

One can compute the multiplication of the $\Rep(H)$ symmetry MPO as follows,
\tikzsetfigurename{repmulti}
\begin{align}
\Br^{\Gamma_i} \Br^{\Gamma_j} &= 
\begin{ZX}
    && && && && && \\
    && && && && &&\zxZ{\chi_{\Gamma_i}} \ar[ld,)] \\
    &&\zxN{}  \ar[rr] \ar[ld,s] && \zxX{}\ar[r] \ar[ld,)] && ... &&\zxX{} \ar[r] \ar[ld,)] \ar[l] &&  \\
    & \zxZ{}\ar[dd] \ar[uuu,3d above,very thick]  && \zxZ{} \ar[dd] \ar[uuu,3d above,very thick]&& \zxN{} && \zxZ{}\ar[dd] \ar[uuu,3d above,very thick] && \zxN{} &\\
    && && && && && \\
    && && && && &&\zxZ{\chi_{\Gamma_j}} \ar[ld,)] \\
    &&\zxN{}  \ar[rr] \ar[ld,s] && \zxX{}\ar[r] \ar[ld,)] && ... &&\zxX{} \ar[r] \ar[ld,)] \ar[l] &&  \\
    & \zxZ{}\ar[dd] \ar[uuu,3d above,very thick]  && \zxZ{} \ar[dd] \ar[uuu,3d above,very thick]&& \zxN{} && \zxZ{}\ar[dd] \ar[uuu,3d above,very thick] && \zxN{} &\\
    && && && && && \\
    && && && && &&
\end{ZX}\stackrel{\text{\eqref{eq:assoc}},\text{\eqref{eq:hopfbialgebra}}}{=}\begin{ZX}
&& && && && &&& \\
&& && && && &&\zxZ{\chi_{\Gamma_i}} \ar[ld,)]&\zxZ{\chi_{\Gamma_j}} \ar[lld,)] \\
&&\zxN{}  \ar[rr] \ar[ld,s] && \zxX{}\ar[r] \ar[ld,)] && ... &&\zxX{} \ar[r] \ar[ld,)] \ar[l] &\zxZ{}&&  \\
& \zxZ{}\ar[dd] \ar[uuu,3d above,very thick]  && \zxZ{} \ar[dd] \ar[uuu,3d above,very thick]&& \zxN{} && \zxZ{}\ar[dd] \ar[uuu,3d above,very thick] && \zxN{} &&\\
&& && && && &&& \\
&& && && && &&&
\end{ZX} \nonumber \\
&= \sum_{k} N_{\Gamma_i,\Gamma_j}^{\Gamma_k} \begin{ZX}
&& && && && && \\
&& && && && &&\zxZ{\chi_{\Gamma_k}} \ar[ld,)] \\
&&\zxN{}  \ar[rr] \ar[ld,s] && \zxX{}\ar[r] \ar[ld,)] && ... &&\zxX{} \ar[r] \ar[ld,)] \ar[l] &&  \\
& \zxZ{}\ar[dd] \ar[uuu,3d above,very thick]  && \zxZ{} \ar[dd] \ar[uuu,3d above,very thick]&& \zxN{} && \zxZ{}\ar[dd] \ar[uuu,3d above,very thick] && \zxN{} &\\
&& && && && && \\
&& && && && &&
\end{ZX}  =\sum_{k} N_{\Gamma_i,\Gamma_j}^{\Gamma_k} \Br^{\Gamma_k}
\label{eq:Brep-mult}
\end{align}
where $N_{\Gamma_i,\Gamma_j}^{\Gamma_k}$ is the structure tensor for the fusion of irreducible representations of the Hopf algebra $H$, i.e.,
\tikzsetfigurename{chi-mult}
\begin{equation}\label{eq:chi_mult}
\begin{ZX}
    \zxZ{\chi_{\Gamma_i}} && \zxZ{\chi_{\Gamma_j}} \\
    & \zxZ{} \ar[lu,)]  \ar[ru,(] \ar[dd]& \\
    &&\\ &&
\end{ZX} = \begin{ZX}
\zxZ{\chi_{\Gamma_i\otimes \Gamma_j}} \ar[dd] \\ \\ 
\end{ZX} = \sum_{k} N_{\Gamma_i,\Gamma_j}^{\Gamma_k} \begin{ZX}
\zxZ{\chi_{\Gamma_k}} \ar[dd] \\ \\
\end{ZX} \ .
\end{equation}
In the above equations, $\sum_k$ denotes a sum over all the irreps of $H$. 

For the case of the group algebra $\CC[G]$, the comultiplication $\comult(g)=g\otimes g$, so that the multiplication of characters factorizes $\chi_V(g) \chi_W(g) = \chi_{V\otimes W}(g)$, which is no longer the case for a general Hopf algebra.

Let us now discuss some local operators that commute with the $\Rep(H)$ symmetry operators $\Br^\Ga$. We find the operators,
\tikzsetfigurename{hopfXXZ}
\begin{equation} \label{eq:RepH-symmetric-ops}
    \lact{Z}^{\alpha}= \tikzsetnextfilename{Zterm} \begin{ZX}
        \zxZ{\alpha S}&&&\\
        &\zxZ{} \ar[u] \ar[d] \ar[lu,)]&&\\
        &&&
    \end{ZX},\qquad (\ract{X} \lact{X})^{a}_{\op} =  \includegraphicsr{}{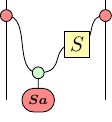}
\end{equation}
with $\al\in{\dl H}$ and $a\in H$, commute with $\Br^\Ga$.
Using these operators, one can define gapped Hamiltonians realizing both a $\Rep(H)$ symmetric gapped phase (paramagnet) as well as one with the $\Rep(H)$ symmetry spontaneously broken (ferromagnet). A fixed point Hamiltonian for the $\Rep(H)$ paramagnetic phase for a periodic chain of $L$ qudits is given by
\tikzsetfigurename{PauliZham}
\begin{equation}\label{eq:RepH-Sym-Ham}
    \sH_{\Sym} = -\frac{1}{|H|}\sum_{j=1}^L\sum_{\Gamma \in \Irr(H)} d_\Gamma \Zl^{\chi_\Gamma }_j = -\sum_{j=1}^L Z^{\phi}_j \stackrel{\eqref{eq:singlesiteprojector}}{=} -\sum_{j=1}^L \begin{ZX}
        &\\
        \zxX{} \ar[u]&\\
        \zxXTri'{} \ar[d]& \\
        &\\[5 pt]
        \zxbase{j}&\\
        &
    \end{ZX},
\end{equation}
whose ground state is the product state $\ket{\unit}^{\otimes L}$
\tikzsetfigurename{RepHsymGS}
\begin{equation}\label{eq:repHsymGS}
    \begin{ZX}
        &&&&\\
        \zxX{} \ar[u]&\zxX{} \ar[u]&...&\zxX{} \ar[u]&
    \end{ZX} \ .
\end{equation}
On the other hand, the fixed point Hamiltonian for the $\Rep(H)$ ferromagnetic phase described by
\begin{equation}\label{eq:repHSSBham00}
    \sH_{\SSB} = -\frac{1}{|H|}\sum_{j=1}^L \sum_{\drep \in \Irr(\dl{H})} d_\drep (\Xr_j \Xl_{j+1})^{\dchi_\drep}_\op = -\sum_{j=1}^L (\Xr_j \Xl_{j+1})_\op^{\wc\haar} =-\frac{1}{\sqrt{|H|}}\sum_{j=1}^L \includegraphicsr{}{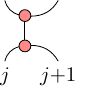}
\end{equation}
Its ground state is given by a generalized cat state,
\tikzsetfigurename{RepHGHZ1}
\begin{equation}\label{eq:genGHZ}
    \ket{\GHX} = \begin{ZX}
        &&&&&\\
        \zxN{} \ar[d,C]&\zxX{} \ar[l] \ar[u] \ar[r]&\zxX{}  \ar[u] \ar[r]&...&\zxX{} \ar[l] \ar[u] \ar[r]&\zxN{} \ar[d,C-]\\
        \zxN{} \ar[r]&&&...&&\zxN{} \ar[ll]
    \end{ZX} = |H|\begin{ZX}
    && && &&& && && \\
    && && &&& && &&\zxXTri'{} \ar[ld,)] \\
    \zxN{} \ar[u] &&\zxN{}  \ar[rr] \ar[ld,s] \ar[u]&& \zxX{}\ar[r] \ar[ld,)] &...&& \zxN{}\ar[u] &&\zxX{} \ar[r] \ar[ld,)] \ar[lll] &&  \\
    & \zxX{} \ar[lu,),3d above,very thick]  && \zxX{} \ar[lu,),3d above,very thick]&&& \zxN{} && \zxX{}\ar[lu,),3d above,very thick] && \zxN{} &\\
    && && &&& && && \\
    && && &&& && && \\
    \end{ZX} \ .
\end{equation}
Following \cite{Bombin23}, we will call this state the \emph{Hopf GHX state} (the ferromagnet with dual $\Rep(\dl{H})$ symmetry in Eq.~\eqref{eq:GHZ} will be called the Hopf GHZ state).
% \footnote{We will call the ferromagnetic ground state of $\Rep(H)$ GHX, in contrast to the GHZ of $\Rep(\dl H)$.}

The entanglement entropy of the Hopf GHX state is $\log (|H|)$, while the ground state degeneracy (with periodic boundary conditions) is given by the number of irreducible representations in the regular representation of $H$. For example, the Hopf GHX state for the fully spontaneously symmetry breaking phase of the following symmetries is given by
\begin{equation}
    \begin{array}{c|c|c|c|c}
        \text{Symmetry} & \Rep(\ZZ_2) \cong \VEC_{\ZZ_2} & \Rep(S_3) & \Rep(D_8) & \Rep(H_8) \cong \Rep(H_8^*) \\ \hline 
        \text{Algebra} & \CC[\ZZ_2] \cong \CC^{\ZZ_2} & \CC[S_3] & \CC[D_8] & \CC[H_8] \cong \CC[H_8^*] \\ \hline 
        \text{Entanglement Entropy}& \log(2) & \log(6) & \log(8) & \log(8)\\ \hline
        \text{Ground state deg.}& 2 & 3 & 5 & 5\\ 
    \end{array}
\end{equation}

Since \tikzsetfigurename{char-sum0} $\sum_{\Gamma \in \Irr(H)} d_\Gamma \begin{ZX}
    \zxZ{\chi_\Gamma} \ar[d] \\
\end{ZX} = |H| \begin{ZX}
    \zxXTri'{} \ar[d]\\
\end{ZX}$, the Hopf GHX state can be decomposed into the basis states labeled by irreducible representations of $H$, 
\begin{align}
    \ket{\GHX} &= \sum_{\Gamma \in \Irr(H)} d_\Gamma  \ket{\chi_{\Gamma}}\\
\tikzsetfigurename{ssb-short-range}
    \ket{\chi_{\Gamma}} &= \begin{ZX}
        && &&& &&& && && \\
        && &&& &&& && &\zxN{}\ar[d,C-] \ar[lllllllllll]& \\
        \zxN{} \ar[uu,3d above,very thick] \ar[u,C] &&\zxrep{$\Gamma$}  \ar[rrr] \ar[ld,)] \ar[ll] &\zxN{} \ar[uu,3d above,very thick]&& \zxrep{$\Gamma$}\ar[r] \ar[ld,)] &...&& \zxN{}\ar[uu,3d above,very thick] &&\zxrep{$\Gamma$}\ar[r] \ar[ld,)] \ar[lll] &&  \\
        & \zxX{} \ar[lu,),3d above,very thick]  &&& \zxX{} \ar[lu,),3d above,very thick]&&& \zxN{} && \zxX{}\ar[lu,),3d above,very thick] && \zxN{} &\\
        && &&& &&& && && \\
        && &&& &&& && && \\
    \end{ZX} = \begin{ZX}
        && && &&& && && \\
        && && &&& && &&\zxZ{\chi_\Gamma} \ar[ld,)] \\
        \zxN{} \ar[u] &&\zxN{}  \ar[rr] \ar[ld,s] \ar[u]&& \zxX{}\ar[r] \ar[ld,)] &...&& \zxN{}\ar[u] &&\zxX{} \ar[r] \ar[ld,)] \ar[lll] &&  \\
        & \zxX{} \ar[lu,),3d above,very thick]  && \zxX{} \ar[lu,),3d above,very thick]&&& \zxN{} && \zxX{}\ar[lu,),3d above,very thick] && \zxN{} &\\
        && && &&& && && \\
        && && &&& && && \\
    \end{ZX}.
\end{align}
Note that each $\ket{\chi_\Ga}$ corresponds to a short-range entangled state. This is because the representation is irreducible, and so the tensor $\Gamma$ corresponds to an injective MPS. Moreover, if $\Ga$ is 1-dimensional, then $\ket{\chi_\Ga}$ is in fact a product state. The overlap between $\ket{\chi_{\Gamma_i}}$ and $\ket{\chi_{\Gamma_j}}$ is given by $\sum_{(\haar)}\chi_{\Gamma_i}(\haar_{(1)}) \chi_{\Gamma_j}(\haar_{(2)})$ and vanishes when $i\neq j$ \cite{larson1971characters}.
From the diagram, it is clear that the virtual bond dimension of the different SSB states $\ket{\chi_\Gamma}$ depends on the dimension of irreducible representation $\Gamma$. We remark that the states $\ket{\chi_\Gamma}$ for general (weak) Hopf algebra is also considered in \cite{jia2024generalized}, and it reduces to the group case in \cite{Albert:2021vts,nat2025repG,chung2025noninv}.
% In particular, if $\Gamma$ is 1-dimension irreducible representation, then the state is a product state.

The $\Rep(H)$ symmetry MPO acting on the states $\ket{\chi_{\Gamma}}$ gives
\begin{equation}
    \Br^{\Gamma_i} \ket{\chi_{\Gamma_j}} = \sum_{k} N_{\Gamma_i,\Gamma_j}^{\Gamma_k} \ket{\chi_{\Gamma_k}}.
\end{equation}
As discussed around \eqref{eq:ZGaXa}, the central elements of $H$ are represented by a scalar multiple of the identity matrix $\mathbbm 1_{d_\Ga}$ in any irrep $\Ga$, and this can be used to build a local order parameter for the $\Rep(H)$ SSB phases. Indeed, we can define such a local operator,
\begin{equation}\label{eq:RepHorderparameter}
    \Xl_j^{[a]} \equiv \sum_{(\wc \haar)}  \Xl_j^{{\wc \haar}_{(1)} a S({\wc \haar}_{(2)})}
\end{equation}
with $a\in H$, and $[a]$ denoting its class sum, which is defined as~\footnote{Since the multiplication is associative, we can write the red node with 3 input legs without ambiguity.}
\tikzsetfigurename{hopfconj} 
\begin{equation} 
\label{eq:cclass}
    [a] = \sum_{(\wc \haar)} {\wc \haar}_{(1)} a S({\wc \haar}_{(2)})  =  \begin{ZX}
        &&\\
        &\zxX{} \ar[u] \ar[d]&\\
        \zxN{}\ar[ru,)]&\zxX{a}&\zxrep{$S$}\ar[lu,(]\\
        &\zxZ{} \ar[lu,)] \ar[ru,(] \ar[d]&\\
        &\zxZTri.{}&
    \end{ZX}
\end{equation}
The class sum $[a]$ defines a central element in the Hopf algebra because it is invariant under the adjoint action of any other elements up to a counit.\footnote{The adjoint action is defined as ${\rm ad}_x(\cdot) = \sum_{(x)}x_{(1)} (\cdot) S(x_{(2)})$, and ${\rm ad}_x([a]) =\wh\counit(x) [a]$.}
Therefore, its representation is proportional to the identity matrix. In particular, ${\Gamma([a]) =\frac{\chi_\Gamma(a)}{d_\Gamma} \mathbbm{1}_{d_\Gamma}}$.
Hence, the operator $\Xl^{[a]}_j$ defined above acts on the states $\ket{\chi_{\Ga}}$ as
\begin{equation}\label{eq:localorderparameteraction}
    \Xl^{[a]}_j \ket{\chi_\Gamma} =\frac{\chi_\Gamma(a)}{d_\Gamma}\ket{\chi_\Gamma}
\end{equation}

Notably, because of the sum in the definition of $[a]$ \eqref{eq:cclass}, $\Xl_j^{[a]}$ is in general a non-invertible matrix. It is also interesting to note that the order parameter for the non-invertible symmetry could be local invertible operator when $a$ is a group-like central element as discussed around \eqref{eq:ZGaXacentral}. We will use such a local order parameter to determine the phase diagram of the $H_8$ Ising model in \cref{sec:H8-Ising}.  

Let us perform a sanity check with the case of the group algebra $H=\CC[G]$ which gives a $\Rep(G)$ symmetry. The conjugacy class $[a]$ reduces to the usual class sum of groups,
\begin{equation}
    [a] = \frac{1}{|G|}\sum_{g\in G} gag^{-1}
\end{equation}
where the normalization factor $\frac{1}{|G|}$ is due to the Haar integral of the group algebra. Then the representation $\Gamma$ of $[a]$ is
\begin{equation}
    \Gamma([a]) = \frac{1}{|G|}\sum_{g\in G} \Gamma(g)\Gamma(a)\Gamma(g^{-1}) = \frac{\chi_\Gamma(a)}{d_\Gamma} \mathbbm{1}_\Gamma
\end{equation}
where we used a corollary of Schur's lemma in the second equality.
Since $[a]$ is in general a summation of multiple elements, the operator $\Xl^{[a]}_j$ is non-invertible.

\subsection{\texorpdfstring{$\Rep (\dl{H})$}{Rep(H*)} symmetry and SSB}
\label{sec:RepHdual-symm}

In the previous subsection, we discussed an MPO representation of $\Rep(H)$ symmetry on a periodic chain of Hopf qudits. Similarly, we can define the $\Rep(\dl{H})$ symmetry MPO for an irrep $\drep \in \Irr(\dl H)$, whose character corresponds to a cocommutative element $\dchi_\drep\in \Cocom(H)$. Note that the local Hilbert space here is still $H$. This is distinct from using a local Hilbert space of $H^*$, which will instead give the MPO in Eq.~\eqref{eq:rephmpo} by inputting the dual Hopf algebra. The MPO is built by comultiplying $\dchi_\drep$ followed by left multiplication into the qudit on each site:
\tikzsetfigurename{rephdualmpo}
\begin{equation}\label{eq:rephdualmpo} \Al^{\drep}=
    \begin{ZX}
        &&&&&&&&&&\\
        &&\zxX{} \ar[u] && \zxX{} \ar[u] &&&& \zxX{} \ar[u]&&\\
        & \zxZ{} \ar[ru,)]\ar[rr]&& \zxZ{}\ar[rr] \ar[ru,)]&& ... &&\zxN{} \ar[l]\ar[ru,s]&&&\\
        \zxX{\dchi_\drep} \ar[ru,)]&& \ar[uu,3d above,very thick]&&\ar[uu,3d above,very thick]&&&&\ar[uu,3d above,very thick]&&\\
        &&&&&&&&&&\\
    \end{ZX} = \left(\prod \lact{X}\right)^{\dchi_\drep}
\end{equation}
Since $\dchi_\drep\in \text{Cocom}(H)$, $\Al^\drep$ commutes with translation symmetry on a closed chain, similarly to \eqref{eq:translation-inv}. 

The multiplication of the $\Rep(\dl{H})$ symmetry MPO is
\begin{equation}\label{eq:RepHd-fusion}
    \Al^{\drep_i}  \Al^{\drep_j} = \Al^{{\drep_i\otimes \drep_j}}  =\sum_{k}N_{\drep_i \drep_j}^{\drep_k} \Al^{\drep_k}
\end{equation}
where $N_{\drep_i \drep_j}^{\drep_k}$ is the structure constant for the fusion of irreducible representations of $\dl{H}$, $\drep_i \in \Irr(\dl{H})$, which is analogous to \eqref{eq:chi_mult}. Similarly, we have
\tikzsetfigurename{cocomH-sum}
\begin{equation}
    \begin{ZX}
        &&\\ &&\\
        &\zxX{} \ar[uu] \ar[ld,(] \ar[rd,)]&\\
        \zxX{\dchi_{\drep_i}}&&\zxX{\dchi_{\drep_j}} 
    \end{ZX} = \begin{ZX}
        \\ \zxX{\dchi_{\drep_i\otimes \drep_j}} \ar[u] 
    \end{ZX} = \sum_{k} N_{\drep_i \drep_j}^{\drep_k}\begin{ZX}
        \\ \zxX{\dchi_{\drep_{k}}} \ar[u] 
    \end{ZX},\quad\sum_{\drep \in \mathrm{Irr}(\dl{H})} d_\drep \begin{ZX}
        \\ \zxX{\dchi_\drep} \ar[u]
    \end{ZX} = |\dl{H}| \begin{ZX}
    \\ \zxZTri.{} \ar[u]
    \end{ZX} \ .
\end{equation}
The symmetry MPO corresponding to the regular representation of $\dl{H}$ is given by,
\tikzsetfigurename{RepHd-reg-MPO}
\begin{equation}
    |\dl{H}|\Al^{\wc\haar}= \sum_{\drep \in\Irr(\dl{H})} d_\drep \Al^\drep  =  |\dl{H}| \begin{ZX}
        &&&&&&&&&&\\
        &&\zxX{} \ar[u] && \zxX{} \ar[u] &&&& \zxX{} \ar[u]&&\\
        & \zxZ{} \ar[ru,)]\ar[rr]&& \zxZ{}\ar[rr] \ar[ru,)]&& ... &&\zxN{} \ar[l]\ar[ru,s]&&&\\
        \zxZTri.{}\ar[ru,)]&& \ar[uu,3d above,very thick]&&\ar[uu,3d above,very thick]&&&&\ar[uu,3d above,very thick]&&\\
        &&&&&&&&&&\\
    \end{ZX} = \left(\prod \lact{X}\right)^{\wc\haar}
\end{equation}
When defined on a dual plaquette of a 2D lattice, this term precisely reproduces the vertex term of the Hopf quantum double model.

The simplest $\Rep(\dl{H})$-invariant local operators are
\tikzsetfigurename{RepHstarinv}
\begin{equation}\label{eq:RepHstarinv}
    \ract{X}^{a}= \tikzsetnextfilename{Xterm}\begin{ZX}
        &&&\\
        \zxX{} \ar[u] \ar[d] \ar[rd,)]&&&\\
        &\zxX{Sa}&&
    \end{ZX},\quad (\ract{Z} \lact{Z})^{\alpha}_\op= \tikzsetnextfilename{ZZterm}
    \begin{ZX}
        &&\zxZ{\alpha}&\\
        &&\zxX{}\ar[rdd,s]\ar[ld,s] \ar[u]&\\
        &\zxrep{$S$}\ar[ld,s]&&\\
        \zxZ{}\ar[uuu]\ar[d]&&&\zxZ{}\ar[uuu]\ar[d]\\
        &&&
    \end{ZX} \ .
\end{equation}
Similar to the previous construction, we can use these operators to write down fixed point Hamiltonians for two gapped phases, one that $\Rep(\dl H)$ symmetric and another that spontaneously breaks the entire symmetry. 
The $\Rep(\dl{H})$ symmetric phase is realized by the gapped Hamiltonian,
\tikzsetfigurename{PauliXham}
\begin{equation}\label{eq:PauliXham}
    \sH'_{\Sym} = -\frac{1}{|H|}\sum_{j=1}^L \sum_{\drep \in \Irr(\dl{H})} d_\drep \Xr^{\dchi_\drep}_j = -\sum_{j=1}^L X^{\wc\haar}_j \stackrel{\eqref{eq:singlesiteprojector}}{=} -\sum_{j=1}^L \begin{ZX}
        &\\
        \zxZTri.{} \ar[u]&\\
        \zxZ{} \ar[d]& \\
        &\\[5 pt]
        \zxbase{j}&\\
        &
    \end{ZX} \ .
\end{equation}
The ground state of $\sH'_{\Sym}$ is the product state of Haar integrals \tikzsetnextfilename{Xproductstate2}\begin{ZX}
    &&&\\
    \zxZTri.{} \ar[u]&\zxZTri.{} \ar[u]& ... &\zxZTri.{} \ar[u]
\end{ZX}. 
On the other hand, for the SSB phase, we have
\tikzsetfigurename{RepHdSSBham}
\begin{equation}\label{eq:RepHdSSBham}
    \sH'_{\SSB} = -\frac{1}{|H|}\sum_{j=1}^L \sum_{\Gamma \in \Irr(H)} d_\Gamma (\ract{Z}_j \lact{Z}_{j+1})^{\chi_\Gamma} = -\sum_{j=1}^L (\ract{Z}_j \lact{Z}_{j+1})^{\cohaar} =-\frac{1}{\sqrt{|H|}}\sum_{j=1}^L \begin{ZX}
        &&&&\\
        &\zxZ{} \ar[dd] \ar[lu,)] \ar[ru,(]&&&\\
        &&&&\\
        &\zxZ{} \ar[ld,(] \ar[rd,)]&&&\\
        &&&&\\[5 pt]
        \zxbase{j}&&\zxbase{j+1}&&\\
    \end{ZX}
\end{equation}
and the ground state is given by the Hopf GHZ state
\tikzsetfigurename{repHstarZZ}
\begin{equation}\label{eq:GHZ}
    \ket{\GHZ} = \tikzsetnextfilename{ZZssbstate}
    \begin{ZX}
        &&&&&&&&&\\
        &\zxZ{} \ar[u] \ar[rr]&&\zxZ{} \ar[u] \ar[rr]&&\cdots&\zxZ{} \ar[l]\ar[u] \ar[ru,(]&&&\\    
        \zxZTri.{} \ar[ru,)]&&&&&&&&&\\
        &&&&&&&&
    \end{ZX} =\tikzsetnextfilename{ZZssbstate2}\begin{ZX}
        &&&&&&&&&&\\
        &&\zxX{} \ar[u] && \zxX{} \ar[u] &&&& \zxX{} \ar[u]&&\\
        & \zxZ{} \ar[ru,)]\ar[rr]&& \zxZ{}\ar[rr] \ar[ru,)]&& ... &&\zxN{} \ar[l]\ar[ru,s]&&&\\
        \zxZTri.{} \ar[ru,)]&& \zxX{} \ar[uu,3d above,very thick]&&\zxX{} \ar[uu,3d above,very thick]&&&&\zxX{} \ar[uu,3d above,very thick]&&\\
        &&&&&&&&&&\\
    \end{ZX}
\end{equation}
The last equation means that the Hopf GHZ state is the symmetrized state by the \tikzsetfigurename{haardecompose}$\Rep(\dl{H})$ symmetry MPOs. Since, $\sum_{\drep \in \Irr(\dl{H})} d_\drep \begin{ZX}
    \\ \zxX{\dchi_\drep} \ar[u]
\end{ZX} = |\dl{H}| \begin{ZX}
    \\ \zxZTri.{} \ar[u]
\end{ZX}$, the Hopf GHZ state of $\Rep(\dl{H})$ can be decomposed as 
\tikzsetfigurename{rephdualGHZ}
\begin{equation}\label{eq:RepHdualGHZ}
    \ket{\GHZ} = \sum_{\drep \in \mathrm{Irr}(\dl{H})} d_\drep \ket{\drep} =\sum_{\drep \in \Irr(\dl{H})} d_\drep  \begin{ZX}
        &&&&&&&&&\\
        &\zxZ{} \ar[u] \ar[rr]&&\zxZ{} \ar[u] \ar[rr]&&\cdots&\zxZ{} \ar[l]\ar[u] \ar[ru,(]&&&\\    
        \zxX{\dchi_\drep} \ar[ru,)]&&&&&&&&&\\
        &&&&&&&&
    \end{ZX} 
\end{equation}
One can define a local order parameter similarly as in \eqref{eq:RepHorderparameter}.

\section{Gauging and Hopf Kramers-Wannier duality}
\label{sec:KW}

In this section, we explain how to gauge the $\Rep(\dl{H})$ symmetry generated by $\Al^\drep$ in \eqref{eq:rephdualmpo} and show that the resulting theory acquires a $\Rep(H)$ symmetry, generated by $ \Br^\Ga $ as defined in \eqref{eq:rephmpo}.\footnote{One can alternatively start with $\Rep(H)$ and gauge it to get $\Rep(\dl{H})$ symmetry which is given by ``flipping'' our gauging map upside down. Here we choose to start with $\Rep(\dl{H})$ symmetry because it reduces to the $\prod X$ symmetry in the ordinary transverse field Ising model for ${H=\CC[\ZZ_2]}$.} 
Moreover, when the Hopf algebra $H$ is self-dual, i.e., isomorphic to its dual $\dl{H}$, one can construct a Hadamard operator that provides an explicit isomorphism between the two symmetries: it transforms $\Rep(H)$ back to $\Rep(\dl{H})$, thereby establishing a duality between the gauged theory and the original theory.

\subsection{Gauging the \texorpdfstring{$\Rep(\dl{H})$}{Rep(H*)} symmetry}
\label{sec:gaugingRepH*}

In the group case, the gauging map has been implemented as a tensor network operator previously in the literature~\cite{HaegemanGauging15,tantivasadakarn_long-range_2022,tantivasadakarn_hierarchy_2022,Lootens24,nat2025repG}. 
The use of ZX calculus in the case of $G=\ZZ_2$ can be found in \Rf{nat2024zxKW}. Going beyond groups, several recent works considered the gauging of non-anomalous categorical (non-invertible) symmetries~\cite{tantivasadakarn_hierarchy_2022,Pace:2024acq,Sahand2025gaugenon,bram2025gauging,blanik2025}. In particular, the Gauss law for $\Rep(G)$ was explicitly discussed in \Rf{Pace:2024acq}.
Here, we will show that the gauging of $\Rep(\dl H)$ symmetry proceeds analogously to the group case, except that the multiplication and comultiplication must be explicitly tracked.

To gauge the $\Rep(\dl H)$ symmetry generated by $\Al^{\drep}$, we introduce a Hopf qudit as a gauge field degree of freedom on each link.
The Hilbert space is thus enlarged from ${\cH_{\rm sites}=H^{\otimes L}}$ to ${\cH_{\rm enlarged} = H^{\otimes L} \otimes H^{\otimes L}}$. Each gauge field qudit is initialized in the unit state $\ket{\unit}$.

Next, we define the Gauss operators,
\tikzsetfigurename{Gaussdef}
\begin{equation}
    G_j^a 
    = ( \Xr_{j-\frac{1}{2}} \Xl_{j} \Xl_{j+\frac{1}{2}})^{a}
    = \tikzsetnextfilename{G_j}
    \begin{ZX}
        &&&&&\\
        \zxX[a=X1]{}\ar[rd,)]\ar[ddd]\ar[u]&&&&\zxX[a=X2]{}\ar[ddd]\ar[u]&\zxX[a=X3]{}\ar[ddd]\ar[u]\\
        &\zxrep{$S$}&&\zxZ[a=Z2, yshift =-4pt]{}\ar[to=X2,IO,), ls=1]\ar[to=X3,3d above,very thick, (]&&\\
        &&\zxZ[a=Z1,xshift=-5pt]{}\ar[lu,)]\ar[ru,(] \ar[d]&&&\\
        &&\zxX[xshift=-6.1pt]{a}&&&\\[5pt]
        \zxbase{j-\tfrac12}&&&&\zxbase{j}&\zxbase{j+\tfrac{1}{2}}\\
    \end{ZX}
    \,.
\end{equation}
The gauge-invariant physical subspace $\cH_{\rm physical}$ is defined by 
\ie 
\cH_{\rm physical} = \left \{\ket{\psi} \in \cH_{\rm enlarged} \, \big | \, G_j^a \ket{\psi } = \counith(a) \ket{\psi },\  \forall j\in\{1,2,\dots, L\}, \forall a\in H \right \} \,.
\fe 
Projection to $\cH_{\rm physical}$ is implemented by the (mutually commuting) projection operators $G_j^\haarc$ (${j=1,\dots,L}$).
To see that these projectors indeed do the job, we first note that for any ${\ket{\psi}\in \cH_{\rm physical}}$, we immediately have ${G_j^\haarc \ket{\psi} = \ket{\psi}}$ by setting ${a=\haarc}$. 
On the other hand, for any $\ket{\psi}$ satisfying ${G_j^\haarc \ket{\psi} = \ket{\psi}}$, we have
\ie 
G_j^a \ket{\psi} = G_j^a G_j^\haarc \ket{\psi} = \counith(a) G_j^\haarc \ket{\psi} = \counith(a) \ket{\psi}\,,
\fe 
and hence ${\ket{\psi}\in \cH_{\rm physical}}$. 
Here, we have used that for ${a,b\in H}$, the Gauss operators satisfy
\ie \label{eq:Gauss-mult}
G_j^a G_j^b = G_j^{ab}\,,\qquad G_j^{a\haarc} = \counith(a) G_j^{\haarc}\,.
\fe 

Let us show that the Gauss operators in fact serve as local symmetry actions of the original symmetry operators $\Al^{\drep}$. We compute
\begingroup
\allowdisplaybreaks
\tikzsetfigurename{KW-kernel-Al}
\begin{align}\label{eq:KW-symkernel}
    \Al^{r} &=
    \tikzsetnextfilename{Ax-1}
    \begin{ZX}
        &&& & &&& & &&&&\\
        &&\zxX{} \ar[u] & & & \zxX{} \ar[u] & & &&& \zxX{} \ar[u]&&\\
        & \zxZ{} \ar[ru,)]\ar[rrr]&&  & \zxZ{}\ar[rrr] \ar[ru,)]&& &  ... &&\zxN{} \ar[l]\ar[ru,s]&&&\\
        \zxX{\dchi_\drep} \ar[ru,)]&& \ar[uu,3d above,very thick]& & &\ar[uu,3d above,very thick]& & &&&\ar[uu,3d above,very thick]&&\\
        &&& \zxX{}\ar[uuuu,3d above,very thick] &&& \zxX{}\ar[uuuu,3d above,very thick] & &&&&\zxX{}\ar[uuuu,3d above,very thick] & \\
    \end{ZX} 
    \stackrel{\eqref{eq:Hopf-rule}}{=}
    \tikzsetnextfilename{Ax-2}
    \begin{ZX}
        &&&&&&&&&&&&&&&&& & &\\
        &&&&\zxX{}\ar[u]&&&&&\zxX{}\ar[u]&&&&&&&& &\zxX{}\ar[u] & &\\
        &&\zxX{} \ar[uu] &&\zxX{}\ar[u]\ar[d]& \zxrep{\tiny $S$}\ar[lu,(]&& \zxX{} \ar[uu] &&\zxX{}\ar[u]&\zxrep{\tiny $S$}\ar[lu,(] &&&&&& \zxX{} \ar[uu]& & \zxX{}\ar[u]&\zxrep{\tiny $S$}\ar[lu,(] & \\
        & \zxZ{} \ar[ru,)]\ar[rr]&&\zxZ{} \ar[ru,)]\ar[rr]&&\zxZ{}\ar[u]\ar[r] & \zxZ{}\ar[rr] \ar[ru,)]&&\zxZ{}\ar[ru,)]\ar[rr]&&\zxZ{}\ar[u]\ar[rr]&&& ... &&\zxZ{} \ar[l]\ar[ru,)] \ar[rr]&& \zxZ{}\ar[ru,)]\ar[r]& \zxN{}\ar[ru,(] & &\\
        \zxX{\dchi_\drep} \ar[ru,)]&& \ar[uu,3d above,very thick]&&\uar\dar&&&\ar[uu,3d above,very thick]&&\uar\dar&&&&&&&\ar[uu,3d above,very thick]& & & &\\
        &&&& \zxX{}\ar[uuu,3d above,very thick] &&&& &\zxX{}\ar[uuu,3d above,very thick]&&&&&&&& & \zxX{}\ar[uuu,3d above,very thick]& &\\
    \end{ZX} 
    \notag
    \\
    & \stackrel{\mathmakebox[\widthof{=}]{\eqref{eq:assoc}}}{=}
    \tikzsetnextfilename{Ax-3}
    \begin{ZX}
        && && && && && && && & && & && && &
        \\
        && && \zxX{}\ar[u] && && \zxX{} \ar[u] && \zxX{}\uar\dar && && & && & && && &
        \\
        && &&\zxN{}\dar\uar &\zxrep{\tiny $S$}\ar[lu,(]&  &\zxZ{}\ar[ru,)]\ar[rrru,(]& && \zxX{}\ar[u] && && & && & &\zxX{}\ar[uu]& &\zxX{}\ar[uu]& &
        \\
        && \zxX{} \ar[uuu] && \zxX{} && \zxZ{}\ar[lu,)]\ar[ur,(] &&  && \zxN{}\ar[dd]\ar[u] &\zxrep{\tiny $S$}\ar[lu,(]& &\zxZ{}\ar[ru,)]\ar[r,(]&  & ... && & \zxZ{}\ar[ru,)]\ar[rrru,(,3d above,very thick] && &\zxX{}\ar[u]& & 
        \\
        &\zxZ{}\ar[ru,)]\ar[rrru,(]& &&  && &&  &&  && \zxZ{}\ar[lu,)]\ar[ru,(] && & && \zxZ{}\ar[l,)]\ar[ru,(] & && && \zxrep{\tiny $S$}\ar[lu,(] & 
        \\
        &\zxZ{} \ar[u] \ar[rrrrr]& && && \zxZ{}\ar[rr] \ar[uu] && &\zxN{}\ar[l]\ar[rrr]& && \zxZ{}\ar[u]\ar[rrr] && &... && \zxZ{} \ar[l]\ar[u] \ar[rrrr] & && &\zxN{}\ar[ru,(]& &
        \\
        \zxX{\dchi_\drep} \ar[ru,)] && \ar[uuu,3d above,very thick] && \zxX{}\ar[uuu,3d above,very thick] && && \ar[uuuuu,3d above,very thick] && \zxX{}\ar[uu,3d above,very thick] && && & && & &\ar[uuuu,3d above, very thick]& &\zxX{}\ar[uuu,3d above,very thick]& &
    \end{ZX}
    \notag
    \\
    &=
    \tikzsetnextfilename{Ax-4}
    \begin{ZX}
        && & && && && && && && & && & && && &
        \\
        && & && \zxX{}\ar[u] && && \zxX{} \ar[u] && \zxX{}\uar\dar && && & && & && &\zxX{}\ar[u]& &
        \\
        && & \zxX{}\ar[uu] && \zxX{}\dar\uar &\zxrep{\tiny $S$}\ar[lu,(]&  &\zxZ{}\ar[ru,)]\ar[rrru,(]& && \zxX{}\ar[u] && && & && & &\zxX{}\ar[uu]& &\zxX{}\ar[u]& \zxrep{\tiny $S$}\ar[lu,(]\ar[rd,(] &
        \\
        && \zxZ{}\ar[ru,)]\ar[rrru,(] &   && && \zxZ{}\ar[lu,)]\ar[ur,(] &&  && \zxN{}\ar[dd]\ar[u] &\zxrep{\tiny $S$}\ar[lu,(]& &\zxZ{}\ar[ru,)]\ar[r,(]&  & ... && & \zxZ{}\ar[ru,)]\ar[rrru,(,3d above,very thick] && &&  & 
        \\
        &\zxZ{}\ar[ru,(]\ar[lu,)] & & &&  && &&  &&  && \zxZ{}\ar[lu,)]\ar[ru,(] && & && \zxZ{}\ar[l,)]\ar[ru,(] & && && & 
        \\
        &\zxZ{} \ar[u] \ar[rrrrr]& & && && \zxZ{}\ar[rr] \ar[uu] && &\zxN{}\ar[l]\ar[rrr]& && \zxZ{}\ar[u]\ar[rrr] && &... &\zxN{} \ar[ru,(]&  & && && &
        \\
        \zxX{\dchi_\drep} \ar[ru,)] && & \ar[uuuu,3d above,very thick] && \zxX{}\ar[uuuu,3d above,very thick] && && \ar[uuuuu,3d above,very thick] && \zxX{}\ar[uu,3d above,very thick] && && & && & &\ar[uuuu,3d above, very thick]& &\zxX{}\ar[uuuu]& &
    \end{ZX}
    = ( \prod_{j=1}^L G_{j} )^{\dchi_\drep}  
\end{align}
\endgroup
In the fourth equality above, we used the cocommutativity of $\dchi_\drep$ and the diagram should be read with periodic boundary conditions. 

Thus, we have shown that by introducing gauge fields initialized in the state $\ket{\unit}$, we can express the original symmetry action as a (sum of) product of local Gauss laws. In the group algebra case, the comultiplication has a single summand, and this reproduces the well-known result for the Gauss law of invertible symmetries.

It is now clear that in the gauge invariant subspace, where ${G_j^a \sim \counith(a) \mathbbm{1}}$, the symmetry acts as 
\begin{align}
    \Al^{r} =\sum_{(x_r)} G_1^{x_{r\,(1)}} \cdots G_L^{x_{r\,(L)}}  \sim \sum_{(x_r)} \wh{\counit}(x_{r\,(1)}) \cdots \wh{\counit}(x_{r\,(L)}) \mathbbm{1} = \wh{\counit}(x_r)\mathbbm{1} = d_r \mathbbm{1}
\end{align}
where $d_\drep$ is the dimension of the irrep $r$ of $\dl{H}$.
In the final step, we used the fact that in the Gauss-constrained physical Hilbert space, ${G_j^a = \counith(a) \mathbbm{1}}$ for all ${a\in H}$.

Among the $\Rep(\dl H)$-symmetric local operators in \cref{eq:RepHstarinv}, the ``$X$'' term acts within $\cH_{\rm physical}$. On the other hand, the ``$ZZ$'' term can be minimally coupled to the gauge field qudits so that it also commutes with the Gauss operators $G_j^\haarc$ and, hence, acts within $\cH_{\rm physical}$,
\tikzsetfigurename{ZZmincpl}
\begin{equation}
    (\Zr_j \Zl_{j+1})^{\alpha}_\op  \xrightarrow{\text{min cpl}} (\Zr_j \Zl_{j+\frac{1}{2}} \Zl_{j+1})^{\alpha}_\op
    = \tikzsetnextfilename{ZZZ}
    \begin{ZX}
        &&\zxZ{\alpha S}&&&\\
        &&\zxX[a=X1]{}\ar[ld,(]\ar[to=X2,)] \ar[u]&&&\\
        &\zxrep{$S$}&&\zxX[a=X2, yshift =4pt]{}\ar[to=Z2,IO,(, ls=1]\ar[to=Z3, )]&&\\
        \zxZ[a=Z1]{}\ar[ru,(]\ar[uuu]\ar[d]&&&&\zxZ[a=Z2]{}\ar[uuu,3d above,very thick]\ar[d]&\zxZ[a=Z3]{}\ar[uuu]\ar[d]\\
        &&&&&\\[5pt]
        \zxbase{j}&&&&\zxbase{j+\tfrac{1}{2}}&\zxbase{j+1}\\
    \end{ZX} \,.
\end{equation}
One can check that indeed,  ${[G_{j'}^{\haarc},(\Zr_j\Zl_{j+\frac12} \Zl_{j+1} )^{\alpha S}_{\op}] = 0}$ for any $j,j'$.
Furthermore, since the gauge qudit is initialized in the state $\ket{\unit}$, the action of the minimally coupled operator on the site qudits is identical to the original uncoupled operator.

To implement the projection from $\cH_{\rm enlarged}$ to $\cH_{\rm physical}$, it is convenient to perform a unitary transformation,
\tikzsetfigurename{Gausssimplify}
\begin{equation}
    U_{ZX}  =\prod_{j=1}^L \ZXgate_{j,j+\frac{1}{2}}^\dagger \prod_{j=1}^L \XZgate_{j-\frac{1}{2},j}
    = \cdots \tikzsetnextfilename{Ucond}
    \begin{ZX}
        \zxN{}\dar&&&&\zxN{}\dar&&\\
        &&\zxX[a=X1]{}\ar[dd]\ar[u]&&&&\zxX[a=X2]{}\uar\ar[dddd]\\
        &\zxrep{$S$}\ar[ld]\ar[ru]&&&&\zxrep{$S$}\ar[ld]\ar[ru]&\\
        \zxZ[a=Z1]{}\ar[uu]\ar[dd]&&\zxX[a=X3]{}\ar[dd]\ar[rrd]&&\zxZ[a=Z2]{}\ar[uu]&&\\[5pt]
        \zxN{}\uar&&&&\zxZ[a=Z3]{}\ar[u]\dar&&\\
        &&&&&&\\[5pt]
        \zxbase{j-1}&&\zxbase{j-\tfrac12}&&\zxbase{\phantom{1}j\phantom{1}}&&\zxbase{j+1}\\
    \end{ZX}
    \cdots 
\end{equation}
All gates in $U_{ZX}$ commute because of associativity and coassociativity. Under this unitary transformation, we find
\begin{align}
    G_j^\haarc &\xrightarrow{U_{ZX}} X_j^\haarc\\
    \sum_{(\alpha)}\Zr_{j}^{S\alpha_{(1)}} \Zl_{j+\frac{1}{2}}^{S\alpha_{(2)}} \Zl_{j+1}^{S\alpha_{(3)}}   &\xrightarrow{U_{ZX}}  \Zl_{j+\frac{1}{2}}^{S\alpha} \\
    \Xr_j^{a} &\xrightarrow{U_{ZX}} (\Xr_{j-\frac{1}{2}} \Xr_{j} \Xl_{j+\frac{1}{2}})^{a}_{\op}
\end{align}
That is, the Gauss law reduces to a single site projector.

Restricting to the subspace where the unitarily transformed Gauss constraints acts as identity, ${X_j^\haarc \sim \mathbbm{1}}$, we find
\begin{equation}
     (\Xr_{j-\frac{1}{2}} \Xr_{j} \Xl_{j+\frac{1}{2}})^{a}_{\op} \sim X_j^\haarc (\Xr_{j-\frac{1}{2}} \Xr_{j} \Xl_{j+\frac{1}{2}})^{a}_{\op} 
    = 
   X_j^\haarc (\Xr_{j-\frac{1}{2}}  \Xl_{j+\frac{1}{2}})^{a}_{\op}
    \sim (\Xr_{j-\frac{1}{2}}  \Xl_{j+\frac{1}{2}})^{a}_{\op}
    \ .
\end{equation}
Thus, upon imposing the Gauss constraints, the 3-site operator simplifies to a 2-site one. The original site qudits get decoupled, and we are only left with the gauge field qudits on links.
This sequence of steps, collectively referred to as ``the gauging map'', defines a linear map $\gaugemap$ from ${H^{\otimes L}}$ on the integer-sites (``vertices"), to the half-integer sites (``edges"). 
The gauging map is implemented by an MPO\footnote{We abuse notation slightly by using KW to denote both the map and the MPO that implements it},
\tikzsetfigurename{gaugingmap}
\begin{equation} \label{eq:KW-MPO}
    \gaugemap=
    \tikzsetnextfilename{gmap-fig1}
    \begin{ZX}
        &&&&&&&&&&&&&&\\
        &&&&\zxX{} \ar[ld,(] \ar[rd,)] \ar[u]&&&&\zxX{} \ar[ld,(] \ar[rd,)] \ar[u]&&&&\zxX{} \ar[ld,(] \ar[rd,)] \ar[u]&&\\
          \cdots&\zxN{}&&\zxrep{$S$} &&\zxN{}&&\zxrep{$S$}&&\zxN{}&&\zxrep{$S$}&&\zxN{}&\cdots\\
        &&\zxZ{} \ar[lu,)] \ar[ru,(] \ar[d] &&&&\zxZ{} \ar[lu,)] \ar[ru,(] \ar[d]&&&&\zxZ{} \ar[lu,)] \ar[ru,(] \ar[d]&&&\\
        &&&&&&&&&&&&&&
    \end{ZX} =
    \tikzsetnextfilename{gmap-fig2}
    \begin{ZX}
        &&&&&&&&&&&&&&\\
        &&&&&&&&&&&&&&\\
        &&\zxZ{} \ar[dd]&&\zxX{} \ar[ld]  \ar[uu] \ar[dd]&&\zxZ{} \ar[dd]&&\zxX{} \ar[ld]  \ar[uu] \ar[dd]&&\zxZ{} \ar[dd]&&\zxX{} \ar[ld]  \ar[uu] \ar[dd]&&\\
        &&&\zxrep{$S$} \ar[ld]& &&&\zxrep{$S$} \ar[ld]& &&&\zxrep{$S$} \ar[ld]& &&\\
        \cdots&&\zxZ{}&&\zxX{} \ar[rrd]&&\zxZ{}&&\zxX{} \ar[rrd]&&\zxZ{}&&\zxX{} \ar[rrd]&&\cdots\\
        && \zxZ{} \ar{llu}  \ar[dd] \ar[u] &&&&   \zxZ{}\ar[dd] \ar[u]&&&& \zxZ{} \ar[dd] \ar[u]&&&&\zxN{}\\
        &&&&\zxX{} \ar[uu]&&&&\zxX{} \ar[uu]&&&&\zxX{} \ar[uu]&&\\\
        &&&&&&&&&&&&&&
    \end{ZX} 
\ .
\end{equation}
In equations, ${\gaugemap = \bra{\counit}_V  U_{ZX} \ket{\unit}_E}$, where ${\ket{\unit}_E = \bigotimes_{j=1}^L \ket{\unit}_{j-\frac12}}$ and ${\bra{\counit}_V = \bigotimes_{j=1}^L \bra{\counit}_j}$. 
The state $ \ket{\unit}_E$ represents the initialization of the gauge qudits in a trivial state, the unitary $U_{ZX}$ implements a change of basis, and finally the contraction with $\bra{\counit}_V$ implements the Gauss constraints, projecting to the physical gauge-invariant subspace.

We remark that if we impose that the gauging map maps back to the original Hilbert space, then we can also realize it by a sequential circuit followed by a projection with a single Hopf qudit ancilla:
\tikzsetfigurename{KW-seq-circ}
\begin{align} \label{eq:KW-seq-circ}
   \gaugemap&= 
   \tikzsetnextfilename{kwseqcirc-fig1}
   \begin{ZX}
        &&&&&&&&&&&&&&\\
        &&\zxX{} \ar[ld] \ar[u] \ar[dd]&&\zxX{} \ar[u]\ar[dd] && \zxX{} \ar[u] \ar[dd]&&  \zxN{}\\
        \cdots&&&\zxrep{$S$} \ar[ld] \ar[ru]& &\zxrep{$S$} \ar[ld] \ar[ru]& &\zxrep{$S$} \ar[ld] \ar[ru]&\cdots &&\\
        &&\zxZ{}\ar[d]&&\zxZ{} \ar[d]&&\zxZ{} \ar[d]&&&&\\
        &&&&&&&&&&&&&&
\end{ZX}  = \tikzsetnextfilename{kwseqcirc-fig2}\begin{ZX}
        &&\zxZ{}   \ar[ddd]&&&&&&&&&&&&\\
        &&&&\zxX{} \ar[u]\ar[ddddd] && &&  \\
        &&&\zxrep{$S$} \ar[ld] \ar[ru]& && && &&\\
        &&\zxZ{}\ar[ddd]&&&&&&&&\\
        &&&&&&\zxX{} \ar[uuuu] \ar[ddddd]&&&&&&&&\\
        &&&&&\zxrep{$S$} \ar[ld] \ar[ru]&&&&&&&&&\\
        &&\zxN{}\ar[ld,)]&&\zxZ{} \ar[ddddddd]&&&&&&&&\\
        &&&&&&&&\zxX{} \ar[uuuuuuu] \ar[ddddd]&&&\\
        &&&&&&&\zxrep{$S$} \ar[ld] \ar[ru]&&&\zxN{}\ar[ru,)]&\\
        &&&&&&\zxZ{} \ar[dddd]&&&&&&\\   
        &&&&&&&&&&\zxX{} \ar[uu] \ar[ddd] &&&&&&\\
        &&&&&&&&&\zxN{} \ar[ld] \ar[ru]&&&&&&&&\\
        &&&&&&&&\zxZ{} \ar[d]&&&&&&\\ 
        &&&&&&&&&&\zxX{}&&&&\\
         &&0&&1&&2& \cdots&L&&0&&&&\\
\end{ZX} \\
&= \bra{\counit}_0  \left(\prod_{j=0}^{L-1}  \ZXgate_{j,j+1}^\dagger\right)\ZXgate_{L,0}\ket{\unit}_0.
\end{align}
where $0$ denotes the ancilla qudit.

The gauging map satisfies the algebraic relations, 
\ie \label{eq:KW-AB}
\gaugemap \ \Al^{\drep} =\counith(\dchi_\drep) \gaugemap = d_{\drep} \gaugemap \,, \qquad 
\Br^{\Gamma} \ \gaugemap  = \chi_\Gamma(\unit) \gaugemap = d_\Gamma \gaugemap \,,
\fe 
where $\Br^\Gamma$ is as defined in \cref{eq:rephmpo}.
The first of these two equalities follows from the fact that the $\Rep(\dl H)$ symmetry operators $\Al^\drep$ act like the identity operator times $d_\drep$ on $\cH_{\rm physical}$ (see \cref{eq:KW-symkernel}), which means that $\Al^\drep$ has been gauged.
This also follows directly from the MPO form of $\gaugemap$ \eqref{eq:KW-MPO}. 
Using the same MPO, one can also show that the second equality in \cref{eq:KW-AB} holds for any irrep $\Gamma$ of $H$.
As a consequence of \cref{eq:KW-AB}, we find
\ie 
{\rm im}\left(\Al^\drep - d_\drep \mathbbm{1}\right) \subseteq {\rm ker} (\gaugemap) \,,
\quad 
{\rm im}(\gaugemap) \subseteq {\rm ker} \left(\Br^\Gamma - d_\Gamma \mathbbm{1}\right) .
\fe 
In particular, the second of the above inclusion relations implies that any state in the image of the gauging map is necessarily in the stabilizer of the action of the $\Rep(H)$ symmetry operators $\Br^\Gamma$.

Moving on from the action of the gauging map on states to that on local operators, we find the transformation of the ``$X$'' and ``$ZZ$'' operators under $\gaugemap$ as
\begin{equation}
    \Xr_j^{a}\xrightarrow{\gaugemap}  
    (\Xr_{j}  \Xl_{j+1})^{a}_\op
        \,,     \qquad
    (\Zr_{j}  \Zl_{j+1})^{\alpha }_\op  \xrightarrow{\gaugemap}
    \Zl_{j+1}^{\alpha} 
    \,.
\end{equation}
In diagrams, we have
\tikzsetfigurename{gaugingmappaulis}
\begin{align}\label{eq:XKWZZKW-diag}
    \begin{ZX}
        &&\\
        \zxX{} \ar[u] \ar[d] \ar[rd,)]&&\\
        &\zxX{Sa}&\\
        \zxbase{j}&&\\
        &&
    \end{ZX} 
    \xrightarrow{\gaugemap}
    \begin{ZX}
        &&&\\
        \zxX{}\ar[u]\ar[ddd]&&&\zxX{}\ar[u]\ar[ddd]\\
        &&\zxrep{$S$}\ar[ru,s]&\\
        &\zxZ{}\ar[luu,s]\ar[ru,s] \ar[d]&&\\
        &\zxX{Sa}&&\\
        \zxbase{j}&&&\zxbase{j+1}\\
        &&&
    \end{ZX} 
    \ , \qquad
    \begin{ZX}
        &&\zxZ{\alpha S}&\\
        &&\zxX{}\ar[rdd,s]\ar[ld,s] \ar[u]&\\
        &\zxrep{$S$}\ar[ld,s]&&\\
        \zxZ{}\ar[uuu]\ar[d]&&&\zxZ{}\ar[uuu]\ar[d]\\
        &&&\\[5 pt]
        \zxbase{j}&&&\zxbase{j+1}\\
        &&&
    \end{ZX} 
    \xrightarrow{\gaugemap} \begin{ZX}
        \zxZ{\alpha S} &\\
        & \zxZ{}\ar[u]\ar[d]\ar[lu,)] \\
        &\\[5 pt]
        &\zxbase{j+1}\\
        &
    \end{ZX} 
    \ .
\end{align}
As we noted in \cref{eq:RepH-symmetric-ops}, the operators on the r.h.s.~commute with the $\Rep(H)$ symmetry operators, $\Br^\Gamma$.
The above fact demonstrates that the gauging map $\gaugemap$ takes $\Rep(\dl{H})$ symmetric local operators to $\Rep(H)$ symmetric local operators. Furthermore, the algebra of these local operators is also preserved under this map, so $\gaugemap$ can be viewed as an isomorphism of local symmetric operator algebras~\cite{Cobanera:2011wn,JiWen20,chatterjee_symmetry_2023,Jones:2024lws,Ma:2024ypm,jones2025op_alg}.

\subsection{Hopf Hadamard gate}

We have seen that the gauging map $\gaugemap$ transforms a local Hamiltonian with $\Rep(\dl{H})$ symmetry to one with $\Rep(H)$ symmetry.
We now specialize to the self-dual case, where ${H \cong \dl{H}}$ through a Hopf algebra isomorphism. 
In this case, one can define a \emph{Hopf Hadamard gate}, which implements such an isomorphism, from $H$ to $\dl{H}$, in Hopf ZX diagrams.
As we will show in \Cref{sec:DKW}, this (unitary) transformation turns the $\Rep(H)$ symmetry operator $\Br$ into the $\Rep(\dl H)$ symmetry operator $\Al$.

For a self-dual Hopf algebra $H$, the isomorphism between $H$ and $\dl{H}$ is captured by the non-degenerate Hadamard form ${\Theta:H \times H \to \CC}$ which is defined by the following properties:\footnote{See \appref{app:isoHandHdual} for more details on this.}
\tikzsetfigurename{hadamardform0}
\begin{equation}\label{eq:ThMultComult}
\begin{ZX}
    &&\zxrep{$\Theta$}\ar[ld,(] \ar[rd,)]&&\\
    &\zxX{} \ar[ld,(] \ar[rd,)] &&\zxN{} \ar[d]&\\
    &&&&
\end{ZX} = \begin{ZX}
    &\zxrep{$\Theta$} \ar[ldd,(] \ar[rrdd,)]&&\zxrep{$\Theta$} \ar[ldd,(,3d above,very thick] \ar[rd,)]& \\
    &&&&\\
    \zxN{} \ar[d]&&\zxN{} \ar[d]&\zxZ{}\ar[ru,N]\ar[d]&\\
    &&&&\\
\end{ZX},\quad \begin{ZX}
&\zxrep{$\Theta$} \ar[ldd,(] \ar[rd,)]&&&\\
&&\zxX{} \ar[ld,(] \ar[rd,)]&&\\
&&&&
\end{ZX} = \begin{ZX}
&\zxrep{$\Theta$} \ar[rdd,)] \ar[ld,(]&&\zxrep{$\Theta$} \ar[lldd,(,3d above,very thick] \ar[rdd,)]& \\
&&&&\\
&\zxZ{} \ar[lu,N] \ar[d]&\zxN{}\ar[d]&& \zxN{} \ar[d]\\
&&&&
\end{ZX}
\end{equation}
The Hadamard form maps the unit to the counit and is compatible with the antipode, i.e.,
\tikzsetfigurename{hadamardform11}
\begin{equation}\label{eq:ThUnitHaarS}
    \begin{ZX}
        &\zxrep{$\Theta$} \ar[ld,(] \ar[rd,)]&\\
        &&\zxX{}
    \end{ZX}=\begin{ZX}
        \zxZ{} \ar[d]&\\ \zxN{}\ar[u]&
    \end{ZX}
    = 
    \begin{ZX}
        &\zxrep{$\Theta$} \ar[ld,(] \ar[rd,)]&\\
        \zxX{}&&
    \end{ZX},\quad \begin{ZX}
    &\zxrep{$\Theta$} \ar[ld,(] \ar[rd,)]&\\
    \zxrep{$S$}\ar[d]&&\zxN{} \ar[d]\\
    &&
    \end{ZX} =  \begin{ZX}
    &\zxrep{$\Theta$} \ar[ld,(] \ar[rd,)]&\\
    \zxN{} \ar[d]&&\zxrep{$S$}\ar[d]\\
    &&
    \end{ZX}
\end{equation}
Since $\Theta$ is non-degenerate, there exists a unique element ${\Theta^{-1}\in H\otimes H}$ \footnote{In Sweedler notation, ${\Theta^{-1} = \sum_{(\Theta^{-1})}\Theta^{-1}_1 \otimes \Theta^{-1}_2}$. \Rf{majid2022zxhopf} refers to $\Theta^{-1}$ as the ``metric'' dual to the bilinear form $\Theta$.}, such that,
\tikzsetfigurename{hadamardform12}
\begin{equation}\label{eq:inverseTheta}
     \begin{ZX}
        &\zxrep{$\Theta$}\ar[ld,(] \ar[rrd,s] &&&\\
        &&&\zxrep{$\Theta^{-1}$} \ar[ru,(]&
    \end{ZX}
    =
    \begin{ZX}
        &\zxN{} \dar& \\[10pt]
        &\zxN{}\uar& 
    \end{ZX} 
    = \begin{ZX}
    &&&\zxrep{$\Theta$}\ar[rd,)] \ar[lld,s] &\\
    &\zxrep{$\Theta^{-1}$} \ar[lu,)]&&&
    \end{ZX}
\end{equation}
Lastly, since the Haar (co)integral are unique, under the Hadamard form, we have
\begin{equation}\label{eq:ThHaar}
     \begin{ZX}
        &\zxrep{$\Theta$} \ar[ld,(] \ar[rd,)]&\\
        &&\zxZTri.{} 
     \end{ZX} = \begin{ZX}
        \zxXTri'{}  \ar[d]&\\ \zxN{}\ar[u]&
    \end{ZX}
    = 
    \begin{ZX}
        &\zxrep{$\Theta$} \ar[ld,(] \ar[rd,)]&\\
        \zxZTri.{} &&
    \end{ZX}
\end{equation}
We further assume the Hadamard form is symmetric, so that it gives rise to a self-duality (we will show this fact later in \eqref{eq:argsymmetricTheta}).\footnote{\label{footnote:symm-Had} It is possible to obtain a non-symmetric Hadamard form, for example, by applying an appropriate Hopf algebra automorphism on one side of the Hadamard form. However, the resulting $\kw$ operator will also differ by the automorphism, and generally results in a $p$-ality~\cite{Lu:2024ytl} with ${p>2}$ rather than a duality transformation. On the other hand, applying the same automorphism on both inputs of the Hadamard form results in a $\kw$ operator which still implements a self-duality, but which differs by conjugating the original self-duality by the automorphism.} 
It then follows that $\Theta^{-1}$ is also symmetric,
\begin{equation}
    \begin{ZX}
        &&\\
        &\zxrep{$\Theta$} \ar[ld,(] \ar[rd,)]&\\
        &&
    \end{ZX}=\begin{ZX}
        &&& \\
        &\zxrep{$\Theta$}\ar[d,C] &&\\
        &\zxN{} \ar[rd,)]\ar[u,C-,3d above,very thick]\ar[ld,(,3d above,very thick] &&\\
        &&&
    \end{ZX},\quad \quad 
    \begin{ZX}
        &&\\
        &\zxrep{$\Theta^{-1}$} \ar[lu,)], \ar[ru,(] &\\
        &&
    \end{ZX}
    =
    \begin{ZX}
        \zxN{}\ar[rd,(]&&\\
        &\zxN{} \ar[ru,(,3d above,very thick]\ar[d,C-] &\\
        &\zxrep{$\Theta^{-1}$}  \ar[u,C,3d above,very thick] &\\
        &&
    \end{ZX}
\end{equation} 
One can modify the Hadamard form by a Hopf algebra automorphism ${\tau:H\rightarrow H}$, defining
\begin{equation}\label{eq:modifiedHad}
    \Theta'(\cdot,\cdot) \equiv \Theta(\cdot,\tau(\cdot)) \,.
\end{equation}
$\Theta'$ is also symmetric if $\tau$ and $\Theta$ satisfy ${\Theta(a,\tau(b)) = \Theta(\tau(a),b)}$ for all ${a,b\in H}$.

\Cref{eq:ThMultComult,eq:ThUnitHaarS} define a so-called type 2 Hadamard form, which exists if and only if $H$ is self-dual~\cite{majid2022zxhopf}. 
% The Hadamard form acts on the elements as follows,
Using this Hadamard form, we introduce the notations,
\tikzsetfigurename{defineThetaelements}
\begin{equation}\label{eq:defThElements}
    \Theta_a \equiv \Theta(a,\cdot) 
    =\begin{ZX}
        \zxZ{\Theta_a} \ar[d]&\\& 
    \end{ZX}
    % \equiv
    =
    \begin{ZX}
        &\zxrep{$\Theta$} \ar[ld,(] \ar[rd,)]&\\
        \zxX{a}&&
    \end{ZX}
    \,,
    \qquad 
    \Theta^{-1}_\chi \equiv \sum_{(\Theta^{-1})}\chi(\Theta^{-1}_{(1)}) \Theta^{-1}_{(2)}
    \equiv
    \begin{ZX}
        &\\
        \zxX{\Theta^{-1}_\chi} \ar[u]&
    \end{ZX} 
    =
    \begin{ZX}
        \zxZ{\chi} \ar[rd,(]&&\\
        &\zxrep{$\Theta^{-1}$} \ar[ru,(]&
    \end{ZX}
    \,,
\end{equation}
generalizing the first two equalities of \cref{eq:ThUnitHaarS}\,.

With the Hadamard form $\Theta$ as defined above, one can define a Hopf Hadamard gate\footnote{We avoid using the word Fourier transform in this context. Although they coincide for finite abelian groups, the Fourier transform exists for any Hopf algebra (e.g. non-Abelian group algebras) while the Hopf Hadamard gate only exists when the Hopf algebra is self-dual.} as follows:
\tikzsetfigurename{hadamard}
\begin{equation}\label{eq:hadamardgatedef}
    \FF\equiv\begin{ZX}
        \\
        \zxHad{} \ar[u]\ar[d]\\
    \end{ZX} = \begin{ZX}
        &\zxrep{$\Theta$} \ar[rrd,s] \ar[ld,(] &&&\\
        &&&\zxX{}\ar[ru,(]&
    \end{ZX}
    \,,
    \qquad 
    \FF^{-1} 
    = \begin{ZX}
    &\zxX{} \ar[rrd,s] \ar[ld,(] &&&\\
    &&&\zxrep{$\Theta^{-1}$}\ar[ru,(]&
    \end{ZX}
\end{equation}
For ${H = \CC[\ZZ_2]}$, ${\FF = \frac{1}{\sqrt{2}}\left(\begin{smallmatrix}
    1&1 \\1 &-1
\end{smallmatrix}\right)}$, which is the usual Hadamard gate for qubits. 
For ${H = \CC[\ZZ_{N}]}$ (${N>2}$), ${(\FF)_{ij} = \frac1{\sqrt N}\omega^{ij} }$ (${i,j\in\ZZ_N}$) with ${\omega = \ee^{2\pi i/N}}$, is the familiar complex Hadamard matrix. If the Hadamard form is modified by automorphism $\tau$ according to \cref{eq:modifiedHad}, the Hopf Hadamard gate satisfies ${\tau \FF = \FF \tau^{-1}}$. From \cref{eq:ThMultComult,eq:hadamardgatedef}, we find that the Hadamard gate satisfies
\tikzsetfigurename{hadamardmapping}
\begin{equation}\label{eq:type2hadgate}
\tag{CC1}
\text{Color change 1:} \qquad
    \begin{ZX}
        &&\\
        &\zxX{} \ar[lu,)] \ar[ru,(]&\\
        &\zxHad{} \ar[u] \ar[d] &\\
        &&
    \end{ZX}=\begin{ZX}
        &&\\
        \zxHad{}\ar[u]&&\zxHad{} \ar[u]\\
        &\zxZ{} \ar[lu,)] \ar[ru,(] \ar[d]&\\
        &&
    \end{ZX}
    \,, 
    \qquad 
    \begin{ZX}
        &&\\
        &\zxZ{} \ar[ld,(] \ar[rd,)] \ar[u]&\\
        \zxHad{} \ar[d]&&\zxHad{} \ar[d]\\
        &&\\
    \end{ZX} = \begin{ZX}
        &&\\
        &\zxHad{} \ar[u] &\\
        &\zxX{} \ar[u] \ar[d,C] &\\
        &\zxN{} \ar[rd,)]\ar[u,C-,3d above,very thick]\ar[ld,(,3d above,very thick] &\\
        &&
    \end{ZX}
    \,.
\end{equation}

Using the definitions above, we note an interesting identity involving $\FF$ and $S$,
\begin{equation}\label{eq:F2=S}
    \begin{ZX}
        \\ \zxHad{}\ar[u]\ar[d] \\ \zxHad{} \ar[d]\\
    \end{ZX}=\begin{ZX}
        &\zxrep{$\Theta$} \ar[ld,(] \ar[rd,s]&&\zxrep{$\Theta$} \ar[ld,s] \ar[rd,s]&&\\
        &&\zxZ{}\ar[d]&&\zxZ{}\ar[d] \ar[ru,(]&\\
        &&\zxZ{}&&\zxZ{}&
    \end{ZX}
    \stackrel{\eqref{eq:ThMultComult}}{=}
    \begin{ZX}
        &&\zxrep{$\Theta$}\ar[d,)] \ar[ld,(]&&\\
        &\zxX{}\ar[ld,(] \ar[rd,s]&\zxZ{}&&\\
        &&\zxZ{} \ar[ru,(] \ar[d]&&\\
        &&\zxZ{}&&
    \end{ZX} 
    \stackrel{\eqref{eq:ThUnitHaarS}}{=} 
    \begin{ZX}
        &\zxX{} \ar[d]&&&\\
        &\zxX{}\ar[ld,(] \ar[rd,s]&&&\\
        &&\zxZ{} \ar[ru,(] \ar[d]&&\\
        &&\zxZ{}&&
    \end{ZX} 
    \stackrel{\eqref{eq:FrobeniusForm}}{=}
    \begin{ZX}
        &\zxZ{}\ar[ld,(] \ar[rd,s]&&&\\
        &&\zxX{} \ar[ru,(] &&\\
    \end{ZX} 
    \stackrel{\eqref{eq:rcgc=S}}{=}
    \begin{ZX}
        \\ \zxrep{$S$} \ar[u] \ar[d] \\
    \end{ZX}
    \,.
\end{equation}
Here, we have made use of the symmetricity of the chosen Hadamard form. Thus, we find that ${\FF^2=S}$. Applying \cref{eq:F2=S} on \cref{eq:type2hadgate}, we find
\begin{equation}\label{eq:type1hadgate}
\tag{CC2}
\text{Color change 2:} \qquad
    \begin{ZX}
        &&\\
        &\zxX{} \ar[ld,(] \ar[rd,)] \ar[u]&\\
        \zxHad{} \ar[d]&&\zxHad{} \ar[d]\\
        &&\\
    \end{ZX} = \begin{ZX}
        &&\\
        &\zxHad{} \ar[u] &\\
        &\zxZ{} \ar[ld,(] \ar[rd,)] \ar[u]&\\
        &&
    \end{ZX}
    \,,\qquad 
    \begin{ZX}
        &&\\
        &\zxZ{} \ar[lu,)] \ar[ru,(]&\\
        &\zxHad{} \ar[u] \ar[d] &\\
        &&
    \end{ZX}=\begin{ZX}
        &&\\
        \zxHad{} \ar[u]\ar[rd,(]&&\zxHad{} \ar[u]\\
        &\zxN{} \ar[ru,(,3d above,very thick]\ar[d,C-] &\\
        &\zxX{} \ar[d] \ar[u,C,3d above,very thick] &\\
        &&
    \end{ZX}
    \,.
\end{equation}
These are the defining conditions for the so-called type 1 Hadamard gate~\cite{majid2022zxhopf}.\footnote{We have thus shown that given a type 2 Hadamard gate associated with a finite-dimensional semi-simple self-dual Hopf algebra, it is automatically also a type 1 Hadamard gate. The fact that ${S^2=\mathbbm{1}}$ for any finite-dimensional semi-simple Hopf algebra over $\CC$ was used in proving the key identity \cref{eq:F2=S}.}

We re-express \cref{eq:defThElements} in terms of the Hopf Hadamard gate for convenience in subsequent calculations,
\tikzsetfigurename{hadamardmapping-state}
\begin{equation}
    % \begin{ZX}
    %   \zxX{a}\\ \zxHad{} \ar[u] \ar[d] \\
    % \end{ZX} = 
    \begin{ZX}
     &\zxX{} \ar[rd,)]&\\
     \zxX{a} \ar[ru,)]&&\zxHad{} \ar[d]\\
     &&
    \end{ZX} = \begin{ZX}
    \zxZ{\Theta_a}\\ \zxN{} \ar[u] \ar[d] \\  
    \end{ZX}\ ,\qquad 
    \begin{ZX}
    &&\\
    \zxZ{\chi} \ar[rd,(] &&\zxHad{} \ar[u]\\
    &\zxZ{} \ar[ru,(] &
    \end{ZX} = \begin{ZX}
        \\ \zxN{} \ar[u] \ar[d] \\ \zxX{\Theta^{-1}_\chi} 
    \end{ZX}.
\end{equation}

The Hadamard form $\Theta$ discussed so far only implements an isomorphism of Hopf algebras. We further require that this isomorphism is compatible with the $\star$-structure so that it lifts to an isomorphism of Hopf $C^\star$ algebras. In equations, this means (see \Cref{app:isoHandHdual})
\tikzsetfigurename{hadamard-star}
\begin{equation}
    \Theta(a^\star,b) = \overline{\Theta(a,S(b)^\star)}.
\end{equation}
The Hadamard gate defined using such a Hadamard form is unitary, since we have that ${\langle \FF(x),\FF(y) \rangle_{\cohaar}=\cohaar(\FF(x)^\star \FF(y)) = \phi(x^\star y) = \langle x, y \rangle_{\cohaar}}$, where the first and the last equal signs are by definition. To see this, we first note that
\begin{equation}\label{eq:star-had-star}
    \cohaar(x^\star y) = \cohaar(\FF(x)^\star \FF(y)) = \cohaar(\FF(\FF(x)^\star) y)  \Leftrightarrow \FF(x^\star)^\star = \FF^{-1}(x)\,.
\end{equation}
Indeed, we find\footnote{We remind the reader that due to the anti-linearity of $\star$, the Hadamard form in the second expression of \cref{eq:star-had-star} acquires a complex conjugation, denoted by the overline.}
 \tikzsetfigurename{hadamard-star-star}
 \begin{equation}
    (\FF(x^\star))^\star = \begin{ZX}
     &&&\\
         &\zxrep{$\overline\Theta$} \ar[ld,(]\ar[rd,s]&&\zxrep{$\star$} \ar[u] \ar[ld,)]\\
         \zxrep{$\star$} \ar[d]&&\zxX{}&\\
         &&&
     \end{ZX} = \begin{ZX}
     &&&&&\\
         &\zxrep{$\Theta$} \ar[ld,(]\ar[rd,s]&&&&\zxrep{$\star$} \ar[u] \ar[ld,)]\\
         &&\zxrep{$S$} \ar[r]&\zxrep{$\star$}\ar[r]&\zxX{}&
     \end{ZX} = \begin{ZX}
         \\ \zxrep{$S$} \ar[u] \ar[d] \\ \zxHad{} \ar[d]\\
     \end{ZX} = \FF^{-1}(x) \,.
 \end{equation}
Therefore, $\FF^{-1} = \FF^\dagger$, i.e.~$\FF$ is a unitary gate. We introduce the following notation:
\tikzsetfigurename{hadamard-dagger}
\ie 
\FF^\dagger = \begin{ZX}
         \\ \zxHad{\tiny $\dagger$} \ar[u] \ar[d] \\
     \end{ZX}
     =
     \begin{ZX}
    &\zxX{} \ar[rrd,s] \ar[ld,(] &&&\\
    &&&\zxrep{$\Theta^{-1}$}\ar[ru,(]&
    \end{ZX} \,.
\fe

\subsection{Self-duality symmetry}
\label{sec:DKW}
In the following, we demonstrate the Hopf Kramers--Wannier duality by placing an additional layer of Hopf Hadamard gates on top of the gauging map. The self-duality symmetry MPO is defined as,
\tikzsetfigurename{KW}
\begin{equation}\label{eq:dualityop}
    \kw= (\FF^\dagger)^{\otimes N} \times \gaugemap =
    \begin{ZX}
        &&&&&&&&&&&&&\\
        &&&&\zxN{}\ar[ru,(]&&&&\zxN{}\ar[ru,(]&&&&\zxN{}\ar[ru,(]&\\
        &&&\zxHad{\tiny$\dagger$}\ar[ur,)]&&&&\zxHad{\tiny$\dagger$}\ar[ur,)]&&&&\zxHad{\tiny$\dagger$}\ar[ur,)]&&\\
        &&&\zxX{} \ar[ld,(] \ar[rd,)] \ar[u]&&&&\zxX{} \ar[ld,(] \ar[rd,)] \ar[u]&&&&\zxX{} \ar[ld,(] \ar[rd,)] \ar[u]&&\\
        \zxN{} &&\zxrep{$S$} &&\zxN{}&&\zxrep{$S$}&&\zxN{}&&\zxrep{$S$}&&&\\
        &\zxZ{} \ar[lu,)] \ar[ru,(] \ar[d] &&&&\zxZ{} \ar[lu,)] \ar[ru,(] \ar[d]&&&&\zxZ{} \ar[lu,)] \ar[ru,(] \ar[d]&&&\\
        &&&&&&&&&&&&&
    \end{ZX}
    =  \begin{ZX}
        &&&&&&&&&&&&&\\
        &&&&\zxN{}\ar[ru,(]&&&&\zxN{}\ar[ru,(]&&&&\zxN{}\ar[ru,(]&\\
        &&&\zxZ{} \ar[ld,(] \ar[rd,)]\ar[ur,)]&&&&\zxZ{} \ar[ld,(] \ar[rd,)] \ar[ur,)]&&&&\zxZ{} \ar[ld,(] \ar[rd,)] \ar[ur,)]&&\\
        \zxN{} &&\zxHad{} &&\zxHad{\tiny$\dagger$}&&\zxHad{}&&\zxHad{\tiny$\dagger$}&&\zxHad{}&&&\\
        &\zxZ{} \ar[lu,)] \ar[ru,(] \ar[d] &&&&\zxZ{} \ar[lu,)] \ar[ru,(] \ar[d]&&&&\zxZ{} \ar[lu,)] \ar[ru,(] \ar[d]&&&\\
        &&&&&&&&&&&&&
    \end{ZX}  
\end{equation}
which reduces to the $\ZZ_2$ Kramers-Wannier MPO for ${H=\CC[\ZZ_2]}$ \cite{aasen2016duality,Chen:2023qst,SeibergShao23,tantivasadakarn_long-range_2022,oshikawa2023kw,shuheng2024KWlsm,arkya2024reps3dual,nat2024zxKW,lu2025strange}. 

The $\Rep(\dl{H})$-symmetric one- and two-body Hopf Pauli operators are mapped by $\kw$ as
\tikzsetfigurename{gaugingmapspaulisa}
\begin{align}
    \ract{X}^{a}=\begin{ZX}
        &\\
        \zxX{} \ar[u] \ar[d] \ar[rd,)]&\\
        &\zxX{Sa}
    \end{ZX}\xrightarrow{\kw}
     \begin{ZX}
        &&&\\
        &&\zxZ{\Theta_{Sa}} \ar[d]&\\
        &&\zxX{}\ar[rdd,s]\ar[ld,s] &\\
        &\zxrep{$S$}\ar[ld,s]&&\\
        \zxZ{}\ar[uuu]\ar[d]&&&\zxZ{}\ar[uuu]\ar[d]\\
        &&&
    \end{ZX} = (\ract{Z}\lact{Z})_{\op}^{\Theta_a}\,,\quad 
    (\ract{Z}\lact{Z})_{\op}^{\Theta_a} =
    \begin{ZX}
        &&\zxZ{\Theta_{Sa}}&\\
        &&\zxX{}\ar[rdd,s]\ar[ld,s] \ar[u]&\\
        &\zxrep{$S$}\ar[ld,s]&&\\
        \zxZ{}\ar[uuu]\ar[d]&&&\zxZ{}\ar[uuu]\ar[d]\\
        &&&
    \end{ZX} \xrightarrow{\kw} 
    \begin{ZX}
        &&&\\
        \zxX{} \ar[u] \ar[d] \ar[rd,)]&&&\\
        &\zxX{S a}&&
    \end{ZX} ,
\end{align}
where $\Theta(a,S(\cdot)) = \Theta(S(a),\cdot) \equiv \Theta_{Sa}$ due to \eqref{eq:ThUnitHaarS}. In equations,
\begin{align}\label{eq:KW-XZmap}
    \Xr^a_j \xrightarrow{\kw} (\ract{Z}_j\lact{Z}_{j+1})_\op^{\Theta_a}  \xrightarrow{\kw} \Xr^a_{j+1} \,,
\end{align}
% In particular, if $a\in \Cocom(H)$ (resp. $\chi \in \Cocom(\dl{H})$), then $\Theta_a\in \Cocom(\dl{H})$ (resp. $\sum_{(\Theta^{-1})}\chi(\Theta^{-1}_{(1)}) \Theta^{-1}_{(2)} \in \Cocom(H)$). 
and in particular, for ${a\in \Cocom(H)}$, we have ${\Theta_a\in \Cocom(\dl{H})}$ (a similar comment applies to ${\chi\in\Cocom(\dl{H})}$ and $\Theta^{-1}_{\chi}$). 
In \cref{eq:KW-XZmap}, we used the following action of the Hopf Hadamard gate on the Hopf Pauli operators:\footnote{The notation ${\cO\xrightarrow{\FF^{\dagger}} \cO'}$ here is shorthand for ${\cO' = \FF^{\dagger} \cO \FF}$.}
\begin{align}
    \Xl^a \xrightarrow{\FF^\dagger}\Zl^{\Theta_a}\xrightarrow{\FF^\dagger} \Xr^{ a}\xrightarrow{\FF^\dagger} \Zr^{\Theta_a}\xrightarrow{\FF^\dagger} \Xl^a 
\end{align}
The simplest $\Rep(\dl{H})$ and $\kw$ symmetric (i.e.~self-dual) Hamiltonian is given by~\footnote{We note the following simplification: $\Theta_{x_r} S = \Theta(x_r,S(\cdot))=\Theta(S(x_r),\cdot)= \Theta_{S x_r}$}
\tikzsetfigurename{self-dual-Ham}
\begin{align}
    \sH_\text{sd} &= -\sum_{j=1}^L \sum_{\drep \in \Irr(\dl{H})} g_\drep \left( \ract{X}^{\dchi_\drep}_j +(\ract{Z}_j\lact{Z}_{j+1})_\op^{\Theta_{\dchi_\drep}}\right)
    \label{eq:HsdH8} \\
    &= -\sum_{j=1}^L \sum_{\drep \in \Irr(\dl{H})} g_\drep \left(\begin{ZX}
        &&&\\
        \zxX{} \ar[u] \ar[dd] \ar[rd,)]&&&\\
        \zxN{}&\zxX{S\dchi_\drep}&&\\
        &&&\\[5 pt]
         \zxbase{j}&&&\\
         &&&
    \end{ZX}
    +\begin{ZX}
        &&&\\
        &&\zxZ{\Theta_{S\dchi_\drep}} \ar[d]&\\
        &&\zxX{}\ar[rdd,s]\ar[ld,s] &\\
        &\zxrep{$S$}\ar[ld,s]&&\\
        \zxZ{}\ar[uuu]\ar[d]&&&\zxZ{}\ar[uuu]\ar[d]\\
        &&&\\[5 pt]
        \zxbase{j}&&&\zxbase{j+1}\\
        &&&
        \end{ZX}\right)
\end{align}
The self-duality of this Hamiltonian follows from \cref{eq:KW-XZmap}.

Conjugation by the Hopf Hadamard gates $\prod_j \FF_j^\dagger$ turns the $\Rep(H)$ symmetry MPOs into the corresponding $\Rep(\dl{H})$ symmetry MPOs,
\tikzsetfigurename{symmetryhadamard}
\begin{equation}\label{eq:FourierSymMPO}
    \Br^{\Theta_{x_r}} = \begin{ZX}
        && && && && && \\
        && && && && &&\zxZ{\Theta_{x_r}} \ar[ld,)] \\
        &&\zxN{}  \ar[rr] \ar[ld,s] && \zxX{}\ar[r] \ar[ld,)] && ... &&\zxX{} \ar[r] \ar[ld,)] \ar[l] &&  \\
        & \zxZ{}\ar[dd] \ar[uu,3d above,very thick]  && \zxZ{} \ar[dd] \ar[uu,3d above,very thick]&& \zxN{} && \zxZ{}\ar[dd] \ar[uu,3d above,very thick] && \zxN{} &\\
        && && && && && \\
        && && && && && \\
    \end{ZX} \xrightarrow{\FF^\dagger} \begin{ZX}
        &&&&&&&&&&&\\
        &&&\zxX{} \ar[u] && \zxX{} \ar[u] &&&& \zxX{} \ar[u]&&\\
        && \zxZ{} \ar[ru,)]\ar[rr]&& \zxZ{}\ar[rr] \ar[ru,)]&& ... &&\zxN{} \ar[l]\ar[ru,s]&&&\\
        &\zxX{ \dchi_\drep} \ar[ru,)]&& \ar[uu,3d above,very thick]&&\ar[uu,3d above,very thick]&&&&\ar[uu,3d above,very thick]&&\\
        &&&&&&&&&&&\\
    \end{ZX} 
    = \Al^{\drep}
    \, 
\end{equation}
 where we note that every ${\drep\in \Irr(\dl{H})}$ has a unique corresponding ${\Gamma\in \Irr(H)}$ given by ${\chi_\Gamma =\Theta_{ \dchi_\drep}}$.
Combining \cref{eq:FourierSymMPO,eq:KW-AB}, we have the following symmetry algebra:
\ie \label{eq:KW-AA}
\kw \ \Al^{\drep} = d_{\drep} \kw = 
\Al^{\drep} \ \kw 
\fe 

Let us next investigate the algebra of the self-duality symmetry operator with itself, $\kw$. In the $\mathbb Z_2$ case, it is known that the self-duality symmetry operator squares to a projector to the $\ZZ_2$ symmetric subspace, composed with lattice translation~\cite{SeibergShao23,nat2024zxKW}. For more general self-dual Hopf algebras, using \eqref{eq:dualityop}, we similarly find
\tikzsetfigurename{KWsquared}
\begin{align}\label{eq:kwsquare}
    T^{-1}(\kw)^2 &= 
    \begin{ZX}
        &&&&&&&&&&&\\
        &&\zxZ{} \ar[u] \ar[ld] \ar[rd]&&&&\zxZ{} \ar[u] \ar[ld] \ar[rd]&&&&&\\
        &&&\zxHad{\tiny$\dagger$}&&\zxHad{}&&\zxHad{\tiny$\dagger$}&&\zxHad{} \ar[ru]&&\\
        \cdots&&&&\zxZ{}\ar[d] \ar[lu] \ar[ru]&&&&\zxZ{}\ar[d] \ar[lu] \ar[ru]&&&\cdots\\
        &&&&\zxZ{}\ar[rd] \ar[ld] &&&&\zxZ{}\ar[rd] \ar[ld]&&&\\
        &&&\zxHad{} \ar[ld]&&\zxHad{\tiny$\dagger$} \ar[rd]&&\zxHad{}\ar[ld]&&\zxHad{\tiny$\dagger$} \ar[rd]&&\\
        &&\zxZ{} \ar[d] \ar[lu] &&&&\zxZ{} \ar[d] &&&&&\\
        &&&&&&&&&&&
    \end{ZX}  \stackrel{\eqref{eq:hopf-frob-cond}}{=} 
    \begin{ZX}
        &&&&&&&&&&&\\
        &&\zxZ{} \ar[u] \ar[ld] \ar[rd]&&&&\zxZ{} \ar[u] \ar[ld] \ar[rd]&&&&&\\
        &&&\zxHad{\tiny$\dagger$} \ar[d]&&\zxHad{}\ar[dd]&&\zxHad{\tiny$\dagger$} \ar[d]&&\zxHad{} \ar[dd] \ar[ru]&&\\
        \cdots&&&\zxZ{} \ar[dd] \ar[rrd]&& &&\zxZ{}\ar[dd] \ar[rrd]&&&&\cdots\\
        &&&&&\zxZ{}\ar[d]&&&&\zxZ{}\ar[d]&&\\
        &&&\zxHad{} \ar[ld]&&\zxHad{\tiny$\dagger$} \ar[rd]&&\zxHad{}\ar[ld]&&\zxHad{\tiny$\dagger$} \ar[rd]&&\\
        &&\zxZ{} \ar[d] \ar[lu] &&&&\zxZ{} \ar[d] &&&&&\\
        &&&&&&&&&&&
    \end{ZX}\stackrel{\eqref{eq:type1hadgate}}{=} \begin{ZX}
        &&&&&&&&\\
        &&&&&\zxHad{\tiny$\dagger$} \ar[u]\ar[d]&&&\\
        &&&&&\zxX{}\ar[ld,(] \ar[rd,)]&&&\\
        &&\cdots&&\zxrep{$S$}\ar[d]& &\zxZ{} \ar[rrd] \ar[lddd,)]&&\\
        &&&&\zxZ{} \ar[llu] \ar[d]&&&&\cdots\\
        &&&&\zxrep{$S$} \ar[rd,(]&&&&\\
        &&&&&\zxX{} \ar[d]&&&\\
        &&&&&\zxHad{} \ar[d]&&&\\
        &&&&&&&&
    \end{ZX} \nonumber\\
    &\stackrel{\eqref{eq:proof-kw-sqaure}}{=} \begin{ZX}
        &&&&&&\\
        ...&&\zxZ{}\ar[l] \ar[rr]&&&...&\\
        &&&\zxX{} \ar[uu,3d above,very thick] \ar[d] \ar[lu]&&&\\
        &&&&&&\\
    \end{ZX}\stackrel{\eqref{eq:assoc}}{=} \begin{ZX}
        && &&& && && && \\
        \zxZ{} \ar[rd,(] \ar[u,C-] \ar[u,C]&& &&& && && &&\\
        &\zxZ{} \ar[rd,(] \ar[rr] &&&... && \zxZ{}\ar[l]\ar[r] \ar[rd,(] && \zxN{}\ar[l] \ar[rd,s] &&&  \\
        && \zxX{}\ar[dd] \ar[uuu,3d above,very thick]  &&& \zxN{}&& \zxX{} \ar[dd] \ar[uuu,3d above,very thick] && \zxX{}\ar[dd] \ar[uuu,3d above,very thick] && \zxN{} \\
        && &&& && && && \\
        && &&& && && && \\
    \end{ZX}
    =|H|\begin{ZX}
        && &&& && && && \\
        \zxZTri'{} \ar[rd,(]&& &&& && && &&\\
        &\zxZ{} \ar[rd,(] \ar[rr] &&&... && \zxZ{}\ar[l]\ar[r] \ar[rd,(] && \zxN{}\ar[l] \ar[rd,s] &&&  \\
        && \zxX{}\ar[dd] \ar[uuu,3d above,very thick]  &&& \zxN{}&& \zxX{} \ar[dd] \ar[uuu,3d above,very thick] && \zxX{}\ar[dd] \ar[uuu,3d above,very thick] && \zxN{} \\
        && &&& && && && \\
        && &&& && && && \\
    \end{ZX}
\end{align}
where $T$ is the translation operator, which shifts ${j\rightarrow j+1}$. More details of the proof are in \cref{eq:proof-kw-sqaure}.\footnote{\begin{ZX}
    &&\zxZ{} \ar[ll,->,>=stealth,pos=0.2] \ar[rr,<-,>=stealth,pos=0.5] \ar[rdd,(,<-,>=stealth,pos=0.5] &&\\
    &&&&\\
    &&&&\\
\end{ZX}
gives the regular representation of the associated Hopf algebra $\aso{H}$, which is isomorphic to $\dl{H}$.}
Further simplification of \cref{eq:kwsquare} gives
\begin{equation}
    T^{-1} (\kw)^2
    =|H|\begin{ZX}
        && &&& && && && \\
        \zxZTri'{} \ar[rd,(]&& &&& && && &&\\
        &\zxZ{} \ar[rd,(] \ar[rr] &&&... && \zxZ{}\ar[l]\ar[r] \ar[rd,(] && \zxN{}\ar[l] \ar[rd,s] &&&  \\
        && \zxX{}\ar[dd] \ar[uuu,3d above,very thick]  &&& \zxN{}&& \zxX{} \ar[dd] \ar[uuu,3d above,very thick] && \zxX{}\ar[dd] \ar[uuu,3d above,very thick] && \zxN{} \\
        && &&& && && && \\
        && &&& && && && \\
    \end{ZX} 
    \stackrel{\eqref{eq:assoH}}{=}
    |H|\begin{ZX}
        &&&&&&&&&&\\
        &&\zxX{} \ar[u] && \zxX{} \ar[u] &&&& \zxX{} \ar[u]&&\\
        & \zxZ{} \ar[ru,)]\ar[rr]&& \zxZ{}\ar[rr] \ar[ru,)]&& ... &&\zxN{} \ar[l]\ar[ru,s]&&&\\
        \zxZTri.{} \ar[ru,)]&& \ar[uu,3d above,very thick]&&\ar[uu,3d above,very thick]&&&&\ar[uu,3d above,very thick]&&\\
        &&&&&&&&&&\\
    \end{ZX}
    = |H| \Al^{\wc\haar} 
    = \sum_{\drep\in \Irr(\dl{H})} d_{\drep} \Al^{\drep}
    \,.
\end{equation}

Thus we conclude that applying $\kw$ twice projects onto the symmetric sector of $\Rep(\dl{H})$, followed by a lattice translation.

Borrowing the terminology used in the continuum, the operator $\kw$ represents a defect that is topological (freely deformable) in the time direction. This is a defect parallel to the spatial dimension, localized to a single time-slice. We can also define a defect for $\kw$ that lies parallel to the time direction, instead.
This is done by gauging the $\Rep(\dl{H}) $ symmetry on one half of an infinite chain, following the recipe discussed in \cref{sec:gaugingRepH*}. For concreteness we consider the self-dual Hamiltonian \eqref{eq:HsdH8}.
We then find the following defect Hamiltonian localized at the $(J-1,J)$ link:\footnote{The derivation largely follows the case of $\ZZ_2$ Kramers-Wannier discussed in Appendix B.2 of \Rf{shuheng2024KWlsm}.}
\ie 
\sH_{\kw}^{(J-1,J)} = -  \sum_{\drep \in \mathrm{Irr}(\dl{H})} g_\drep \left [
\sum_{j\neq J} \left (\ract{X}^{\dchi_\drep}_j +(\ract{Z}_{j-1}\lact{Z}_{j})_\op^{\Theta_{\dchi_\drep}} \right )
+ \FF^\dagger_{J} (\ract{Z}_{J-1} \lact{Z}_{J})_\op^{\Theta_{\dchi_\drep}} \FF_{J}
 \right ]
\fe 
The last term is depicted diagrammatically as
\tikzsetfigurename{defectHam}
\ie \label{eq:def-Ham-sd}
\FF^\dagger_{J} (\ract{Z}_{J-1} \lact{Z}_{J})_\op^{\Theta_{\dchi_\drep}} \FF_{J}
=
\tikzsetnextfilename{ZHZH}
\begin{ZX}
    &&&\\
    &&\zxZ{\Theta_{S\dchi_\drep}} \ar[d]&\\
    &&\zxX{}\ar[rdd,s]\ar[ld,s] &\\
    &\zxrep{$S$}\ar[ld,s]&&\zxHad{\tiny$\dagger$}\ar[uu]\dar\\
    \zxZ{}\ar[uuu]\ar[dd]&&&\zxZ{}\ar[u]\ar[d]\\
    \uar\dar &&&\zxHad{}\dar\\
    &&&\\[5pt]
    \zxbase{J-1}&&&\zxbase{J} 
\end{ZX} \ .
\fe 

One can check the defect movement operator is given by the unitary,
\tikzsetfigurename{defectMO}
\ie \label{eq:def-movement-op}
U_{J,J+1} =\FF_{J+1}^\dagger \ZXgate_{J,J+1}\FF_{J+1}\FF_J =  
\tikzsetnextfilename{UJ1}
\begin{ZX}
    &&\\
    \zxN{}&&\zxHad{\tiny$\dagger$}\ar[u]\\
    &&\zxX[]{}\ar[dd]\ar[u]\\
    \zxZ[]{}\ar[uuu]\ar[d]\ar[rru]&&\\
    \zxHad{}\ar[d]&&\zxHad{}\ar[d]\\
    &&\\[5 pt]
    \zxbase{J}&&\zxbase{J+1}
\end{ZX}
\,.
\fe
In the case ${H = \CC[\ZZ_2]}$, this reduces to $U_{J,J+1} = CZ_{J,J+1} H_J$ in agreement with~\cite{aasen2016duality,shuheng2024KWlsm} (see also \cite{SeibergShao23,Chen:2023qst,Tantivasadakarn:2025txn} which uses an equivalent unitary up to conjugation). The operator $U_{J,J+1}$ shifts the defect from the link $(J-1,J)$ to the link $(J,J+1)$,
i.e., ${\sH_{\kw}^{(J,J+1)} = U_{J,J+1} \sH_{\kw}^{(J,J+1)} U_{J,J+1}^\dagger}$.

This movement operator is intimately related to the sequential circuit discussed in \cref{eq:KW-seq-circ}. One can check that the $\kw$ MPO can be decomposed into a (unitary) sequential circuit on ${L+1}$ sites, followed by a projection on the ancilla qudit. Indeed, we find
\ie \label{eq:Dkw-seq-circ}
\kw = \prescript{}{0}{\bra{\cohaar}} \left( \prod_{j=0}^{L-1}  U_{j,j+1}^\dagger \right) \left( \FF_L^\dagger \ZXgate_{L,0} \right) \ket{\unit}_0
\fe 
where the subscript $0$ represents the ancilla, which is initialized in the unit state and projected into the cointegral state. This agrees with the expression from the sequential circuit of the gauging map in \cref{eq:KW-seq-circ} after applying the onsite basis transformation with $\FF^\dagger$.

One can interpret \cref{eq:Dkw-seq-circ} as pair-creating a defect and an anti-defect, moving one of them around the periodic chain, and finally annihilating them. The non-invertibility of the symmetry requires that we need to project onto the vacuum fusion channel.

\subsection{MPO fusion and \texorpdfstring{$F$}{F}-symbols}\label{sec:fusionFsym}
Fusion rules of non-invertible symmetry MPOs are obtained by composing the MPOs on a periodic chain. If the MPOs admit local fusion in a finite region (the ``on-site'' condition), the three-leg fusion/splitting tensors yield lattice $F$-symbols, encoding associativity and determining the fusion category~\cite{jose2023mpoclas,kansei2024nispt,Meng:2024nxx}, which applies for $\Rep(\dl H)$ symmetry MPOs. For non–``onsite'' MPOs one can still extract fusion/splitting tensors
via a weaker zipping rule~\cite{lu2025strange}, but reconstructing $F$-symbols is generally subtler; for instance, the Frobenius–Schur indicator of the Kramers-Wannier duality MPO cannot be fixed.

\paragraph{\texorpdfstring{$\Rep(\dl H)$}{Rep(H*)} fusion tensors and \texorpdfstring{$F$}{F}-symbols}
The $\Rep(\dl H)$ symmetry MPO tensors are of the form
\tikzsetfigurename{Atensor}
\begin{equation}
    \begin{ZX}
        &&&\\
        &&\zxX{} \ar[u]&\\
        &\zxrep{$r$} \ar[l,blue] \ar[rr,blue] \ar[ru,)] &&\\
        &&\zxN{} \ar[uu,3d above, very thick]&
    \end{ZX}
\end{equation}
where $r\in \Irr(\dl H)$. Since the bond dimension is equal to the dimension of the irrep $d_r$, (which is the same as its quantum dimension as an object of $\Rep(H)$) the MPO is ``on-site''~\cite{Meng:2024nxx}. The fusion of two such MPO tensors is as follows~\cite{jose2023mpoclas,kansei2024nispt,Meng:2024nxx}:
\tikzsetfigurename{MPOdecompose}
\begin{equation}\label{eq:MPOdecompose}
    \begin{tikzpicture}
        \tikzset{baseline=(current bounding box.center)}
        % Draw horizontal arrows
        \draw[thick, black, ->-=.3,->-=0.8] (-1,0.5) -- (1,0.5);
        \draw[thick, black, ->-=.3,->-=0.8] (-1,-0.5) -- (1,-0.5);
        % Draw vertical arrows
        \draw[thick,gray, ->-=0.55,->-=0.12,->-=0.95] (0,-1) -- (0,1);
        % Draw nodes
        \filldraw[fill=gray!30, draw=black, thick] (0,0.5) circle (0.15);
        \filldraw[fill=gray!30, draw=black, thick] (0,-0.5) circle (0.15);
        % Labels
        \node at (0.6,0.8) {$\mathcal{O}_x$};
        \node at (0.6,-0.2) {$\mathcal{O}_y$};
    \end{tikzpicture}  = \sum_{z\in \Irr(\cC)} \sum_{1\le\mu \le N_{xy}^z} \ \begin{tikzpicture}
        \tikzset{baseline=(current bounding box.center)}
        % Draw horizontal arrows
        \draw[thick, black, ->-=.3,->-=0.8] (-1,0) -- (1,0);
        % Draw vertical arrows
        \draw[thick,gray, ->-=0.25,->-=0.95] (0,-0.5) -- (0,0.5);
        % Draw nodes
        \filldraw[fill=gray!30, draw=black, thick] (0,0) circle (0.15);
        \draw[thick, black] (1,0) -- (1,0.5);
        \draw[thick, black, ->-=.6] (1,0.5) -- (1.5,0.5);
        \draw[thick, black] (1,0) -- (1,-0.5);
        \draw[thick, black, ->-=.6] (1,-0.5) -- (1.5,-0.5);
        %left
        \draw[thick, black] (-1,0) -- (-1,0.5);
        \draw[thick, black, -<-=.6] (-1,0.5) -- (-1.5,0.5);
        \draw[thick, black] (-1,0) -- (-1,-0.5);
        \draw[thick, black, -<-=.6] (-1,-0.5) -- (-1.5,-0.5);
        % Labels
        \filldraw[fill=white, draw=black, thick] (1,0) circle (0.08);
        \filldraw[fill=black, draw=black, thick] (-1,0) circle (0.08);
        \node at (0.6,0.3) {$\mathcal{O}_z$};
        \node[left] at (-1.0,0.0) {$(\phi_{xy}^z)_\mu$};   \node[right] at (1.0,0.0) {$(\bar{\phi}_{xy}^z)_\mu$};
        \node at (1,-0.7) {};
    \end{tikzpicture}
\end{equation}
where $x,y,z$ are simple objects in the fusion category $\cC$, and $N_{xy}^z$ is the fusion structure constant. $(\phi_{xy}^z)_\mu$ and $(\bbphi_{xy}^z)_\mu$ are fusion and split 3-leg tensors respectively. In particular, the fusion of two $\Rep(\dl H)$ symmetry MPO tensors then has the decomposition,
\tikzsetfigurename{Atensorfuse}
\begin{equation}\label{eq:RepHfusion}
    \begin{ZX}
        &&&\\
        &&\zxX{} \ar[u]&\\
        &\zxrep{$r_i$} \ar[l,blue] \ar[rr,blue] \ar[ru,)] &&\\
        &&\zxX{} \ar[uu,3d above, very thick]&\\
        &\zxrep{$r_j$} \ar[l,blue] \ar[rr,blue] \ar[ru,)] &&\\
        &&\zxN{} \ar[uu,3d above, very thick]&
    \end{ZX} 
    =
    \begin{ZX}
        &&&&\\
        &&&&\zxX{}\ar[u]\\
        &&&\zxX{} \ar[ru,)]&\\
        &\zxrep{$r_i$} \ar[l,blue] \ar[rr,blue] \ar[rru,)] &&&\\
        &&&&\\
        &\zxrep{$r_j$} \ar[l,blue] \ar[rr,blue] \ar[rruuu,3d above, very thick,(] &&&\\
        &&&&\zxN{} \ar[uuuuu,3d above, very thick]
    \end{ZX}
    =
    \sum_{k,\mu}\begin{ZX}
        &&&&&\\
        &&&\zxX{} \ar[u]&&\\
        &\zxrep{$T_{ij}^{k,\mu}$}\ar[lu,(,blue] \ar[ld,),blue]&\zxrep{$r_k$} \ar[l,blue] \ar[rr,blue] \ar[ru,)] &&\zxrep{$(T^\vee)_{ij}^{k,\mu}$}\ar[ru,),blue] \ar[rd,(,blue]&\\
        &&&\zxN{} \ar[uu,3d above, very thick]&&
    \end{ZX}
\end{equation}
where $r_k \subset r_i \otimes r_j$ and $\mu$ is the multiplicity of $r_k$ inside the decomposition. $T_{ij}^{k,\mu}$ and $(T^\vee)_{ij}^{k,\mu}$ are the fusion and split tensors and they further satisfy the orthogonality and completeness conditions,
\tikzsetfigurename{AtensorOrth}
\begin{equation}
    \begin{ZX}
        &&&&\\
        &\zxrep{$(T^\vee)_{ij}^{k,\mu}$}\ar[rr,C'=0.5,blue] \ar[rr,C.=0.5,blue] \ar[l,blue]&&\zxrep{$T_{ij}^{k',\mu'}$} \ar[r,blue]&\\
        &&&&
    \end{ZX} = \delta_{k,k'} \delta_{\mu,\mu'} \begin{ZX}
    \zxN{} \ar[rrr,blue] &&& \zxN{}\end{ZX} ,\quad
     \sum_{k,\mu}\begin{ZX}
        &&&&&\\
        &&&&&\\
        &\zxrep{$T_{ij}^{k,\mu}$}\ar[lu,(,blue] \ar[ld,),blue] \ar[rrr,blue]&&&\zxrep{$(T^\vee)_{ij}^{k,\mu}$}\ar[ru,),blue] \ar[rd,(,blue]&\\
        &&&&&
    \end{ZX} = \begin{ZX}
        \zxN{} \ar[rrrrr,blue] &&&&&\\
                \zxN{}  &&&&&\\
                        \zxN{}  &&&&&\\
                                \zxN{} \ar[rrrrr,blue] &&&&&
    \end{ZX}
\end{equation} 
The $T_{ij}^{k,\mu}$ is the intertwiner between the tensor product of irreps $r_i,r_j$ and irrep $r_k$. Then the $F$ symbols are given by
\begin{equation}
    T_{ij}^{m,\mu} T_{mk}^{l,\nu} = F^{ijk}_{l,(m,\mu,\nu),(n,\rho,\sigma)} T_{jk}^{n,\rho} T_{in}^{l,\sigma}.
\end{equation}
Since the tensor $T_{ij}^{k,\mu}$ depends on specific matrix form of the representations, the $F$ symbols are basis dependent.

\paragraph{Fusion of \texorpdfstring{$\Rep(\dl H)$}{Rep(H*)} and \texorpdfstring{$\kw$}{Dkw} tensors}
The $\kw$ MPO is not ``on-site'', since its quantum dimension $\sqrt{|H|}$ is not equal to the bond dimension $|H|$. In general, for the non-``on-site'' MPO realizations, the decomposition of MPO fusion is no longer valid, as the fusion/split 3-leg tensors do not form a complete basis, and one needs to use the alternative ``zipping condition'' to find the fusion tensors \cite{lu2025strange}. However, in the current case, the \eqref{eq:MPOdecompose} still holds, because $\Al^\drep\times \kw =d_r \kw$, the fusion outcome is still $\kw$, so effectively we only need to decompose the $\Rep(\dl H)$ part which is realized in an ``on-site'' way.

We first consider the fusion between the MPO tensors of a $\Rep(\dl H)$ element, $\Al^\drep$, and $\kw$.
\tikzsetfigurename{Akwtensorfuse}
\begin{equation}
    \begin{ZX}
        &&&&\\
        &&&\zxX{} \ar[u]&\\
        &&\zxrep{$r$} \ar[ll,blue] \ar[rr,blue] \ar[ru,)] &&\\
        &&&\zxHad{\tiny$\dagger$} \ar[uu,3d above, very thick]&\\
        &&&\zxX{} \ar[u] \ar[rd,)]&\\
        &&\zxrep{$S$} \ar[ru,)]&&\\
        &\zxZ{} \ar[d] \ar[lu,)] \ar[ru,(]&&&\\
        &&&&
    \end{ZX} = \begin{ZX}
        &&&&&&\\
        &\zxrep{$\Theta_r$} \ar[l,blue] \ar[rrrr,blue] &&&&\zxrep{$\Theta_r$} \ar[r,blue] &\\
        &\zxrep{$S$}\ar[u]&&&\zxHad{\tiny$\dagger$} \ar[uu,3d above, very thick]&&\\
        &&&&\zxX{} \ar[u] \ar[rd]&&\\
        &\zxZ{} \ar[lu] \ar[uu]&&\zxrep{$S$} \ar[ru,)]&&\zxZ{} \ar[rd] \ar[uuu]&\\
        &&\zxZ{} \ar[d] \ar[lu] \ar[ru,(]&&&&\\
        &&&&&&
    \end{ZX} = \sum_\mu  \begin{ZX}
    &&&&&&\\
    &\zxrep{$\Theta_r$} \ar[l,blue] \ar[r,blue] &\zxend{$\mu$}&\zxend{$\mu$}&&\zxrep{$\Theta_r$} \ar[r,blue] \ar[ll,blue]&\\
    &\zxrep{$S$}\ar[u]&&&\zxHad{\tiny$\dagger$} \ar[uu,3d above, very thick]&&\\
    &&&&\zxX{} \ar[u] \ar[rd]&&\\
    &\zxZ{} \ar[lu] \ar[uu]&&\zxrep{$S$} \ar[ru,)]&&\zxZ{} \ar[rd] \ar[uuu]&\\
    &&\zxZ{} \ar[d] \ar[lu] \ar[ru,(]&&&&\\
    &&&&&&
    \end{ZX},
    \quad \text{where }\begin{ZX}
        &\zxrep{$\Theta_r$} \ar[l,blue] \ar[r,blue] \ar[d] &\\
        &&
    \end{ZX} \equiv \begin{ZX}
        &&\zxrep{$\Theta$} \ar[ld,(] \ar[rd,)]&\\
        &\zxrep{$r$}\ar[l,blue] \ar[r,blue]&&
    \end{ZX} 
\end{equation}
We find the fusion tensor,
\tikzsetfigurename{Akwftensor}
\begin{equation} \label{eq:r-kw-fusion-tensor}
\begin{tikzpicture}
    \tikzset{baseline=(current bounding box.center)}
    \draw[thick, black, ->-=.7] (-0.5,0.25) -- (0,0.25);
    \draw[thick, black, ->-=.7] (-0.5,-0.25) -- (0,-0.25);
    \draw[thick, black] (0,0.25) -- (0,-0.25);
    \draw[thick, black, ->-=.7] (0,0) -- (0.5,0);
    \filldraw[fill=black, draw=black, thick] (0,0) circle (0.05);
    \node[left] at (-0.5,0.25) {$r$};
    \node[left] at (-0.5,-0.25) {$\kw$};
    \node[left] at (0,0) {$\mu$};
    \node[right] at (0.5,0) {$\kw$};
\end{tikzpicture} \equiv    \begin{ZX}
        &\zxrep{$\Theta_r$} \ar[l,blue] \ar[r,blue] \ar[d] &  \zxend{$\mu$}\\
        &\zxrep{$S$} \ar[d]&\\
        &\zxZ{} \ar[lu] \ar[rd]&\\
        &&\zxN{}
    \end{ZX} = \begin{ZX}
    \zxend{$\mu$}&\zxrep{$\Theta_r$} \ar[l,blue] \ar[r,blue] \ar[dd] &  \\
        &&\\
        &\zxZ{} \ar[lu] \ar[rd]&\\
        &&\zxN{}
    \end{ZX}
\end{equation}
such that the zipping condition is satisfied, where we used $\Gamma(S(\cdot)) = \Gamma(\cdot)^\intercal$ for $\Gamma \in \Irr(H)$. 
Similarly, fusing these MPO tensors in the reverse order, $\kw \times \Al^\drep$, we get
\begin{equation}
    \begin{ZX}
        &&&&&&\\
        &&&&\zxHad{\tiny$\dagger$} \ar[u,3d above, very thick]&&\\
        &&&&\zxX{} \ar[u] \ar[rd,)]&&\\
        &&&\zxrep{$S$} \ar[ru,)]&&&\\
        &&\zxZ{}  \ar[lu,)] \ar[ru,(]&&&&\\
        &&\zxX{} \ar[u]&&&&\\
        &\zxrep{$r$} \ar[l,blue] \ar[rr,blue] \ar[ru,)]&& &&&\\
        &&\zxN{} \ar[uu,3d above, thick]&&&&
    \end{ZX}  =\begin{ZX}
        &&&&&&\\
        &&&&\zxHad{\tiny$\dagger$} \ar[u,3d above, very thick]&&\\
        &&&&\zxX{} \ar[u] \ar[rd]&&\\
        &\zxX{} \ar[lu]&&\zxrep{$S$} \ar[ru,)]&&\zxX{} \ar[rd]&\\
        &&\zxZ{}  \ar[lu] \ar[ru,(]&&&\zxrep{$S$} \ar[u]&\\
        &&&&&&\\
        &\zxrep{$r$} \ar[uuu] \ar[l,blue] \ar[rrrr,blue]&& &&\zxrep{$r$} \ar[uu] \ar[r,blue]&\\
        &&\zxN{} \ar[uuu,3d above, thick]&&&&
    \end{ZX}  =\sum_\nu\begin{ZX}
    &&&&&&\\
    &&&&\zxHad{\tiny$\dagger$} \ar[u,3d above, very thick]&&\\
    &&&&\zxX{} \ar[u] \ar[rd]&&\\
    &\zxX{} \ar[lu]&&\zxrep{$S$} \ar[ru,)]&&\zxX{} \ar[rd]&\\
    &&\zxZ{}  \ar[lu] \ar[ru,(]&&&\zxrep{$S$} \ar[u]&\\
    &&&&&&\\
    &\zxrep{$r$} \ar[uuu] \ar[l,blue] \ar[rr,blue]&&\zxend{$\nu$} &\zxend{$\nu$}&\zxrep{$r$} \ar[l,blue] \ar[uu] \ar[r,blue]&\\
    &&\zxN{} \ar[uuu,3d above, thick]&&&&
    \end{ZX}  
\end{equation}
The corresponding fusion tensor is
\tikzsetfigurename{kwAfusiontensor}
\begin{equation}\label{eq:kw-r-fusion-tensor}
\begin{tikzpicture}
    \tikzset{baseline=(current bounding box.center)}
    \draw[thick, black, ->-=.7] (-0.5,0.25) -- (0,0.25);
    \draw[thick, black, ->-=.7] (-0.5,-0.25) -- (0,-0.25);
    \draw[thick, black] (0,0.25) -- (0,-0.25);
    \draw[thick, black, ->-=.7] (0,0) -- (0.5,0);
    \filldraw[fill=black, draw=black, thick] (0,0) circle (0.05);
    \node[left] at (-0.5,0.25) {$\kw$};
    \node[left] at (-0.5,-0.25) {$r$};
    \node[left] at (0,0) {$\nu$};
    \node[right] at (0.5,0) {$\kw$};
\end{tikzpicture} \equiv    \begin{ZX}
        &&\\
        &\zxX{} \ar[dd] \ar[rd] \ar[lu]&\\
        &&\\
        &\zxrep{$r$} \ar[l,blue] \ar[r,blue]  &  \zxend{$\nu$}\\
        &&
    \end{ZX} \ .
\end{equation}
Notably, the associativity of MPO fusion is no longer a strict equality but an isomorphism given by the $F$-symbol. 

On the lattice, the $F$-symbol can be obtained by comparing two sequences of fusion, in particular, $(\Al^{r_i}\times \kw) \times \Al^{r_j}$, $\Al^{r_i}\times (\kw \times \Al^{r_j}) $ differ by the $F$-symbol, 
\begin{equation}
\begin{tikzpicture}
    \tikzset{baseline=(current bounding box.center)}
    \draw[thick, black, ->-=.7] (-1.5,0.5) -- (-0.5,0.5);
    \draw[thick, black, ->-=.7] (-1.5,-0.5) -- (-0.5,-0.5);
    \draw[thick, black] (-0.5,-0.5) -- (-0.5,0.5);
    \draw[thick, black, ->-=.7] (-0.5,0) -- (0.5,0);
    \draw[thick, black, ->-=.7] (-1.5,-1) -- (0.5,-1);
    \draw[thick, black] (0.5,0) -- (0.5,-1);
    \draw[thick, black, ->-=.7] (0.5,-0.5) -- (1.5,-0.5);
    \filldraw[fill=black, draw=black, thick] (-0.5,0) circle (0.08);
    \filldraw[fill=black, draw=black, thick] (0.5,-0.5) circle (0.08);
    \node[left] at (-1.5,0.5) {$r_i$};
    \node[left] at (-1.5,-0.5) {$\kw$};
    \node[left] at (-1.5,-1) {$r_j$};
    \node[left] at (-0.5,0) {$\mu$};
    \node[left] at (0.5,-0.5) {$\nu$};
    \node[above] at (0,0) {$\kw$};
    \node[above] at (1.2,-0.5) {$\kw$};
\end{tikzpicture}  = \ \sum_{v,\rho,\sigma} (F_{\kw}^{r_i \kw r_j})_{(\kw,\mu,\nu),(\kw,\rho,\sigma)} \begin{tikzpicture}
\tikzset{baseline=(current bounding box.center)}
\draw[thick, black, ->-=.7] (-1.5,0.5) -- (0.5,0.5);
\draw[thick, black, ->-=.7] (-1.5,0) -- (-0.5,0);
\draw[thick, black] (-0.5,0) -- (-0.5,-1);
\draw[thick, black, ->-=.7] (-0.5,-0.5) -- (0.5,-0.5);
\draw[thick, black, ->-=.7] (-1.5,-1) -- (-0.5,-1);
\draw[thick, black] (0.5,0.5) -- (0.5,-0.5);
\draw[thick, black, ->-=.7] (0.5,0) -- (1.5,0);
\filldraw[fill=black, draw=black, thick] (-0.5,-0.5) circle (0.08);
\filldraw[fill=black, draw=black, thick] (0.5,0) circle (0.08);
\node[left] at (-1.5,0.5) {$r_i$};
\node[left] at (-1.5,0) {$\kw$};
\node[left] at (-1.5,-1) {$r_j$};
\node[left] at (-0.5,-0.5) {$\rho$};
\node[left] at (0.5,0) {$\sigma$};
\node[below] at (0,-0.5) {$\kw$};
\node[above] at (1.2,0) {$\kw$};
\end{tikzpicture}
\end{equation}
where
\begin{equation}\label{eq:bicharacterF}
    F^{r_i \kw r_j}_{\kw,(\kw,\mu,\nu),(\kw,\rho,\sigma)} = \Theta((r_i)_{\mu, \rho},(r_j)_{\sigma,\nu})
\end{equation}
which reduce to the ordinary bicharacter for the Tambara-Yamagami fusion category for the abelian group case, i.e., when ${H=\CC[G]}$ for an abelian group $G$. To be specific,
\tikzsetfigurename{Fmove}
\begin{align}
    \begin{tikzpicture}
        \tikzset{baseline=(current bounding box.center)}
        \draw[thick, black, ->-=.7] (-1.5,0.5) -- (-0.5,0.5);
        \draw[thick, black, ->-=.7] (-1.5,-0.5) -- (-0.5,-0.5);
        \draw[thick, black] (-0.5,-0.5) -- (-0.5,0.5);
        \draw[thick, black, ->-=.7] (-0.5,0) -- (0.5,0);
        \draw[thick, black, ->-=.7] (-1.5,-1) -- (0.5,-1);
        \draw[thick, black] (0.5,0) -- (0.5,-1);
        \draw[thick, black, ->-=.7] (0.5,-0.5) -- (1.5,-0.5);
        \filldraw[fill=black, draw=black, thick] (-0.5,0) circle (0.08);
        \filldraw[fill=black, draw=black, thick] (0.5,-0.5) circle (0.08);
        \node[left] at (-1.5,0.5) {$r_i$};
        \node[left] at (-1.5,-0.5) {$\kw$};
        \node[left] at (-1.5,-1) {$r_j$};
        \node[left] at (-0.5,0) {$\mu$};
        \node[left] at (0.5,-0.5) {$\nu$};
        \node[above] at (0,0) {$\kw$};
        \node[above] at (1.2,-0.5) {$\kw$};
    \end{tikzpicture}  = \begin{ZX}
        \zxend{$\mu$} &\zxrep{$\Theta_{r_i}$} \ar[l,blue] \ar[rr,blue] \ar[dd] &&  \\
        &&&\\
        &\zxZ{} \ar[lu] \ar[rd]&&\\
        &&\zxX{} \ar[rd] \ar[dd]&\\ 
        &&&\\
        &&\zxrep{$r_j$} \ar[ll,blue] \ar[r,blue]  &\zxend{$\nu$} 
    \end{ZX} = \begin{ZX}
        \zxend{$\mu$} \ar[rr,blue]& & \zxrep{$\Theta_{r_i}$} \ar[r,blue]  &\zxrep{$\Theta_{r_i}$} \ar[r,blue]& \\
        \zxN{} \ar[rd] & & & & \\
        & \zxX{} \ar[rrd] \ar[ddd]& && \\
        &&&\zxZ{}\ar[rd] \ar[uuu] &\\
        &&&&\\
        \zxN{}\ar[r,blue] & \zxrep{$r_j$} \ar[r,blue] &\zxrep{$r_j$} \ar[uuuuu,3d above, very thick] \ar[rr,blue] &&\zxend{$\nu$} 
    \end{ZX} \nonumber \\ 
    = \sum_{\rho,\sigma}\begin{ZX}
        \zxend{$\mu$} \ar[r,blue] & \zxrep{$\Theta_{r_i}$} \ar[r,blue] & \zxend{$\rho$} \\
        \zxend{$\sigma$} \ar[r,blue] & \zxrep{$ r_j$} \ar[r,blue] \ar[u] & \zxend{$\nu$} \\
    \end{ZX}\begin{ZX}
        \zxend{$\rho$} &&\zxrep{$\Theta_{r_i}$} \ar[ll,blue] \ar[r,blue] \ar[ddd] &  \\
        &&&\\
        &\zxX{} \ar[lu] \ar[rd] \ar[ddd]&&\\
        &&\zxZ{} \ar[rd] &\\ 
        &&&\\
        &\zxrep{$r_j$} \ar[l,blue] \ar[rr,blue] & &\zxend{$\sigma$} 
    \end{ZX} =\sum_{\rho,\sigma} \begin{ZX}
    &\zxrep{$\Theta$} \ar[ld,(] \ar[rd,)]& \\
    \zxrep{$(r_i)_{\mu\rho}$} &&\zxrep{$(r_j)_{\sigma\nu}$} 
    \end{ZX}\begin{tikzpicture}
    \tikzset{baseline=(current bounding box.center)}
    \draw[thick, black, ->-=.7] (-1.5,0.5) -- (0.5,0.5);
    \draw[thick, black, ->-=.7] (-1.5,0) -- (-0.5,0);
    \draw[thick, black] (-0.5,0) -- (-0.5,-1);
    \draw[thick, black, ->-=.7] (-0.5,-0.5) -- (0.5,-0.5);
    \draw[thick, black, ->-=.7] (-1.5,-1) -- (-0.5,-1);
    \draw[thick, black] (0.5,0.5) -- (0.5,-0.5);
    \draw[thick, black, ->-=.7] (0.5,0) -- (1.5,0);
    \filldraw[fill=black, draw=black, thick] (-0.5,-0.5) circle (0.08);
    \filldraw[fill=black, draw=black, thick] (0.5,0) circle (0.08);
    \node[left] at (-1.5,0.5) {$r_i$};
    \node[left] at (-1.5,0) {$\kw$};
    \node[left] at (-1.5,-1) {$r_j$};
    \node[left] at (-0.5,-0.5) {$\rho$};
    \node[left] at (0.5,0) {$\sigma$};
    \node[below] at (0,-0.5) {$\kw$};
    \node[above] at (1.2,0) {$\kw$};
    \end{tikzpicture}
\end{align}
where we used the property of the representation as algebra homomorphism and inserted the resolution of identity \tikzsetfigurename{identityresolve}$\begin{ZX}
    \zxN{} \ar[rr,blue] && \zxN{}
\end{ZX} = \sum_\rho \begin{ZX}
    \zxN{}\ar[r,blue] & \zxend{$\rho$} &\zxend{$\rho$}\ar[r,blue]&\zxN{}
\end{ZX}$ in the last equation.

We close this subsection with a consistency check. Since $\kw$ implements a duality, its fusion with itself should include the identity. Now, consider the following sequence of $F$-moves:

 \tikzsetfigurename{fsymbol}
\begin{align}\label{eq:argsymmetricTheta}
&\begin{ZX}
    i&\zxN{} \ar[rd,),red] &&&&&&\\
    \DD&\zxN{} \ar[rrr]&\mu&\nu&\zxN{} \ar[rd,)]&&& \\ 
    j&\zxN{} \ar[r,red]& \zxN{} \ar[ru,(,red]&&&\zxN{} \ar[rr,dashed]&&\zxN{}\\
    \DD &\zxN{} \ar[rrr]&&&\zxN{} \ar[ru,(]&&&\\
    &&&&&&&
\end{ZX} = F_{\DD,(\DD,\mu,\nu),(\DD,\rho,\sigma)}^{i\DD j} \begin{ZX}
i&\zxN{} \ar[r,red] &\zxN{}  \ar[rd,),red]&&&&&\\
\DD&\zxN{} \ar[rrr]&\rho&\sigma&\zxN{} \ar[rd,)]&&& \\ 
j&\zxN{} \ar[ru,(,red]& &&&\zxN{} \ar[rr,dashed]&&\zxN{}\\
\DD &\zxN{} \ar[rrr]&&&\zxN{} \ar[ru,(]&&&\\
&&&&&&&
\end{ZX}= F_{\DD,(\DD,\mu,\nu),(\DD,\rho,\sigma)}^{i\DD j} \begin{ZX}
i&\zxN{} \ar[rrrr,red] &&&&\zxN{}  \ar[ddddd,C-=0.5,red]&&\\
\DD&\zxN{} \ar[rrr]&&&\zxN{} \ar[rd,)]&&& \\ 
j&\zxN{} \ar[r,red]&\zxN{}\ar[rd,),red] &&&\zxN{} \ar[rr,dashed]&&\zxN{}\\
\DD &\zxN{} \ar[rrr]&\sigma&\rho&\zxN{} \ar[ru,(]&&&\\
&\zxN{} \ar[ru,(,red]&&&&&&\\
&\zxN{}\ar[u,C,red]&&&&\zxN{} \ar[llll,red]&&
\end{ZX}\\
&= F_{\DD,(\DD,\mu,\nu),(\DD,\rho,\sigma)}^{i\DD j} (F^{-1})_{\DD,(\DD,\sigma,\rho),(\DD,\nu,\mu)}^{j\DD i}\begin{ZX}
i&\zxN{} \ar[rrrr,red] &&&&\zxN{}  \ar[ddddd,C-=0.5,red]&&\\
\DD&\zxN{} \ar[rrr]&&&\zxN{} \ar[rd,)]&&& \\ 
j&\zxN{} \ar[rd,),red]& &&&\zxN{} \ar[rr,dashed]&&\zxN{}\\
\DD &\zxN{} \ar[rrr]&\nu&\mu&\zxN{} \ar[ru,(]&&&\\
&\zxN{} \ar[r,red] &\zxN{} \ar[ru,(,red]&&&&&\\
&\zxN{}\ar[u,C,red]&&&&\zxN{} \ar[llll,red]&&
\end{ZX} = F_{\DD,(\DD,\mu,\nu),(\DD,\rho,\sigma)}^{i\DD j} (F^{-1})_{\DD,(\DD,\sigma,\rho),(\DD,\nu,\mu)}^{j\DD i} \begin{ZX}
i&\zxN{} \ar[rd,),red] &&&&&&\\
\DD&\zxN{} \ar[rrr]&\mu&\nu&\zxN{} \ar[rd,)]&&& \\ 
j&\zxN{} \ar[r,red]& \zxN{} \ar[ru,(,red]&&&\zxN{} \ar[rr,dashed]&&\zxN{}\\
\DD &\zxN{} \ar[rrr]&&&\zxN{} \ar[ru,(]&&&\\
&&&&&&&
\end{ZX} \nonumber
\end{align}
Here, ${i,j\in \Rep(\dl H)}$ are topological, i.e., their defects can be moved by unitary operators,
% where we use the fact that $i,j\in \Rep(\dl H)$ are topological and whose defects can be moved by unitary operators. 
and we have made use of the fusion tensors in \cref{eq:r-kw-fusion-tensor,eq:kw-r-fusion-tensor}.
The equality above implies that
\begin{align}
&F_{\DD,(\DD,\mu,\nu),(\DD,\rho,\sigma)}^{i\DD j} (F^{-1})_{\DD,(\DD,\sigma,\rho),(\DD,\nu,\mu)}^{j\DD i}  = 1 \Rightarrow  \Theta((r_i)_{\mu, \rho},(r_j)_{\sigma,\nu})  \Theta^{-1}((r_j)_{\sigma, \nu},(r_i)_{\mu,\rho}) = 1 \nonumber\\
& \Rightarrow  \Theta((r_i)_{\mu, \rho},(r_j)_{\sigma,\nu}) =  \Theta((r_j)_{\sigma, \nu},(r_i)_{\mu,\rho}) \,.
\end{align}
This is consistent with our assumption, in defining $\kw$, that the Hadamard form $\Theta$ is symmetric. This diagrammatic manipulation is also reminscent of the proof in the original Tambara-Yamagami fusion category of abelian group with bicharacter being symmetric, where the explicit pentagon equations are solved~\cite{tambara1998tensor}. When the Hadamard form $\Theta$ is non-symmetric, the associated $\kw$ operator can no longer implement a duality (see related discussion in \cref{footnote:symm-Had}). Since we expect the construction of $\kw$ to go through for any self-dual finite-dimensional Hopf $C^\star$-algebra, we conjecture that any such Hopf algebra must admit a symmetric Hadamard form.

\subsection{Categorical description of the self-duality symmetry}
\label{sec:categorical-selfduality}

As noted previously, the MPO \eqref{eq:dualityop} generalizes Kramers--Wannier duality: when $H$ is a self-dual Hopf algebra, it implements the self-duality obtained by gauging the $\Rep(H)$ symmetry, followed by a Hadamard isomorphism (since we are in the continuum, we drop the distinction between $\Rep(H)$ and $\Rep(\dl{H}))$ in this subsection). Under the renormalization group flow, if the lattice translation flows to the identity in the infrared (IR) field theory, then the lattice $\kw$ symmetry MPO flows to the $\cD$ line in the $\ZZ_2$ extension of the $\Rep(\dl{H})$ fusion category. The fusion rules of the latter fusion category are 
\begin{equation}\label{eq:Ez2RepH}
    \cD \times \cD = \sum_{a\in \Irr(H)} d_a \cL_a,\quad \cD\times \cL_a = \cL_a \times \cD = d_a \cD,\quad \cL_a\times \cL_b = \sum_{c} N_{ab}^c \cL_c
\end{equation}
where $d_a$ is the quantum dimension of ${a\in \Irr(H))}$, and $N_{ab}^c$ follows from \eqref{eq:RepHd-fusion}. 
The fusion rules \eqref{eq:Ez2RepH} realize a particular $\ZZ_2$-extension of $\Rep(H)$ in which the nontrivial $\ZZ_2$-graded component consists of a single simple object of quantum dimension $\sqrt{|H|}$. 
For these fusion rules, there are (in general) multiple gauge-inequivalent associators ($F$-symbols), yielding distinct fusion categories. In general, solving the pentagon equations directly is difficult, since $\Rep(H)$ contains simple objects with quantum dimension larger than $1$. Consequently, the $\ZZ_2$-extended fusion category is not multiplicity-free; for example, ${\cD \times \cL_a = d_a \cD}$. 
For self-dual Hopf algebras of small order (e.g., the 8-dimensional Kac--Paljutkin algebra $H_8$), the $F$-symbols can be obtained by solving the pentagon equations explicitly~\cite{choi2023self}. It is then natural to consider how the lattice self-duality operator \eqref{eq:dualityop} matches with the duality defect in the IR fusion category symmetry. 

According to \Rf{lu2025sdset}, the $\ZZ_2$-extensions of $\Rep(H)$ fusion categories are classified by the different $\ZZ_2$ symmetry-enriched topological orders of ${\Rep(D(H))\cong Z(\Rep(H))}$. In particular, there are 3 layers of data: 
\begin{enumerate}
    \item   ${\rho: \ZZ_2^\text{em}\rightarrow \Aut(\Rep(D(H)))}$, which represents a braided autoequivalence, permuting certain ``electric'' anyons with ``magnetic'' anyons.
    \item $H^2_{[\rho]}(\ZZ_2^{\text{em}},\mathcal{A})$, which represents a symmetry fractionalization class, valued in the abelian anyons $\mathcal{A}$ within $\Rep(D(H))$ .
    \item  $H^3(\ZZ_2^{\text{em}},U(1))$, the Frobenius-Schur (FS) indicator of the duality defect.
\end{enumerate}
The first two layers of data can be matched with the lattice self-duality symmetry operator \eqref{eq:dualityop}, consistent with the $F$ symbols discussed in \secref{sec:fusionFsym}. The FS indicator is not well-defined for the lattice symmetry operator since it is not self-adjoint ${\kw^\dagger = T^{-1} \kw \neq \kw}$, where $T$ is the operator implementing translation by one site to the right. 
Nevertheless, for any given microscopic Hamiltonian, one can discuss the FS indicator of the IR symmetry that $\kw$ flows to. The choice of lattice Hamiltonian would affect the IR fate of the translation operator, which in turn affects the FS indicator of the symmetry emanating from $\kw$ in the IR~\cite{shuheng2024KWlsm}.

Before discussing the $\ZZ_2$ extension of more general $\Rep(H)$ categories, let us first discuss the case of Tambara-Yamagami categories $\TY(A,\chi,\epsilon)$, where $H$ is taken to be the group algebra of an abelian group $A$. The $\TY(A,\chi,\epsilon)$ fusion category describes the self-duality under gauging the abelian symmetry, described by $A$. 
Its fusion rules are given by
\begin{equation}
    \cN\times \cN = \sum_{g\in A} g,\quad g\times \cN = \cN \times g = \cN,\quad g\times h =gh.
\end{equation}
where $\cN$ is the self-duality non-invertible line with quantum dimension $\sqrt{|A|}$. 
The nontrivial $F$ symbols are ${F^{g\cN h}_{\cN} = F_{h}^{\cN g \cN}= \chi(g,h)}$ and ${F_{\cN,gh}^{\cN \cN \cN} = \frac{\epsilon}{\sqrt{|A|}} \chi(g,h)}$, where ${\chi:A\times A\to U(1)}$ is a symmetric non-degenerate bicharacter and $\epsilon$ is the FS indicator.
$\chi$ is determined by the layer-(1.) data, which corresponds to an automorphism of the group $A$. There is no non-trivial symmetry fractionalization class appearing in the layer-(2.) data. However, there are two choices of layer-(3.) data that sets ${\epsilon = \pm 1}$. 
The bicharacter data can be matched at the lattice level, i.e.~whether the duality defect is combined with the non-trivial group automorphism or not. As a consistency check, in the case of an abelian group algebra, \cref{eq:bicharacterF} reduces to the choice of a symmetric bicharacter $\chi$ for the TY category.

Moving on to the more general case, for the self-duality under gauging $\Rep(H)$ symmetry, there could be a non-trivial layer-(2.) fractionalization class data. In fact, this is the case for the self-duality arising from gauging $\Rep(H_8)$. There are, in total, 8 different $\ZZ_2$-extensions of $\Rep(H_8)$ with each layer contributing a factor of 2~\cite{choi2023self,lu2025sdset}. 
Indeed, our lattice construction admits four inequivalent symmetric Hadamard forms, distinguished by the automorphisms of $H_8$, which yield four distinct $\kw$ MPOs. 
Modifying any specific choice of Hadamard form by one of the automorphisms of $H_8$ (which form a $\ZZ_2 \times \ZZ_2$ group~\cite{Sage_2012}) using \eqref{eq:modifiedHad} leads to a different Hadamard gate and, hence, a modification of the self-duality MPO \eqref{eq:dualityop}. Modulo the ambiguity of the FS indicator of the self-duality symmetry discussed above, we thus have a lattice realization for each of the 8 IR fusion category symmetries described above.

The categorical structure of $\ZZ_2$-extension of $\Rep(H_\text{sd})$, for self-dual of Hopf algebras $H_\text{sd}$ and with a single non-trivial object in the $\ZZ_2$-graded sector, was studied in \Rf{davydov2013z}. Our Hadamard form $\Theta$, Hadamard gate $\FF$ and FS indicator correspond to the choice of $\gamma$, $\Gamma$ and $\lambda$ in their paper. The self-duality under gauging a generalized symmetry is also studied in the context of subfactor theory \cite{evans1994orbifold,liu_jones-wassermann_2019,liu2020lifting,evans2023tambara,Liu:2024vsq,Yu:2025iqf}.

\section{Hopf Ising models}\label{sec:HopfIsing}

In the preceding sections, we have constructed a non-invertible Kramers-Wannier symmetry operator $\kw$ for any \textit{self-dual} Hopf algebra $H$, generalizing the construction for Abelian groups. This operator implements the gauging of an anomaly-free, on-site (in the sense of \cite{Meng:2024nxx}) $\Rep(\dl H)$ symmetry, followed by a unitary transformation to turn the dual $\Rep(H)$ symmetry back into $\Rep(\dl H)$. 
In this section, we discuss a family of $\Rep(\dl H)$-symmetric local Hamiltonians on a tensor product Hilbert space that we dub ``Hopf Ising models''. They are constructed from the Hopf Pauli operators introduced in \cref{sec:hopf-pauli}. The action of the associated $\kw$ operator exchanges different local terms in the Hamiltonian. 
While these Hamiltonians are well-defined for any (finite-dimensional semisimple) Hopf $C^\star$-algebra, it is only the self-dual ones for which the operator $\kw$ can be defined, so we choose to focus on these.

Let us consider the family of $\Rep(\dl H)$-symmetric Hamiltonians on a periodic chain of $L$ $H$-qudits defined by
\ie \label{eq:Hopf-Ising}
    \sH(J_\bullet,h_\bullet) = 
    -\sum_{j=1}^{L} \left[ 
    \sum_{\Gamma \in \Irr(H)} J_{\Gamma} \left (\Zr_j \Zl_{j+1} \right )^{\chi_\Gamma} + 
    \sum_{g \in \Irr(\dl{H})} h_{g}\,  \Xr^{x_g}_j  \right] .
\fe 
Here, the first term involves a sum over irreducible representations of $H$, while the second term over those of $\dl H$. The notation reflects the fact that characters of irreps of $\dl H$ can be identified with elements of $H$ itself.\footnote{We use the notation introduced below \cref{eq:Pauli-XX}.}
In general, the coefficients $J_{\Gamma}$ and $h_g$ are valued in $\CC$, but in order for $\sH$ to be Hermitian, we need that ${J_{\Gamma^\star} =  \overline{J_\Gamma}}$ and ${h_{g^\star} =  \overline{h_{g}}}$ for every $\Ga$ and $g$.
Following \cref{eq:RepHstarinv}, we recall that all the terms in this Hamiltonian commute with the $\Rep(\dl H)$ symmetry given by the operators $\Al^r$, defined in \cref{eq:rephdualmpo}.

Let us comment on how this somewhat abstract construction reduces to some well-known models for the case of the group algebra, ${H= \CC[G]}$. 
We denote ${\Rep(\CC[G])}$ simply as $\Rep(G)$, and the dual Hopf algebra ${\dl H = \CC^G}$ has the associated representation category, $\Rep(\CC^G) \cong \VEC_G$. In this case, the above Hamiltonian reduces to
\ie \label{eq:G-Potts}
    \sH_G = 
    -\sum_{j=1}^{L} \left[ 
    \sum_{ \Gamma \in\Irr(G) }  J_{\Gamma}\,  \Tr_{\Ga} \left (Z^{\Ga^\star}_j  Z^\Ga_{j+1} \right ) + 
    \sum_{g \in G} h_{g}\,  \Xr^{g}_j  \right] ,
\fe 
where ${(Z^\Ga)_{\al\bt} = \sum_{h \in G} \Ga(h)_{\al\bt} \ket{h}\bra{h}}$ and ${\Xr^g =  \sum_{h \in G} \ket{h g^{-1}}\bra{h} }$ are the group-valued Pauli Z and Pauli X operators, respectively (cf.~\Rfs{Albert:2021vts,tantivasadakarn_hierarchy_2022,nat2025repG}).

For the Abelian group ${G = \ZZ_n}$, we can further simplify the above Hamiltonian to
\ie \label{eq:Zn-Ham}
\sH_{\ZZ_n} = 
-\sum_{j=1}^{L} \left[ 
\sum_{m=0}^{n-1} J_{m}\,  (Z^\dagger_j Z_{j+1})^m + 
 h_{m}\,  X^m_j  \right] ,
\fe 
where $Z$ and $X$ now denote $\ZZ_n$ Pauli Z and Pauli X operators, respectively. The choice of coupling constants ${J_1 = J_{n-1} = J} $ and ${h_1 = h_{n-1} = h}$, with $h,J\in\RR$, and all other couplings vanishing leads to the usual (non-chiral) $\ZZ_n$ clock model. 
For $n>2$, we could let ${J_m\in\CC}$ such that ${J_1 = \overline{J_{n-1}} = J \ee^{i\theta}} $ and ${h_1 = \overline{h_{n-1}} = h \ee^{i\varphi}} $ with ${h,J\in\RR}$ and ${\phi,\varphi \in \RR/2\pi\ZZ}$. This is referred to as the chiral $\ZZ_n$ clock model~\cite{ostlund1981incommensurate,Fendley:2012vv}. 
Lastly, the choice of ${J_m = J}$ and $ {h_m = h}$ for all $m$ defines the $n$-state Potts model. 

The phase diagrams of these lattice models have been studied in great detail, especially in the statistical mechanical setting. Here, we recall some of the lessons from these studies. The $n$-state Potts and the (non-chiral) $\ZZ_n$ clock models are equivalent for ${n=2,3}$, with the critical point between ordered and disordered phase being described by Ising (${c=\frac12}$) and 3-state Potts (${c=\frac45}$) conformal field theories (CFTs) respectively. 
For ${n=4}$, the clock and Potts models are not equivalent. However, they can be mapped to different one-parameter families in the phase diagram of the Ashkin-Teller model~\cite{Ashkin:1943zza}.
The quantum Ashkin-Teller Hamiltonian on a periodic chain of $L$ sites, with two qubits per site, is
    \begin{equation} 
    \sH_{AT}(J,U) = - J \sum_{j=1}^L \left ( \si^z_j \si^z_{j+1} + \tau^z_j \tau^z_{j+1} + U \si^z_j \tau^z_j \si^z_{j+1} \tau^z_{j+1} \right ) - (1-J) \sum_{j=1}^L \left ( \si^x_j + \tau^x_j + U \si^x_j \tau^x_j \right )\,.
    \end{equation}
    Here $\si$ and $\tau$ correspond to Pauli operators acting on the two qubits on each site. $\sH_{AT}$ is unitarily equivalent to $H_{\ZZ_4}$ with ${J_1 = J_3 = J,\ J_2 = JU}$ and ${h_1=h_3 =1-J,\ h_2 = (1-J)U}$ (see e.g.~\Rf{Tsui:2017ryj} for an explicit expression of the unitary). The same unitary maps the $\ZZ_4$ clock model to the one-parameter family $\sH_{AT}(J,0)$ of two decoupled Ising models~\cite{kohmoto1981hamiltonian}.\footnote{As a side remark, the Hamiltonian $\sH_{AT}(\frac12,0)$ realizes a $\Rep(H_8)$ symmetry in the IR, implemented by a simultaneous Kramers-Wannier transformation acting on each Ising chain~\cite{Thorngren2019iar}. However, this symmetry transformation mixes with lattice translation in the UV. As a result, it is also possible for this symmetry operator to flow to the anomalous TY$(\ZZ_2 \times \ZZ_2)$ symmetry in the IR for a different choice of Hamiltonian~\cite{shuheng2024KWlsm} (see Sec.~\ref{sec:categorical-selfduality} for further discussion). This is to be contrasted with the $\Rep(H_8)$ symmetry realized in the $H_8$ Ising model~\cref{eq:H8-Ham}.\label{footnote:Ashkin-Teller-H8}} On the other hand, the (non-chiral) 4-state Potts model corresponds to $\sH_{AT}(J,1)$~\cite{Kadanoff:1980PhRvB,Aoun:2024lmd}.\footnote{It is interesting to note that $\sH_{AT}$ is identical to the Hamiltonian $\sH_G$ in \cref{eq:G-Potts} for ${G=\ZZ_2\times\ZZ_2}$, with couplings ${J_{(1,0)} = J_{(0,1)} = J}$, ${J_{(1,1)} = JU}$, ${h_{(1,0)} = h_{(0,1)} = h }$ and ${h_{(1,1)} = hU}$.}
As such, both the $\ZZ_4$ clock and the $4$-state Potts models undergo direct continuous phase transitions as a function of the coupling $J$. However, the IR effective field theory of the two critical points correspond to different points on the orbifold branch of the compact boson CFT (${c=1}$)~\cite{Ginsparg:1987eb}. An exactly marginal operator deforms one into the other. 
For ${n\geq 5}$, the Potts and clock models have distinct phase diagrams. While the clock model has a two-step transition with an intermediate gapless phase~\cite{Jose:1977gm}, the Potts counterpart has a direct first-order transition~\cite{baxter1973potts}. 
The latter are known to be proximate to complex conformal fixed points~\cite{Ma:2018euv}. 

Returning to the general case of the Hopf Ising Hamiltonians \eqref{eq:Hopf-Ising}, we note that setting ${h_g = \frac{d_g}{|H|}}$, where $d_g$ is the dimension of the irrep $g$ of $\dl H$, and all ${J_\Ga = 0}$ reproduces the Hamiltonian \eqref{eq:PauliXham}. This realizes a $\Rep(\dl H)$-symmetric phase with a unique gapped ground state.
On the other hand, setting ${J_\Ga = \frac{d_\Ga}{|H|}}$ and all ${h_g = 0}$ recovers the Hamiltonian \eqref{eq:RepHdSSBham}, which realizes a gapped phase with the $\Rep(\dl H)$ symmetry fully spontaneously broken, with $|\Irr(H)|$ degenerate ground states. 
Interpolating between these two, we get the Hamiltonian,
\ie \label{eq:Hopf-Potts}
\sH_{\rm HP}(J) &= -\frac{1}{|H|}\sum_{j=1}^{L} \Big[ 
J \sum_{\Gamma \in \Irr(H)} d_{\Gamma}  (\Zr_j \Zl_{j+1}  )^{\chi_\Gamma} + (1-J) 
\sum_{g \in \Irr(\dl H)} d_{g}\,  \Xr^{\dchi_g}_j  \Big]
\\
&=  - \sum_{j=1}^{L} \left(
J  (\Zr_j \Zl_{j+1}  )^{\cohaar} + (1-J)  X^{\haarc}_j  \right)
,
\fe 
which realizes an ordered (Hopf-ferromagnet) phase, with $|H|$ degenerate ground states, for ${0< J \ll \frac12 }$ and a disordered (Hopf-paramagnet) phase, with a unique symmetric ground state, for ${J \gg \frac12}$. 
Details of intermediate phases and transitions possibly depend on the choice of $H$.
By analogy with the $n$-state Potts model, we refer to $\sH_{\rm HP}$ as the Hopf-Potts model. In \cref{sec:H8-Ising}, we will discuss the phase diagram of a Hopf Ising model associated with the Hopf algebra $H_8$, which will include a 1-parameter family representing an $H_8$ Potts model.
% For other choices of the values of $J_\Ga$ and $h_g$, the Hamiltonian is referred to as an anisotropic Hopf Potts model.

Finally, we focus on the self-dual Hopf algebras, for which there is a Hopf Kramers Wannier duality, as discussed in \cref{sec:KW}. This is implemented by the operator $\kw$, which acts as (see \cref{eq:KW-XZmap})
\ie \label{eq:Dkw-action}
\kw \Xr^{\dchi_g}_j =     (\Zr_{j} \Zl_{j+1}  )^{\Theta_{\dchi_g} } \kw 
  \,, \qquad 
\kw  (\Zr_{j-1} \Zl_j  )^{\Theta_{\dchi_g}} =   \Xr^{\dchi_g}_j  \kw \,.
\fe 
We recall that $\dchi_g \in {\rm Cocom}(H) $ and, hence, $\Theta_{\dchi_g} \in {\rm Cocom}(\dl{H}) $.  
As a result of \cref{eq:Dkw-action}, the action of $\kw$ on the corresponding Hopf Ising Hamiltonian \eqref{eq:Hopf-Ising} amounts to swapping the respective pairs of coupling constants, $(J_\bullet,h_\bullet)$. The Hamiltonian realized by setting ${J_\bullet = h_\bullet}$ is self-dual, i.e.~commutes with $\kw$.

Specializing to the Hopf-Potts Hamiltonian $\sH_{\rm HP}$ \eqref{eq:Hopf-Potts}, the action of $\kw$ exchanges the ferromagnetic and paramagnetic terms, i.e.,
\ie \label{eq:Dkw-HP}
\kw \, \sH_{\rm HP}(J) = \sH_{\rm HP}(1-J)\, \kw \,,
\fe 
with $J=\frac12$ being self-dual, i.e.~${[\kw,\sH_{\rm HP}(\frac12)] = 0}$.
Let us note that for the $\ZZ_n$-symmetric Hamiltonian \eqref{eq:Zn-Ham}, there is a well-known Kramers-Wannier duality~\cite{Fateev:1985mm,thorngren_fusion_2021,ParayilMana:2024txy}
whose action on $\ZZ_n$-symmetric operators is given by
\ie 
\mathsf{D}_{\ZZ_n} Z_{j-1}^\dagger Z_{j} = X_j \mathsf{D}_{\ZZ_n} \,, 
\qquad 
\mathsf{D}_{\ZZ_n} X_{j} = Z_{j}^\dagger Z_{j+1} \mathsf{D}_{\ZZ_n} \,. 
\fe 
This has the effect of interchanging $J_m$ with $h_m$ for each $m$. The more general construction discussed above reduces to this case upon setting ${H=\CC[\ZZ_n]}$.

\subsection{\texorpdfstring{$H_8$}{H8} Ising model and self-duality}
\label{sec:H8-Ising}

For the rest of this section, we specialize to the smallest self-dual semisimple Hopf $C^\star$-algebra that is not a group algebra or its dual, namely, the 8-dimensional Kac-Paljutkin algebra, $H_8$.

The defining data of $H_8$ can be found in \cref{app:H8data}. It has a four-dimensional group-like Hopf sub-algebra, isomorphic to $\CC[\ZZ_2\times \ZZ_2]$. In the $x,y,z$ basis, it is spanned by the four elements, $\{1,x,y,xy\}$.
Of the four non-trivial irreps of $H_8$, three are the one-dimensional ``sign'' reps associated with the $\CC[\ZZ_2\times \ZZ_2]$ sub-algebra mentioned above. We denote them as ${\mathbf{s}}_n $, with ${n=1,2,3}$.
The remaining nontrivial irrep is two-dimensional, denoted $\two$. 
The irreps of $H_8$ and their characters are summarized in \cref{eq:H8-irreps} and \cref{eq:H8-characters}, respectively.

We remark that the fusion category $\Rep(H_8)$ can also be thought of as the subcategory generated by simples $\{1,\psi_1,\psi_2,\psi_1\psi_2, \sigma_1\sigma_2\}$ in the fusion category $\text{Ising}^2$, where $\sigma_i$ has quantum dimension $\sqrt{2}$ for each copy $i=1,2$, and $\psi_i$'s are invertible. Nevertheless, on the lattice, the Ising fusion category can only be realized by an extension by spatial symmetries, such as lattice translation or reflection~\cite{SeibergShao23,nat2024zxKW,Pace25}. Therefore, it is possible to realize a different symmetry which mixes with lattice translation, and will flow to $\Rep(H_8)$ in the continuum for an appropriate Hamiltonian~\cite{Seifnashri24} (for example, by simultaneously enacting the Kramers-Wanniers duality in two copies of the Ising chain, see \cref{footnote:Ashkin-Teller-H8}). However, this same symmetry can also flow to a different IR fusion category for a different choice of UV Hamiltonian\footnote{This is due to the fact that internal symmetries that are extended non-trivially by lattice translation do not have a well-defined Frobenius-Schur indicator~\cite{shuheng2024KWlsm}.}. 
In contrast, our lattice realization of $\Rep(H_8)$ does not mix with lattice translation, and therefore, for any choice of Hamiltonian, it will flow to the fusion category $\Rep(H_8)$ as long as the full symmetry acts faithfully in the IR.

We consider the following family of Hamiltonians, each of which commutes with the $\Rep(\dl H_8)$ symmetry operators $\Al^r$ (see \cref{eq:rephdualmpo}):
\ie \label{eq:H8-Ham}
\sH_{H_8} = 
- J \sum_{j=1}^L\left (K \sum_{n=1}^3  (\Zr_j \Zl_{j+1} )^{\chi_{{\mathbf{s}}_n}} +  (\Zr_j \Zl_{j+1}  )^{\chi_\two} \right ) 
- (1-J) \sum_{j=1}^L\left (K \sum_{n=1}^3  \Xr^{\dchi_{{\mathbf{s}}_n}}_j  +  \Xr^{\dchi_{\two}}_j  \right )
\fe 
Here, ${\{\dchi_{{\mathbf{s}}_n}, \dchi_{\two}\}}$ are elements of $H_8$ that are related to the characters ${\{\chi_{{\mathbf{s}}_n}, \chi_\two \}}$ of the irreps ${\{ {\mathbf{s}}_n, \two \}}$, respectively, by the Hopf Hadamard transformation, i.e.,
\ie 
\dchi_{{\mathbf{s}}_n} = \Th^{-1}_{\chi_{{\mathbf{s}}_n}}\,, \qquad  \dchi_\two = \Th^{-1}_{ \chi_\two} \,.
\fe 
One can observe readily that $\sH_{H_8}$ commutes with $\kw$ at ${J=0.5}$, for any value of $K$. For ${K=0.5}$, we obtain the $H_8$ Potts model, as defined in \cref{eq:Hopf-Potts}, up to an overall scale. 
We note that $\sH_{H_8}$ is Hermitian because the irreps of $H_8$ satisfy ${\Gamma^\star = \Gamma}$ and, hence, ${\chi_\Gamma^\star = \chi_\Gamma}$. 

As a further simplification, the irreps of $H_8$ are all self-dual\footnote{One way to see if the representation $\Gamma$ is self-dual is from its Frobenius-Schur indicator. For semisimple and cosemisimple Hopf algebras, the representations are self-dual if and only if their Frobenius–Schur indicators are $\pm 1$ (the FS indicator for an irrep, self-dual or not, is valued in $\{0,1,-1\}$)~\cite{linchenko2000frobenius,kashina2006higher,ng2005higher}. According to \Rf{ng2005higher}, the simple objects in all four Tambara-Yamagami fusion categories of $\ZZ_2\times \ZZ_2$, including $\Rep(H_8)$, are all self-dual, i.e.~$\chi_\Gamma(S(\cdot)) = \chi_\Gamma(\cdot), \forall  \Gamma \in \Irr(H_8)$.}~\cite{ng2005higher}, i.e., each irrep $\Gamma$ is isomorphic to its dual irrep $\Gamma^\vee$, defined by $\Gamma^\vee(\cdot) = \Gamma(S(\cdot))^\intercal$ (see \Cref{app:def} for more details). In that case, the corresponding character satisfies ${\chi_\Gamma(S(\cdot)) = \chi_{ \Gamma}(\cdot)}$ (see \eqref{eq:dualHchar}).

We can express the family of Hamiltonians \eqref{eq:H8-Ham} more explicitly by choosing a specific basis. Unlike group algebras, there is no canonical basis. Here we the ``$D_8$'' basis (see \cref{app:H8data}, including other bases discussed in the literature) .  We can identify the eight-dimensional Hilbert space associated with $H_8$ with that of three qubits, and label the corresponding Pauli operators as $X_{j,\al}, Y_{j,\al}, Z_{j,\al}$ where ${\al=1,2,3}$ is the label for the three on-site qubits. 
The $\Xr $ operators are given by
\ie 
\Xr^{\dchi_{{\mathbf{s}}_1}}_j &= P^+_{j,2} X_{j,3} + P^-_{j,2} Y_{j,3}\,, 
%\\
\qquad 
& \Xr^{\dchi_{{\mathbf{s}}_2}}_j &= P^+_{j,2} X_{j,3} - P^-_{j,2} Y_{j,3}\,, 
\\
\Xr^{\dchi_{{\mathbf{s}}_3}}_j &= X_{j,2}\,, 
%\\
\qquad & \Xr^{\dchi_{\two}}_j &= 2\, X_{j,1} P^+_{j,2} P^+_{j,3} 
\,,
\fe 
while the $\Zr \Zl$ operators are given by
\ie 
 (\Zr_j \Zl_{j+1}  )^{\chi_{{\mathbf{s}}_1}} &= P^-_{j,2} Z_{j,3} Z_{j+1,1} Z_{j+1,3} + Z_{j,1} P^+_{j,2} Z_{j,3} Z_{j+1,1} Z_{j+1,3}   \,,\\
 (\Zr_j \Zl_{j+1}  )^{\chi_{{\mathbf{s}}_2}} &= P^+_{j,2} Z_{j,3} Z_{j+1,3} + Z_{j,1} P^-_{j,2} Z_{j,3} Z_{j+1,3} \,, \\
 (\Zr_j \Zl_{j+1}  )^{\chi_{{\mathbf{s}}_3}} &= Z_{j,1} Z_{j+1,1} \,, \\
 (\Zr_j \Zl_{j+1}  )^{\chi_{\two}} &= 
2\, P^0_{j,1} Y_{j,2} P^1_{j,3} P^0_{j+1,1} Y_{j+1,2} P^1_{j+1,3}
+ 2\, P^0_{j,1} Z_{j,2} P^0_{j,3} P^0_{j+1,1} Z_{j+1,2} P^0_{j+1,3} \\
&\hspace{22.5pt} + P^1_{j,1} Z_{j,2}  P^1_{j+1,1} Z_{j+1,2} X_{j+1,3}
+ P^1_{j,1} Y_{j,2} Z_{j,3} P^1_{j+1,1} Z_{j+1,2} Y_{j+1,3} 
\,.
\fe 
In writing the above expressions, we have defined the following projection operators: 
\ie \label{eq:proj-ops}
P^\pm = \frac{1\pm X}{2} \,,
\qquad P^{0} = \frac{1+Z}{2} \,,
\qquad P^1 = \frac{1-Z}{2} \,.
\fe

As discussed around~\eqref{eq:ZGaXa}, the Hopf Pauli operators $\Zr^\Ga$ and $\Xl^a$ may commute only up to a phase. The simplest instance occurs when $a$ is a central group-like element, which can then be used to construct the order parameter. In $H_8$, the nontrivial central group-like element is $r^2$ (equal to $xy$ in the $xyz$ basis). Indeed, we find ${\Zr^\Gamma \Xl^{r^2}=\ee^{i\phi_\Ga}\,\Xl^{r^2}\Zr^\Gamma}$, with ${\ee^{i\phi_\Ga} = \frac{\chi_\Ga(r^2)}{d_\Ga}}$ nontrivial, and equal to $-1$, only for the two-dimensional irrep, ${\Ga = \two}$.
Accordingly, $\Xl^{r^2}$ anticommutes with the Hopf Pauli Z operator associated with the 2d irrep of $H_8$, while commuting with all the 1d irreps; this is reflected in the character table \eqref{eq:d8chitab}. On a $3$-qubit local Hilbert space we realize $\Xl^{r^2}_j=X_{j,2}$, which serves as the order parameter for the 2d irrep. In the numerical investigation, we use the $\Rep(H_8^*)$ presentation of the symmetry. Hence, the corresponding order parameter is obtained by applying the Hadamard, giving ${Z_{j,1} = \FF^{\dagger} X_{j,2} \FF}$. We will momentarily employ $Z_{j,1}$ as a local order parameter, and use it to define a corresponding magnetization.

\subsection{Phase diagram of \texorpdfstring{$H_8$}{H8} Ising model}

\begin{figure}
    \centering
    \includegraphics[width=0.7\linewidth]{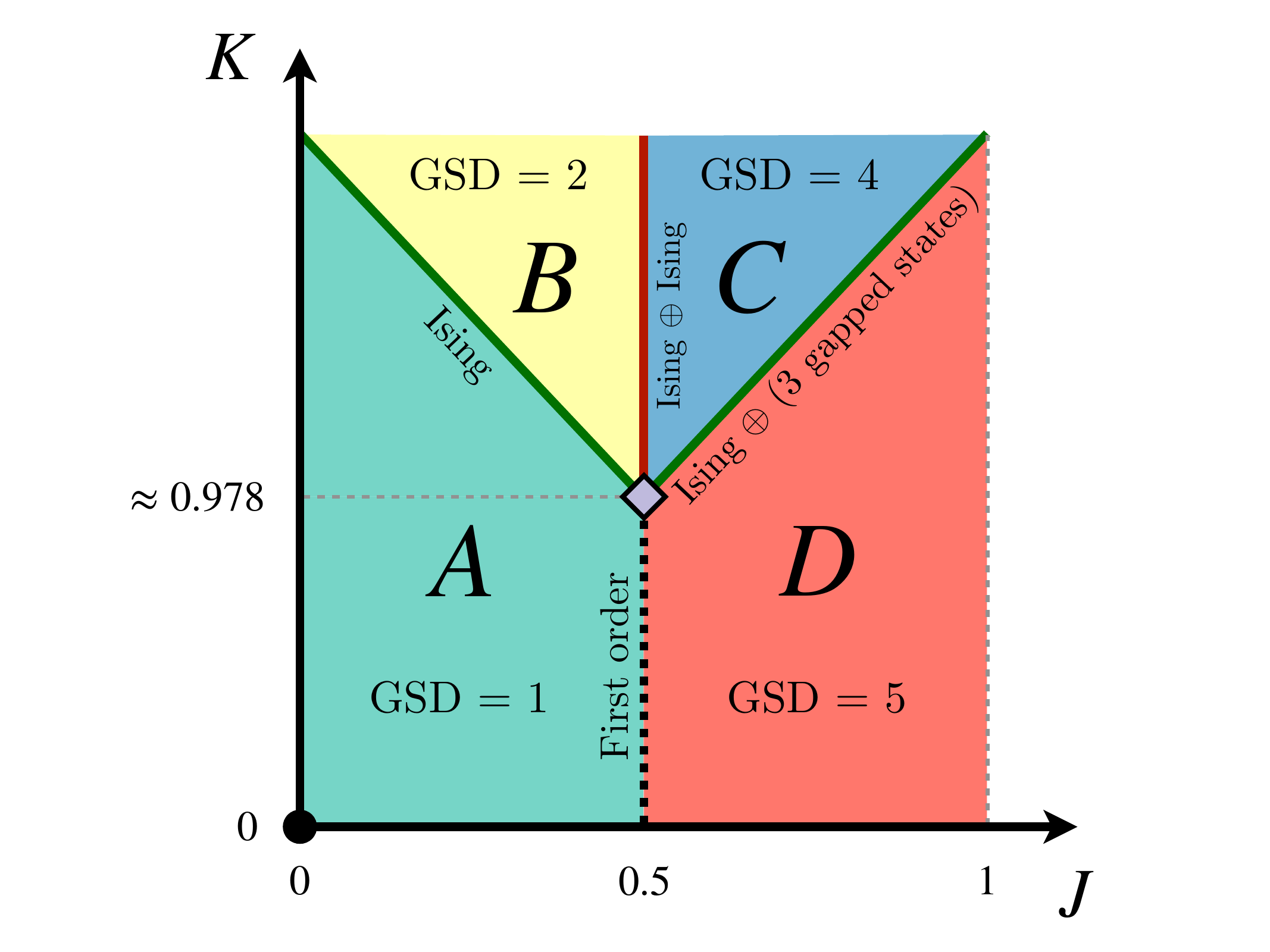}
    \caption[]{Schematic phase diagram of the $H_8$ Ising model $\sH_{H_8}$. The self-dual line corresponds to $J=0.5$.
        }
    \label{fig:schematic-phase-diagram}
\end{figure}

The $(J,K)$ parameter space of $\sH_{H_8}$ realizes a number of different gapped phases, each reflecting different spontaneous symmetry breaking (SSB) patterns of $\Rep(H_8)$. 
We have already discussed the ${J\approx 0}$ and ${J\approx 1}$ limits, which correspond to a complete $\Rep(H_8)$ SSB phase and a symmetric gapped phase, respectively. 
There is a rich phase diagram for intermediate values of ${J \approx 0.5}$, with a strong dependence on the ``anisotropy'' parameter $K$.
The phase diagram has the schematic form shown in \Cref{fig:schematic-phase-diagram}.

First, we use exact diagonalization on small system sizes to identify the ground state degeneracies in the different gapped phases. We find four distinct phases with ground state degeneracies of $1,2,4,5$; let us name them A, B, C and D, respectively.
As we tune $J$ through the self-dual ${J=0.5}$ line, for fixed $K$ slice, we find
\begin{enumerate}[(a)]
    \item  for ${0\leq K \ll 1}$, there is a direct transition from the symmetric phase (A) with one ground state to the full SSB phase with 5-fold degenerate ground state (D); and
    \item for ${K\gg 1}$, there are intermediate gapped phases that spontaneously break $\Rep(H_8)$ down to different subsymmetries---first, a transition to a phase that preserves the invertible $\ZZ_2\times \ZZ_2$ sub-symmetry (B), followed by one that preserves its diagonal $\ZZ_2$ subsymmetry (C), and finally to the completely SSB phase.
\end{enumerate}

\begin{figure}[b!]
    \begin{center}
        \begin{subfigure}{0.46\textwidth}
            \includegraphics[width=\textwidth]{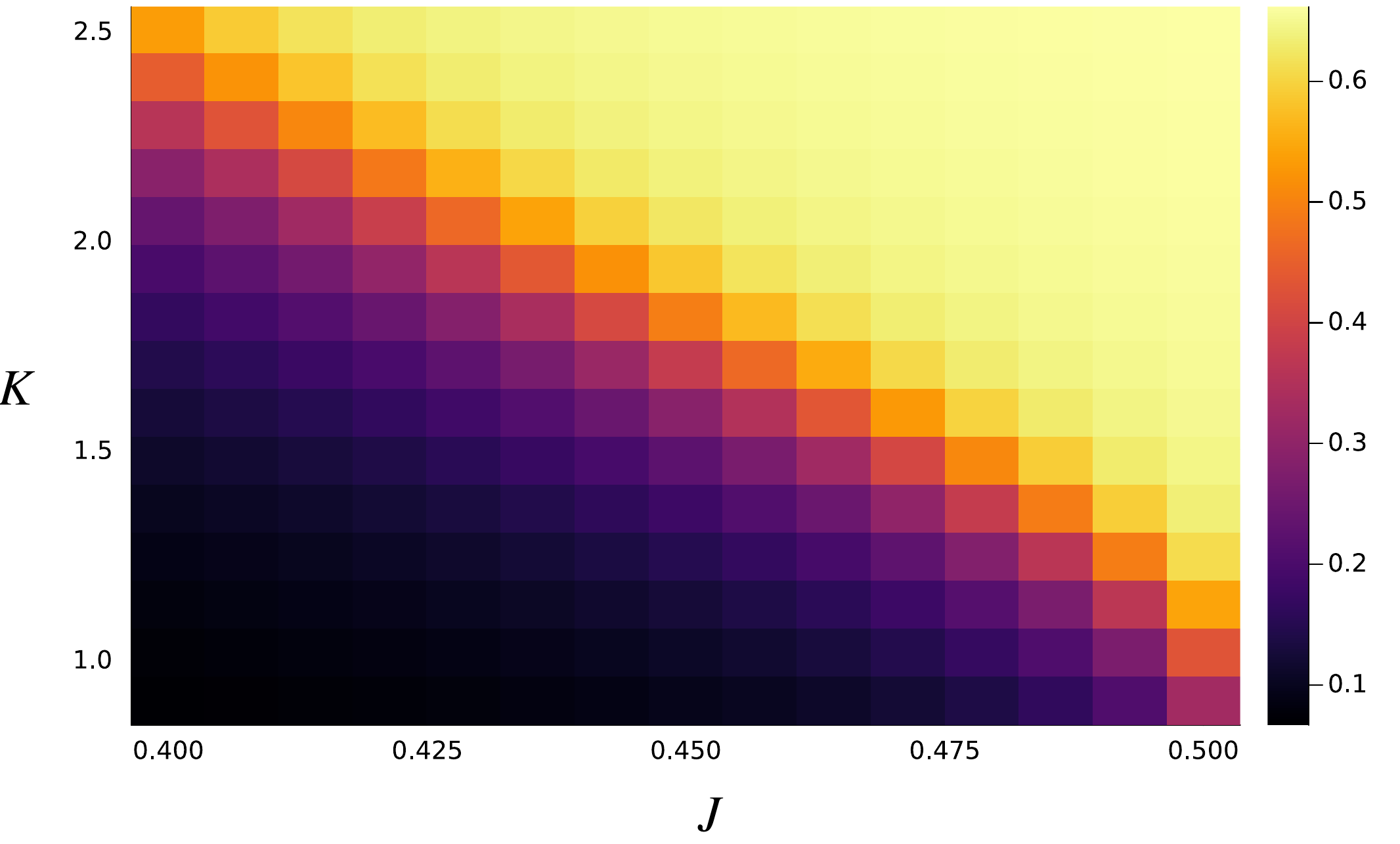}
            \caption{}
            \label{fig:ZIIbinder-phase-diagram}
        \end{subfigure}
        \hspace{0.05\textwidth} 
        \begin{subfigure}{0.425\textwidth}
            \includegraphics[width=\textwidth]{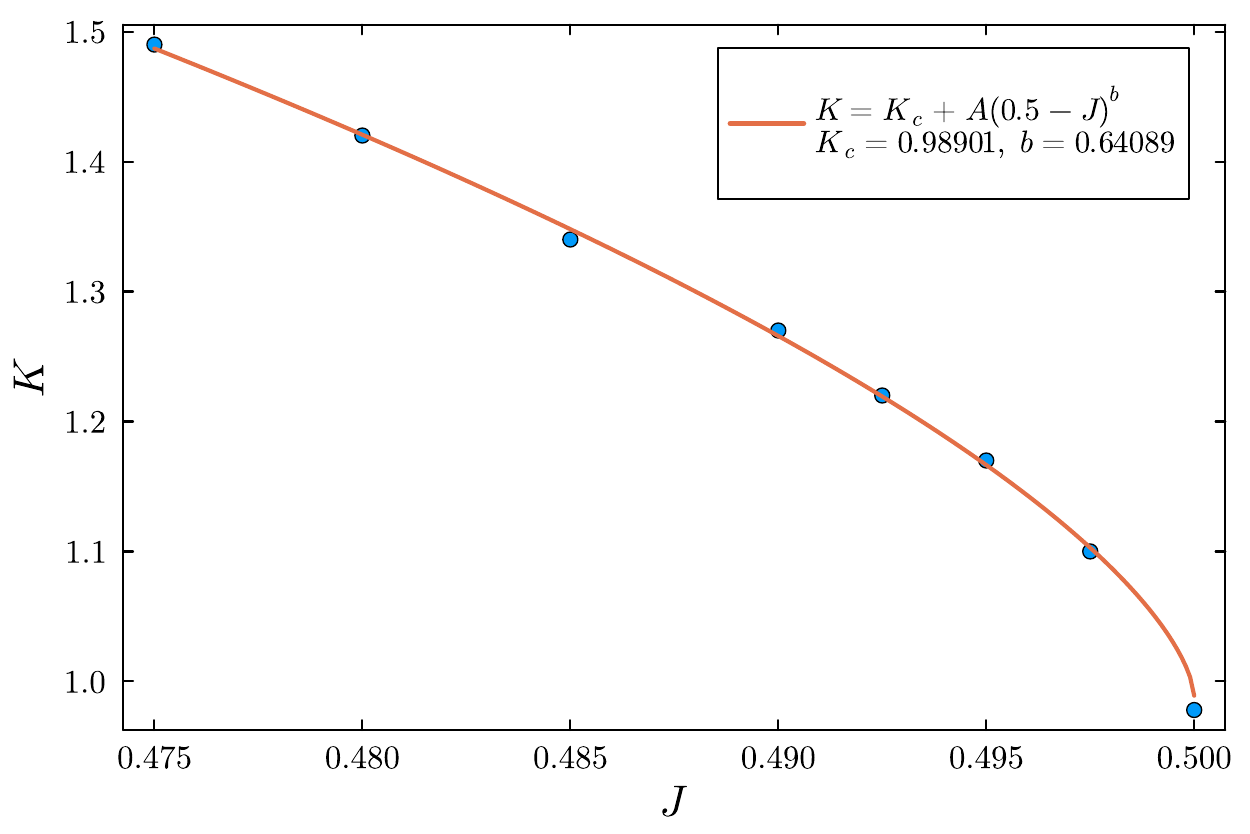}
            \caption{}
            \label{fig:ZIIbinder-phase-boundary}
        \end{subfigure}
    \end{center}
    \caption{
    (a) Plot of the Binder cumulant $B_m$ computed in the ground state of $\sH_{H_8}$ with ${L=40}$ qudits, ${J\in[0.4,0.5]}$, ${K\in[0.9,2.5]}$, and open boundary conditions.
    (b) Phase boundary obtained from the crossing of Binder cumulant as a function of $K$ and $L$ for different values of ${J\in[0.475,0.5]}$ (blue circles), along with a nonlinear fit (red line).
    } 
\end{figure}

Before investigating more fine-grained information about the phase diagram, we note that gapped phases are mapped to each other under topological operations such as discrete gauging. Moreover, such an operation also preserves correlation functions of local operators which remain local under the duality. In particular, this holds for the action of $\kw$, since it is constructed from a gauging map. Therefore, we conclude that $\kw$ maps between phases A and D, as well as between phases B and C, with the intervening phase boundaries also getting mapped to their respective partners under $\kw$. Because of this simplification, it is largely enough to restrict our attention to the left half of the phase diagram \Cref{fig:schematic-phase-diagram}.\footnote{Two continuous transitions mapped to each other by $\kw$ need not have the same number of decoupled gapped ground states, however. This is consistent with the fact that a discrete gauging does not necessarily preserve global properties even though it preserves correlation function of local operators in the sense we discuss above.}
Finally, since the phase transition from C $\to$ D proceeds via the breaking of a $\ZZ_2$ symmetry, the local low-energy properties of that transition are expected to be described by the Ising CFT. By the duality argument, we expect this to also hold for its dual partner A $\to$ B.

With these general expectations set, we now turn to numerical methods to shed further light on the nature of the phase transitions in \cref{fig:schematic-phase-diagram}. 
We study the ground state wavefunction using finite MPS with the ITensor library~\cite{ITensor,itensor-r0.3}, as well as infinite MPS (with three qubits per unit cell) using the MPSKit library~\cite{Van_Damme_MPSKit_2025}. This is to extract complementary information about the ground state and the low-lying excited states. The MPS was optimized using DMRG (Density Matrix Renormalization Group), while the iMPS was optimized by the VUMPS (Variational Uniform Matrix Product State) algorithm, before switching to gradient descent for small tolerance to accelerate convergence~\cite{HauruGradientDescent}.

There are four main observables we are interested in:
\begin{enumerate}[(1)]
    \item Local order parameter ${Z_{j,1}}$, and the associated magnetization ${m = \frac1L \sum_{j=1}^L Z_{j,1}}$; see discussion below \cref{eq:proj-ops}.
    \item Binder cumulant ${B_m = \frac{\langle m^4\rangle}{\langle m^2\rangle^2}}$ associated with the above order parameter. 
    \item Bipartite entanglement entropy whose scaling with subsystem size encodes the central charge at continuous transitions described by a CFT in the low-energy limit.
    \item Ground state degeneracy and the finite size scaling of the gap above the ground state(s).
\end{enumerate}

\begin{figure}
    \begin{center}
        \begin{subfigure}{0.46\textwidth}
            \includegraphics[width=\textwidth]{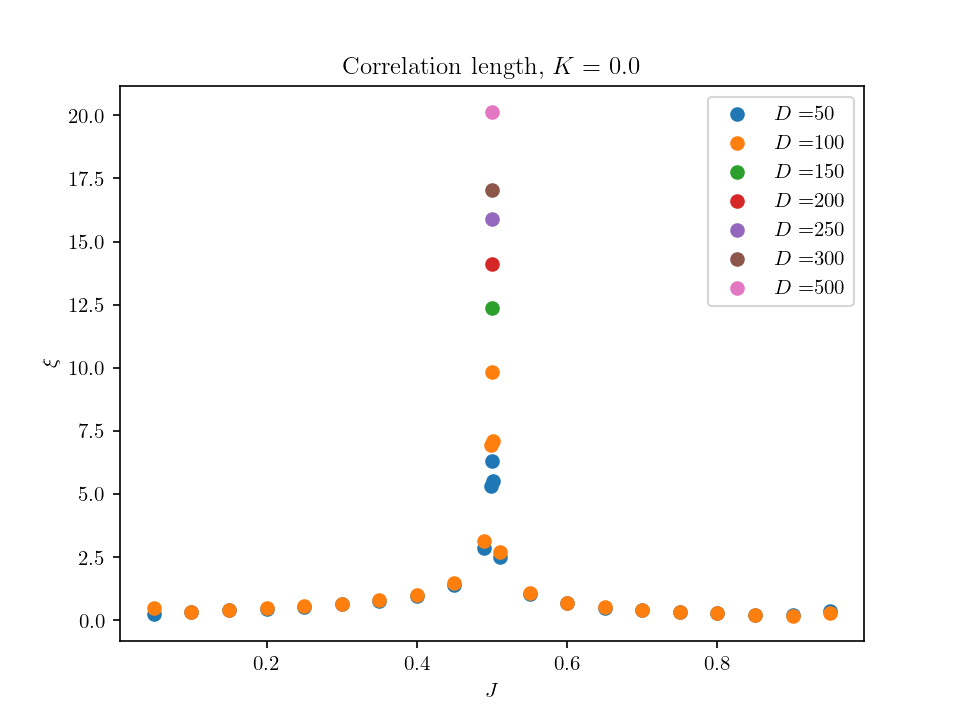}
            \caption{}
        \end{subfigure}
        \hspace{0.05\textwidth} 
        \begin{subfigure}{0.46\textwidth}
            \includegraphics[width=\textwidth]{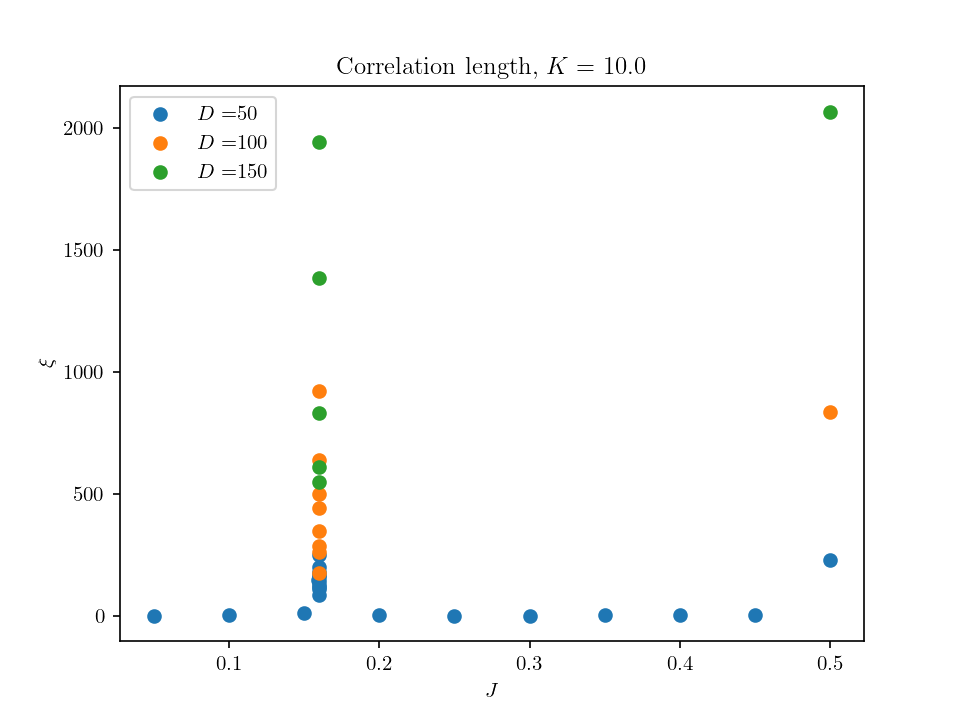}
            \caption{}
        \end{subfigure}
    \end{center}
    \caption{
        Correlation length $\xi$, defined by the second largest eigenvalue of the transfer matrix associated with the iMPS ground state, plotted along two different constant $K$ slices: (a) ${K=0}$ and (b) ${K=10}$. Different MPS bond dimensions ($D$) are shown in different colors.
    } \label{fig:corr-length}
\end{figure}

\begin{figure}[t]
    \begin{center}
    \begin{subfigure}{0.48\textwidth}
        \includegraphics[width=\textwidth]{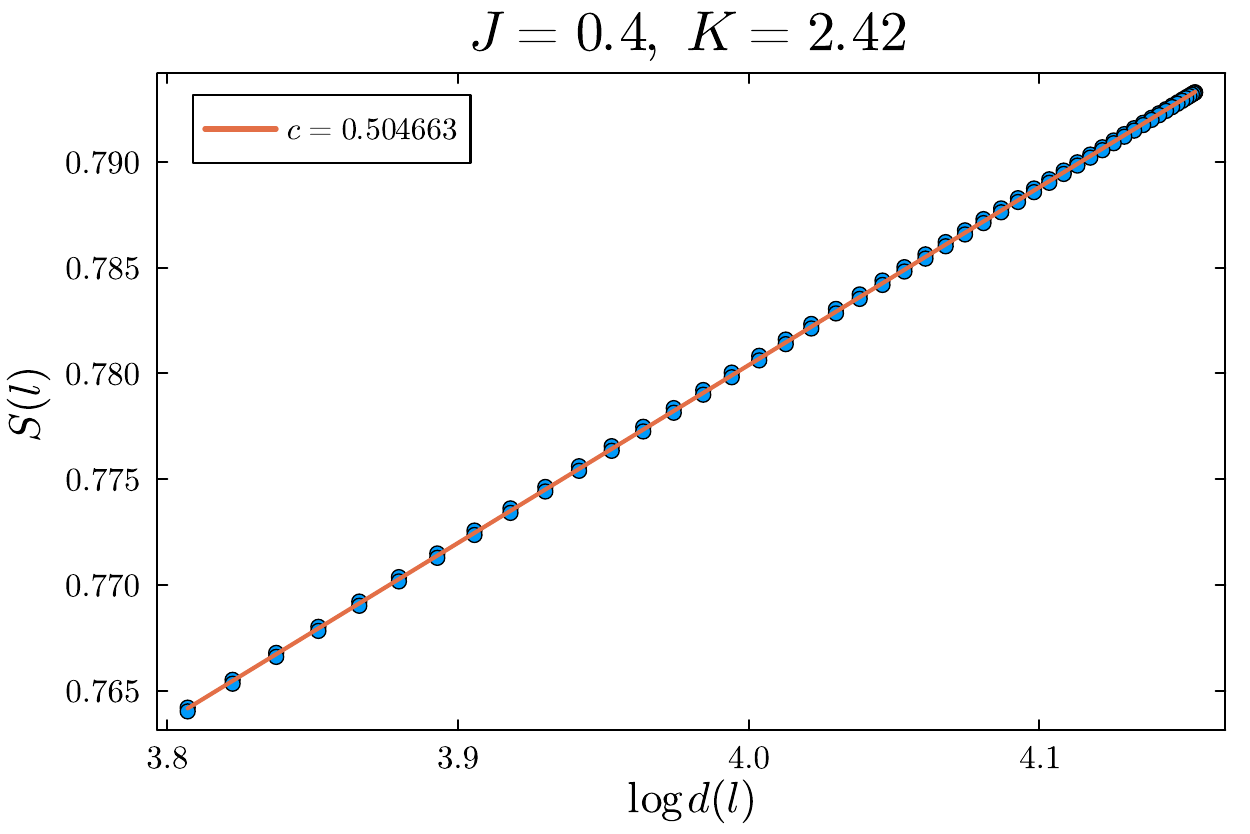}
        \caption{}
        \label{fig:ccfit-Ising1}
    \end{subfigure}
    \hspace{0.05\textwidth} 
    \begin{subfigure}{0.43\textwidth}
        \includegraphics[width=\textwidth]{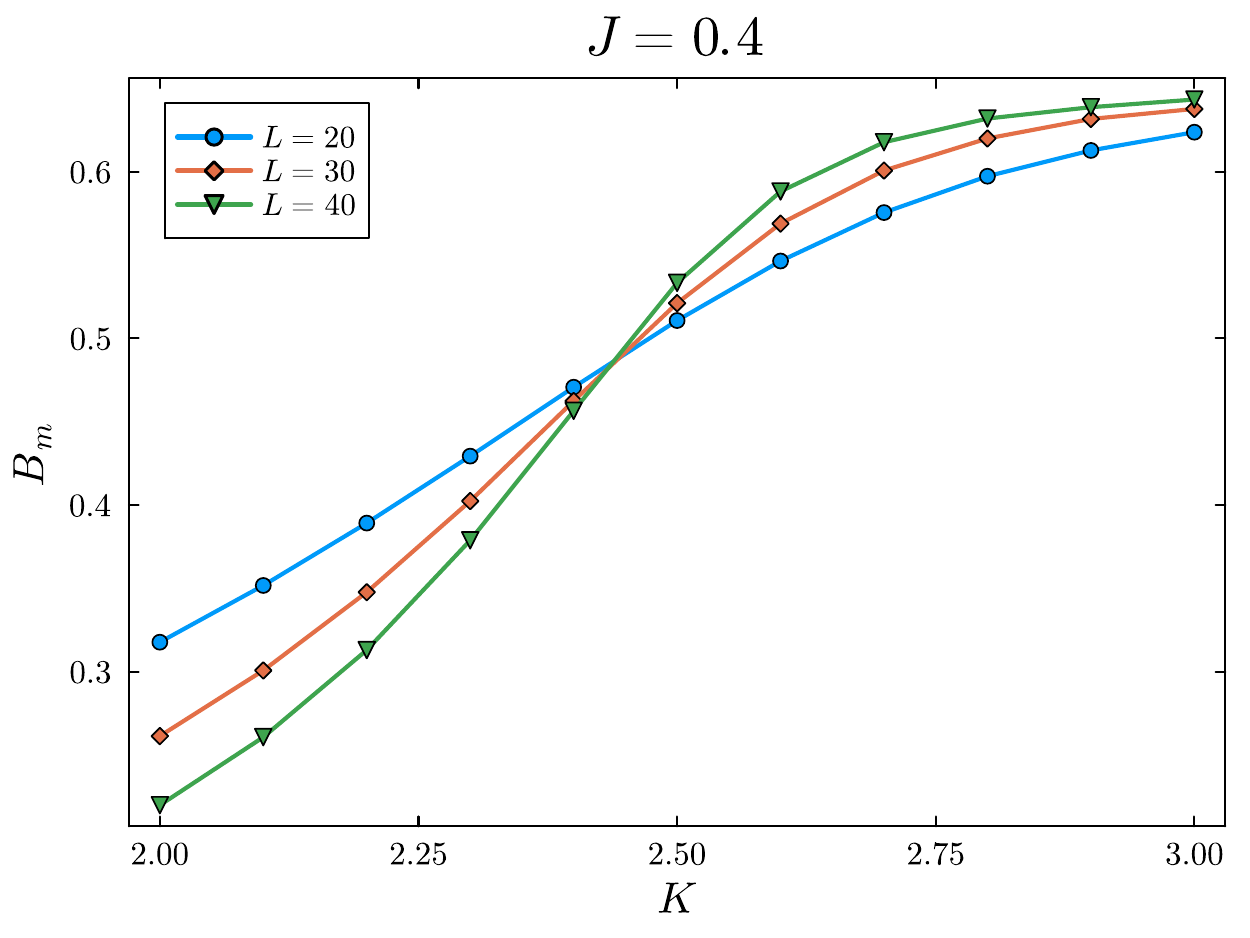}
        \caption{}
        \label{fig:Binder-Ising1}
    \end{subfigure}
    \end{center}
    \caption{(a) Bipartite entanglement entropy in the ground state MPS of a chain with  ${L=200}$ qudits, ${J = 0.4}$, ${K = 2.42}$ and open boundary conditions. The horizontal axis plots $\log d(l)$, where $l$ is the number of qudits in the subsystem and ${d(l) = \frac{2 L}{\pi}  \sin\left ( \frac{\pi l }{L} \right ) }$. The red line shows a linear fit to the Calabrese-Cardy formula ${S(l) = \frac{c}{6}\log d(l) + c_1}$ with the value of $c$ extracted in the inset. (b) Binder cumulant for ${L=20,30,40}$ with $J=0.4$ and ${K\in[2.35,2.5]}$.} 
    \label{fig:Ising1}
\end{figure}

In \Cref{fig:ZIIbinder-phase-diagram}, we plot a heatmap for the Binder cumulant $B_m$ 
in the MPS ground state of $\sH_{H_8}$ for a chain of ${L=40}$ qudits with open boundary conditions. 
The shape of the phase boundary between phases A and B, near the multicritical point, is shown in \Cref{fig:ZIIbinder-phase-boundary}. 
The locations of the phase transitions along constant $K$ slices of the phase diagram are reflected in the divergence of the correlation length obtained from the iMPS ground state, \Cref{fig:corr-length}.
We compute the central charge on the phase boundary between the phases A and B, by fitting the bipartite entanglement entropy in the MPS ground state to the Calabrese-Cardy formula. One example is shown in \Cref{fig:ccfit-Ising1}, demonstrating that the central charge is approximately equal to $\frac12$, consistent with the prediction that the A $\to$ B transition is described by the Ising CFT.
\Cref{fig:Binder-Ising1} shows the crossing of the Binder cumulants computed for different system sizes, ${L=20,30,40}$ at ${J=0.4}$. This allows us to locate the phase transition on the constant $J$ slice to ${K=2.42}$. Similar computations were used in obtaining \Cref{fig:ZIIbinder-phase-boundary}.

\begin{figure}[t!]
    \begin{center}
    \begin{subfigure}{0.45\textwidth}
        \includegraphics[width=\textwidth]{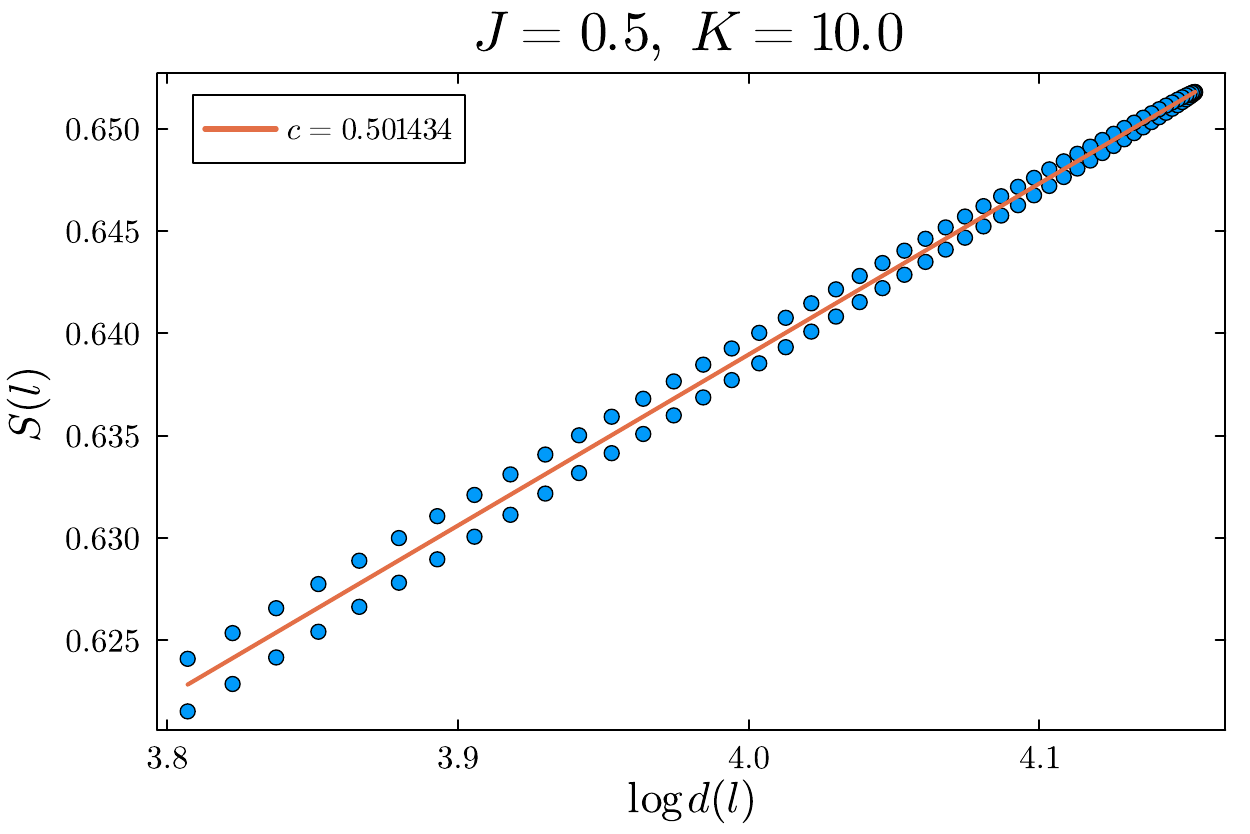}
        \caption{}
        \label{fig:ccfit-Ising2}
    \end{subfigure}
    \hspace{0.05\textwidth} 
    \begin{subfigure}{0.4\textwidth}
        \includegraphics[width=\textwidth]{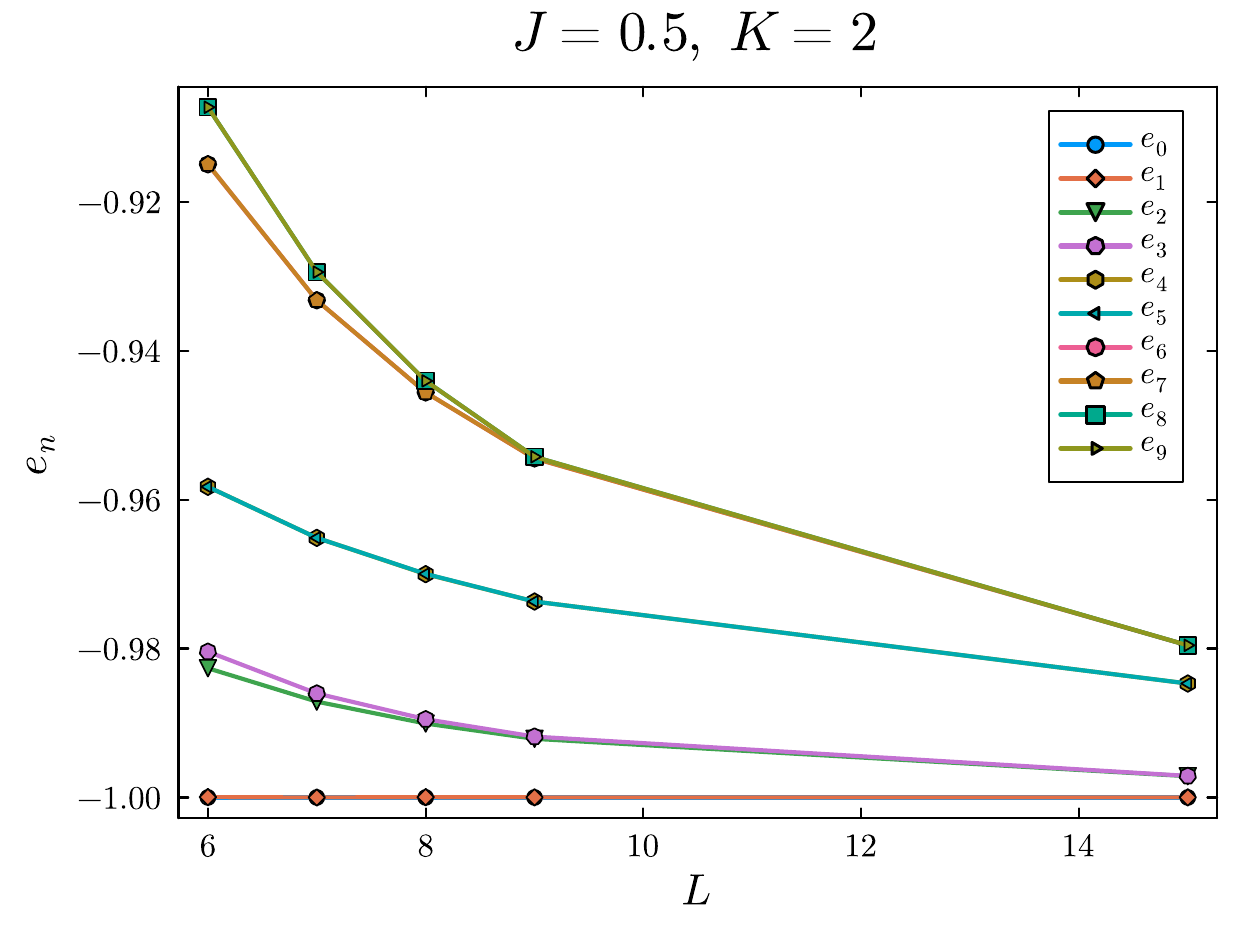}
        \caption{}
        \label{fig:Ising2-GSD}
    \end{subfigure}
    \end{center}
    \caption{(a) Bipartite entanglement entropy and Calabrese-Cardy fit similar to \Cref{fig:ccfit-Ising1} with ${L=200}$, ${J=0.5}$, ${K=10}$ and open boundary conditions. 
    (b) Smallest 10 energy eigenvalues, normalized as ${e_n = \frac{E_n}{|E_0|}}$, for ${J=0.5}$, ${K=2}$, and ${L=6,7,8,9,15}$ qudits with periodic boundary conditions. The lines are guides to the eye.
    }
    \label{fig:Ising2}
\end{figure}

The central charge obtained from fitting bipartite entanglement entropy at the B $\to$ C phase boundary is shown in \Cref{fig:ccfit-Ising2}. This is consistent with the Ising universality class. However, this transition differs from the one in \Cref{fig:Ising1} due to the existence of two degenerate ground states, rather than a single (symmetric) one. 
In fact, we find an almost exact 2-fold degeneracy for \textit{low-energy} eigenstates, not just the ground state.
We demonstrate this numerically in \cref{fig:Ising2-GSD}, where we show the $10$ lowest energy eigenvalues at ${J=0.5}$ and ${K=0}$, for periodic chains of length ${L=6,7,8,9}$. The degeneracies are $(2,2,2,4)$, which are exactly 2 times the degeneracies of the Ising CFT spectrum, where these energy eigenstates correspond to the operators ${\mathbbm{1}, \sigma,\varepsilon,L_{\pm 1} \sigma}$ via the state-operator correspondence.
Using exact diagonalization on small systems, we find that only the $\Rep(\dl{H})$ symmetry operator corresponding to the 2-dimensional irrep of $H_8$ acts non-trivially on the ground state subspace. Indeed, we find the matrix elements of the symmetry operators $\Al^{{\mathbf{s}}_n}$ and $\Al^{\two}$ to be $\begin{psmallmatrix}
    1 & 0\\0 & 1
\end{psmallmatrix}$ and $\begin{psmallmatrix}
    2 & 0\\0 & -2
\end{psmallmatrix}$ respectively.

Using the terminology coined by \Rf{Bhardwaj:2024qrf}, we call the gapless state realized at the B $\to$ C transition a gapless spontaneous symmetry breaking (gSSB) state, which breaks the $\Rep(\dl{H_8})$ symmetry down to its group-like subsymmetry $\ZZ_2\times\ZZ_2$.

\begin{figure}[t]
    \begin{center}
        \begin{subfigure}{0.45\textwidth}
            \includegraphics[width=\textwidth]{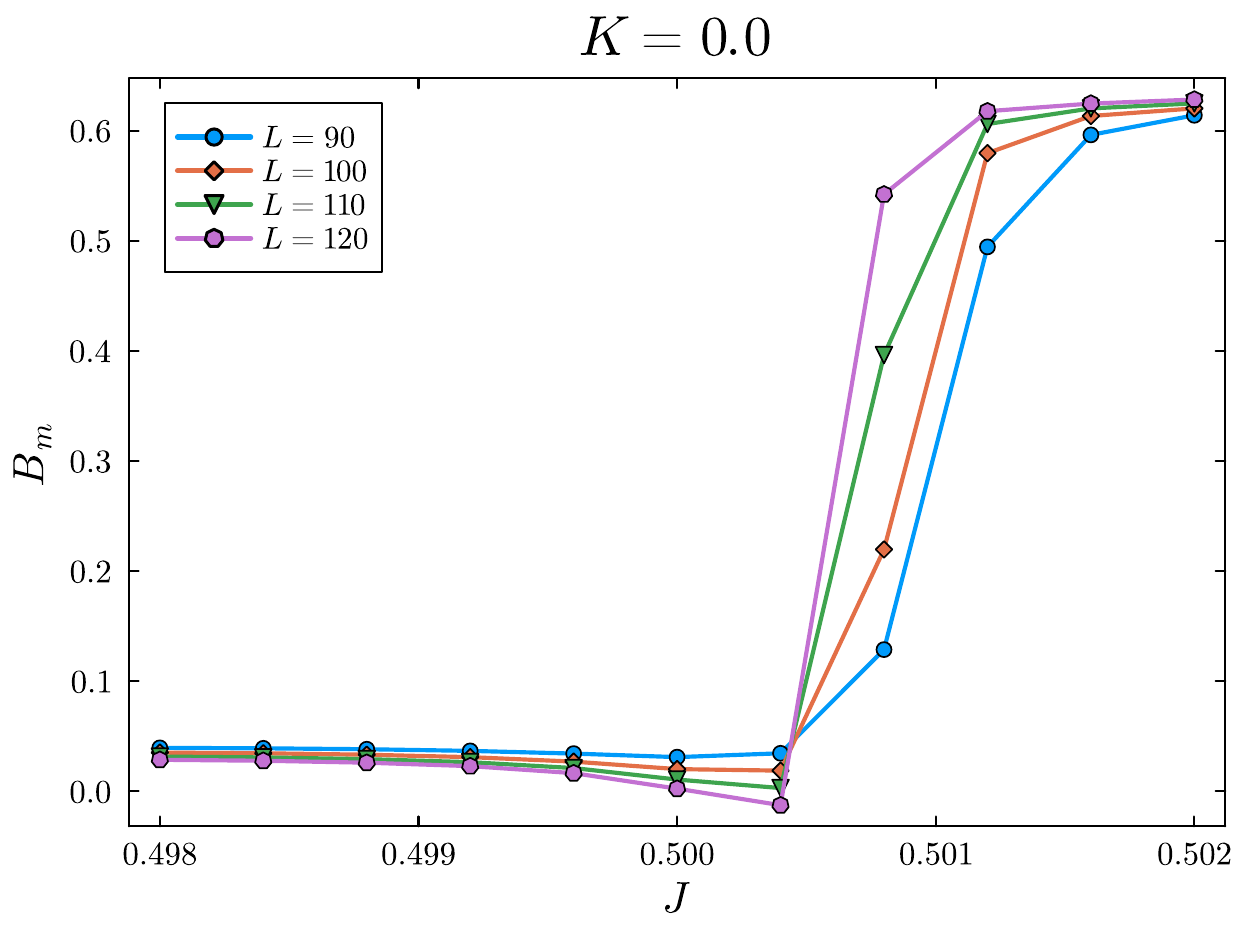}
            \caption{}
            \label{fig:binder-J-1/2}
        \end{subfigure}
        \hspace{0.05\textwidth} 
        \begin{subfigure}{0.45\textwidth}
            \includegraphics[width=\textwidth]{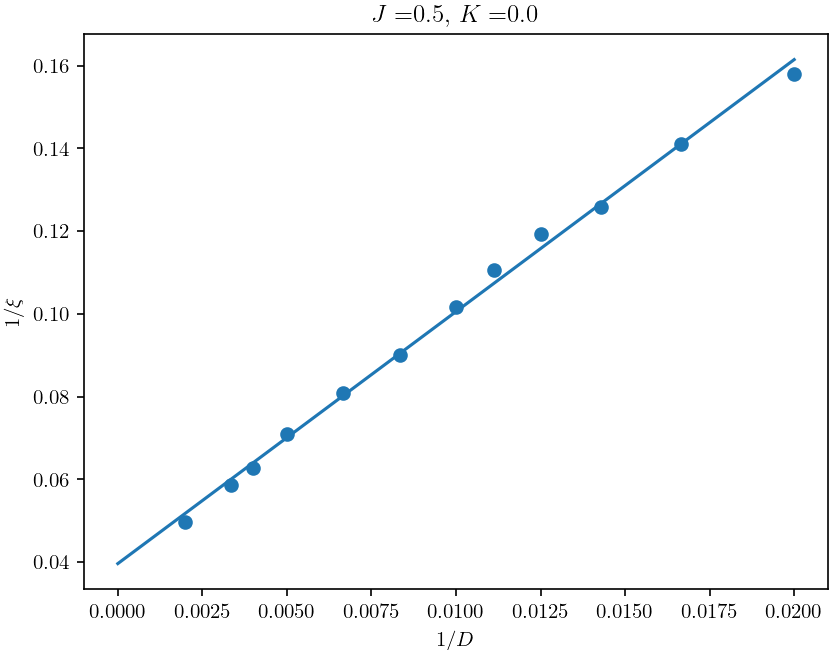}
            \caption{}
            \label{fig:1/xi-1/D-J-1/2}
        \end{subfigure}
    \end{center}
    \caption{(a) Binder cumulant near the A $\to$ D direct phase transition at ${K=0}$. 
    (b) Inverse correlation length $\xi^{-1}$ versus $D^{-1}$ for different iMPS bond dimensions $D$, at ${J = 0.5}$, ${K= 0}$. The numerical data is shown with blue dots, while the blue line is a linear fit.}
    \label{fig:weakly-first}
\end{figure}

\begin{figure}
    \begin{center}
        \begin{subfigure}{0.45\textwidth}
                \includegraphics[width=\textwidth]{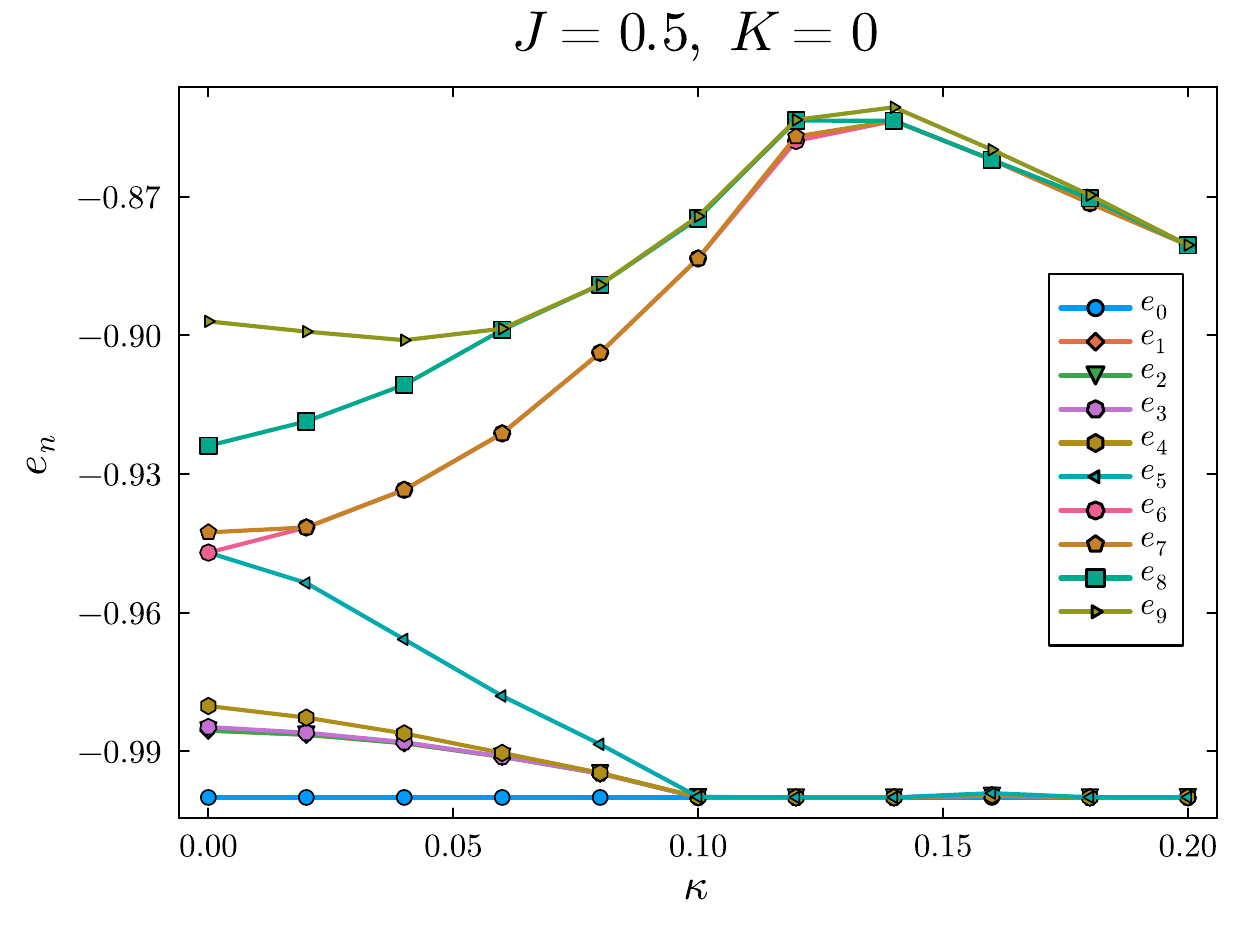}
                \caption{}
                \label{fig:evals-OBF-K=0}
            \end{subfigure}
            \hspace{0.05\textwidth} 
            \begin{subfigure}{0.45\textwidth}
                \includegraphics[width=\textwidth]{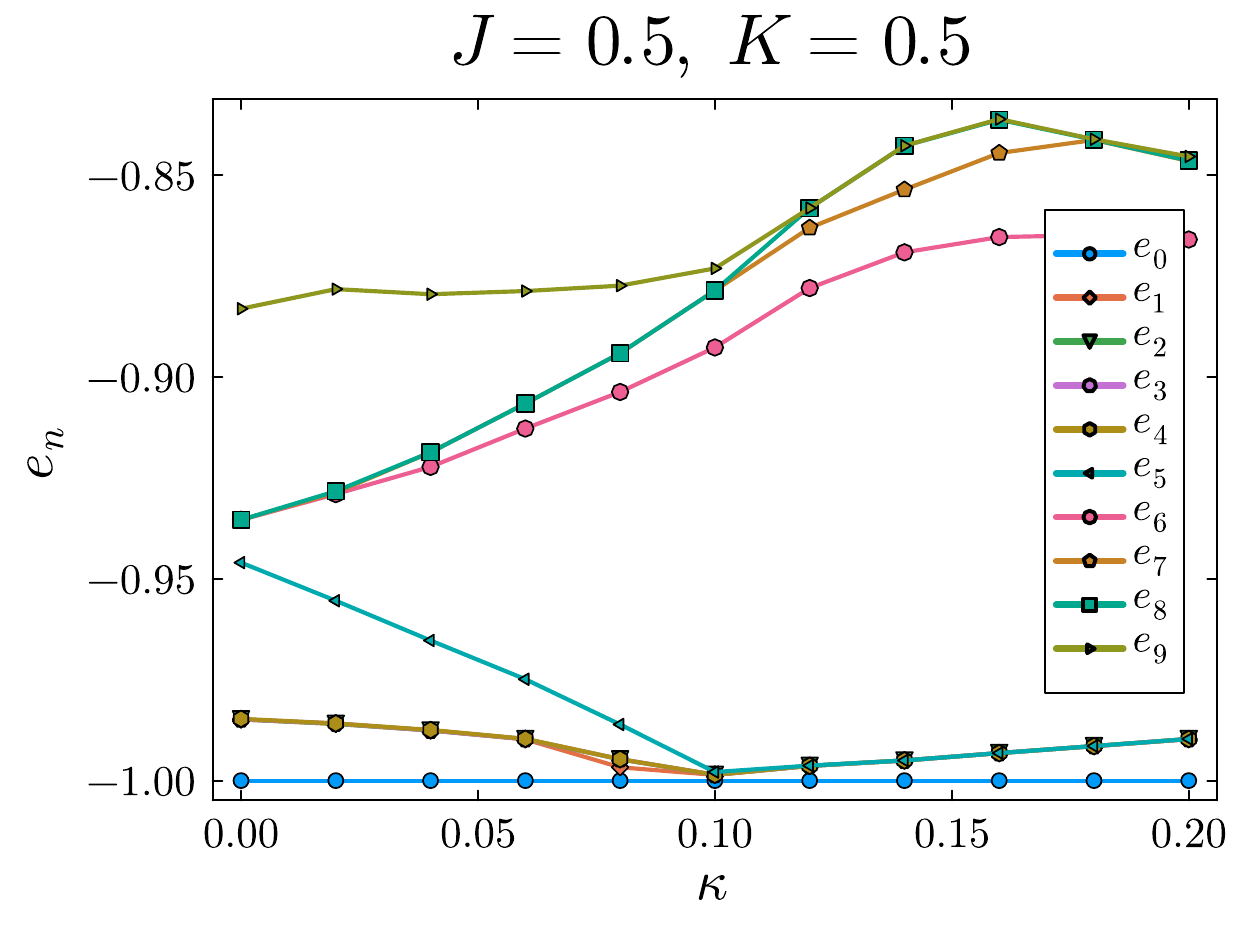}
                \caption{}
                \label{fig:evals-OBF-K=0.5}
            \end{subfigure}
    \end{center}
    \caption{Smallest 10 energy eigenvalues, normalized as in \Cref{fig:Ising2-GSD}, for a periodic chain of ${L=6}$ qudits, with ${J=0.5}$, ${\ofcoeff\in[0,0.2]}$ and (a) ${K=0}$, (b) ${K=0.5}$. The lines are guides to the eye.}
    \label{fig:evals-OBF}
\end{figure}

Let us now comment on the nature of the direct A $\to $ D phase transition, at ${J=0.5}$ for ${0 \le K\lesssim 1}$.
The Binder cumulant computed in the MPS ground state becomes negative right at the transition as the system size is increased. This is a well-known signature of a first-order transition~\cite{vollmayr1993finite}. 
Moreover, we observe an approximate linear fitting of the inverse of the correlation length $\xi^{-1}$ of the iMPS ground state with the inverse of the bond dimension $D^{-1}$,
\begin{equation}
    \frac{1}{\xi} \approx  \frac{1}{\xi_0} + \frac{C}{D} \,,
\end{equation}
with $\xi_0 \approx 25$ for $K=0$ (see \Cref{fig:1/xi-1/D-J-1/2}). 
The above features are consistent with the A $\to$ D transition being first-order. 
As such, we expect the ground state at the transition to be $6$-fold degenerate (for periodic boundary conditions), due to the 5 ferromagnetic ground states and the 1 paramagnetic ground state becoming degenerate. To obtain this GSD numerically, we compute the low-energy spectrum of $\sH_{H_8}$ using DMRG at ${J=0.5}$, ${K=0}$ on a finite-size periodic chain. Performing finite size scaling on the low energy spectrum of this Hamiltonian turns out to be prohibitively expensive, even for modestly large systems.
However, upon adding the following O'Brien-Fendley type term~\cite{fendley2018tricritical} to the above Hamiltonian, with a small coefficient $\ofcoeff$, we find a $6$-fold degenerate ground state: 

\begin{multline}
    \delta \sH_{OF} = \ofcoeff \sum_{j=1}^L\Big [K \sum_{n=1}^3  (\Zr_j \Zl_{j+1} )^{\chi_{{\mathbf{s}}_n}}\Xr_{j+2}^{\chi_{{\mathbf{s}}_n}} +  (\Zr_j \Zl_{j+1}  )^{\chi_\two}\Xr_{j+2}^{\chi_{\two}} \\
    + K \sum_{n=1}^3  \Xr_{j}^{\chi_{{\mathbf{s}}_n}} (\Zr_{j+1} \Zl_{j+2} )^{\chi_{{\mathbf{s}}_n}} +  \Xr_{j}^{\chi_{\two}}(\Zr_{j+1} \Zl_{j+2}  )^{\chi_\two} \Big ] 
\end{multline}
Note that $\delta H_{OF}$ is invariant under the Hopf Kramers-Wannier duality, $\kw$. In \Cref{fig:evals-OBF}, we show the low-energy spectrum for ${L=6}$ with ${\ofcoeff\in[0,0.2]}$ and (a) ${K=0}$, (b) ${K=0.5}$. 
For ${K=0}$, we find that the system is in a gapped phase with $6$ almost exactly degenerate ground states for ${\ofcoeff\gtrsim 0.1}$. (This degeneracy is lifted slightly for ${K =0.5}$, but it is likely a finite-size effect.)
In fact, when ${\ofcoeff=0.25}$ and ${J=K=0.5}$, the Hamiltonian,
\begin{multline} \label{eq:H-fr-free}
    \sH_{H_8} + \delta \sH_{OF} 
    = -\frac L2 +  
    \frac14 \sum_{j=1}^L \Big[ \left(1 - \Xr^{\haarc}_{j-1}\right)\left(1 - (\Zr_j \Zl_{j+1})^{\cohaar}\right) 
    \\
    + \left(1 - (\Zr_j \Zl_{j+1})^{\cohaar}\right)\left(1 - \Xr^{\haarc}_{j+2}\right)\Big]
\end{multline}
is frustration-free,
with exactly 6 degenerate gapped ground states. These include the fixed point ground state of the paramagnetic phase, $\ket{\haar}^{\otimes L}$ (see \cref{eq:PauliXham}) and the 5 fixed point ground states of the ferromagnetic phase $\ket{r}$, for $r\in \Irr(H_8^*)$ (see \cref{eq:RepHdualGHZ}).
Based on the numerical evidence in \Cref{fig:evals-OBF}, we conclude that these are the only ground states of Hamiltonian \eqref{eq:H-fr-free} and that they have an $O(1)$ gap to excited states.
This is consistent with the ground state degeneracy of a gapped phase that fully spontaneously breaks the $\Rep(\dl{H_8})$ self-duality symmetry $\kw$. 
A more thorough investigation of the phase diagram on the self-dual ${(K,\ofcoeff)}$ plane is an interesting question to explore in future work.

\begin{figure}[t]
    \begin{center}
        \begin{subfigure}{0.45\textwidth}
            \includegraphics[width=\textwidth]{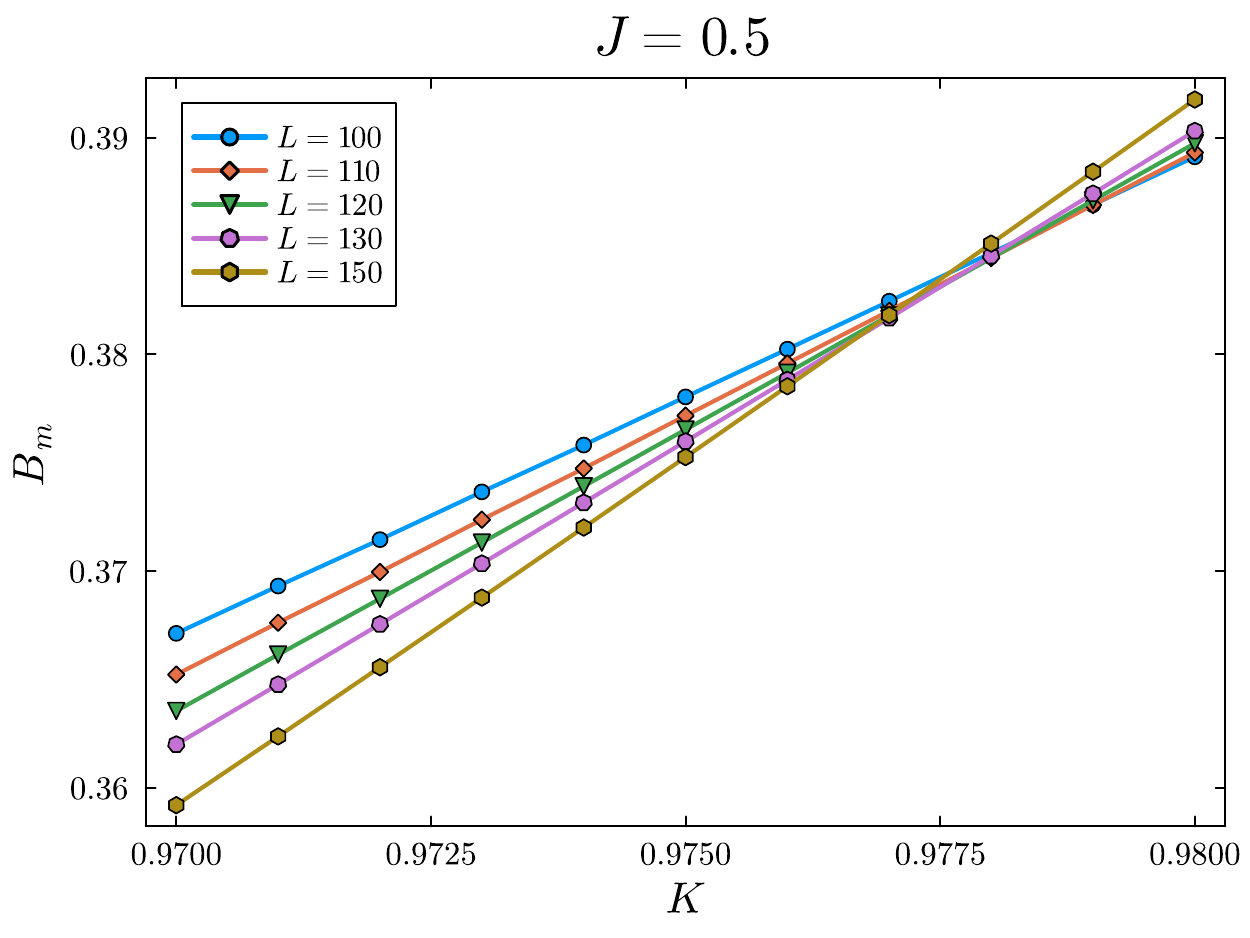}
            \caption{}
            \label{fig:binder-K-0.978}
        \end{subfigure}
        \hspace{0.05\textwidth} 
        \begin{subfigure}{0.45\textwidth}
            \includegraphics[width=\textwidth]{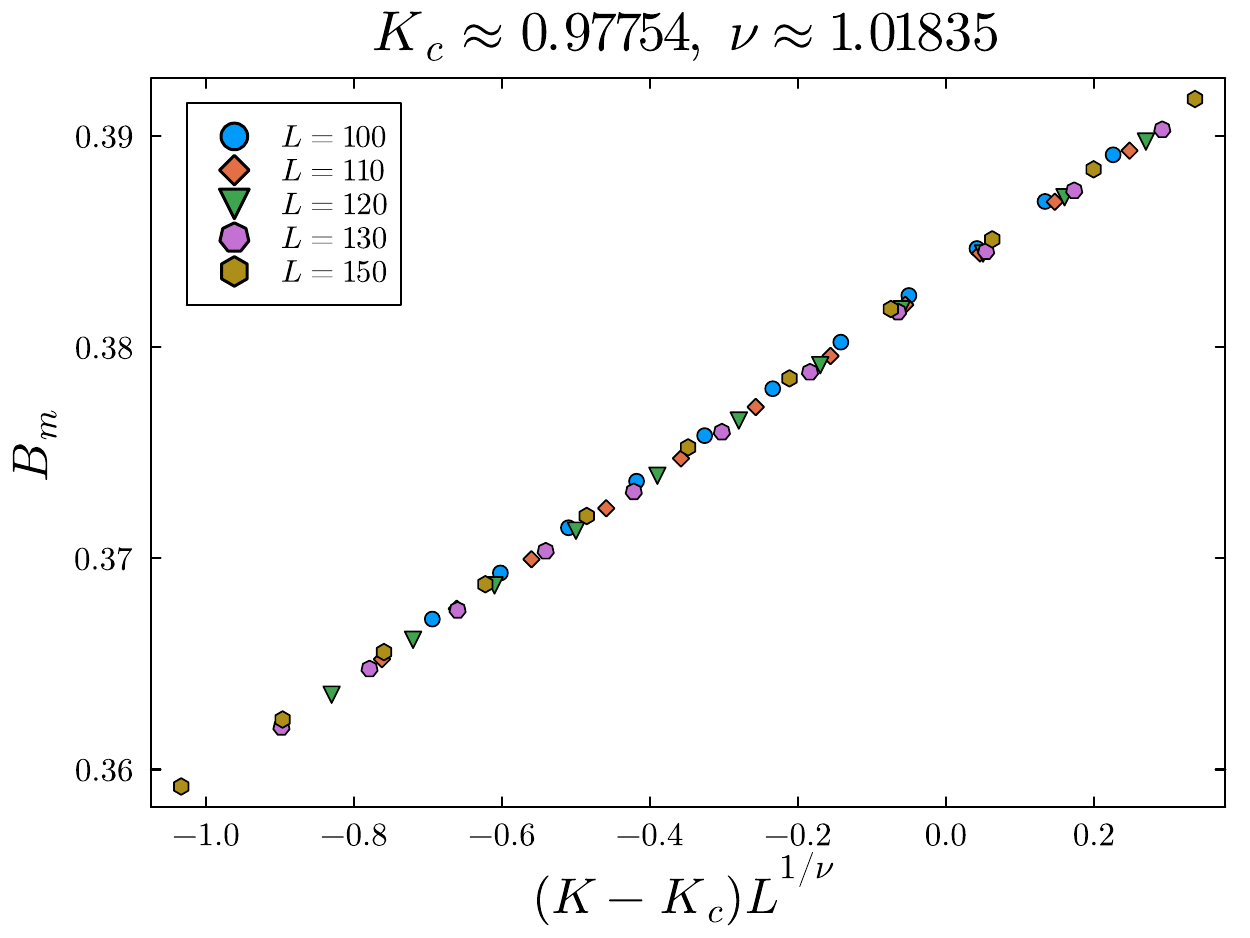}
            \caption{}
            \label{fig:bindercollapse-K-0.978}
        \end{subfigure}
    \end{center}
    \caption{(a) Binder cumulant near the multicritical point on the ${J=0.5}$ line for $L$ between $90$ and $150$. 
    (b) Collapse of the Binder cumulant data to the scaling form ${B_m\propto (K-K_c)^{1/\nu}}$, predicting the location of the multicritical point at ${K_c \approx 0.978}$ and the correlation length critical exponent as ${\nu\approx 1.02}$.}
    \label{fig:multi-Binder}
\end{figure}

\begin{figure}[t!]
    \begin{center}
        \begin{subfigure}{0.45\textwidth}
            \includegraphics[width=\textwidth]{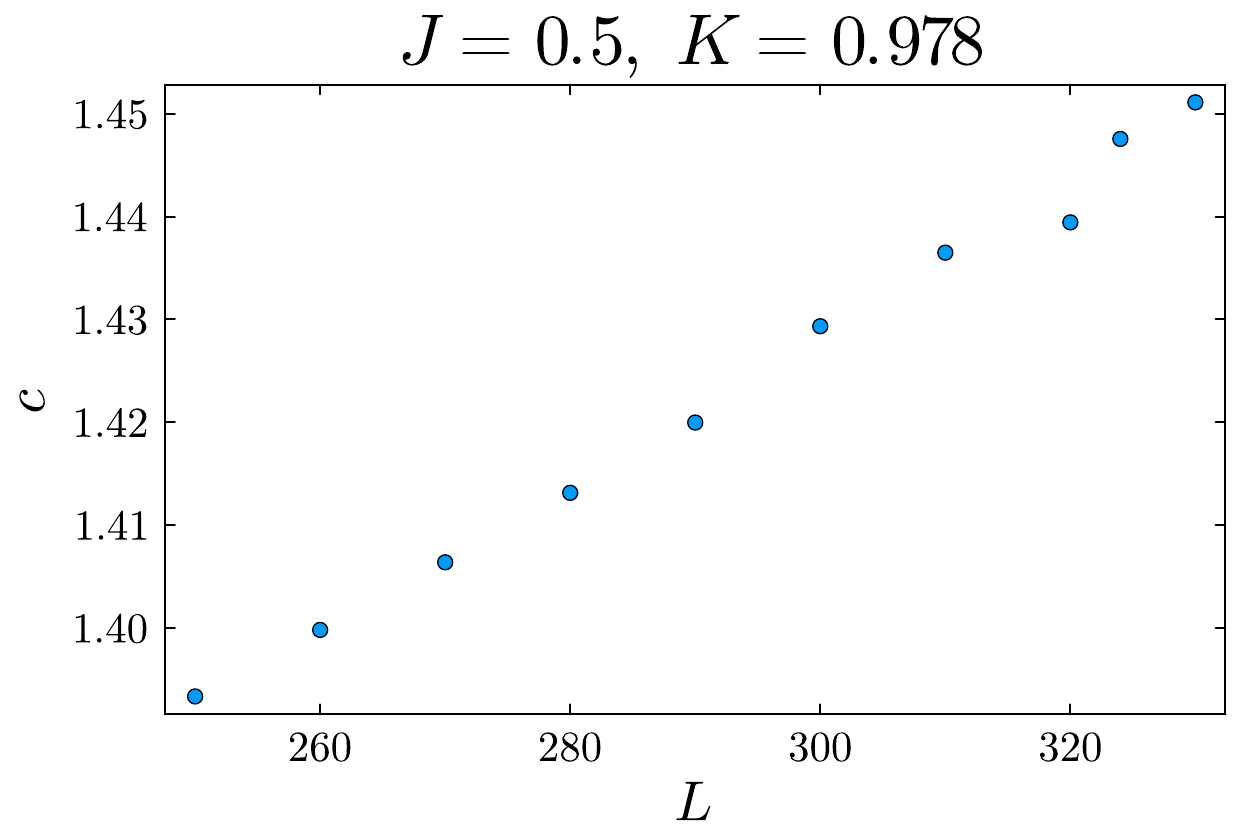}
            \caption{}
            \label{fig:cc-vs-N}
        \end{subfigure}
        \hspace{0.05\textwidth} 
        \begin{subfigure}{0.45\textwidth}
            \includegraphics[width=\textwidth]{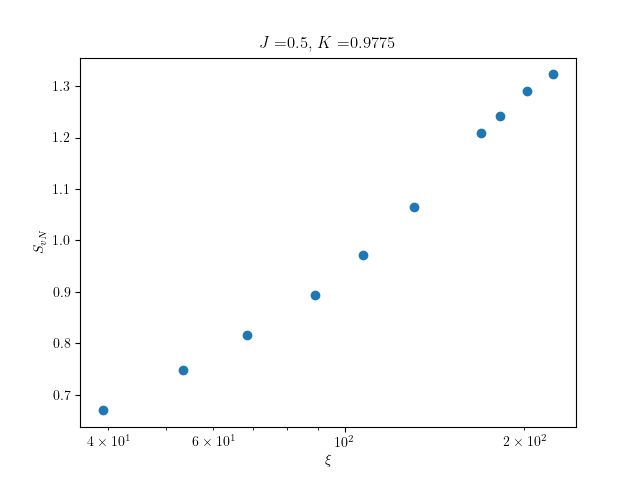}
            \caption{}
            \label{fig:cc-VUMPS}
        \end{subfigure}
    \end{center}
    \caption{(a) Central charge computed from the MPS ground state with open boundary conditions close to the multicritical point, for ${L\in [250,330]}$. (b) Entanglement entropy vs correlation length from VUMPS for bond dimension ${D\in[100,500]}$ displays a slight non-linear scaling.
    }
\end{figure}

As the value of $K$ is increased along the ${J=0.5}$ self-dual line, we find a multicritical point where all 4 gapped phases meet. This transition is located numerically at ${K_c \approx 0.978}$, using the collapse of the Binder cumulant $B_m$; see \Cref{fig:multi-Binder}. 
The critical exponent associated to the corresponding correlation length is ${\nu \approx 1.02}$, which suggests that the corresponding local order parameter $Z_{j,1}$ flows to a scaling dimension ${\Delta = 2 - \frac{1}{\nu} \approx 1.02}$ operator in the IR limit.
The scaling collapse of the Binder cumulant suggests that this is a continuous transition, but we cannot conclusively pin down the universality class. 
The scaling of bipartite entanglement entropy with subsystem size in the MPS ground state is consistent with the Calabrese-Cardy formula. However, the central charge obtained from fitting the entanglement entropy to this scaling law increases with system size $L$, as shown in \Cref{fig:cc-vs-N} for ${J=0.5,\, K=0.978}$ and $L$ between 250 and 330. Our computational resources appear to be insufficient for accessing the thermodynamic limit. 
We complement this with a computation of the central charge from the scaling of entanglement entropy in the iMPS ground state with the correlation length: this is shown in Fig~\ref{fig:cc-VUMPS}. A slight non-linear scaling can be seen. Extracting the central charge from the slope towards higher correlation lengths using the formula ${S= \frac{c}{6}\log \xi}$ gives ${c\approx 3}$, starkly contrasting the value obtained from the MPS. It is therefore very likely that our system size and bond dimension are currently not large enough to make accurate predictions about the ground state.

To conclude, our numerical results provide some (but not in any way conclusive) evidence that the low-energy theory is described by a CFT, enriched with both $\Rep(H_8)$ and self-duality $\kw$ symmetries, and warrants a more thorough numerical study.

\section{\texorpdfstring{$\Rep(H)$}{Rep(H)}-symmetric gapped phases from \texorpdfstring{$H$}{H}-comodule algebras}
\label{sec:H-gapped-phases}

In the previous sections, we assumed the local Hilbert space is the Hopf algebra $H$ itself and constructed symmetry MPOs and symmetric Hamiltonians using the Hopf Pauli matrices. However, there are more exotic symmetry gapped phases that cannot be realized by simple Hopf Pauli matrices alone. Thus, while the Hopf ZX calculus provides a compact and transparent description of local operators, it is not designed to parameterize fully general MPOs on general finite-dimensional Hilbert spaces, since generic MPO algebras typically realize pre-bialgebra structures \cite{molnar2022MPO,yuhan2025mpo}.

In this section, we now assume an arbitrary finite dimensional Hilbert space, and construct more general $\Rep(H)$ symmetry operators and the $\Rep(H)$-symmetric gapped phases using the structure of $H$-modules and $H$-comodule algebras acting on this space \cite{inamura_lattice_2022}. After reviewing the basic notions of $H$-modules and $H$-comodule algebras, we revisit the construction of symmetry operators and Hamiltonians of gapped phases in \Rf{inamura_lattice_2022} using the graphical calculus. We also construct the corresponding order parameters and fixed point ground states. We further specialize to the $H_8$ case and construct all 6 symmetric gapped phases using the $H_8$-comodule algebras and analyze their properties. 

% However, Hopf ZX calculus is sufficient for constructing MPO realizations of anomaly-free non-invertible symmetries on the lattice. Conversely, the algebra generated by such on-site anomaly-free symmetry MPOs is expected to recover the underlying Hopf algebra $H$.
 
Before diving into the mathematical formalism, we briefly provide the relevant background on comodule algebras relevant to the construction of topological phases.
Given an anomaly-free non-invertible symmetry described by $\Rep(H)$, symmetric gapped phases are classified by module categories over $\Rep(H)$~\cite{Thorngren2019iar,Komargodski:2020mxz}, e.g. a single indecomposable object gives a fully symmetric (trivial or SPT) phase, while $m$ indecomposable objects indicate spontaneous symmetry breaking with $m$ degenerate ground states. For lattice constructions, it is convenient to work with the algebraic counterpart of this classification, namely $H$-comodule algebras. Indecomposable exact $\Rep(H)$-module categories are in one-to-one correspondence with $H$-comodule algebras up to equivariant Morita equivalence \cite{andruskiewitsch2007module}. Fixed-point lattice Hamiltonians and their ground states realizing these phases can be constructed from Frobenius $H$-comodule algebras \cite{inamura_lattice_2022}.

Every Hopf algebra $H$ admits at least two canonical $H$-comodule algebras: the trivial algebra $\CC$, corresponding to a fully symmetric gapped phase, and $H$ itself, corresponding to a fully symmetry-broken phase. More generally, inequivalent $H$-comodule algebras give rise to distinct symmetric gapped phases. Some of these are subalgebras of $H$ and describe partially symmetry-broken phases, while others are not subalgebras of $H$ and cannot be characterized by a simple symmetry-breaking pattern, but can be diagnosed by the string order parameters.

Frobenius $H$-comodule algebras also characterize gapped boundaries of the quantum double $D(H)$. As a result, the classification of $\Rep(H)$-symmetric gapped phases coincides with the classification of gapped boundaries of $D(H)$ \cite{zhian2023double,kita2012gappedbdy,kong2013EMdual,zhenghan2016gapbdy,zhenghan2017defect,yuting2018gapbdy,dong2025gapbdy}. Complementary perspectives on fusion category symmetries and their symmetric gapped phases have also been developed using operator algebraic methods \cite{nill1997hopfchain,Szlachanyi93,Jones:2023ptg,jones2025op_alg}.

Our Hopf and comodule algebra construction parallels the MPO framework based on fusion and module categories for anomaly-free non-invertible symmetries \cite{lootens_dualities_2022,Lootens24,delcamp2024dual}. However, it is particularly well suited to lattice realizations, since the tensor-product Hilbert space is built in from the outset. In this approach, the duality operator is determined directly by the Hopf algebra data together with a single additional isomorphism, which is straightforward to construct. By contrast, extracting the corresponding duality operator from bimodule category data is considerably more involved \cite{Lootens24}.

\subsection{\texorpdfstring{$H$}{H}-module and \texorpdfstring{$H$}{H}-comodule}
\label{sec:Hmod-Hcomod}
 A (simple) left $H$-module is the generalization of a(n irreducible) representation of a group (we leave the formal definition to \cref{app:def}). In the graphical notation, we represent the left $H$-module $M$ by a blue line, while $H$ is represented by a black line as before. An element $m\in M$ has the following map, called the $H$-action:
\tikzsetfigurename{H-mod-action}
\begin{equation}
\Gamma: H\otimes M \rightarrow M,\quad (h,m) \mapsto h\cdot m ,\qquad 
\begin{ZX}
    \zxNone{}&&\zxN{} &\\
    &&&\\
    &&\zxrep{$\Gamma$} \ar[dd,blue, on background layer] \ar[uu,blue, on background layer]&\\
        &&&\\
    \zxNone{} \ar[zxarrow,rruu,)]&&\zxNone{} &
\end{ZX}
\end{equation}
This map satisfies the associativity condition,
\begin{equation}\label{eq:mod_asso}
    (hg)\cdot m =h\cdot(g\cdot m),\qquad
    \begin{ZX}
        \zxNone{}&&&&&\\
        &&&&&\\
        &&&&\zxrep{$\Gamma$} \ar[uu,blue] \ar[dddd,blue]&\\
        &&&&&\\
        &&&&&\\
        &\zxX{} \ar[zxarrow,rrruuu,)]&&&&\\
        \zxNone{} \ar[ru,)]&&\zxNone{} \ar[lu,(]&&\zxNone{} &
    \end{ZX}=\begin{ZX}
        \zxNone{}&&&&&\\
        &&&&&\\
        &&&&\zxrep{$\Gamma$} \ar[uu,blue]&\\
        &&&&&\\
        &&&&\zxrep{$\Gamma$} \ar[dd,blue] \ar[uu,blue]&\\
        &&&&&\\
        \zxNone{} \ar[zxarrow,rrrruuuu,)]&&\zxNone{} \ar[zxarrow,rruu,)]&&\zxNone{} &
    \end{ZX} 
\end{equation}
$H$ acts on the tensor product of the $H$-modules $M_1$ and $M_2$ as,
\tikzsetfigurename{tensorprod}
\begin{equation}\label{eq:tensorprodirreps}
\sum_{(h)} (h_{(1)}\cdot m_1) \otimes (h_{(2)}\cdot m_2),\qquad 
\begin{ZX}
    \zxNone{}&&\zxN{} &&\\
    &&&&\\
    &&\zxrep{$\Gamma_1$} \ar[dddd,blue, on background layer] \ar[uu,blue, on background layer]&&\zxrep{$\Gamma_2$} \ar[dddd,blue, on background layer] \ar[uu,blue, on background layer]\\
        &&&&\\
    \zxZ{} \ar[dd] \ar[zxarrow,rruu,)] \ar[zxarrow,rrrruu,(]&&\zxNone{} &&\\
    &&&&\\
    &&&&
\end{ZX}
\end{equation}
For a group algebra ${H=\CC[G]}$, $\comult(g)=g\otimes g$, therefore, the tensor product of irreps is the same as tensor product of the corresponding matrices. But for a general Hopf algebra, it involves the non-trivial comultiplication structure. 
Note that a left $H$-module can be turned into a right $H$-module by the antipode,
\tikzsetfigurename{right-H-mod}
\begin{equation}
    m\cdot h \equiv S(h)m,\quad \begin{ZX}
        &&&&\\
        &&&&\\
        &\zxrep{$\Gamma$} \ar[uu,blue] \ar[dddd,blue]&&&\\
        \zxrep{$S$}\ar[zxarrow,ru,)] \ar[rrddd,N]&&&&\\
        &&&&\\
        &&&&\\
        &&&&
    \end{ZX}
\end{equation}

% Similar to the left $H$-modules, we can define the right $H$-comodule, using the coaction $\delta$ and denote the vector space by a red line,
Dual to a left $H$-module is a right $H$-comodule $V$, which will be denoted graphically by a red line. The comodule struture is defined using the coaction $\delta$
\tikzsetfigurename{H-comod}
\begin{equation}
    % \delta: V\rightarrowtail V\otimes H,
    \delta: V\rightarrow V\otimes H,
    \quad v\mapsto \sum v_{(0)}\otimes v_{(1)}, \quad 
    \begin{ZX}
        \zxN{}&&&&\\
        &&&&\\
        \zxrep{$\delta$} \ar[dd,red] \ar[uu,red] \ar[zxdrrow,rruu,(]&&&&\\
        &&&&\\
        \zxN{}&&&&
    \end{ZX}
\end{equation}
The coaction satisfies the coassociativity condition,
\tikzsetfigurename{H-comod-asso}
\begin{equation}\label{eq:comodule-coasso}
    (\id_V\otimes \Delta) \circ \delta = (\delta\otimes \id_H)\circ \delta,\quad \begin{ZX}
        &&&&\\
        &&\zxZ{} \ar[lu,)] \ar[ru,(]&&\\
        &&&&\\
        \zxrep{$\delta$} \ar[dd,red] \ar[uuu,red] \ar[zxdrrow,rruu,(] &&&&\\
        \zxN{}&&&&\\
        &&&&
    \end{ZX} = \begin{ZX}
    &&&&\\
    &&&&\\
    \zxrep{$\delta$} \ar[zxdrrow,rruu,(] \ar[uu,red]&&&&\\
    &&&&\\
    \zxrep{$\delta$} \ar[dd,red]  \ar[uu,red]  \ar[zxdrrow,rrrruuuu,(]&&&&\\
    &&&&\\
    &&&&
    \end{ZX} \,.
\end{equation}
A morphism of right $H$-comodules ${f:V\rightarrow V'}$ is a linear map satisfying
\tikzsetfigurename{H-comod-mor}
\begin{equation}\label{eq:comodule_mor}
    (f\otimes \id_H) \circ \delta_V = \delta_{V'}\circ f,\quad 
    \begin{ZX}
        &&&&\\
        &&&&\\
        \zxrep{$f$} \ar[uu,red]&&&&\\
        &&&&\\
        \zxrep{$\delta_V$} \ar[dd,red] \ar[uu,red] \ar[zxdrrow,rruuuu,(]&&&&\\
        &&&&\\
        \zxN{}&&&&
    \end{ZX} = \begin{ZX}
    &&&&\\
    &&&&\\
    \zxrep{$\delta_{V'}$} \ar[uu,red] \ar[zxdrrow,rruu,(]&&&&\\
    &&&&\\
    \zxrep{$f$} \ar[dd,red] \ar[uu,red] &&&&\\
    &&&&\\
    \zxN{}&&&&
    \end{ZX}
\end{equation}

\subsection{\texorpdfstring{$\Rep(H)$}{Rep(H)} symmetry operator on a chain}

We recall that $\Rep(H)$ is the tensor category of finite-dimensional left $H$-modules, which we referred to earlier as representations of $H$. For any $H$-module $\Gamma$, we would like to construct a corresponding symmetry operator on a periodic chain of qudits. 

% In general, the local Hilbert space can be isomorphic to the underlying vector space of an $H$-comodule $V$, i.e. $\CC^{|V|}$.
Let us consider the local Hilbert space associated with each site of the chain to be isomorphic to the underlying vector space of an $H$-comodule $V$, i.e. $\CC^{|V|}$.
The $\Rep(H)$ symmetry MPO is constructed out of the tensor,
% The $\Rep(H)$ symmetry MPO tensor is given by
\begin{equation}\label{eq:repHmpocomod}
    \tikzname{repHmpocomod}
    \begin{ZX}
        & & & &&\\
        &  &  \zxrep{$\Gamma$} \ar[ll,blue] \ar[rrr,blue] && &\\
        & & & &&\\
        &  &   & & \zxN{} \ar[zxarrow,lluu,N] &\\
        & & &  \zxrep{$\delta$} \ar[dd,red] \ar[uuuu,red] \ar[zxdrrow,ru,(] &&\\
        & & & &&\\
        & & & &&
    \end{ZX}
\end{equation}
where the blue line corresponds to the virtual bond of the $\Rep(H)$ MPO and its dimension is equal to the dimension of the $H$-module $d_\Gamma$, the red line corresponds to the comodule $V$ with dimension $|V|$. Contracting the $\Rep(H)$ MPO tensors on a closed chain produces 
\tikzsetfigurename{rephmpo2}
\begin{equation}
    \Br_V^{\Gamma}=
    \begin{ZX}
        && && && && && \\
        \zxN{}\ar[rrrrr,blue]&& && &&... && &&\zxN{} \ar[lll,blue] \\
        \zxN{} \ar[u,C,blue]&&\zxrep{$\Gamma$} \ar[ll,blue] \ar[rr,blue] \ar[zxarrowr,rrdd,N] && \zxrep{$\Gamma$}\ar[r,blue] \ar[zxarrowr,rrdd,N] && ... &&\zxrep{$\Gamma$} \ar[rr,blue] \ar[zxarrowr,rrdd,N] \ar[l,blue] && \zxN{} \ar[u,C-,blue]\\
        && && && && && \\
        && && && && && \\
        &&& \zxrep{$\delta$} \ar[dd,red] \ar[uuuuu,red] \ar[zxdrrow,ru,(]   && \zxrep{$\delta$} \ar[dd,red] \ar[uuuuu,red] \ar[zxdrrow,ru,(]&& \zxN{} && \zxrep{$\delta$} \ar[dd,red] \ar[uuuuu,red] \ar[zxdrrow,ru,(] &\\
        && && && && && \\
        && && && && && \\
    \end{ZX} = 
    \begin{ZX}
        && && && && && \\
        &\zxrep{$\Gamma$} \ar[u,C,blue] \ar[u,C-,blue]& && && && && \\
        \zxN{} &&\zxX{} \ar[lu,)] \ar[rr] \ar[zxarrowr,rrdd,N] && \zxX{}\ar[r] \ar[zxarrowr,rrdd,N] && ... &&\zxN{}  \ar[rrdd,N] \ar[l] && \\
        && && && && && \\
        && && && && && \\
        &&& \zxrep{$\delta$} \ar[dd,red] \ar[uuuuu,red] \ar[zxdrrow,ru,(]   && \zxrep{$\delta$} \ar[dd,red] \ar[uuuuu,red] \ar[zxdrrow,ru,(] && \zxN{} && \zxrep{$\delta$} \ar[dd,red] \ar[uuuuu,red] \ar[zxdrrow,ru,(] &\\
        && && && && && \\
        && && && && && \\
    \end{ZX}
    =\begin{ZX}
        && && && && && \\
        &\zxrep{$\chi_\Gamma$} & && && && && \\
        \zxN{} &&\zxX{} \ar[lu,)] \ar[rr] \ar[zxarrowr,rrdd,N] && \zxX{}\ar[r] \ar[zxarrowr,rrdd,N] && ... &&\zxN{}  \ar[rrdd,N] \ar[l] && \\
        && && && && && \\
        && && && && && \\
        &&& \zxrep{$\delta$} \ar[dd,red] \ar[uuuuu,red] \ar[zxdrrow,ru,(]   &&\zxrep{$\delta$} \ar[dd,red] \ar[uuuuu,red] \ar[zxdrrow,ru,(] && \zxN{} && \zxrep{$\delta$} \ar[dd,red] \ar[uuuuu,red] \ar[zxdrrow,ru,(] &\\
        && && && && && \\
        && && && && && \\
    \end{ZX}
\end{equation}
which is due to the associativity condition of the $H$-module \eqref{eq:mod_asso},
\tikzsetfigurename{repHmult}
\begin{equation}\label{eq:repsmult}
    \begin{ZX}
        & \zxrep{$\Gamma$} \ar[l,blue] \ar[ddr,N] \ar[r,blue]& &\zxrep{$\Gamma$}\ar[l,blue] \ar[ddr,N] \ar[r,blue]&\\
        &&&&\\
        &&&&\\
    \end{ZX} =\begin{ZX}
        & \zxrep{$\Gamma$} \ar[l,blue] \ar[dr,N] \ar[r,blue] && \\
        &&\zxX{} \ar[dr,)] \ar[dl,(]& \\
        &&&
    \end{ZX} 
\end{equation}

Upon fusing two MPOs, using \cref{eq:comodule-coasso,eq:chi_mult}, we get
\begin{equation}
    \Br_V^{\Gamma_i} \times \Br_V^{\Gamma_j} = \sum_k N_{\Gamma_i,\Gamma_j}^{\Gamma_k} \Br_V^{\Gamma_k} 
\end{equation}
where $N_{\Gamma_i,\Gamma_j}^{\Gamma_k}$ is the structure constant. The regular representation corresponds to taking the $H$-module to be $H$ itself and the $H$-action is given by the multiplication tensor of $H$
\tikzsetfigurename{repreg}
\begin{equation}
    \begin{ZX}
        &   \zxrep{Reg} \ar[l,blue] \ar[ddr,N] \ar[r,blue] & \\
        &&\\
        &&
    \end{ZX} = \begin{ZX}
        %&&\zxX{} \ar[ll,->,>=stealth,pos=0.2] \ar[rr,<-,>=stealth,pos=0.5] \ar[rdd,(,<-,>=stealth,pos=0.5] &&\\
        &&\zxX{} \ar[ll] \ar[rr] \ar[rdd,N] &&\\
        &&&&\\
        &&&&\\
    \end{ZX}
\end{equation}
Then the corresponding symmetry MPO is
\tikzsetfigurename{rephmporeg}
\begin{equation}
    \Br_V^\cohaar= \begin{ZX}
        && && && && && \\
        &\zxX{} \ar[u,C] \ar[u,C-]& && && && && \\
        \zxN{} &&\zxX{} \ar[lu,)] \ar[rr] \ar[zxarrowr,rrdd,N] && \zxX{}\ar[r] \ar[zxarrowr,rrdd,N] && ... &&\zxN{}  \ar[rrdd,N] \ar[l] && \\
        && && && && && \\
        && && && && && \\
        &&& \zxrep{$\delta$} \ar[dd,red] \ar[uuuuu,red] \ar[zxdrrow,ru,(]    && \zxrep{$\delta$} \ar[dd,red] \ar[uuuuu,red] \ar[zxdrrow,ru,(]  && \zxN{} && \zxrep{$\delta$} \ar[dd,red] \ar[uuuuu,red] \ar[zxdrrow,ru,(]  &\\
        && && && && && \\
        && && && && && \\
    \end{ZX}
    =|H|\begin{ZX}
        && && && && && \\
        &\zxXTri'{} & && && && && \\
        \zxN{} &&\zxX{} \ar[lu,)] \ar[rr] \ar[zxarrowr,rrdd,N] && \zxX{}\ar[r] \ar[zxarrowr,rrdd,N] && ... &&\zxN{}  \ar[rrdd,N] \ar[l] && \\
        && && && && && \\
        && && && && && \\
        &&&\zxrep{$\delta$} \ar[dd,red] \ar[uuuuu,red] \ar[zxdrrow,ru,(]   && \zxrep{$\delta$} \ar[dd,red] \ar[uuuuu,red] \ar[zxdrrow,ru,(]  && \zxN{} && \zxrep{$\delta$} \ar[dd,red] \ar[uuuuu,red] \ar[zxdrrow,ru,(]   &\\
        && && && && && \\
        && && && && && \\
    \end{ZX} = |H|\begin{ZX}
        && && && && && \\
        && && && && &&\zxXTri'{} \ar[ld,)] \\
        &&\zxN{}  \ar[rr]  && \zxX{}\ar[r]  && ... &&\zxX{} \ar[r]  \ar[l] &&  \\
        & \zxrep{$\delta$} \ar[dd,red] \ar[uuu,red] \ar[zxdrrow,ru] &&  \zxrep{$\delta$} \ar[dd,red] \ar[uuu,red] \ar[zxdrrow,ru,(] && \zxN{} && \zxrep{$\delta$} \ar[dd,red] \ar[uuu,red] \ar[zxdrrow,ru,(]  && \zxN{} &\\
        && && && && && \\
        && && && && && \\
    \end{ZX} 
\end{equation}
Since the cointegral can be decomposed as
\begin{equation}
    \cohaar = \frac{1}{|H|}\sum_{\Gamma\in \text{Irr}(H)} d_\Gamma \chi_\Gamma,
\end{equation}
where $d_\Gamma$ is the dimension of simple $H$-module $\Gamma$, we have
\begin{equation}
    \Br_V^\cohaar = \frac{1}{|H|}\sum_{\Gamma\in \text{Irr}(H)} d_\Gamma \Br_V^\Gamma \,.
\end{equation}

When $V$ is chosen to be the Hopf algebra $H$ itself, the local Hilbert space associated with each site is isomorphic to the underlying vector space of the Hopf algebra $H$, i.e., $\CC^{|H|}$. Moreover, the local Hilbert space has a canonical right $H$-coaction given by the comultiplication $\Delta$. 
This defines a canonical self-comodule structure on a Hopf algebra. 
The $\Rep(H)$ symmetry MPO from the above definitions then reduces to
\tikzsetfigurename{repHmpo1}
\begin{equation}\label{eq:repHmpotensor}
    \begin{ZX}
        & & & &&\\
        &  &  \zxrep{$\Gamma$} \ar[zxarrowr,rrdd,N] \ar[ll,blue] \ar[rrr,blue] && &\\
        & & & &&\\
        &  &   & & \zxN{} &\\
        & & &  \zxZ{} \ar[dd] \ar[uuuu,3d above,very thick] \ar[ru,(] &&\\
        & & & &&\\
        & & & &&
    \end{ZX} \,,
\end{equation}
thereby reproducing \eqref{eq:rephmpo}, ${\Br_{V=H}^{\Gamma} = \Br^{\Gamma}}$.

\subsection{\texorpdfstring{$\Rep(H)$}{Rep(H)} symmetric gapped phases: fixed point Hamiltonians}
\label{sec:fp-Hams-comod}

Commuting projector Hamiltonians realizing $\Rep(H)$ symmetric gapped phases were constructed using $H$-comodule algebras in \Rf{inamura_lattice_2022}. Here, we review that construction and construct corresponding ground state wave functions as well as (string) order parameters.
$\Rep(H)$-symmetric gapped phases are classified by the indecomposable module categories over $\Rep(H)$, or equivalently, by indecomposable $H$-comodule algebras up to equivariant Morita equivalence~\cite{andruskiewitsch2007module} (Thm 1.25). 
For the group algebra $\CC[G]$, its comodule algebras are given by $\CC[K]^{\psi}$, where $K$ is a subgroup of $G$ and ${[\psi]\in H^2(K,U(1))}$ is a 2-cocycle twist. This gives the classification of module categories over $\Rep(G)$~\cite{ostrik_module_2003,etingof2016tensor}.

Here, we will work with right $H$-comodule algebras. A right $H$-comodule algebra consists of the triple $(V,m_V,\eta_V)$, where $V$ is a right $H$-comodule, $m_V$ is a multiplication structure defined on $V$,
\begin{equation}
    m_V:V \otimes V \rightarrow V,\quad \tikzsetnextfilename{multA}
    \begin{ZX}
        & \zxN{}  &\\
        & \zxamult{} \ar[u,red] &  \\
        \zxN{} \ar[ru, bend left,red]  & 
        &  \zxN{} \ar[lu, bend right,red]
    \end{ZX}
\end{equation}
% and,
and $\eta_V$ is the unit, which satisfies
\tikzsetfigurename{comod-unit}
\begin{align}
    \begin{ZX}
        &   & \zxN{} \ar[d,red] \\
        &   & [\zxWRow] \\
        &   & \zxamult{} \ar[u,red]  \\
        & \zxamult{} \ar[ru, bend left,red]  & 
        \zxN{} \ar[u,red]\ar[d,red]\\
        & [\zxWRow] &
    \end{ZX}
    =
    \begin{ZX}
        & \zxN{} \ar[d,red] &   \\ 
        & [\zxWRow] &   \\ 
        & \zxamult{} \ar[u,red]  &   \\ 
        & \zxN{} \ar[u,red]\ar[d,red]    & \zxamult{} \ar[lu, bend right,red]  \\
        & [\zxWRow] &
    \end{ZX}
    =
    \begin{ZX}
        \zxN{} \ar[d,red] \\
        [\zxWRow] \\
        \zxN{} \ar[u,red] \ar[d,red] \\
        [\zxWRow] \\
        \zxN{} \ar[u,red] 
    \end{ZX} \ .
\end{align}
We use pink dots to represent the multiplication and the unit of the comodule algebra.
The multiplication is compatible with the $H$-coaction of the underlying comodule $V$,
% And it is compatible with the $H$-coaction, 
\tikzsetfigurename{comod-alg-coaction}
\begin{equation}\label{eq:comod-alg-coaction}
    \begin{ZX}
        &&\\
        &&\\
        &\zxN{}\ar[uu,red] \ar[zxdrrow, ruu,(]&\\
        & \zxN{}  &\\
        & \zxamult{} \ar[uu,red] &  \\
        \zxN{} \ar[ru, bend left,red]  & 
        &  \zxN{} \ar[lu, bend right,red]
    \end{ZX} = \begin{ZX}
    &&&&\\
    & \zxN{}  &&&\\
    & \zxamult{} \ar[uu,red] &  &\zxX{} \ar[uu]&\\
    &&&&\\
    \zxN{} \ar[ruu, bend left,red] \ar[d,red] \ar[rrruu,zxdrrow,S]  & 
    &  \zxN{} \ar[luu, bend right,red] \ar[d,red] \ar[ruu,zxdrrow,C-] &&\\
    &&&&
    \end{ZX},\quad \begin{ZX}
    &&\\
    &&\\
    \zxN{} \ar[zxdrrow, ruu,(]&&\\
    &&\\
    \zxamult{} \ar[uuuu,red]&&
    \end{ZX} = \begin{ZX}
    &&\\
    &&\\
    &&\\
    &&\\
    \zxamult{} \ar[uuuu,red]&&\zxX{} \ar[uuuu]
    \end{ZX} \ .
\end{equation}

We can further equip this $H$-comodule algebra with a comultiplication $\Delta_V$ and a counit $\epsilon_V$, such that they satisfy the Frobenius relation,
\tikzsetfigurename{comod-Frobenius}
\begin{equation}\label{eq:comod-alg-frob}
    \begin{ZX}
        &&\\
        \zxamult{} \ar[u,red] \ar[dd,red] \ar[rd,red]&&\\
        &\zxamult{} \ar[uu,red] \ar[d,red]&\\
        &&
    \end{ZX} = \begin{ZX}
        &&\\
        &\zxamult{} \ar[u,red] \ar[dd,red]&\\
        \zxamult{}\ar[uu,red] \ar[d,red] \ar[ru,red]&&\\
        &&\\
        &&
    \end{ZX} = \begin{ZX}
        &&\\
        &\zxamult{} \ar[lu,red] \ar[ru,red] \ar[d,red]&\\
        &\zxamult{} \ar[ld,red] \ar[rd,red]&\\
        &&
    \end{ZX} \ ,
 %    \quad \begin{ZX}
    % &\zxN{}  & \\
    % & & \\
    % &\zxamult{} \ar[uu,red] \ar[dd,C-,red] \ar[dd,C,red] & \\
    % &   &\\
    % &\zxamult{} \ar[dd,red]   &   \\
    % &  &  \\
    % &\zxN{}  &  
    % \end{ZX} = \begin{ZX}
    % \zxN{} \ar[dddd,red]\\
    % \\
    % \\
    % \\
    % \end{ZX}.
\end{equation}
and the separable condition,\footnote{When the separability holds up to a scalar pre-factor not equal to 1, a Frobenius algebra is called \textit{special}. Appropriate rescaling of the counit and the comultiplication can turn a special Frobenius algebra into a separable one.}
\tikzsetfigurename{comod-separable}
\begin{equation}\label{eq:comod-alg-separable}
    \begin{ZX}
    &\zxN{}  & \\
    & & \\
    &\zxamult{} \ar[uu,red] \ar[dd,C-,red] \ar[dd,C,red] & \\
    &   &\\
    &\zxamult{} \ar[dd,red]   &   \\
    &  &  \\
    &\zxN{}  &  
    \end{ZX} = \begin{ZX}
    \zxN{} \ar[dddd,red]\\
    \\
    \\
    \\
    \end{ZX} \ .
\end{equation}
% One can rescale the counit and comultiplication of the special Frobenius algebra to make it separable. 
The coalgebra structure defined by $\Delta_V$ and $\epsilon_V$ is also compatible with the $H$-coation,
\tikzsetfigurename{comod-coprod}
\begin{equation}\label{eq:comod-coalg-coaction}
    \begin{ZX}
        &&&&\\
        &&&&\\
        &\zxamult{} \ar[dd,red] \ar[lu,),red] \ar[ru,(,red]&&&\\
        &&&&\\
        &\zxN{}  \ar[d,red]\ar[zxdrrow,rruuu,(]&&&\\
        &&&&
    \end{ZX} = \begin{ZX}
    &&&&\\
    &&&\zxX{} \ar[u]&\\
    &&&&\\
    \zxN{}  \ar[uuu,red]\ar[zxdrrow,rrruu,S]&&\zxN{}  \ar[uuu,red]\ar[zxdrrow,ruu,(]&&\\
    &\zxamult{} \ar[d,red] \ar[lu,),red] \ar[ru,(,red]&&&\\
    &&&&
    \end{ZX}
\end{equation}
Thus, the quintuple $(V,m_V,\eta_V,\Delta_V,\epsilon_V)$ forms a separable Frobenius $H$-comodule algebra.

It follows from \cref{eq:comod-alg-coaction,eq:comod-coalg-coaction} that $\Delta_V \circ m_V$ satisfies the ``pull through'' condition,
\tikzsetfigurename{comod-commutation}
\begin{equation}
\begin{ZX}
    &&&&\\
    &&&\zxX{} \ar[u]&\\
    &&&&\\
    \zxN{}  \ar[uuu,red]\ar[zxdrrow,rrruu,S]&&\zxN{}  \ar[uuu,red]\ar[zxdrrow,ruu,(]&&\\
    &\zxamult{} \ar[dd,red] \ar[lu,),red] \ar[ru,(,red]&&&\\
    &&&&\\
    &\zxamult{} \ar[ld,(,red] \ar[rd,),red]&&&\\
    &&&&
\end{ZX} = \begin{ZX}
&&&&\\
&&&&\\
&\zxamult{} \ar[dd,red] \ar[lu,),red] \ar[ru,(,red]&&&\\
&&&&\\
&\zxN{}  \ar[dd,red]\ar[zxdrrow,rruuu,(]&&&\\
&&&&\\
&\zxamult{} \ar[ld,(,red] \ar[rd,),red]&&&\\
&&&&
\end{ZX} = \begin{ZX}
&&&&\\
&\zxamult{} \ar[lu,),red] \ar[ru,(,red]&&&\\
& &&&\\
& \zxamult{} \ar[uu,red] &  &\zxX{} \ar[uuu]&\\
&&&&\\
\zxN{} \ar[ruu, bend left,red] \ar[d,red] \ar[rrruu,zxdrrow,S]  & 
&  \zxN{} \ar[luu, bend right,red] \ar[d,red] \ar[ruu,zxdrrow,C-] &&\\
&&&&
\end{ZX} \ .
\end{equation}
Due to this pull through condition and \eqref{eq:repsmult}, the tensor ${\Delta_V\circ m_V}$ commutes with the $\Rep(H)$ symmetry MPO~\eqref{eq:repHmpocomod}. Therefore, the tensor $\Delta_V \circ m_V: V\otimes V \rightarrow V\otimes V$ gives a 2-body interaction term that commutes with the $\Rep(H)$ symmetry operator. 
% One can show this tensor commutes with the $\Rep(H)$ symmetry MPO tensor using \eqref{eq:repsmult}. Therefore, the tensor $\Delta_V \circ m_V: V\otimes V \rightarrow V\otimes V$ gives the 2-body interaction term that commutes with the $\Rep(H)$ symmetry. 
% Furthermore, it gives a commuting projector Hamiltonian due to the Frobenius relation \eqref{eq:comod-alg-frob},
Furthermore, these terms are mutually commuting due to the Frobenius relation \eqref{eq:comod-alg-frob} and they are projectors due to the separable condition \eqref{eq:comod-alg-separable}. Thus we have a commuting projector Hamiltonian~\cite{inamura_lattice_2022},
\tikzsetfigurename{commuting-projector-Ham}
\begin{equation}
    \sH_V =-\sum_j \begin{ZX}
        &&&&\\
        &\zxamult{} \ar[dd,red] \ar[lu,),red] \ar[ru,(,red]&&&\\
        &&&&\\
        &\zxamult{} \ar[ld,(,red] \ar[rd,),red]&&&\\
        &&&&\\[5 pt]
        \zxbase{j}&&\zxbase{j+1}&&
    \end{ZX}
\end{equation}
where $j$ is a lattice site index. 

A ground state of $\sH_V$ is given by the generalized GHZ state of the comodule algebra $V$,
\tikzsetfigurename{VGHZ1}
\begin{equation}\label{eq:VGHZ1}
    \begin{ZX}
        &&&&&\\
        \zxN{} \ar[d,C,red]&\zxamult{} \ar[l,red] \ar[u,red] \ar[r,red]&\zxamult{}  \ar[u,red] \ar[r,red]&...&\zxamult{} \ar[l,red] \ar[u,red] \ar[r,red]&\zxN{} \ar[d,C-,red]\\
        \zxN{} \ar[r,red]&&&...&&\zxN{} \ar[ll,red]
    \end{ZX}  = \begin{ZX}
    &&&&&\\
    &\zxN{} \ar[lu,),red]&\zxamult{} \ar[l,red] \ar[u,red] \ar[r,red]&...&\zxamult{} \ar[l,red] \ar[u,red] \ar[rd,N,red]&\\
    &&&&&\zxamult{} \ar[d,C,red] \ar[d,C-,red] \\
    &&&&&
    \end{ZX}
\end{equation}

The tensor \tikzsetfigurename{VGHZ10}$\begin{ZX}
\\ \zxamult{} \ar[u] \ar[d,C,red] \ar[d,C-,red] \\
\end{ZX}$ corresponds to the character of the regular representation of the algebra $V$. Decomposing it into simple modules produces a sum over the corresponding characters, similar to \cref{eq:chardecompose}. This way, we find a decomposition of the generalized GHZ state into short-range entangled states, labeled by characters of the simple $V$-modules.

It is also instructive to construct a string order parameter which detects whether a given state is in the gapped phase associated with the comodule algebra $V$. Such an operator is given by the product of the 2-body projectors appearing in the Hamiltonian $\sH_V$\footnote{The four-legged tensors appearing in \cref{eq:Oij-stringop} can be split in two three-legged tensors in an unambiguous manner since the (co)multiplication is (co)associative.},
\tikzsetfigurename{orderparameter}
\begin{equation}\label{eq:Oij-stringop}
    \cO_{i,j} = \begin{ZX}
        &&&&\\
        \zxamult{} \ar[u,red] \ar[d,red] \ar[r,red]&\zxamult{} \ar[u,red] \ar[d,red] \ar[r,red]&...&\zxamult{} \ar[u,red] \ar[d,red] \ar[l,red]&\\
        &&&&\\[5 pt]
        \zxbase{i} &&&\zxbase{j}&\\
    \end{ZX}
\end{equation}
where $i,j$ are lattice site indices. 
Such an operator will give string long-range order only if in the phase.

To realize different $\Rep(H)$-symmetric gapped phases on the same local Hilbert space, we embed the comodule algebras $V$ into a common space. 
The Hopf algebra $H$ itself serves as a natural choice and we fix $H$ as the local Hilbert space. There is a comodule inclusion morphism ${\iota: V\rightarrow H}$ and projection ${p: H \rightarrow V}$, such that $p\circ \iota = \id_{V}$ and
\tikzsetfigurename{cocom-mor2}
\begin{equation}
    \begin{ZX}
        & & & &&\\
        & & & &&\\
        & & &  \zxZ{} \ar[dd] \ar[uu] \ar[ruu,(] &&\\
        & & & &&\\
        & & & \zxrep{$\iota$} \ar[dd,red]&&\\
        & & & &&\\
        & & & &&
    \end{ZX} 
    =   
    \begin{ZX}
        & & & &&\\
        & & & &&\\
        & & & \zxrep{$\iota$} \ar[uu] &&\\
        & & & &&\\
        & & & \zxN{}\ar[dd,red] \ar[uu,red] \ar[zxdrrow,ruuuu,(] &&\\
        & & & &&\\
        & & & &&
    \end{ZX}\ ,
    \quad 
    \begin{ZX}
    & & & &&\\
    & & & &&\\
    & & &  \zxN{} \ar[dd,red] \ar[uu,red] \ar[zxdrrow,ruu,(] &&\\
    & & & &&\\
    & & & \zxrep{$p$} \ar[dd]&&\\
    & & & &&\\
    & & & &&
    \end{ZX} 
    =
    \begin{ZX}
    & & & &&\\
    & & & &&\\
    & & & \zxrep{$p$} \ar[uu,red] &&\\
    & & & &&\\
    & & & \zxZ{}\ar[dd] \ar[uu] \ar[ruuuu,(] &&\\
    & & & &&\\
    & & & &&
    \end{ZX} \ .
\end{equation}
% Then we build the commuting projector Hamiltonian using the 2-body term,
With $H$ as the local Hilbert space, the commuting projector Hamiltonian associated with $V$ is constructed out of a modified 2-body term, as
\tikzsetfigurename{commuting-projector}
\begin{equation}\label{eq:HcomoduleHam}
\t{\sH}_V =-\sum_j \begin{ZX}
&&&&\\
\zxrep{$\iota$}\ar[u]&&\zxrep{$\iota$} \ar[u]&&\\
&\zxamult{} \ar[dd,red] \ar[lu,),red] \ar[ru,(,red]&&&\\
&&&&\\
&\zxamult{} \ar[ld,(,red] \ar[rd,),red]&&&\\
\zxrep{$p$}\ar[d]&&\zxrep{$p$} \ar[d]&&\\
&&&&\\[5 pt]
\zxbase{j}&&\zxbase{j+1}&&\\
\end{ZX}
\end{equation}
where $j$ is a lattice site index. The commuting projector Hamiltonian $\t{\sH}_V$ commutes with the $\Rep(H)$ symmetry MPO \eqref{eq:repHmpotensor}. The ground state \eqref{eq:VGHZ1} can be included into the new Hilbert space by $\iota$ so that it is a ground state of $\t{\sH}_V$,
\tikzsetfigurename{generalGHZ}
\begin{equation}
    \begin{ZX}
        &&&&&\\
        &\zxrep{$\iota$}\ar[u]&\zxrep{$\iota$}\ar[u]&&\zxrep{$\iota$}\ar[u]&\\
        \zxN{} \ar[d,C,red]&\zxamult{} \ar[l,red] \ar[u,red] \ar[r,red]&\zxamult{}  \ar[u,red] \ar[r,red]&...&\zxamult{} \ar[l,red] \ar[u,red] \ar[r,red]&\zxN{} \ar[d,C-,red]\\
        \zxN{} \ar[r,red]&&&...&&\zxN{} \ar[ll,red]
    \end{ZX}  = \begin{ZX}
        &&&&&\\
        \zxrep{$\iota$}\ar[u]&&\zxrep{$\iota$}\ar[u]&&\zxrep{$\iota$}\ar[u]&\\
        &\zxN{} \ar[lu,),red]&\zxamult{} \ar[l,red] \ar[u,red] \ar[r,red]&...&\zxamult{} \ar[l,red] \ar[u,red] \ar[rd,N,red]&\\
        &&&&&\zxamult{} \ar[d,C,red] \ar[d,C-,red] \\
        &&&&&
    \end{ZX} \ .
\end{equation}
% And the operator that has string long-range order is given by,
The string order parameter \eqref{eq:Oij-stringop} similarly becomes
\tikzsetfigurename{orderparameter2}
\begin{equation}
    \t{\cO}_{i,j} = \begin{ZX}
        &&&&\\
        \zxrep{$\iota$}\ar[u]&\zxrep{$\iota$}\ar[u]&&\zxrep{$\iota$}\ar[u]&\\
        \zxamult{} \ar[u,red] \ar[d,red] \ar[r,red]&\zxamult{} \ar[u,red] \ar[d,red] \ar[r,red]&...&\zxamult{} \ar[u,red] \ar[d,red] \ar[l,red]&\\
        \zxrep{$p$}\ar[d]&\zxrep{$p$}\ar[d]&&\zxrep{$p$}\ar[d]&\\
        &&&&\\[5 pt]
        \zxbase{i} &\zxbase{i+1}&&\zxbase{j}&\\
    \end{ZX}
    \ .
\end{equation}

Let us make some comments about this construction:
\begin{enumerate}
    \item The multiplication of the $H$-comodule algebra is, in general, different from that of any subalgebra of $H$. An example is the non-trivial SPT phase in the case of the group algebra. Indeed, for the group algebra $\CC[G]$, the twisted group algebra $\CC[K]^\psi$, where ${K\subset G}$ and ${\psi\in H^2(K,U(1))}$, is a $\CC[G]$-comodule algebra. However, due to the non-trivial twist $\psi$, it is not isomorphic to any subalgebra of $\CC[G]$. Using $\psi\in H^2(K,U(1))$ as an input gives an $H$-SPT phase labeled by $\psi$.

    \item The construction using comodule algebras embedded in the Hopf algebra $H$ closely parallels the constructions in \Rfs{hung2025CFTstrange,hung2025CFTlatt} using the strange correlator. The primary differences are that our discussion is purely 1+1d and is formulated in terms of algebraic rather than categorical data. In particular, the commuting projector Hamiltonian \eqref{eq:HcomoduleHam} parallels their construction, which employs different Frobenius algebras while sharing the same module. While in our construction, the Frobenius $H$-comodule algebra $V$ has a natrual action on $H$, which is explicitly given by
\begin{equation}
    \rho: V\otimes H \rightarrow H,\quad  v\otimes h \mapsto \iota (m_V(v \otimes p(h))) \, .
\end{equation}
Frobenius $H$-comodule algebras $V$ correspond to module categories over $\Rep(H)$, which in turn classify the gapped boundaries of the quantum double of Hopf algebra $H$~\cite{buerschaper2013qdalgebra,zhian2023double}.
\end{enumerate}

\subsection{\texorpdfstring{$\Rep(H)$}{Rep(H)} ferromagnet and paramagnet revisited}

Two special cases of the above construction reproduce the $\Rep(H)$ spontaneous symmetry breaking Hamiltonian $\sH_{\SSB}$, defined in \cref{eq:repHSSBham00}, and the $\Rep(H)$-symmetric Hamiltonian $\sH_{\Sym}$, defined in \cref{eq:RepH-Sym-Ham}. These are fixed point Hamiltonians for the $\Rep(H)$-ferromagnet and paramagnet phases respectively.
These Hamiltonians can be obtained by setting the $H$-comodule algebra $V$ in the above construction to $H$ and $\CC$, respectively.

\paragraph{\texorpdfstring{$\Rep(H)$}{Rep(H)} ferromagnetic phase} Here we take the $H$-comodule algebra to be $H$ itself, in the sense that multiplication and unit are given by the Hopf algebra multiplication $\mult$ and unit $\unit$, respectively. 
According to \secref{sec:hopf-frob}, we can construct the comultiplication and counit that satisfies the Frobenius relation \eqref{eq:hopf-frob-cond}. Hence, the commuting projector Hamiltonian is given by,
\tikzsetfigurename{RepHssbham}
\begin{equation}\label{eq:repHSSBham}
\sH_{\SSB} = -\sum_j \frac{1}{\sqrt{H}}\begin{ZX}
    &&&&\\
    &\zxX{} \ar[dd] \ar[lu,)] \ar[ru,(]&&&\\
    &&&&\\
    &\zxX{} \ar[ld,(] \ar[rd,)]&&&\\
    &&&&\\[5 pt]
    \zxbase{j}&&\zxbase{j+1}&&\\
    &&&&
\end{ZX}
\end{equation} 
which is the same as \eqref{eq:repHSSBham00}. A ground state of $\sH_{\SSB}$ is given by the generalized GHZ state,
\tikzsetfigurename{RepHGHZ2}
\begin{equation}
    \begin{ZX}
        &&&&&\\
        \zxN{} \ar[d,C]&\zxX{} \ar[l] \ar[u] \ar[r]&\zxX{}  \ar[u] \ar[r]&...&\zxX{} \ar[l] \ar[u] \ar[r]&\zxN{} \ar[d,C-]\\
        \zxN{} \ar[r]&&&...&&\zxN{} \ar[ll]
    \end{ZX} = |H|\begin{ZX}
    &&&&&\\
&\zxN{} \ar[lu,)]&\zxX{} \ar[l] \ar[u] \ar[r]&...&\zxX{} \ar[l] \ar[u] \ar[rd,N]&\\
    &&&&&\zxXTri.{}
    \end{ZX} \ .
\end{equation}
The full ground state space is degenerate with the degeneracy given by the number of simple $\dl{H}$-modules (i.e., irreducible representation of $\dl{H}$), since $\haar = \sum_{\chi\in \mathrm{Irr}(\Rep(\dl{H}))} \chi$. The entanglement entropy of the Hopf GHZ state is given by the reduced density matrix $\rho_A = \frac{1}{|H|} \mathds{1}$, $S=\log(|H|)$. The operator that exhibits string long-range order is given by,
\tikzsetfigurename{orderpara}
\begin{equation}
    \cO_{i,j} = \begin{ZX}
        &&&&\\
        \zxX{} \ar[u] \ar[d] \ar[r]&\zxX{} \ar[u] \ar[d] \ar[r]&...&\zxX{} \ar[u] \ar[d] \ar[l]&\\
        i &i+1&&j&
    \end{ZX}
\end{equation}

\paragraph{\texorpdfstring{$\Rep(H)$}{Rep(H)} paramagnetic phase} In this case, we take the $H$-comodule algebra to be the trivial algebra $\CC$. From the general construction, we find the comodule inclusion morphism $\iota$ which embeds $\CC$ to the identity element in $H$ and corresponding projection $p$. The commuting projector Hamiltonian is then 
\tikzsetfigurename{RepHsym}
\begin{equation}
    \sH_{\Sym} = -\sum_j \begin{ZX}
        &&\\
        \zxX{} \ar[u]&\zxX{} \ar[u]&\\
        \zxXTri'{} \ar[d]&\zxXTri'{} \ar[d]&\\
        &&\\[5 pt]
    \zxbase{j}&\zxbase{j+1}&\\
    \end{ZX} \ ,
\end{equation}
which recovers \cref{eq:RepH-Sym-Ham}. 
This Hamiltonian has a unique ground state,
\tikzsetfigurename{RepHsymGS2}
\begin{equation}
    \begin{ZX}
        &&&&\\
        \zxX{} \ar[u]&\zxX{} \ar[u]&...&\zxX{} \ar[u]&
    \end{ZX}
\end{equation}
The associated order parameter is given by the following string order parameter, which is essentially a truncated $\Rep(H)$ symmetry MPO:
\tikzsetfigurename{RepHsymOP}
\begin{equation}
    \cO_{i,j} = \begin{ZX}
        && && && && && \\
        \zxend{$\mu$} &&\zxrep{$\Gamma$} \ar[ll,blue] \ar[rr,blue] \ar[zxarrowr,rrdd,N] && \zxrep{$\Gamma$}\ar[r,blue] \ar[zxarrowr,rrdd,N] && ... &&\zxrep{$\Gamma$} \ar[rr,blue] \ar[zxarrowr,rrdd,N] \ar[l,blue] && \zxend{$\nu$} \\
        && && && && && \\
        && && && && && \\
        &&& \zxZ{}\ar[dd] \ar[uuuu,3d above,very thick] \ar[ru,(]   && \zxZ{} \ar[dd] \ar[uuuu,3d above,very thick] \ar[ru,(]&& \zxN{} && \zxZ{}\ar[dd] \ar[uuuu,3d above,very thick]\ar[ru,(] &\\
        && && && && && \\
        && && && && && \\[5 pt]
        && &\zxbase{i}& &\zxbase{i+1}& && &\zxbase{j}& \\
    \end{ZX} \ .
\end{equation}
Here, the blue dots at the left and right ends of the blue virtual bonds represent $d_\Gamma$-dimensional vectors $\mu,\nu$, which can be chosen freely.

\subsection{\texorpdfstring{$\Rep(H_8)$}{Rep(H8)} symmetric gapped phases}
In this subsection, we construct fixed point Hamiltonians for all the $\Rep(H_8)$ symmetric gapped phases using comodule algebras. It is known that there are six module categories over $\Rep(H_8)$ \cite{ostrik_module_2003,meir2012module,choi2023self,Diatlyk:2023fwf}, which in turn correspond to six equivariantly Morita inequivalent $H_8$-comodule algebras $V$. Two of the six gapped phases are the fully symmetric phase, corresponding to ${V=\CC}$, and the fully SSB phase, corresponding to ${V=H_8}$, following the above construction. 
Three of the remaining four gapped phases correspond to the comodule algebras formed by Hopf subalgebras, $\CC[\ZZ_2^x\times \ZZ_2^y],\CC[\ZZ_2^\text{diag}], \CC[\ZZ_2^x]$. The last one corresponds to ${V=A_{xy}^q}$, which is not a Hopf subalgebra of $H_8$~\cite{van2025h}.  
The unbroken symmetry and ground state degeneracy in each of these gapped phases (labeled by the corresponding comodule algebra) are as follows:
\begin{center}
    \begin{tabular}{c|c|c}
        Gapped phase & Unbroken symmetry & Ground state degeneracy\\\hline \hline
         $\CC$ & $\Rep(H_8)$ & 1 \\ \hline  
      $\CC[\langle 1,xy \rangle]$ & $\ZZ_2^{\mathbf{s}_1}\times \ZZ_2^{\mathbf{s}_2}$ & 2 \\ \hline 
         $\CC[\langle 1,x \rangle]$ & $\ZZ_2^\text{diag}$ & 2 \\ \hline 
         $\CC[\langle 1,x,y,xy \rangle]$ & $\ZZ_2^\text{diag}$ & 4 \\ \hline 
         $A_{xy}^q$ & $1$ & 4 \\ \hline 
      $H_8$ & $1$ & 5 \\ \hline
    \end{tabular}
\end{center}
Here, $x,y$ are the group-like elements of $H_8$ generating the $\CC[\ZZ_2^x],\CC[\ZZ_2^y]$ group algebras. 

As discussed in prior subsections, a comodule algebra can have its own multiplication map, distinct from that of the Hopf algebra. This is the case for $A_{xy}^q$. It is spanned by the basis elements $\{1,xy,z,xz\} \subset H_8$ (see \Cref{app:H8data} for details on the notation), its $H_8$-coaction follows from the $H_8$ comultiplication, and the multiplication is given by
\begin{equation}\label{eq:axymult}
    \begin{array}{c|cccc}
        m & 1 & xy& z& xz \\ \hline
        1&1 & xy & z & xz \\
        xy&xy & -1 & -xz & z \\
        z&z & -xz & -\frac{1}{\sqrt{2}}(1-xy) & \frac{1}{\sqrt{2}}(xy+1) \\
        xz &xz & z & \frac{1}{\sqrt{2}}(xy+1) & \frac{1}{\sqrt{2}}(1-xy) \\
    \end{array}
\end{equation}
which satisfies the Frobenius relation in \cref{eq:comod-alg-frob}.
For completeness, the comultiplication is given by
\begin{align}\label{eq:axycomult}
    &\comult(1) =\frac{1}{4} ( 1 \otimes 1-xy \otimes xy + \frac{xz \otimes z}{\sqrt{2}} + \frac{z \otimes xz}{\sqrt{2}} + \frac{xz \otimes xz}{\sqrt{2}} - \frac{z \otimes z}{\sqrt{2}} ),\; \\
    &\comult(xy) =\frac{1}{4} (1 \otimes xy + xy \otimes 1 + \frac{xz \otimes z}{\sqrt{2}} + \frac{z  \otimes xz}{\sqrt{2}} - \frac{xz \otimes xz}{\sqrt{2}} + \frac{z \otimes z}{\sqrt{2}}),\; \\
    &\comult(z) = \frac{1}{4} (1 \otimes z + z \otimes 1+xy \otimes xz + xz \otimes xy) ,\; \\
    &\comult(xz) = \frac{1}{4} (1 \otimes xz + xz \otimes 1 -xy \otimes z - z \otimes xy).
\end{align}
One can check ${\mult \circ \comult = \id}$ and the other conditions in \eqref{eq:comod-alg-frob} are satisfied, so that $A_{xy}^q$ is a unital Frobenius algebra. 

Using the algebra structure, one can construct $A_{xy}^q$-modules. $A_{xy}^q$ has symmetric Frobenius form and the characters are,
\begin{equation}\label{eq:Axy-char}
\begin{array}{c|c|c|c|c}
    & 1 & xy & z & xz \\ \hline
    \chi_1 & 1 & 1 & 1 & 1 \\ \hline
    \chi_2 & -i & i & -i & i \\ \hline
    \chi_3 & \ee^{\frac{5 i \pi}{8}} & \ee^{-\frac{5 i \pi}{8}} & \ee^{-\frac{3 i \pi}{8}} & \ee^{\frac{3 i \pi}{8}} \\ \hline
    \chi_4 & \ee^{-\frac{7 i \pi}{8}} & \ee^{\frac{7 i \pi}{8}} & \ee^{\frac{i \pi}{8}} & \ee^{-\frac{i \pi}{8}}
\end{array}
\end{equation}
Neither $A_{xy}^q$ nor $\CC[\langle 1,x \rangle]$ appears in the parameter space we explored for the phase diagram of the Hopf-Ising model Fig.~\ref{fig:schematic-phase-diagram} and relies on the comodule algebra construction. 
The $H_8$-comodule algebra $A_{xy}^q$ is analogous to the twisted subgroup algebras $\CC[K]^\psi$ in the case of $\CC[G]$ (see remark 1 at the end of \Cref{sec:fp-Hams-comod}), in that it is a comodule algebra that does not constitute a subalgebra.

One may wonder about the gapped phase constructed using $\CC[\langle 1,x,y,xy \rangle]^\psi$, where $\psi$ is the non-trivial 2-cocycle twist of $H^2(\ZZ_2^x \times \ZZ_2^y,U(1))$. This phase is equivalent to the fully symmetric gapped phase $\CC$. Indeed, one can check there is no non-trivial interface mode between them. Mathematically, these two comodule algebras are said to be \textit{equivariantly Morita equivalent} \cite{andruskiewitsch2007module}.

Since $H_8$ is self-dual, one can conjugate the Hamiltonians constructed using the comodule algebras above by the Hopf Hadamard gate $\FF^\dagger$ to get the corresponding $\Rep(\dl{H_8})$-symmetric Hamiltonians. Such a construction is applicable for any self-dual Hopf algebra. The Hamiltonian is given by
\tikzsetfigurename{RepHdcomodHam}
\begin{equation}
\t{\sH}_V =-\sum_j \begin{ZX}
    &&&&\\
    \zxHad{\tiny$\dagger$} \ar[u]&&\zxHad{\tiny$\dagger$} \ar[u]&&\\
    \zxrep{$\iota$}\ar[u]&&\zxrep{$\iota$} \ar[u]&&\\
    &\zxamult{} \ar[dd,red] \ar[lu,),red] \ar[ru,(,red]&&&\\
    &&&&\\
    &\zxamult{} \ar[ld,(,red] \ar[rd,),red]&&&\\
    \zxrep{$p$}\ar[d]&&\zxrep{$p$} \ar[d]&&\\
    \zxHad{} \ar[d]&&\zxHad{} \ar[d]&&\\
    &&&&\\[5 pt]
    \zxbase{j}&&\zxbase{j+1}&&\\
\end{ZX}
\end{equation}
which commutes with the $\Rep(\dl{H}_8)$ symmetry operators $\Al^r$ due to \cref{eq:FourierSymMPO}.

In the following, we list the 6 $\Rep(H_8)$-symmetric gapped phases constructed from the $H_8$-comodule algebras, along with the action of the symmetry operators $\Br^\Gamma$ on the ground state subspace. 
We will use the $xyz$ basis of $H_8$, namely $\{1,x,y,xy,z,xz,yz,xyz\}$. 
The $\Rep(H_8)$ symmetry MPOs are constructed using \eqref{eq:rephmpo} from the characters of $H_8$, which we list below:
\begin{equation}\label{eq:H8-char}
    \begin{array}{c|c|c|c|c|c|c|c|c}
        &1 & x & y & xy & z & xz& yz&xyz \\ \hline
        \chi_\dsi&1 & 1 & 1 & 1 & 1 & 1 & 1 & 1 \\ \hline
        \chi_{{\mathbf{s}}_1}&1 & -1 & -1 & 1 & -i & i & i & -i \\ \hline
        \chi_{{\mathbf{s}}_2}&1 & -1 & -1 & 1 & i & -i & -i & i \\ \hline
        \chi_{{\mathbf{s}}_3}&1 & 1 & 1 & 1 & -1 & -1 & -1 & -1 \\ \hline
        \chi_\two&2 & 0 & 0 & -2 & 0 & 0 & 0 & 0 
    \end{array}
\end{equation}
For the fixed point Hamiltonian for each symmetric gapped phase, there may be multiple degenerate ground states, which are labeled by the characters of the corresponding comodule algebra. 
For $A_{xy}^q$ the characters are listed in \cref{eq:Axy-char}, while the characters of the remaining comodule algebras, being $H_8$ subalgebras, can be read off from \cref{eq:H8-char}.

\subsubsection{\texorpdfstring{$V=\CC$}{V=C} with 1 ground state}

The projection $p: a \mapsto \CC$ is given by the counit and $\iota$ is the unit, which leads to the Hamiltonian,
\begin{equation}
    \sH_{\CC} = - \sum_j \ket{\unit}\bra{\unit}_{j} \otimes \ket{\unit}\bra{\unit}_{j+1} = -\sum_j (P^I P^{II} P^{III})_j  (P^I P^{II} P^{III})_{j+1} \,.
\end{equation}
where $\unit$ is the identity element of $H_8$. 
Since the local Hilbert space can be represented by 3 qubits, we have introduced the projectors ${P^{a}_j = (1+Z^{a}_j)/2}$, ${a\in\{I,II,III\}}$, acting on the $a$-th qubit on site $j$. 
The ground state of $\sH_{\CC}$ is $\otimes_j \ket{\unit}_j$. 
It is an eigenstate of the symmetry MPOs with eigenvalues as follows:
\begin{equation}
    \begin{array}{c|c|c|c|c}
        \chi_\dsi & \chi_{{\mathbf{s}}_1} &\chi_{{\mathbf{s}}_2} &\chi_{{\mathbf{s}}_3} & \chi_\two \\ \hline
        1 & 1& 1& 1& 2
    \end{array}
\end{equation}
The entire $\Rep(H_8)$ symmetry is unbroken in this phase.

\subsubsection{\texorpdfstring{$V=\CC[\gel{1,xy}]$}{V=C[<1,xy>]} with 2 ground states}
The projection $p: H_8 \rightarrow \{1,xy\}$ is given by
\begin{equation}
    p=\begin{pmatrix}
        1&0&0&0&0&0&0&0 \\
        0&0&0&1&0&0&0&0 \\
    \end{pmatrix}
\end{equation}
and $\iota$ is the transpose of $p$. The Hamiltonian,
\begin{align}
    \sH_{\CC[\gel{1,xy}]} = - \sum_j &P^I_j (1+Z^{II}Z^{III})_j P^I_{j+1} (1+Z^{II}Z^{III})_{j+1} \\
    &+P^I_j (X^{II}X^{III}-Y^{II}Y^{III})_j P^I_{j+1} (X^{II}X^{III}-Y^{II}Y^{III})_{j+1},
\end{align}
has two degenerate ground states. The matrix elements of the symmetry MPOs in the ground state subspace are as follows:
\begin{equation}
    \begin{array}{c|c|c|c|c}
        \chi_\dsi & \chi_{{\mathbf{s}}_1} &\chi_{{\mathbf{s}}_2} &\chi_{{\mathbf{s}}_3} & \chi_\two \\ \hline
        \left(
        \begin{array}{cc}
            1 & 0 \\
            0 & 1 \\
        \end{array}
        \right)& \left(
        \begin{array}{cc}
            1 & 0 \\
            0 & 1 \\
        \end{array}
        \right)&\left(
        \begin{array}{cc}
            1 & 0 \\
            0 & 1 \\
        \end{array}
        \right)&\left(
        \begin{array}{cc}
            1 & 0 \\
            0 & 1 \\
        \end{array}
        \right)&\left(
        \begin{array}{cc}
            0 & 2 \\
            2 & 0 \\
        \end{array}
        \right)
    \end{array}
\end{equation}
In this partial SSB phase, the group-like subsymmetry, ${\ZZ_2\times \ZZ_2}$ generated by $\Br^{{\mathbf{s}_1}}$ and $\Br^{{\mathbf{s}_2}}$ is unbroken.

\subsubsection{\texorpdfstring{$V=\CC[\gel{1,x}] \cong \CC[\gel{1,y}]$}{V=C[<1,x>] or C[<1,y>]} with 2 ground states}

The comodule algebras, $\CC[\gel{1,x}]$ and $\CC[\gel{1,y}]$, are equivariantly Morita equivalent, so they describe the same gapped phase. 
The projection $p: H_8 \rightarrow \CC[\gel{1,x}]$ is given by
\begin{equation}
    p=\begin{pmatrix}
        1&0&0&0&0&0&0&0 \\
        0&1&0&0&0&0&0&0 \\
    \end{pmatrix}
\end{equation}
and $\iota$ is the transpose of $p$. 
Following \eqref{eq:HcomoduleHam} and using the multiplication \eqref{eq:H8multxyz}, we have the Hamiltonian,
\begin{equation}
    \sH_{\CC[\gel{1,x}]} = - \sum_j (P^I P^{II} X^{III})_{j} \otimes (P^I P^{II} X^{III})_{j+1} \,.
\end{equation}
Its two degenerate ground states realize an SSB phase on qubit $III$ in each unit cell, owing to the $X_j^{III}X_{j+1}^{III}$ interaction.
The matrix elements of the symmetry MPOs in the ground state subspace are as follows:
\begin{equation}
    \begin{array}{c|c|c|c|c}
        \chi_\dsi & \chi_{{\mathbf{s}}_1} &\chi_{{\mathbf{s}}_2} &\chi_{{\mathbf{s}}_3} & \chi_\two \\ \hline
        \left(
        \begin{array}{cc}
            1 & 0 \\
            0 & 1 \\
        \end{array}
        \right)& \left(
        \begin{array}{cc}
            0 & 1 \\
            1 & 0 \\
        \end{array}
        \right)&\left(
        \begin{array}{cc}
            0 & 1 \\
            1 & 0 \\
        \end{array}
        \right)&\left(
        \begin{array}{cc}
            1 & 0 \\
            0 & 1 \\
        \end{array}
        \right)&\left(
        \begin{array}{cc}
            1 & 1 \\
            1 & 1 \\
        \end{array}
        \right)
    \end{array}
\end{equation}
In this partial SSB phase, the $\ZZ_2^{\rm diag}$ symmetry generated by $\Br^{{\mathbf{s}_3}}$ is unbroken.  Moreover, the ground state subspace satisfies $ {\dsi + \Br^{\mathbf{s}_1} = \dsi + \Br^{\mathbf{s}_2} =  \Br^{\mathbf 2}}$. We contrast this with the SSB pattern discussed in \cref{sec:comod-Cxy}, where the only nontrivial relation satisfied by the symmetry operators is ${\Br^{\mathbf{s}_3}=\dsi}$, i.e.~that the $\ZZ_2^{\rm diag}$ symmetry is unbroken.

\subsubsection{\texorpdfstring{$V=\CC[\gel{1,x,y,xy}]$}{V=C[<1,x,y,xy>]} with 4 ground states}\label{sec:comod-Cxy}
The projection $p: H_8 \rightarrow \{1,x,y,xy\}$ is given by
\begin{equation}
    p=\begin{pmatrix}
        1&0&0&0&0&0&0&0 \\
        0&1&0&0&0&0&0&0 \\
        0&0&1&0&0&0&0&0 \\
        0&0&0&1&0&0&0&0 \\
    \end{pmatrix}
\end{equation}
and $\iota$ is the transpose of $p$. The Hamiltonian is given by
\begin{equation}
    \sH_{\CC[\gel{1,x,y,xy}]} = - \sum_j P^I_j P^I_{j+1} (1+X^{II}_j X^{II}_{j+1}) (1+X^{III}_j X^{III}_{j+1}) \,.
\end{equation}
It has four degenerate ground states, realizing decoupled SSB phases on the qubits $II$ and $III$ in each unit cell. The matrix elements of the symmetry MPOs in the ground state subspace are as follows:
\begin{equation}
    \begin{array}{c|c|c|c|c}
        \chi_\dsi & \chi_{{\mathbf{s}}_1} &\chi_{{\mathbf{s}}_2} &\chi_{{\mathbf{s}}_3} & \chi_\two \\\hline
        \left(
        \begin{smallmatrix}
            1 & 0 & 0 & 0 \\
            0 & 1 & 0 & 0 \\
            0 & 0 & 1 & 0 \\
            0 & 0 & 0 & 1
        \end{smallmatrix}
        \right)&
        \left(
        \begin{smallmatrix}
            0 & 1 & 0 & 0 \\
            1 & 0 & 0 & 0 \\
            0 & 0 & 0 & 1 \\
            0 & 0 & 1 & 0 \\
        \end{smallmatrix}
        \right)&
        \left(
        \begin{smallmatrix}
            0 & 1 & 0 & 0 \\
            1 & 0 & 0 & 0 \\
            0 & 0 & 0 & 1 \\
            0 & 0 & 1 & 0 \\
        \end{smallmatrix}
        \right)&
        \left(
        \begin{smallmatrix}
            1 & 0 & 0 & 0 \\
            0 & 1 & 0 & 0 \\
            0 & 0 & 1 & 0 \\
            0 & 0 & 0 & 1
        \end{smallmatrix}
        \right)&
        \left(
        \begin{smallmatrix}
 0 & 0 & 1 & 1 \\
0 & 0 & 1 & 1 \\
1 & 1 & 0 & 0 \\
1 & 1 & 0 & 0 
        \end{smallmatrix}
        \right)
    \end{array}
\end{equation}
In this partial SSB phase, the $\ZZ_2^{\rm diag}$ symmetry generated by $\Br^{{\mathbf{s}_3}}$ is unbroken.

\subsubsection{\texorpdfstring{$V=A_{xy}^q$}{V=Aqxy} with 4 ground states}
The projection $p: H_8 \rightarrow \{1,xy,z,xz\}$ is given by
\begin{equation}
    p=\begin{pmatrix}
        1&0&0&0&0&0&0&0 \\
        0&0&0&1&0&0&0&0 \\
        0&0&0&0&1&0&0&0 \\
        0&0&0&0&0&1&0&0 \\
    \end{pmatrix}
\end{equation}
and $\iota$ is the transpose of $p$. 
The Hamiltonian $\sH_{A_{xy}^q}$ is constructed using \eqref{eq:axymult} and \eqref{eq:axycomult}, and is quite complicated and not very insightful to represent using $\ZZ_2$ Pauli operators. It has four degenerate ground states labeled by the characters listed in \cref{eq:Axy-char}. 
The matrix elements of the symmetry MPOs in the ground state subspace are as follows:
\begin{equation}
    \begin{array}{c|c|c|c|c}
        \chi_\dsi & \chi_{{\mathbf{s}}_1} &\chi_{{\mathbf{s}}_2} &\chi_{{\mathbf{s}}_3} & \chi_\two \\ \hline
        \left(
        \begin{smallmatrix}
            1 & 0 & 0 & 0 \\
            0 & 1 & 0 & 0 \\
            0 & 0 & 1 & 0 \\
            0 & 0 & 0 & 1
        \end{smallmatrix}
        \right)&\left(
        \begin{smallmatrix}
                1 & 0 & 0 & 0 \\
                0 & 1 & 0 & 0 \\
                0 & 0 & 0 & 1 \\
                0 & 0 & 1 & 0 
        \end{smallmatrix}
        \right)&\left(
        \begin{smallmatrix}
                0 & 1 & 0 & 0 \\
                1 & 0 & 0 & 0 \\
                0 & 0 & 1 & 0 \\
                0 & 0 & 0 & 1 
        \end{smallmatrix}
        \right)&\left(
        \begin{smallmatrix}
    0 & 1 & 0 & 0 \\
    1 & 0 & 0 & 0 \\
    0 & 0 & 0 & 1 \\
    0 & 0 & 1 & 0 
        \end{smallmatrix}
        \right)&\left(
        \begin{smallmatrix}
    0 & 0 & 1 & 1 \\
    0 & 0 & 1 & 1 \\
    1 & 1 & 0 & 0 \\
    1 & 1 & 0 & 0 
        \end{smallmatrix}
        \right)
    \end{array}
\end{equation}
In this phase, all simple objects of $\Rep(H_8)$ have non-trivial matrix elements in the ground state subspace. However the symmetry is not completely SSB since the ground state subspace satisfies ${(\dsi - \Br^{\mathbf{s}_1})(\dsi - \Br^{\mathbf{s}_2}) = 0}$. We contrast this with the SSB pattern discussed in \cref{sec:comod-H8}, where there are no nontrivial relations between the symmetry operators.

\subsubsection{\texorpdfstring{$V=H_8$}{V=H8} with 5 ground states}\label{sec:comod-H8}

For the case of ${V=H_8}$, $p,\iota$ are identity maps. The Hamiltonian is constructed using the $H_8$ multiplication map and the $\aso{H_8}$ comultiplication map. It has five degenerate ground states, realizing complete spontaneous breaking of the $\Rep(H_8)$ symmetry. 
The matrix elements of the symmetry MPOs in the ground state subspace are as follows:
\begin{align}
    \begin{array}{c|c|c|c|c}
        \chi_\dsi & \chi_{{\mathbf{s}}_1} &\chi_{{\mathbf{s}}_2} &\chi_{{\mathbf{s}}_3} & \chi_\two \\ \hline
        \left(
        \begin{smallmatrix}
            1 & 0 & 0 & 0 & 0 \\
            0 & 1 & 0 & 0 & 0 \\
            0 & 0 & 1 & 0 & 0 \\
            0 & 0 & 0 & 1 & 0 \\
            0 & 0 & 0 & 0 & 1
        \end{smallmatrix}
        \right)
        &
        \left(
        \begin{smallmatrix}
            0 & 0 & 1 & 0 & 0 \\
            0 & 0 & 0 & 1 & 0 \\
            1 & 0 & 0 & 0 & 0 \\
            0 & 1 & 0 & 0 & 0 \\
            0 & 0 & 0 & 0 & 1
        \end{smallmatrix}
        \right)
        &
        \left(
        \begin{smallmatrix}
            0 & 1 & 0 & 0 & 0 \\
            1 & 0 & 0 & 0 & 0 \\
            0 & 0 & 0 & 1 & 0 \\
            0 & 0 & 1 & 0 & 0 \\
            0 & 0 & 0 & 0 & 1
        \end{smallmatrix}
        \right)
        &
        \left(
        \begin{smallmatrix}
            0 & 0 & 0 & 1 & 0 \\
            0 & 0 & 1 & 0 & 0 \\
            0 & 1 & 0 & 0 & 0 \\
            1 & 0 & 0 & 0 & 0 \\
            0 & 0 & 0 & 0 & 1
        \end{smallmatrix}
        \right)
        &
        \left(
        \begin{smallmatrix}
            0 & 0 & 0 & 0 & 1 \\
            0 & 0 & 0 & 0 & 1 \\
            0 & 0 & 0 & 0 & 1 \\
            0 & 0 & 0 & 0 & 1 \\
            1 & 1 & 1 & 1 & 0
        \end{smallmatrix}
        \right)
    \end{array}
\end{align}

\section{Conclusion \& Outlook}
\label{sec:conclusion}
In this work, we have developed a unified algebraic and graphical framework for realizing anomaly-free non-invertible symmetries in $1+1$ dimensions on a tensor product Hilbert space using Hopf algebra-valued qudits. We explicitly constructed MPOs representing both the $\Rep(H)$ symmetry and the dual $\Rep(\dl{H})$ symmetry using ZX calculus. This diagrammatic language allowed us to define Hopf Pauli operators and derive algebraic relations through graphical rewriting, greatly simplifying the manipulation of tensor contractions. Furthermore, we defined local order parameters and constructed fixed-point Hamiltonians for both the symmetric paramagnetic and the spontaneously symmetry-broken ferromagnetic phases.

Using this framework, we generalized the Kramers-Wannier duality to a Hopf Kramers-Wannier duality, which exists for any finite-dimensional semisimple self-dual Hopf algebra. By implementing a finite-depth circuit, a projection which performs the gauging, and lastly a canonical Hopf Hadamard gate which rotates us back to the original symmetry basis. This lattice duality exchanges order and disorder operators. We studied the fusion rules, defects, and $F$-symbols associated with this duality, matching the lattice construction with the fusion categorical data, though the Frobenius-Schur indicator is not well-defined. This confirms that the self-duality arising from gauging a non-invertible symmetry can be realized on a tensor product Hilbert space at the cost of mixing with lattice translation symmetry. 

We constructed the self-dual Hopf-Ising model, a family of $\Rep(\dl{H})$-symmetric Hamiltonians, and numerically studied the phase diagram for the specific case of the Kac-Paljutkin algebra $H_8$. Our results identified four distinct gapped phases separated by Ising critical lines and a first-order transition, which converge at a multicritical point.

Finally, we provided a general construction for the commuting projector Hamiltonians of $\Rep(H)$ symmetric gapped phases using the formalism of $H$-comodule algebras~\cite{inamura_lattice_2022}. We applied this method to explicitly construct fixed-point Hamiltonians for all six gapped phases associated with $\Rep(H_8)$ symmetry in $H_8$-spin chains, recovering the full classification of module categories over $\Rep(H_8)$.

We close with a few comments about open questions for future exploration.
\begin{enumerate}
    \item A central theme of this work is to understand which non-invertible symmetries can be realized on tensor-product Hilbert spaces. A recent operator-algebraic result shows that a fusion category $\mathcal C$ can be realized as a categorical symmetry of a tensor-product quasi-local algebra without mixing with lattice translation if and only if $\mathcal C$ is integral \cite{jones2025op_alg}. Our constructions indicate that this criterion can be relaxed if one allows the UV lattice realization of the symmetry to mix with lattice translation, as in the generalized Kramers–Wannier duality where the lattice self-duality symmetry operator mixed with translation.
    In the IR limit, if the translation acts trivially, this realizes a $\mathbb Z_2$ extension of $\Rep(H)$, which is generally not integral but only weakly integral.

    However, our construction is limited to situations with a single duality defect. More general scenarios arise when gauging a sub-symmetry or performing a non-maximal gauging, in which case multiple duality defects can appear. In such situations, additional invertible symmetry lines invariant under the duality can fuse with the duality operator, leading to several distinct duality defects that are permuted by invertible lines. For example, gauging the algebra object $\mathcal A = 1 \oplus ab \oplus \mathcal N$ in $\Rep(D_8)$ or $\Rep(H_8)$ produces two duality defects $\mathcal D_1,\mathcal D_2$ with fusion rules
\[
\mathcal D_i \otimes \mathcal D_i = 1 \oplus ab \oplus \mathcal N,\quad 
\mathcal N \otimes \mathcal D_i = \mathcal D_1 \oplus \mathcal D_2,\quad 
\mathcal D_i \otimes a = a \otimes \mathcal D_i = \mathcal D_{\bar i},
\]
where the $\mathbb Z_2$ invertible lines $a$ or $b$ permute the defects $\mathcal D_1$ and $\mathcal D_2$ \cite{lu2025sdset}.

A closely related example appears in the self-duality of $\Rep(S_3)$ \cite{arkya2024reps3dual}. The category $\Rep(S_3)$ contains two invertible lines $\mathbf 1,\mathbf 1'$ and a non-invertible line $\mathbf 2$ with fusion rules
\[
\mathbf 2 \otimes \mathbf 2 = \mathbf 1 \oplus \mathbf 1' \oplus \mathbf 2,\quad 
\mathbf 1' \otimes \mathbf 1' = \mathbf 1,\quad 
\mathbf 2 \otimes \mathbf 1' = \mathbf 1' \otimes \mathbf 2 = \mathbf 2.
\]
Gauging the algebra object $\mathcal A = \mathbf 1 + \mathbf 2$ yields a self-duality whose extended symmetry contains two duality defects with fusion rules
\[
\mathcal D_i \otimes \mathcal D_i = \mathbf 1 \oplus \mathbf 2,\quad 
\mathbf 1' \otimes \mathcal D_i = \mathcal D_i \otimes \mathbf 1' = \mathcal D_{\bar i},
\]
which coincide with those of $SU(2)_4$. Depending on the $F$-symbols, the resulting fusion category can be $SU(2)_4$ or $\mathrm{JK}_4$ \cite{arkya2024reps3dual,eck2024reps3dual,barkeshli_symmetry_2019,Levaillant:2015nia}. Since $\mathbb C^{S_3}$ has no nontrivial Hopf subalgebras, such self-dualities cannot be realized using the Hopf-algebraic gauging procedure developed in this work. More generally, cases with multiple duality defects fall outside our present framework.

Motivated by these examples, it is natural to speculate that allowing the UV lattice realization of a symmetry to mix with lattice translation may substantially enlarge the class of non-invertible symmetries that can be realized on tensor-product Hilbert spaces. In particular, weakly integral fusion categories, which are known to admit $\mathbb Z_2^n$ gradings \cite{GNO2008Nilpotent,2019arXiv191102633G}, appear to provide a promising class of candidates. It would be interesting to explore whether and how more general weakly integral fusion categories can be realized through lattice gauging constructions on tensor-product Hilbert spaces \cite{GNO2008Nilpotent,natale2015weakly,dong2019class}.

   \item A further class of non-invertible symmetries that would be interesting to construct is one that generalizes the $\Rep(D_8)$ symmetry that arises from the off-diagonal gauging of $\ZZ_2\times \ZZ_2$~\cite{Seifnashri24}. Namely, for any $H$ (not necessarily self-dual), consider two spin chains each with local Hilbert space $H$, one with $\Rep(H)$ symmetry, and the other with $\Rep(\dl{H})$ symmetry. The operator $({\kw^H}^\dagger \otimes \kw^{\dl{H}}) \times \textsc{SWAP}$, where \textsc{SWAP} is the interlayer swap operation, realizes a duality of $\Rep(H) \times \Rep(\dl{H})$. This duality does not mix with translation on the lattice. Moreover, the Hopf algebra cluster state~\cite{jia2024generalized} will be invariant under this duality.

    \item The Hopf-Ising model admits off-diagonal ``$Z^\dagger Z$"-type terms when the input Hopf algebra is not the group algebra. Unlike group algebras, there is no canonical basis for general Hopf algebras. Thus, it is natural to ask whether there exists a basis in which the Hopf-Ising model is stoquastic, or whether there is an obstruction for certain Hopf algebras. If such a basis exists, one should be able to construct a corresponding 2D classical statistical mechanics model. 
    It turns out that $H_8$ Ising model is almost stoquastic in the ``MPO" basis, in the sense that it is stoquastic whenever the ferromagnetic terms from $\chi_{{\mathbf{s}}_1}$ and $\chi_{{\mathbf{s}}_2}$ (defined in \cref{app:MPO basis}) appear with equal weight.
    Such a basis would also be helpful for studying sign-problem-free quantum Monte Carlo (QMC) simulation of Hopf-algebra-valued spin chain, as well as the confinement transition of Hopf algebra quantum doubles in 2+1d \cite{Sandvik:1999noc}. 
    We note that the Hamiltonian \eqref{eq:H8-Ham} satisfies the stoquasticity criterion mentioned above, hence it would be interesting to complement our DMRG results with QMC simulations. This may shed further light on the nature of the multicritical point in \cref{fig:schematic-phase-diagram}.
    More generally, a sign problem-free basis does not necessitate a stoquastic Hamiltonian, as was emphasized recently in \Rf{shackleton2025twistedquantumdoublessign}, which showed that twisted quantum doubles of groups are sign problem-free despite the lack of stoquasticity. It is interesting to consider whether similar arguments apply to Hopf quantum doubles, more generally.
    
    \item The multicritical point of the $H_8$-Ising model \eqref{eq:H8-Ham} is numerically challenging to resolve using standard tensor network tools. On the one hand, this motivates the development of an improved DMRG algorithm that explicitly incorporates non-invertible symmetries (see, e.g., \Rfs{Lootens:2024gfp,Devos:2025yoj} for recent progress). Moreover, we did not explore the full parameter space of possible couplings. It would be interesting to see whether tuning these couplings could land us in the remaining two phases of $\Rep(H_8)$.
    
    On the other hand, the physics of such multicritical points is rich and suggests natural generalizations to other self-dual Hopf algebras. Much as the $\mathbb{Z}_n$ clock and Potts models admit informative large-$n$ limits, $H_8$ represents the first member of an infinite family of self-dual Hopf algebras. For instance, Masuoka~\cite{Masuoka95p} constructed $p+1$ self-dual Hopf algebras of order $p^3$ for each odd prime $p$. It would be interesting to investigate the nature of the corresponding phase transitions in the large-$p$ regime.

    \item Since the Hopf algebra $H_8$ is group-theoretical, phase transitions in a model with $\Rep(H_8)$ symmetry can be mapped, by gauging suitable subgroups, to those in another model that has an anomalous $D_8$ symmetry, with an appropriate choice of 3-cocycle. By contrast, there are non-group-theoretical Hopf algebras~\cite{Nikshych08,natale2010hopf}, for whom the associated $\Rep(H)$ non-invertible symmetries therefore cannot be understood as arising from gauging some invertible symmetry. The smallest such example has dimension 36. A symmetry-protected topological phase associated with this Hopf algebra was recently constructed using a non-on-site symmetry realization~\cite{lu2025inispt}. It would be interesting to investigate the nature of phase transitions in lattice models with such symmetries. Notably, this 36-dimensional Hopf algebra is self-dual~\cite{cuadra2017orders,galindo2024integral}.

    \item 
    A systematic study of the algebra of $\Rep(H_8)$-symmetric local operators (also known as bond algebra~\cite{Cobanera:2011wn}) is interesting for several reasons. 
    The generators of the $\ZZ_2$-symmetric bond algebra in the 1+1d quantum Ising model, $X_j$ and $Z_j Z_{j+1}$ can be used to construct U(1) charges. These charges do not commute but they generate non-Abelian Lie algebra, with the Lie bracket defined by the standard matrix commutator. This Lie algebra is known as the Onsager algebra~\cite{onsager1944crystal}; it played a key role in the exact solution of the 2d classical Ising model.
    It would be interesting to explore similar structures for other self-dual Hopf algebras (see \Rf{Vernier:2018han} for the case of ${H=\CC[\ZZ_n]}$).
    Furthermore, we note that the Hopf Kramers--Wannier duality operator can be viewed as a bounded-spread automorphism of the $\Rep(H)$ symmetric local operator algebra and thus provides an example of a quantum cellular automaton (QCA) within the $\Rep(H)$ symmetric subalgebra. It would be interesting to extend the operator-algebraic notion of QCA index to $\Rep(H)$-symmetric subalgebras for self-dual Hopf algebras and to compute the corresponding invariants for the Hopf Kramers-Wannier operator, such as the GNVW index~\cite{Gross:2011yvb,Jones:2023ptg,Jones:2023imy,Ma:2024ypm,Jones:2024lws,jones2025op_alg}.

    \item The 1+1d quantum Ising model can be fermionized to produce a free fermion model. Dually, the fermion model can be turned into the Ising model by summing over spin structures, i.e.~by a Jordan-Wigner transformation. There is an analogous treatment of $\ZZ_n$ clock models using parafermions~\cite{Fradkin:1980th}. 
    It is natural to consider non-Abelian generalizations of parafermions in the context of self-dual Hopf algebras beyond Abelian groups (see also \cite{Albert:2021vts}, which formulates a generalized Jordan-Wigner transformation without requiring self-duality). Relatedly, since the Quon language provides a unifying framework for the ZX calculus and matchgate constructions \cite{Liu:2016xph,Kang:2025stq,Feng:2025elo}, it would be interesting to investigate possible connections between our self-dual Hopf-algebraic structures and the Quon formalism, as well as their interpretation in terms of string–genus relations and alterfold topological field theories \cite{Liu:2023dhj}. A separate related direction could be to relate our construction to fermionic non-invertible symmetries \cite{davydov2013z,ingo2012fermion,kansei2023fermion}. 

\end{enumerate}

\section*{Acknowledgments}
We thank Xie Chen, Pranay Gorantla, Kansei Inamura, Yuhan Liu, Zhengwei Liu, Andr\'as Moln\'ar, Ingo Runkel, Xicheng Wang, Zhiyuan Wang, Xiao-Gang Wen, Fu Xu, Mingru Yang, and Xinping Yang for helpful discussions, and Kansei Inamura, Zhengwei Liu, and Zhiyuan Wang for helpful comments on the manuscript.
D.C.L. is supported by the Simons Collaboration on Ultra-Quantum Matter, which is a grant from the Simons Foundation (grant No. 651440).
A.C. was partially supported by NSF DMR-2022428 and by the Simons Collaboration on Ultra-Quantum Matter, which is a grant from the Simons Foundation (651446, Wen).
Part of this work was completed during the KITP program ``Generalized Symmetries in Quantum Field Theory: High Energy Physics, Condensed Matter, and Quantum Gravity,'' which is supported in part by grant NSF PHY-2309135 to the KITP. 
The computations in this paper were run on the FASRC Cannon cluster supported by the FAS Division of Science Research Computing Group at Harvard University.

\appendix

\section{\texorpdfstring{$H_8$}{H8} data}\label{app:H8data}
In the literature, the Kac--Paljutkin Hopf algebra $H_8$ is presented in various bases. In this appendix, we tabulate these different bases and the required change of basis transformation between them. (see \appref{app:lineartrans} for more details on the basis transformation)

\begin{equation}
    M: H_8^{xyz} \rightarrow H_8^v,\quad a_i \mapsto M_{i,i'}v_{i'}
\end{equation}
And the other tensors are transformed as,
\begin{align}
     &\mult_{ij}^k M_{k,k'} = M_{i,i'}M_{j,j'}\mult^{v,k'}_{i',j'}\\
    &\comult_k^{i,j} M_{i,i'} M_{j,j'} = M_{k,k'} \comult^{v,i',j'}_{k'}\\
    &S^v = M^{-1}SM\\
    &F^v_{ij} = (M^{-1})_{i,i'}F_{i',j'} (M^{-1})_{k,j'} = M^{-1} F (M^{-1}) ^\intercal
\end{align}

\subsection{\texorpdfstring{$x,y,z$}{x,y,z} basis}
The generators $x,y,z$ appears in \cite{Masuoka95} and have the following multiplication,
\begin{equation}\label{eq:H8multxyz}
    x^2=y^2=1,\,  xy =yx, \, zx=yz,\, zy = xz,\, z^2 = \frac{1}{2}(1+x+y-xy)
\end{equation}
unit is given by $\eta = 1$. The comultiplication,
\begin{equation}
    \comult(1)=1\otimes 1,\, \comult(x)=x\otimes x,\, \comult(y)=y\otimes y,\, \comult(z)= \frac{1}{2}(1\otimes 1 +y\otimes 1 +1\otimes x- y\otimes x)z\otimes z
\end{equation}
and the counit is given by $\wh \epsilon = \sum_a \delta_a$, where $\delta_a(b) = \delta_{a,b}$. The antipode is given by,
\begin{equation}
    S(x)=x,\, S(y)=y,\, S(z)=z. 
\end{equation} 
Note that $S(xz) = S(z)S(x)=zx = yz $, the first equal sign follows from \eqref{eq:antipoderule}. The haar integral is given by $\wc\haar = \sum_{a\in H_8} a $ and cointegral is given by $\phi= \delta_1$. 
The $\star$ operation gives,
\begin{equation}
    x^\star = x,\, y^\star =y,\, z^\star=z^{-1}=\frac{1}{2}(z+zx+zy-zxy)=z^3.
\end{equation}

Taking the basis $B_{xyz}=\{1,x,y,xy,z,xz,yz,xyz\}$, all the maps can represented using tensors of different ranks, 
\begin{equation}
     \mult(a_i \otimes a_j) = \sum_k \mult_{ij}^k a_k,\quad \comult(a_i) = \sum_{j,k} a_j\otimes a_k,\quad S(a_i) = S_{ij} a_j 
\end{equation}
and, for example,
\begin{equation}
    \eta =(1,0,0,0,0,0,0,0)^\intercal, \quad \wh\epsilon = (1,1,1,1,1,1,1,1)
\end{equation}
Following \appref{app:lineartrans}, the $\ZZ_2\times \ZZ_2$ automorphism of $H_8$ is generated by,
\begin{equation}
    \tau_1 = \left(
    \begin{array}{cccccccc}
        1 & 0 & 0 & 0 & 0 & 0 & 0 & 0 \\
        0 & 1 & 0 & 0 & 0 & 0 & 0 & 0 \\
        0 & 0 & 1 & 0 & 0 & 0 & 0 & 0 \\
        0 & 0 & 0 & 1 & 0 & 0 & 0 & 0 \\
        0 & 0 & 0 & 0 & 0 & 0 & 0 & 1 \\
        0 & 0 & 0 & 0 & 0 & 0 & 1 & 0 \\
        0 & 0 & 0 & 0 & 0 & 1 & 0 & 0 \\
        0 & 0 & 0 & 0 & 1 & 0 & 0 & 0 \\
    \end{array}
    \right),\quad \tau_2 = \left(
    \begin{array}{cccccccc}
        1 & 0 & 0 & 0 & 0 & 0 & 0 & 0 \\
        0 & 0 & 1 & 0 & 0 & 0 & 0 & 0 \\
        0 & 1 & 0 & 0 & 0 & 0 & 0 & 0 \\
        0 & 0 & 0 & 1 & 0 & 0 & 0 & 0 \\
        0 & 0 & 0 & 0 & -\frac{1}{2} & \frac{1}{2} & \frac{1}{2} & \frac{1}{2} \\
        0 & 0 & 0 & 0 & \frac{1}{2} & \frac{1}{2} & -\frac{1}{2} & \frac{1}{2} \\
        0 & 0 & 0 & 0 & \frac{1}{2} & -\frac{1}{2} & \frac{1}{2} & \frac{1}{2} \\
        0 & 0 & 0 & 0 & \frac{1}{2} & \frac{1}{2} & \frac{1}{2} & -\frac{1}{2} \\
    \end{array}
    \right)
\end{equation}
and its action on the generators,
\begin{align}
    &\tau_1: z \mapsto xyz,\\
    &\tau_2: x \mapsto y,\, y\mapsto x,\, z\mapsto \frac{1}{2}(-z+xz+yz+xyz).   
\end{align} 
which matches $\tau_2$ and $\tau_4$ in \Rf{Sage_2012}.
% \cite{shi2016H8charaut}. 
Taking $\{\delta_1,\delta_x,\delta_y,\delta_{xy},\delta_z,\delta_{xz},\delta_{yz},\delta_{xyz}\}$ as basis for $\dl{H_8}$, the isomorphism between $H_8$ and $H_8^*$ is given by, $a\mapsto \Theta_{a,a'}\delta_{a'}$ and,
\begin{equation}
    \Theta_{a,a'} = \left(
    \begin{array}{cccccccc}
        1 & 1 & 1 & 1 & 1 & 1 & 1 & 1 \\
        1 & -1 & -1 & 1 & -i & i & i & -i \\
        1 & -1 & -1 & 1 & i & -i & -i & i \\
        1 & 1 & 1 & 1 & -1 & -1 & -1 & -1 \\
        1 & -i & i & -1 & -\sqrt{2} & 0 & 0 & \sqrt{2} \\
        1 & i & -i & -1 & 0 & -i \sqrt{2} & i \sqrt{2} & 0 \\
        1 & i & -i & -1 & 0 & i \sqrt{2} & -i \sqrt{2} & 0 \\
        1 & -i & i & -1 & \sqrt{2} & 0 & 0 & -\sqrt{2} \\
    \end{array}
    \right)
\end{equation}
Combining with the automorphism of $H_8$, we have 4 different isomorphisms between $H_8$ and $H_8^*$.

One can construct the Hopf Hadamard gate according to \eqref{eq:hadamardgatedef},
\begin{equation}
    \FF = \frac{1}{\sqrt{8}}\left(
    \begin{array}{cccccccc}
        1 & 1 & 1 & 1 & 1 & 1 & 1 & 1 \\
        1 & -1 & -1 & 1 & -i & i & i & -i \\
        1 & -1 & -1 & 1 & i & -i & -i & i \\
        1 & 1 & 1 & 1 & -1 & -1 & -1 & -1 \\
        1 & i & -i & -1 & -\sqrt{2} & 0 & 0 & \sqrt{2} \\
        1 & -i & i & -1 & 0 & i \sqrt{2} & -i \sqrt{2} & 0 \\
        1 & -i & i & -1 & 0 & -i \sqrt{2} & i \sqrt{2} & 0 \\
        1 & i & -i & -1 & \sqrt{2} & 0 & 0 & -\sqrt{2} \\
    \end{array}
    \right)
\end{equation}
Similarly, one can construct other $\FF$ by conjugating with the automorphisms. One can check that $\FF$ is unitary and ${\FF^2 =S}$.

The irreducible representations are
\begin{equation}\label{eq:H8-irreps}
    \begin{array}{c|c|c|c|c|c|c|c|c}
        & 1&x&y&xy &z &xz&yz&xyz \\ \hline
        \Gamma_{\dsi} & 1 & 1 & 1 & 1 & 1 & 1 & 1 & 1 \\ \hline
        \Gamma_{{\mathbf{s}}_1} &   1 & -1 & -1 & 1 & -i & i & i & -i \\ \hline
        \Gamma_{{\mathbf{s}}_2} &   1 & -1 & -1 & 1 & i & -i & -i & i \\ \hline
        \Gamma_{{\mathbf{s}}_3} &   1 & 1 & 1 & 1 & -1 & -1 & -1 & -1 \\ \hline
    
        \Gamma_\two &\left(
        \begin{smallmatrix}
            1 & 0 \\
            0 & 1 \\
        \end{smallmatrix}
        \right)
        &
        \left(
        \begin{smallmatrix}
            0 & -1 \\
            -1 & 0 \\
        \end{smallmatrix}
        \right)
        &
        \left(
        \begin{smallmatrix}
            0 & 1 \\
            1 & 0 \\
        \end{smallmatrix}
        \right)
        &
        \left(
        \begin{smallmatrix}
            -1 & 0 \\
            0 & -1 \\
        \end{smallmatrix}
        \right)
        &
        \left(
        \begin{smallmatrix}
            -1 & 0 \\
            0 & 1 \\
        \end{smallmatrix}
        \right)
        &
        \left(
        \begin{smallmatrix}
            0 & -1 \\
            1 & 0 \\
        \end{smallmatrix}
        \right)
        &
        \left(
        \begin{smallmatrix}
            0 & 1 \\
            -1 & 0 \\
        \end{smallmatrix}
        \right)
        &
        \left(
        \begin{smallmatrix}
            1 & 0 \\
            0 & -1 \\
        \end{smallmatrix}
        \right)
    \end{array}
\end{equation}
and the characters are,
\begin{equation}\label{eq:H8-characters}
    \begin{array}{c|c|c|c|c|c|c|c|c}
        & 1&x&y&xy &z &xz&yz&xyz \\ \hline
        \chi_\dsi & 1 & 1 & 1 & 1 & 1 & 1 & 1 & 1 \\ \hline
        \chi_{{\mathbf{s}}_1} & 1 & -1 & -1 & 1 & -i & i & i & -i \\ \hline
        \chi_{{\mathbf{s}}_2} & 1 & -1 & -1 & 1 & i & -i & -i & i \\ \hline
        \chi_{{\mathbf{s}}_3} & 1 & 1 & 1 & 1 & -1 & -1 & -1 & -1 \\ \hline
        \chi_\two&2 & 0 & 0 & -2 & 0 & 0 & 0 & 0 
    \end{array}
\end{equation}
The conjugacy classes are,
\begin{equation}
    \{1, \frac{1}{2}(x+y),xy,\frac12(z+xyz),\frac12(xz+yz)\}
\end{equation}

\subsection{``\texorpdfstring{$D_8$}{D8}'' basis} There is a basis such that the multiplication of $H_8$ is the same as the group algebra of $D_8$ but the comultiplication and antipodes are modified due to the pseudo-twists (not Drinfeld-twist) \cite{nikshych1998k0ring,COHEN2006hopfalgebra}. Although the irreps are the same as $\Irr(D_8)$ due to the algebra part, the tensor product of the irreps are different \eqref{eq:tensorprodirreps} since the comultiplication is modified, therefore the category of its representations gives $\Rep(H_8)$ instead of $\Rep(D_8)$.

We choose the generators to be $r,s$ with multiplication is given by the relations $r^4 =s^2=1, srs = r^{-1}$\footnote{An equivalent presentation often used in the literature (for example, in \cite{Masuoka95}) is with generators $a=s$, $b=sr$, $c=r^2$ satisfying
\begin{align*}
   & a^2 = b^2 =c^2 =1, ac=ca, ab=ba, ba = abc,\\
   &\Delta(a) = a\frac{1+c}{2} \otimes a + a\frac{1-c}{2} \otimes b, \Delta(b) = b\frac{1+c}{2} \otimes b + b\frac{1-c}{2} \otimes a, \Delta(c) = c\otimes c,\\
   & S(a)= a\frac{1+c}{2} + b\frac{1-c}{2}, S(b) = b\frac{1+c}{2} + a\frac{1-c}{2}, S(c)=c.
\end{align*}
}. Taking the basis to be $B_{D_8} = \{1,r,r^2,r^3,s,rs,r^2s ,r^3s \}$, we can relate it to the $x,y,z$ basis by the basis transformation $B_{D_8} = M_{8x} B_{xyz}$, where,
\begin{equation}
    M_{8x} = 
\begin{pmatrix}
1&0&0&0&0&0&0&0\\
0&\frac{1+i}{2}&0&\frac{1-i}{2}&0&0&0&0\\
0&\frac{1-i}{2}&0&\frac{1+i}{2}&0&0&0&0\\
0&0&1&0&0&0&0&0\\
0&0&0&0& s\,\omega^{-5}& s\,\omega^{5}& c\,\omega^{-1}& c\,\omega\\
0&0&0&0& c\,\omega^{3}& c\,\omega^{-3}& s\,\omega^{-1}& s\,\omega\\
0&0&0&0& s\,\omega^{-1}& s\,\omega& c\,\omega^{3}& c\,\omega^{-3}\\
0&0&0&0& c\,\omega^{-1}& c\,\omega& s\,\omega^{-5}& s\,\omega^{5}
\end{pmatrix},
\end{equation}
with $\omega = e^{i\pi/8}, c =1/\sqrt{2} \cos(\pi/8), s= 1/\sqrt{2} \sin(\pi/8)$. The comultiplication and antipode in the $D_8$ basis are,
\begin{align}\label{eq:D8comult}
    \comult(r) = \frac{1}{2}(r\otimes r+r\otimes r^3 +r^3 \otimes r -r^3 \otimes r^3) \\
    \comult(s) = \frac{1}{2}(s\otimes s+ s\otimes r s +r^2s \otimes s-r^2s\otimes r s)
\end{align}
and
\begin{equation}
    S(r)= r, \quad S(s) = \frac{1}{2}(s+rs+r^2s-r^3s)
\end{equation}
The automorphisms are given by
\begin{align}
    &\tau_1: s\mapsto r^2s, \\
    &\tau_2: r\mapsto r^{-1},\quad s\mapsto r^{-1}s \, ,
\end{align}
which can be viewed as a $\ZZ_2^2$ subgroup of $\Aut(\CC[D_8]) \cong D_8$ which is also compatible with the distinct comultiplication and antipode structure of $H_8$.

Interestingly, the isomorphism between $H_8$ and $H_8^*$ in this basis is given by
\begin{equation}
    \Theta^{D_8}=\left(
\begin{array}{cccccccc}
1 & 1 & 1 & 1 & 1 & 1 & 1 & 1 \\
1 & -1 & 1 & -1 & -i & i & -i & i \\
1 & 1 & 1 & 1 & -1 & -1 & -1 & -1 \\
1 & -1 & 1 & -1 & i & -i & i & -i \\
1 & -i & -1 & i & \ee^{3i\pi/4} & \ee^{-3i\pi/4} & \ee^{-i\pi/4} & \ee^{i\pi/4} \\
1 & i & -1 & -i & \ee^{-3i\pi/4} & \ee^{3i\pi/4} & \ee^{i\pi/4} & \ee^{-i\pi/4} \\
1 & -i & -1 & i & \ee^{-i\pi/4} & \ee^{i\pi/4} & \ee^{3i\pi/4} & \ee^{-3i\pi/4} \\
1 & i & -1 & -i & \ee^{i\pi/4} & \ee^{-i\pi/4} & \ee^{-3i\pi/4} & \ee^{3i\pi/4} \\
\end{array}
\right)
\end{equation}
which is isomorphic to the complex Hadamard matrix of $\ZZ_8$. Combining with the automorphism yields 4 different $\Theta^{D_8}$ transformation between $H_8$ and $H_8^*$. And the Hopf Hadamard gate is given by,
\begin{equation}
    \FF^{D_8} = \frac{1}{\sqrt{8}}\left(
\begin{array}{cccccccc}
1 & 1 & 1 & 1 & 1 & 1 & 1 & 1 \\
1 & -1 & 1 & -1 & i & -i & i & -i \\
1 & 1 & 1 & 1 & -1 & -1 & -1 & -1 \\
1 & -1 & 1 & -1 & -i & i & -i & i \\
1 & -i & -1 & i & \ee^{3 i\pi/4} & \ee^{-3 i\pi/4} & \ee^{-i\pi/4} & \ee^{i\pi/4} \\
1 & i & -1 & -i & \ee^{-3 i\pi/4} & \ee^{3 i\pi/4} & \ee^{i\pi/4} & \ee^{-i\pi/4} \\
1 & -i & -1 & i & \ee^{-i\pi/4} & \ee^{i\pi/4} & \ee^{3 i\pi/4} & \ee^{-3 i\pi/4} \\
1 & i & -1 & -i & \ee^{i\pi/4} & \ee^{-i\pi/4} & \ee^{-3 i\pi/4} & \ee^{3 i\pi/4} \\
\end{array}
\right)
\end{equation}
One can check $\FF^{D_8} \cdot \FF^{D_8} = S^{D_8}$.

The irreducible representations are,
\begin{equation}
    \begin{array}{c|c|c|c|c|c|c|c|c}
        & 1&r&r^2&r^3 &s &rs&r^2s&r^3s \\ \hline
        \Gamma_\dsi &1&1&1&1&1&1&1&1\\ \hline
        \Gamma_{{\mathbf{s}}_1} &1&-1&1&-1&1&-1&1&-1\\ \hline
        \Gamma_{{\mathbf{s}}_2} &1&-1&1&-1&-1&1&-1&1\\ \hline
        \Gamma_{{\mathbf{s}}_3} &1&1&1&1&-1&-1&-1&-1\\ \hline
        \Gamma_\two &\left(
        \begin{smallmatrix}
            1 & 0 \\
            0 & 1 \\
        \end{smallmatrix}
        \right)
        &
        \left(
        \begin{smallmatrix}
            0 & i \\
            i & 0 \\
        \end{smallmatrix}
        \right)
        &
        \left(
        \begin{smallmatrix}
            -1 & 0 \\
            0 & -1 \\
        \end{smallmatrix}
        \right)
        &
        \left(
        \begin{smallmatrix}
            0 & -i \\
            -i & 0 \\
        \end{smallmatrix}
        \right)
        &
        \left(
        \begin{smallmatrix}
            -1 & 0 \\
            0 & 1 \\
        \end{smallmatrix}
        \right)
        &
        \left(
        \begin{smallmatrix}
            0 & -i \\
            i & 0 \\
        \end{smallmatrix}
        \right)
        &
        \left(
        \begin{smallmatrix}
            1 & 0 \\
            0 & -1 \\
        \end{smallmatrix}
        \right)
        &
        \left(
        \begin{smallmatrix}
            0 & i \\
            -i & 0 \\
        \end{smallmatrix}
        \right)
    \end{array}
\end{equation}
and the characters are,
\begin{equation}\label{eq:d8chitab}
    \begin{array}{c|c|c|c|c|c|c|c|c}
        & 1&r&r^2&r^3 &s &rs&r^2s&r^3s \\ \hline
        \chi_\dsi &1 & 1 & 1 & 1 & 1 & 1 & 1 & 1 \\ \hline
        \chi_{{\mathbf{s}}_1}&1 & -1 & 1 & -1 & -1 & 1 & -1 & 1 \\ \hline
        \chi_{{\mathbf{s}}_2}&1 & -1 & 1 & -1 & 1 & -1 & 1 & -1 \\ \hline
        \chi_{{\mathbf{s}}_3}&1 & 1 & 1 & 1 & -1 & -1 & -1 & -1 \\ \hline
        \chi_\two&2 & 0 & -2 & 0 & 0 & 0 & 0 & 0 
    \end{array}
\end{equation}
We note that although the irreps are the same as the group algebra $\CC[D_8]$ since it has an identical multiplication, the tensor decomposition of the irreps involves the comultiplication \eqref{eq:D8comult}, which is different from the group algebra. It follows the $F$-symbols are different.

The conjugacy classes are,
\begin{equation}
    \{1, \frac{1}{2}(r+r^3),r^2,\frac12(s+r^2s),\frac12(rs+r^3 s)\}
\end{equation}
Moreover, the comultiplication can be summarized as Pauli matrices, $\comult_{j}^{i a} \equiv (\Lambda^a)_{ij}$,
\begin{align}
    \Lambda^1 = P_+ P_+ P_+,\quad \Lambda^r = P_+ \frac{\sigma^0-i \sigma^2}{2}P_-,\quad \Lambda^{r^2} = P_+ P_- P_+,\quad \Lambda^{r^3} =   P_+ \frac{\sigma^0+i \sigma^2}{2}P_-\\
    \Lambda^s = P_- (\sigma^{00}+\sigma^{03}+\sigma^{31}-i\sigma^{32})/4,\quad \Lambda^{rs} = P_- (\sigma^{00}-\sigma^{03}+\sigma^{31}+i\sigma^{32})/4, \\
    \Lambda^{r^2s} = P_- (\sigma^{00}+\sigma^{03}-\sigma^{31}+i\sigma^{32})/4,\quad \Lambda^{r^3s} =   P_- (\sigma^{00}-\sigma^{03}-\sigma^{31}-i\sigma^{32})/4
\end{align}
where $P_\pm = \frac{\sigma^0 \pm \sigma^3}{2}$. So the $\Rep(H_8)$ symmetry MPO tensor is given by,
\begin{equation}
    \sum_a (\Lambda^a)_{ij} [\Gamma(a)]_{mn}
\end{equation}
where $i,j$ are indices for the physical Hilbert space, while $m,n$ are for the virtual dimension. Depending on the representation $\Gamma$, the bond dimension could be larger than 1, but always equal to the quantum dimension, i.e. the symmetry is realized ``on-site''.

\subsection{``MPO'' basis}
\label{app:MPO basis}

This basis is dual to the original basis presented by Kac \& Paljutkin \cite{kac1966finite}. It is called the MPO basis because the comultiplication has the natural structure used to define the matrix product operator \cite{Meng:2024nxx}. The MPO basis has the basis, $B_\text{MPO} = \{e_1,e_2,e_3,e_4,v_{11},v_{12},v_{21},v_{22}\}$, which is related to $xyz$ basis by $B_\text{MPO} = M_{nx} B_{xyz}$, where
\begin{equation}
    M_{nx}=\left(
\begin{array}{cccccccc}
 1 & 1 & 1 & 1 & \sqrt{2} & 0 & 0 & \sqrt{2} \\
 1 & -1 & -1 & 1 & -\sqrt{2} & 0 & 0 & \sqrt{2} \\
 1 & -1 & -1 & 1 & \sqrt{2} & 0 & 0 & -\sqrt{2} \\
 1 & 1 & 1 & 1 & -\sqrt{2} & 0 & 0 & -\sqrt{2} \\
 1 & i & -i & -1 & 0 & \sqrt{2} & \sqrt{2} & 0 \\
 1 & -i & i & -1 & 0 & \sqrt{2} & -\sqrt{2} & 0 \\
 1 & -i & i & -1 & 0 & -\sqrt{2} & \sqrt{2} & 0 \\
 1 & i & -i & -1 & 0 & -\sqrt{2} & -\sqrt{2} & 0 \\
\end{array}
\right).
\end{equation}
The multiplication table is given by,

\begin{equation}
\begin{array}{|c||c|c|c|c|c|c|c|c|}
\hline
 m&e_1 & e_2 & e_3 & e_4 & v_{11} & v_{12} & v_{21} & v_{22} \\ \hhline{|=||=|=|=|=|=|=|=|=|}
 e_1&e_1 & e_2 & e_3 & e_4 & v_{11} & v_{12} & v_{21} & v_{22} \\ \hline
 e_2&e_2 & e_1 & e_4 & e_3 & v_{22} & -i v_{21} & i v_{12} & v_{11} \\\hline
 e_3&e_3 & e_4 & e_1 & e_2 & v_{22} & i v_{21} & -i v_{12} & v_{11} \\\hline
 e_4&e_4 & e_3 & e_2 & e_1 & v_{11} & -v_{12} & -v_{21} & v_{22} \\\hline
 v_{11}&v_{11} & v_{22} & v_{22} & v_{11} & e_1 + e_4 & 0 & 0 & e_2 + e_3 \\\hline
  v_{12}&v_{12} & i v_{21} & -i v_{21} & -v_{12} & 0 & e_1 - e_4 & -i (e_2 - e_3) & 0 \\\hline
 v_{21}& v_{21} & -i v_{12} & i v_{12} & -v_{21} & 0 & i (e_2 - e_3) & e_1 - e_4 & 0 \\\hline
  v_{22}&v_{22} & v_{11} & v_{11} & v_{22} & e_2 + e_3 & 0 & 0 & e_1 + e_4 \\
 \hline
\end{array}
\end{equation}
and the comultiplication is given by,
\begin{equation}
    \comult(e_i) = e_i \otimes e_i,\quad \comult(v_{ij}) = \frac{1}{\sqrt{2}} \sum_k v_{ik}\otimes v_{kj} 
\end{equation}
The antipode acts as,
\begin{equation}
    S(e_i)=e_i,\quad S(v_{ij}) = v_{ji}
\end{equation}
The (co)unit and Haar (co)integral are given by,
\begin{equation}
    \unit=e_1,\quad \counith = \sum_i\delta_{e_i}+\sqrt{2}(\delta_{v_{11}}+\delta_{v_{22}}),\quad \haarc =\frac{1}{8}(\sum_i{e_i}+\sqrt{2}({v_{11}}+{v_{22}})),\quad \cohaar=\delta_{e_1}
\end{equation}
The Hadamard form is given by,
\begin{equation}
  \Theta^\text{MPO}= 
\left(
\begin{array}{cccccccc}
 1 & 1 & 1 & 1 & \sqrt{2} & 0 & 0 & \sqrt{2} \\
 1 & -1 & -1 & 1 & \sqrt{2} & 0 & 0 & -\sqrt{2} \\
 1 & -1 & -1 & 1 & -\sqrt{2} & 0 & 0 & \sqrt{2} \\
 1 & 1 & 1 & 1 & -\sqrt{2} & 0 & 0 & -\sqrt{2} \\
 \sqrt{2} & \sqrt{2} & -\sqrt{2} & -\sqrt{2} & 0 & 0 & 0 & 0 \\
 0 & 0 & 0 & 0 & 0 & 2 \ee^{3 i\pi/4} & 2 \ee^{-3 i\pi/4} & 0 \\
 0 & 0 & 0 & 0 & 0 & 2 \ee^{-3 i\pi/4} & 2 \ee^{3 i\pi/4} & 0 \\
 \sqrt{2} & -\sqrt{2} & \sqrt{2} & -\sqrt{2} & 0 & 0 & 0 & 0 \\
\end{array}
\right)
\end{equation}
and the Hadamard gate is given by,
\begin{equation}
    \FF^\text{MPO} = \frac{1}{2\sqrt{2}}\left(
\begin{array}{cccccccc}
 1 & 1 & 1 & 1 & \sqrt{2} & 0 & 0 & \sqrt{2} \\
 1 & -1 & -1 & 1 & -\sqrt{2} & 0 & 0 & \sqrt{2} \\
 1 & -1 & -1 & 1 & \sqrt{2} & 0 & 0 & -\sqrt{2} \\
 1 & 1 & 1 & 1 & -\sqrt{2} & 0 & 0 & -\sqrt{2} \\
 \sqrt{2} & -\sqrt{2} & \sqrt{2} & -\sqrt{2} & 0 & 0 & 0 & 0 \\
 0 & 0 & 0 & 0 & 0 & 2 \ee^{3 i\pi/4} & 2 \ee^{-3 i\pi/4} & 0 \\
 0 & 0 & 0 & 0 & 0 & 2 \ee^{-3 i\pi/4} & 2 \ee^{3 i\pi/4} & 0 \\
 \sqrt{2} & \sqrt{2} & -\sqrt{2} & -\sqrt{2} & 0 & 0 & 0 & 0 \\
\end{array}
\right)
\end{equation}
The characters of $\Irr(H_8)$ in the ``MPO'' basis are,
\begin{equation}
    \begin{array}{c|c|c|c|c|c|c|c|c}
      & e_1 & e_2 & e_3 & e_4 & v_{11} & v_{12} & v_{21} & v_{22} \\
\hline
\chi_\dsi & 1 & 1 & 1 & 1 & \sqrt{2} & 0 & 0 & \sqrt{2} \\\hline
\chi_{{\mathbf{s}}_1} & 1 & -1 & -1 & 1 & -\sqrt{2} & 0 & 0 & \sqrt{2} \\\hline
\chi_{{\mathbf{s}}_2} & 1 & -1 & -1 & 1 & \sqrt{2} & 0 & 0 & -\sqrt{2} \\\hline
\chi_{{\mathbf{s}}_3} & 1 & 1 & 1 & 1 & -\sqrt{2} & 0 & 0 & -\sqrt{2} \\\hline
\chi_\two & 2 & 0 & 0 & -2 & 0 & 0 & 0 & 0 
\end{array}
\end{equation}
and the characters of $\Irr(\dl H_8)$ are,
\begin{equation}
    \begin{array}{c|c|c|c|c|c|c|c|c}
      & \delta_{e_1} & \delta_{e_2} & \delta_{e_3} & \delta_{e_4} 
      & \delta_{v_{11}} & \delta_{v_{12}} & \delta_{v_{21}} & \delta_{v_{22}} \\
\hline
\dl\chi_\dsi & 1 & 0 & 0 & 0 & 0 & 0 & 0 & 0 \\\hline
\dl\chi_{{\mathbf{s}}_1} & 0 & 0 & 1 & 0 & 0 & 0 & 0 & 0 \\\hline
\dl\chi_{{\mathbf{s}}_2} & 0 & 1 & 0 & 0 & 0 & 0 & 0 & 0 \\\hline
\dl\chi_{{\mathbf{s}}_3} & 0 & 0 & 0 & 1 & 0 & 0 & 0 & 0 \\\hline
\dl\chi_\two & 0 & 0 & 0 & 0 & \tfrac{1}{\sqrt{2}} & 0 & 0 & \tfrac{1}{\sqrt{2}} 
\end{array}
\end{equation}
An interesting property of the MPO basis is that the single-site Hopf Pauli operators of $H_8$ are stoquastic. Moreover, using \eqref{eq:RepHstarinv}, one can show that the Hopf Pauli operators of $ZZ$- and $X$-types are stoquastic for $\chi_{{\mathbf{s}}_1}+\chi_{{\mathbf{s}}_2},\chi_{{\mathbf{s}}_3},\chi_\two$ and similar for the duals. However, the individual operators for $\chi_{{\mathbf{s}}_1}$ and $\chi_{{\mathbf{s}}_2}$ are not.

\section{Definitions}\label{app:def}
This appendix collects definitions, conventions, and basic facts used throughout the paper. Its purpose is to fix notation and terminology, which may vary across the literature, and to provide a self-contained reference for technical material. Most of the material reviewed here is standard, which can be found in \cite{swedler1969hopf,montgomery1993hopf,klimyk2012quantum}.
\begin{definition}[Left $H$-module]\label{def:left-H-module}
Let $H$ be a Hopf algebra over $\CC$. A (left) $H$-module is a
$\CC$-vector space $M$ together with a $\CC$-linear action
\[
\Gamma \colon H \otimes M \to M,\qquad (h,m)\mapsto h\cdot m,
\]
such that for all $h,g\in H$ and $m\in M$,
\begin{equation}
\unit \cdot m = m,\qquad (hg)\cdot m = h\cdot(g\cdot m).
\end{equation}
where $\unit$ is the unit element in $H$. A morphism of left $H$-modules is a $\CC$-linear map $f\colon M \to N$ such that
\begin{equation}
f(h\cdot m) = h\cdot f(m)\qquad\forall\,h\in H,\;m\in M.
\end{equation}
We write $\Rep(H)$ for the category of finite-dimensional left $H$-modules.
\end{definition}

\paragraph{Monoidal structure on $\Rep(H)$.}
Let $\Delta(h) = \sum h_{(1)}\otimes h_{(2)}$ and let $\wh \counit$ be the counit of $H$.
Then $\Rep(H)$ is a monoidal category with:
\begin{itemize}
  \item \emph{Tensor product.} If $M,N$ are left $H$-modules, their tensor
  product $M\otimes N$ is a left $H$-module with action
  \begin{equation}
  h\cdot(m\otimes n)
  := \sum (h_{(1)}\cdot m)\otimes (h_{(2)}\cdot n).
  \end{equation}
  \item \emph{Unit object.} The unit object is the trivial module
  $\CC$ with action
  \begin{equation}
  h\cdot \lambda := \wh\counit(h)\,\lambda,\qquad h\in H,\;\lambda\in \CC.
  \end{equation}
\end{itemize}

If the antipode $S$ of $H$ is bijective and $M$ is finite-dimensional, then the dual vector space $\dl M = \operatorname{Hom}_\CC(M,\CC)$ becomes a left $H$-module via
\begin{equation}\label{eq:dualHmod}
(h\cdot f)(m) := f\bigl(S(h)\cdot m\bigr),
\end{equation}
making $\Rep(H)$ rigid.

Equivalent, $\Gamma$ as an algebra homomorphism, $\Gamma: H\rightarrow \mathrm{End}(M)$,
\begin{equation}
    \Gamma(e)=\mathbbm{1},\quad \Gamma(hg)_{ij} = \Gamma(h)_{ik}\Gamma(g)_{kj}.
\end{equation}
and the dual $\Gamma^\vee$ is given by,
\begin{equation}\label{eq:dualrep}
    \Gamma^\vee(h)= \Gamma(S(h))^\intercal,
\end{equation}
which follows from \eqref{eq:dualHmod}. Finally, their characters are related by,
\begin{equation}\label{eq:dualHchar}
    \chi_{\Gamma^\vee}(h) = \chi_{\Gamma}(S(h))
\end{equation}

For $C^\star$ Hopf algebras, unitary representations $\Gamma$ are those that satisfy
\begin{equation}
    \Gamma(h^\star) = \Gamma(h)^\dagger \,.
\end{equation}

\paragraph{Examples.}
\begin{itemize}
  \item The \emph{trivial module} $\CC$: $h\cdot \lambda = \wh\counit(h)\lambda$.
  \item The \emph{regular module} $H$: action by left multiplication
  \begin{equation}
  h\cdot x := hx,\qquad h,x\in H.
  \end{equation}
  \item If $A$ is an algebra, an \emph{$H$-module algebra} structure on $A$
  is a left $H$-module structure such that
  \begin{equation}
  h\cdot(ab) = \sum (h_{(1)}\cdot a)(h_{(2)}\cdot b),\qquad
  h\cdot 1_A = \epsilon(h)\,1_A.
  \end{equation}
\end{itemize}

\begin{definition}[Right $H$-comodule]\label{def:right-H-comodule}
Let $H$ be a Hopf algebra over a field $\CC$ with comultiplication
$\Delta\colon H\to H\otimes H$ and counit $\wh\counit\colon H\to \CC$.
A \emph{right $H$-comodule} is a $\CC$-vector space $V$ together with a
$\CC$-linear coaction
\begin{equation}
\delta\colon V \longrightarrow V\otimes H
\end{equation}
satisfying
\begin{align}
(\id_V\otimes \Delta)\circ \delta &= (\delta\otimes \id_H)\circ \delta,
\\
(\id_V\otimes \epsilon)\circ \delta &= \id_V.
\end{align}
In Sweedler notation, we write
\begin{equation}
\delta(v) = \sum v_{(0)}\otimes v_{(1)},\qquad v\in V.
\end{equation}
\end{definition}

\begin{definition}[Morphisms of comodules]\label{def:comodule-morphisms}
Let $(V,\delta_V)$ and $(W,\delta_W)$ be right $H$-comodules.
A morphism of right $H$-comodules is a $\CC$-linear map $f\colon V\to W$
such that
\begin{equation}
(f\otimes \id_H)\circ \delta_V = \delta_W\circ f.
\end{equation}
That is, $f$ intertwines the coactions.
\end{definition}

\paragraph{Monoidal structure on right comodules.}
The category of right $H$-comodules is monoidal with:
\begin{itemize}
  \item \emph{Tensor product.} If $(V,\delta_V)$ and $(W,\delta_W)$ are
  right $H$-comodules, the tensor product $V\otimes W$ is a right $H$-comodule
  with coaction
  \begin{equation}
  \delta_{V\otimes W}(v\otimes w)
  := \sum (v_{(0)}\otimes w_{(0)})\otimes v_{(1)}w_{(1)}.
  \end{equation}
  \item \emph{Unit object.} The unit object is $\CC$ with coaction
  \begin{equation}
  \delta(\lambda) := \lambda\otimes \unit,\qquad \lambda\in \CC.
  \end{equation}
  where $\unit$ is the unit element in $H$.
\end{itemize}

\paragraph{Examples.}
\begin{itemize}
  \item The \emph{trivial comodule} $\CC$ with coaction
  $\delta(\lambda) = \lambda\otimes \unit$.
  \item The \emph{regular comodule} $H$ with coaction given by the
  comultiplication:
  \begin{equation}
  \delta(h) := \Delta(h),\qquad h\in H.
  \end{equation}
\end{itemize}
One can similar define left $H$-comodules.

\paragraph{Finite-dimensional duality.}
If $H$ is finite-dimensional, the category of right $H$-comodules is
canonically equivalent to the category of left $\dl{H}$-modules. Given a
right $H$-comodule $(V,\delta)$ and $f\in \dl{H}$, define
\begin{equation}
f\cdot v := (\id_V\otimes f)\bigl(\delta(v)\bigr)
= \sum v_{(0)}\, f\bigl(v_{(1)}\bigr),\qquad v\in V.
\end{equation}
This action makes $V$ a left $\dl{H}$-module. Similarly, left
$H$-comodules correspond to right $\dl{H}$-modules.

\begin{definition}[Hopf module]\label{def:Hopf-module}
Let $H$ be a Hopf algebra over $\CC$, with comultiplication
$\Delta(h)=\sum h_{(1)}\otimes h_{(2)}$, counit $\wh \counit$, and antipode $S$.
% \medskip
% \noindent
% \textbf{(1) Left-left Hopf module.}
A \emph{left--left $H$-Hopf module} is a vector space $M$ which is both
\begin{itemize}
    \item a left $H$-module $(M,\cdot)$, with action 
    $h\cdot m$,
    \item a left $H$-comodule $(M,\delta_L)$, with coaction
    $\delta_L(m)=\sum m_{(-1)}\otimes m_{(0)}$,
\end{itemize}
such that the following compatibility condition holds for all $h\in H$,
$m\in M$:
\begin{equation}\label{eq:left-left-Hopf-module}
\delta_L(h\cdot m)
  =\sum h_{(1)}\,m_{(-1)}\ \otimes\ h_{(2)}\cdot m_{(0)}.
\end{equation}
One can similar define right-right/right-left/left-right Hopf module.
\end{definition}

\section{Miscellaneous Proofs}\label{app:lineartrans}
\tikzsetfigurename{diagrams}
\subsection{Data of dual Hopf algebra}
Starting from the Hopf algebra $H$, we derive the corresponding data of the dual Hopf algebra $\dl{H}$
Take the basis $a_i \in H$, and $\delta_{a_i} \in \dl{H}$, such that the canonical pairing $\langle \dl{H},H \rangle \rightarrow k$,
\begin{equation}
    \langle \delta_{a_i},a_j\rangle \mapsto \delta_{a_i}(a_j) = \delta_{i,j}
\end{equation}
Recall that the multiplication and co-multiplication in $H$ is given as,
\begin{equation}
\label{eq:appHdata}
    \mult(a_i, a_j) = \sum_k \mult_{ij}^k a_k,\quad \comult(a_k)=\sum_{i,j}\comult_k^{ij} a_i \otimes a_j
\end{equation}
The comultiplication of $\dl{H}$ is defined by,
\begin{equation}
    \langle \dl{\comult}(\delta_c),a\otimes b\rangle\equiv\langle \delta_c,ab\rangle
\end{equation}
Suppose $\dl{\comult}(\delta_{a_i}) = \sum_{lm} \tilde{\comult}_i^{lm} \delta_{a_l} \otimes \delta_{a_m}$. We take the basis $\delta_{a_i}$ and $a_j, a_k$, and compute
\begin{align}
    &\langle \dl{\comult}(\delta_{a_i}),a_j \otimes a_k\rangle = \langle \delta_{a_i},a_j a_k\rangle %= \langle \delta_{a_i},\sum_l \mult^l_{jk}a_l\rangle 
    = \sum_l \mult^l_{jk}\langle \delta_{a_i},a_l\rangle =\sum_l \mult^l_{jk} \delta_{i,l}=\mult_{jk}^i
\end{align}
On the other hand, we also have
 \begin{align}
\langle \dl{\comult}(\delta_{a_i}),a_j \otimes a_k\rangle =\sum_{lm}\tilde{\comult}_i^{lm} \delta_{l,j}\delta_{m,k} = \tilde{\comult}_{i}^{jk}
 \end{align}   
This shows that
 \begin{align}
 \label{eq:dlHcomult}
    \dl{\comult}(\delta_{a_i}) = \sum_{jk} \mult_{jk}^i \delta_{a_j} \otimes \delta_{a_k}
\end{align}
Similarly, the multiplication in $\dl{H}$ is defined as,
\begin{align}
    \langle \dl{\mult}(\delta_a\otimes \delta_b),c\rangle = \langle \delta_a\otimes \delta_b,\comult(c) \rangle
\end{align}
An analogous computation gives
\begin{align}
\label{eq:dlHmult}
    \dl{\mult}(\delta_{a_i}\otimes \delta_{a_j}) = \sum_{k} \comult_{k}^{ij}\delta_{a_k}
\end{align}

\subsection{Hopf algebra isomorphism between \texorpdfstring{$H$}{H} and \texorpdfstring{$\dl H$}{H*}}\label{app:isoHandHdual}
We assume the data of the Hopf algebra and its dual in \eqref{eq:appHdata} \eqref{eq:dlHcomult}, \eqref{eq:dlHmult}, 

For self-dual Hopf algebras, the Hadamard form $\Theta: H\otimes H \rightarrow \CC$ induces the isomorphism $\Fiso$,
\begin{equation}
    \Fiso: H\rightarrow \dl{H}, \quad a\mapsto \Theta_{a,a'} \delta_{a'}
\end{equation}
From the commutative diagrams,
\tikzsetfigurename{commutativediags}
\begin{equation}\label{eq:dualL}
    \begin{tikzpicture}[>=Straight Barb, baseline=(current bounding box.center)]
        \matrix (m) [matrix of math nodes, column sep=2.5em, row sep=2.5em] {
            H & H \otimes H \\
            \dl{H} & \dl{H} \otimes \dl{H} \\
        };
        \draw[->] (m-1-1) -- (m-1-2) node[midway, above] {$\Delta$};
        \draw[->] (m-1-1) -- (m-2-1) node[midway, left] {$\Fiso$};
        \draw[->] (m-1-2) -- (m-2-2) node[midway, right] {$\Fiso \otimes \Fiso$};
        \draw[->] (m-2-1) -- (m-2-2) node[midway, below] {$\dl{\Delta}$};
    \end{tikzpicture},\quad \quad \begin{aligned}
        &k\rightarrow \comult_k^{i,j} i \otimes j \rightarrow \comult_k^{i,j} \Theta_{i,i'}\Theta_{j,j'} \delta_{i'}\otimes \delta_{j'} \\
        & k \rightarrow \Theta_{k,k'}\delta_{k'} \rightarrow \Theta_{k,k'} \mult_{i',j'}^{k'} \delta_{i'}\otimes \delta_{j'}\\
        &\Rightarrow \comult_k^{i,j} \Theta_{i,i'}\Theta_{j,j'} = \Theta_{k,k'} \mult_{i',j'}^{k'}
    \end{aligned}
\end{equation}
\begin{equation}\label{eq:duall}
    \begin{tikzpicture}[>=Straight Barb, baseline=(current bounding box.center)]
        \matrix (m) [matrix of math nodes, column sep=2.5em, row sep=2.5em] {
            H \otimes H & H \\
            \dl{H} \otimes \dl{H} & \dl{H} \\
        };
        % arrows
        \draw[->] (m-1-1) -- (m-1-2) node[midway, above] {$m$};
        \draw[->] (m-1-1) -- (m-2-1) node[midway, left] {$\Fiso \otimes \Fiso$};
        \draw[->] (m-1-2) -- (m-2-2) node[midway, right] {$\Fiso$};
        \draw[->] (m-2-1) -- (m-2-2) node[midway, below] {$\dl{m}$};
    \end{tikzpicture}, \quad \quad \mult_{ij}^k \Theta_{k,k'} = \Theta_{i,i'}\Theta_{j,j'}\comult^{i',j'}_{k'}
\end{equation}
Therefore, \eqref{eq:dualL} and \eqref{eq:duall} give the consistency equation that the $\Theta$ should satisfy. For $H_8$, we have 4 solutions corresponding to the automorphism group of $H_8$, $\operatorname{Aut}(H_8)\cong \ZZ_2\times \ZZ_2$.

We can further check the antipode is compatible with the $\Fiso$,
\tikzsetfigurename{commutativediagsb}
\begin{equation}\label{eq:duals}
    \begin{tikzpicture}[>=Straight Barb, baseline=(current bounding box.center)]
        \matrix (m) [matrix of math nodes, column sep=2.5em, row sep=2.5em] {
            H  & H \\
            \dl{H}  & \dl{H} \\
        };
        % arrows
        \draw[->] (m-1-1) -- (m-1-2) node[midway, above] {$S$};
        \draw[->] (m-1-1) -- (m-2-1) node[midway, left] {$\Fiso$};
        \draw[->] (m-1-2) -- (m-2-2) node[midway, right] {$\Fiso$};
        \draw[->] (m-2-1) -- (m-2-2) node[midway, below] {$\dl{S}$};
    \end{tikzpicture}, \quad \quad S_{a,a'}\Theta_{a',a''} =\Theta_{a,a'}S^\intercal_{a',a''}.
\end{equation}
We further require the $\Fiso$ to be a $C^\star$ Hopf algebra isomorphism and involution is compatible under $\Fiso$, therefore 
\begin{equation}\label{eq:dualinv}
    \begin{tikzpicture}[>=Straight Barb, baseline=(current bounding box.center)]
        \matrix (m) [matrix of math nodes, column sep=2.5em, row sep=2.5em] {
            H  & H \\
            \dl{H}  & \dl{H} \\
        };
        % arrows
        \draw[->] (m-1-1) -- (m-1-2) node[midway, above] {$\star$};
        \draw[->] (m-1-1) -- (m-2-1) node[midway, left] {$\Fiso$};
        \draw[->] (m-1-2) -- (m-2-2) node[midway, right] {$\Fiso$};
        \draw[->] (m-2-1) -- (m-2-2) node[midway, below] {$\dl{\star}$};
    \end{tikzpicture}, \quad \quad I_{a,a'}\Theta_{a',a''} = \overline{\Theta}_{a,a'}I^\dagger_{a',s} S^\intercal_{s,a''},
\end{equation}
where $\star: p_aa \mapsto \overline{p_a} I_{a,a'}a'$, $\dl \star: q_a \delta_a \mapsto \overline{q_a} I^\dagger_{a,a'}S^\intercal_{a',a''}\delta_{a''}$. 

\subsection{Basis transformation}
Similarly, we can check the basis transformation, given by $M: H\rightarrow H$ and $a_i \mapsto M_{i,i'}v_{i'}$.
Upon the change of bases, we have,
\begin{equation}
    \comult_k^{i,j} M_{i,i'} M_{j,j'} = M_{k,k'} \comult^{z,i',j'}_{k'},\quad \mult_{ij}^k M_{k,k'} = M_{i,i'}M_{j,j'}\mult^{z,k'}_{i',j'}
\end{equation}
where $\mult^{z,k'}_{i',j'}, \comult^{z,i',j'}$ are the structure constants for the (co)multiplication of $H$ in the $v_i$ bases induced by $M$. With these definitions of $\comult^z,\mult^z$, the following diagrams commute,
\begin{equation}
    \begin{tikzpicture}[>=Straight Barb, baseline=(current bounding box.center)]
        \matrix (m) [matrix of math nodes, column sep=2.5em, row sep=2.5em] {
            H & H \otimes H \\
            H & H \otimes H \\
        };
        \draw[->] (m-1-1) -- (m-1-2) node[midway, above] {$\Delta$};
        \draw[->] (m-1-1) -- (m-2-1) node[midway, left] {$M$};
        \draw[->] (m-1-2) -- (m-2-2) node[midway, right] {$M \otimes M$};
        \draw[->] (m-2-1) -- (m-2-2) node[midway, below] {$\Delta^z$};
    \end{tikzpicture},\quad \quad \begin{tikzpicture}[>=Straight Barb, baseline=(current bounding box.center)]
        \matrix (m) [matrix of math nodes, column sep=2.5em, row sep=2.5em] {
            H \otimes H & H \\
            H \otimes H & H \\
        };
        \draw[->] (m-1-1) -- (m-1-2) node[midway, above] {$m$};
        \draw[->] (m-1-1) -- (m-2-1) node[midway, left] {$M \otimes M$};
        \draw[->] (m-1-2) -- (m-2-2) node[midway, right] {$M$};
        \draw[->] (m-2-1) -- (m-2-2) node[midway, below] {$\mult^z$};
    \end{tikzpicture}.
\end{equation}
For the transformation between $F$ and $F^z$, we use the inverse basis transformation,
\begin{equation}
    M^{-1}: H \rightarrow H,\quad v_i \mapsto (M^{-1})_{i,i'}h_{i'}
\end{equation}
Then the $M^{-1}$ induces,
\begin{equation}
    F^z_{ij} = (M^{-1})_{i,i'}F_{i',j'} (M^{-1})_{k,j'},
\end{equation}
such that the following diagram commutes,
\tikzsetfigurename{commutativediagc}
\begin{equation}
    \begin{tikzpicture}[>=Straight Barb, baseline=(current bounding box.center)]
        \matrix (m) [matrix of math nodes, column sep=2.5em, row sep=2.5em] {
            H & H \\
            \dl{H} & \dl{H} \\
        };
        \draw[<-] (m-1-1) -- (m-1-2) node[midway, above] {$M^{-1}$};
        \draw[->] (m-1-1) -- (m-2-1) node[midway, left] {$F$};
        \draw[->] (m-1-2) -- (m-2-2) node[midway, right] {$F^z$};
        \draw[->] (m-2-1) -- (m-2-2) node[midway, below] {$\dl{(M^{-1})}$};
    \end{tikzpicture}, \quad \quad 
\end{equation}
where the induced dual morphism $\dl{(M^{-1})}$ is given by $(M^{-1})^\intercal$ follows from the definitions\footnote{A \href{https://math.stackexchange.com/questions/4933473/origin-of-notation-of-a-dual-space}{useful StackExchange answer} on the dual of a vector space.}. In the matrix form,
\begin{equation}
    F^z = M^{-1} F (M^{-1})^\intercal
\end{equation}
For basis transformation of unit and counit, we first consider the unit, $\CC \xrightarrow{1_\cH} H, 1\mapsto 1_\cH^i h_i$, where $1_\cH^i$ is the coefficient. Similarly, $\CC \xrightarrow{1_\cV} H, 1\mapsto 1_\cV^i v_i$, where $1_\cV^i$ is the coefficient. We define,
\begin{equation}
    1_\cH^i M_{i,i'}= 1_\cV^{i'} 
\end{equation}
such that the following diagram commutes, 
\begin{equation}
    \begin{tikzpicture}[>=Straight Barb, baseline=(current bounding box.center)]
        \matrix (m) [matrix of math nodes, row sep=2.5em, column sep=2.5em] {
            \CC & H \\
            & H \\
        };
        \draw[->] (m-1-1) -- (m-1-2) node[midway, above] {$1_\cH$};
        \draw[->] (m-1-1) -- (m-2-2) node[midway, left] {$1_\cV$};
        \draw[->] (m-1-2) -- (m-2-2) node[midway, right] {$M$};
    \end{tikzpicture}.
\end{equation}
Similarly, for counit, $H \xrightarrow{\epsilon_\cH} \CC, h_i\mapsto \epsilon_\cH^j \delta_{h_j}(h_i)=\epsilon_\cH^j\delta_{i,j}$, and $H \xrightarrow{\epsilon_\cV} \CC, v_i\mapsto \epsilon_\cV^j \delta_{v_j}(v_i)=\epsilon_\cV^j\delta_{i,j}$. Therefore, we have,
\begin{equation}
    \epsilon_\cH^i = \epsilon_\cV^j \delta_{v_j}(M_{i,i'}v_{i'}) = M_{ij}\epsilon_\cV^j
\end{equation}
and the following diagram commutes,
\tikzsetfigurename{commutativediagd}
\begin{equation}
    \begin{tikzpicture}[>=Straight Barb, baseline=(current bounding box.center)]
        \matrix (m) [matrix of math nodes, row sep=2.5em, column sep=2.5em] {
            H& \CC \\
            H & \\
        };
        \draw[->] (m-1-1) -- (m-1-2) node[midway, above] {$\epsilon_\cH$};
        \draw[->] (m-1-1) -- (m-2-1) node[midway, left] {$M$};
        \draw[->] (m-2-1) -- (m-1-2) node[midway, right] {$\epsilon_\cV$};
    \end{tikzpicture}.
\end{equation}
Finally, for antipode, $H \xrightarrow{S} H$, with $h_i\mapsto S_{ij}h_j$, we define,
\begin{equation}
    S_{ij} = M_{ii'}S^z_{i' j'} (M^{-1})_{j'j},
\end{equation}
or in matrix form $S^z = M^{-1}SM$, where $S^z$ is the antipode in $v_i$ bases induced by $M$. Therefore, the following diagram commutes,
\begin{equation}
    \begin{tikzpicture}[>=Straight Barb, baseline=(current bounding box.center)]
        \matrix (m) [matrix of math nodes, row sep=2.5em, column sep=2.5em] {
            H & H \\
            H & H \\
        };
        \draw[->] (m-1-1) -- (m-1-2) node[midway, above] {$M$};
        \draw[->] (m-1-1) -- (m-2-1) node[midway, left] {$S$};
        \draw[->] (m-1-2) -- (m-2-2) node[midway, right] {$S^z$};
        \draw[->] (m-2-1) -- (m-2-2) node[midway, below] {$M$};
    \end{tikzpicture}.
\end{equation}

\subsection{Proof of \eqref{eq:kwsquare}}
\tikzsetfigurename{proof-KW}
\ie \label{eq:proof-kw-sqaure}
    \tikzsetnextfilename{KWpf-1}
    \begin{ZX}
        &&&&&&&&\\
        &&&&&\zxHad{\tiny$\dagger$} \ar[u]\ar[d]&&&\\
        &&&&&\zxX{}\ar[ld,(] \ar[rd,)]&&&\\
        &&\cdots&&\zxrep{$S$}\ar[d]& &\zxZ{} \ar[rrd] \ar[lddd,)]&&\\
        &&&&\zxZ{} \ar[llu] \ar[d]&&&&\cdots\\
        &&&&\zxrep{$S$} \ar[rd,(]&&&&\\
        &&&&&\zxX{} \ar[d]&&&\\
        &&&&&\zxHad{} \ar[d]&&&\\
        &&&&&&&&
    \end{ZX}
    & \stackrel{\mathmakebox[\widthof{\eqref{eq:rcgc=S}}]{\eqref{eq:rcgc=S}}}{=}    
    \tikzsetnextfilename{KWpf-2}
    \begin{ZX}
        &&&&&& & &&\\
        &&&\zxHad{\tiny$\dagger$} \ar[u]\ar[d]&&& & &&\\
        &&&\zxX{}\ar[ld,(] \ar[rrrrdd,)]&&& & &&\\
        \cdots&&\zxrep{$S$}\ar[d]& & & & &&&\\
        &&\zxZ{} \ar[llu] \ar[d]&&&&& \zxZ{} \ar[rrd] \ar[lddd,)] &&\\
        &&\zxrep{$S$} \ar[rd,(]&&\zxX{}&&&&&\cdots\\
        &&&\zxX{} \ar[d] \ar[ru,s]&&\zxrep{$S$} \ar[lu,(] \ar[rd,(]&&&\\
        &&&\zxHad{} \ar[d]&&&\zxZ{}&&&\\
        &&&&&&&&&
    \end{ZX}
    \stackrel{\eqref{eq:hopf-frob-cond},
    % \eqref{eq:assoc},
    \eqref{eq:antipoderule}}{=}
    \tikzsetnextfilename{KWpf-3}
    \begin{ZX}
        &&&&& &&\\
        &&\zxHad{\tiny$\dagger$} \ar[u]\ar[d]&&& &&\\
        &&\zxX{} \ar[rrdddddd,)]&&& &&\\
        \cdots& & & & && &\\
        & &\zxrep{$S$} \ar[llu,(]&&&&&\\
        &\zxZ{} \ar[ru,(] \ar[rd,(] \ar[ruuu,),3d above,very thick]&&&&&&\\
        &&\zxX{} \ar[d] \ar[rd,)]&&& &\\
        &&\zxHad{} \ar[d]&\zxrep{$S$}&&&&\\
        &&&&\zxZ{} \ar[lu,)] \ar[rrd] &&&\\
        &&&&&&&\cdots
    \end{ZX}
    \stackrel{\eqref{eq:hopfbialgebra}}{=}  
    \tikzsetnextfilename{KWpf-4}
    \begin{ZX}
        &&&&& &&\\
        &&\zxHad{\tiny$\dagger$} \ar[u]\ar[d]&&& &&\\
        &&\zxX{} \ar[rrdddddd,)]&&& &&\\
        \cdots& & & & && &\\
        & &\zxrep{$S$} \ar[llu,(]&&&&&\\
        &\zxX{}  \ar[d] \ar[ruuu,),3d above,very thick]&\zxX{}\ar[u]&&&&&\\
        &\zxZ{} \ar[ru] \ar[d] & \zxZ{} \ar[d] \ar[u] \ar[lu,3d above,very thick]&&& &\\
        &\zxHad{} \ar[d]&\zxrep{$S$}&&&&&\\
        &&&&\zxZ{} \ar[llu,)] \ar[rrd] &&&\\
        &&&&&&&\cdots
    \end{ZX}
    \\
    &\stackrel{\mathmakebox[\widthof{\eqref{eq:rcgc=S}}]{\eqref{eq:antipoderule},\eqref{eq:assoc}}}{=}
    \tikzsetnextfilename{KWpf-5}
    \begin{ZX}
        &&&&&& &&\\
        &&\zxHad{\tiny$\dagger$} \ar[u]\ar[d]&&&& &&\\
        &&\zxX{} \ar[rrrddd,)]&&&& &&\\
        \cdots& & & && && &\\
        & &\zxX{}\ar[llu,(]&&&&&&\\
        & &&\zxrep{$S$}\ar[lu,(]&&\zxX{}&&&\\
        &\zxZ{} \ar[ruuuu,),3d above,very thick] \ar[rru,(] \ar[d] & \zxN{} &&\zxrep{$S$} \ar[ru,)]&& &\\
        &\zxHad{} \ar[d]&& && \zxZ{} \ar[lu,)] \ar[uu,(]&&&\\
        &&&&&\zxZ{} \ar[llluuuu,),3d above,very thick] \ar[u] \ar[rrd]  &&&\\
        &&&&&&&&\cdots
    \end{ZX}
\stackrel{\eqref{eq:Hopf-rule},\eqref{eq:Id},\eqref{eq:antipoderule}}{=}
    \tikzsetnextfilename{KWpf-6}
    \begin{ZX}
        &&&&&&\\
        &&\zxHad{\tiny$\dagger$} \ar[u]\ar[dd]&&&&\\
        \cdots& \zxrep{$S$} \ar[l] \ar[rr,3d above,very thick]  & &\zxX{} \ar[r]& \zxrep{$S$}\ar[r]&\cdots &\\
        & &\zxZ{} \ar[ru,(] \ar[d]&&&&\\
        & &\zxHad{} \ar[d]&&&&\\
        &&&&&&
    \end{ZX}
    \stackrel{\eqref{eq:type1hadgate}}{=}
    \tikzsetnextfilename{KWpf-7}
    \begin{ZX}
        &&&&&&\\
        &&&\zxN{}\ar[u]\ar[dd]&&&\\
        \cdots&\zxHad{\tiny$\dagger$}\ar[l]\ar[r] &  \zxZ{} \ar[rr,3d above,very thick] & &\zxHad{} \ar[l]\ar[r]  &\cdots\\
        &&&\zxX{} \ar[lu,)] \ar[d] &&&\\
        &&& & &&
    \end{ZX}
\fe

Due to the translation invariance of the MPO, the $\FF^{\dagger},\FF$ on the virtual bond cancel in the last step.

\hypersetup{linkcolor=brn} % to color backref links
\bibliography{bib}
\bibliographystyle{ytphys}

\end{document}